\documentclass[a4paper,10pt,titlepage,english]{book}
\pdfoutput=1
\newif\iffrontmatter\frontmattertrue
\usepackage{MyThesisD}

\title{FOR-ARIXV}

\usepackage{enumerate,multirow}

\newcommand{\Hrule}{\noalign{\hrule height1pt}}
\newcommand{\Vrule}{\vrule width 1pt}
\newcommand{\SUSYbreaking}{\cancel{\text{SUSY}}}
\newcommand{\TO}{\text{--}}
\newcommand{\Hc}{\text{H.c.}}
\newcommand{\starline}{\par\begin{center}*\qquad*\qquad*\end{center}\par}
\makeatletter
\def\@@withpackage[#1]#2{{\tt #2~\!\!#1}}
\def\@withpackage#1{{\tt #1}}
\def\withpackage{\@ifnextchar[{\@@withpackage}{\@withpackage}}
\makeatother
\newcommand{\YUKAWA}{Yukawa}
\newcommand{\NAMBUGOLDSTONE}{Nambu--Goldstone}
\def\MSbar{\overline{\rm MS}}
\renewcommand{\Re}{\mathop{\mathrm{Re}}}

\newcommand{\invfb}{\un{fb^{-1}}}
\newcommand{\invpb}{\un{pb^{-1}}}
\newcommand{\pmat}[1]{\begin{pmatrix}#1\end{pmatrix}}
\newcommand{\abssq}[1]{\left|#1\right|^2}
\newcommand{\trans}[1]{#1\raisebox{1.0ex}{\tiny$\mathrm T$}}
\newcommand{\bQ}{\bar Q}

\newcommand{\HnouZ}{\Tilde H\s u^0}
\newcommand{\HnodZ}{\Tilde H\s d^0}
\newcommand{\HnouP}{\Tilde H\s u^+}
\newcommand{\HnodM}{\Tilde H\s d^-}
\newcommand{\TEN}{\vc{10}}
\newcommand{\TENbar}{\overline{\vc{10}}}
\newcommand{\EIGHT}{\vc{8}}
\newcommand{\FIVE}{\vc{5}}
\newcommand{\FIVEbar}{\overline{\vc{5}}}
\newcommand{\THREE}{\vc{3}}
\newcommand{\THREEbar}{\overline{\vc{3}}}
\newcommand{\TWO}{\vc{2}}
\newcommand{\ONE}{\vc{1}}
\newcommand{\ONEbar}{\overline{\vc{1}}}
\newcommand{\neut}[1][1]{\Tilde\chi^0_#1}
\newcommand{\chgP}[1][1]{\Tilde\chi^+_#1}
\newcommand{\chgM}[1][1]{\Tilde\chi^-_#1}
\newcommand{\chgPM}[1][1]{\Tilde\chi^\pm_#1}

\newcommand{\smu}{\Tilde{\mu}}
\newcommand{\smuL}{\Tilde{\mu}\s L}
\newcommand{\smuR}{\Tilde{\mu}\s R}

\newcommand{\snumu}{\Tilde\nu_\mu}
\newcommand{\bino}{\Tilde b}
\newcommand{\wino}{\Tilde w}
\newcommand{\smuRc}{\Tilde \mu\s R^c{}}
\newcommand{\muRc}{\mu\s R^c{}}
\newcommand{\vu}{v\s u}
\newcommand{\vd}{v\s d}

\newcommand{\oneloopforthree}[1]{\Big\langle\!\Big\langle #1 \Big\rangle\!\Big\rangle\stx{for $i=3$}}

\usepackage{mypdfpages}
\def\eataletter#1{\expandafter\expandafter\expandafter\eataletterr#1}
\def\eataletterr#1{}

\begin{document}
\pdfbookmark[0]{Front Cover}{bookmark:first}
\makeatletter
\let\@afterindenttrue\@afterindentfalse
\@afterindenttrue
\makeatother

\setcounter{page}{901}
\def\thepage{A}
\includepdf[noautoscale=true,offset=72 -72,pages={1}]{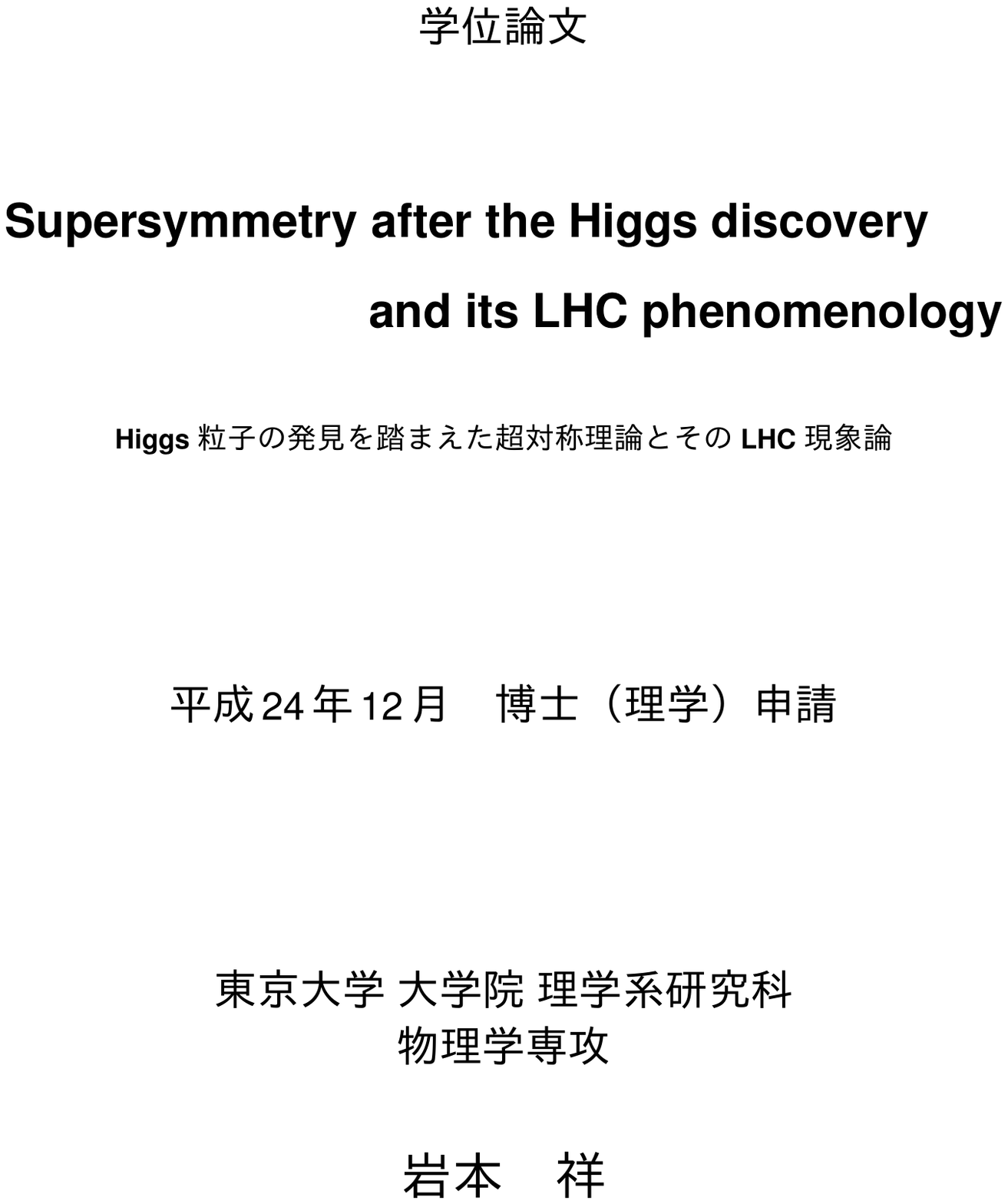}
\def\thepage{B}
\clearpage

\frontmatter\setcounter{page}{1001}
\makeatletter
\def\thepage{\protect\eataletter{\@roman\c@page}}
\makeatother
\includepdf[noautoscale=true,offset=72 -72,pages={1}]{jap/dd.pdf}

\clearpage
\thispagestyle{empty}
\mbox{}
\clearpage

\pdfbookmark[0]{Preface}{bookmark:preface}
\newcommand{\phvmn}{\usefont{T1}{phv}{m}{n}}
\newcommand{\phvbn}{\usefont{T1}{phv}{b}{n}}
\newcommand{\phvbsc}{\usefont{T1}{phv}{b}{sc}}
\makeatletter
\def\thepage{\protect\eataletter{\@roman\c@page}}
\makeatother

\chapter*{{\phvbsc{Preface}}}
\label{chap:preface-first}
{\phvmn
{\phvbn This dissertation is based on the following works by Author accomplished during the Ph.~D.~course at the Graduate School of Science, the University of Tokyo.}
\begin{itemize}{\small
 \item [{\phvbn[A]}]
   M.~Endo, K.~Hamaguchi, S.~Iwamoto, and N.~Yokozaki,
   {\em {Higgs Mass and Muon Anomalous Magnetic Moment in Supersymmetric Models with Vector-Like Matters}},
   \href{http://dx.doi.org/10.1103/PhysRevD.84.075017}{Phys.\ Rev.\ {\bfseries D84} (2011) 075017}
   {\ttfamily [\href{http://arxiv.org/abs/1108.3071}{arXiv:1108.3071}]}.
 \item [{\phvbn[B]}]
   M.~Endo, K.~Hamaguchi, S.~Iwamoto, and N.~Yokozaki,
   {\em {Higgs mass, muon g-2, and LHC prospects in gauge mediation models with vector-like matters}},
   \href{http://dx.doi.org/10.1103/PhysRevD.85.095012}{Phys.\ Rev.\  {\bfseries D85} (2012) 095012}
   {\ttfamily [\href{http://arxiv.org/abs/1112.5653}{arXiv:1112.5653}]}.
 \item [{\phvbn[C]}]
   M.~Endo, K.~Hamaguchi, S.~Iwamoto, and N.~Yokozaki,
   {\em {Vacuum Stability Bound on Extended GMSB Models}},
   \href{http://dx.doi.org/10.1007/JHEP06(2012)060}{JHEP {\bfseries 1206} (2012) 060}
   {\ttfamily [\href{http://arxiv.org/abs/1202.2751}{arXiv:1202.2751}]}.
 \item [{\phvbn[D]}]
   M.~Endo, K.~Hamaguchi, K.~Ishikawa, S.~Iwamoto, and N.~Yokozaki,
   {\em {Gauge Mediation Models with Vectorlike Matters at the LHC}},
   \href{http://dx.doi.org/10.1007/JHEP01(2013)181}{JHEP {\bfseries 1301} (2013) 181}
   {\ttfamily [\href{http://arxiv.org/abs/1212.3935}{arXiv:1212.3935}]}.
}\end{itemize}
{\phvbn An article related to this topic by Author was published in conference proceedings:}
\begin{itemize}{\small
 \item [{\phvbn[E]}]
   S.~Iwamoto,
   {\em {Muon g-2 anomaly and 125 GeV Higgs: Extra vector-like quark and LHC prospects}},
   \href{http://dx.doi.org/10.1063/1.4742079}{AIP Conf.\ Proc.\  {\bfseries 1467} (2012) 57--61}
  {\ttfamily [\href{http://arxiv.org/abs/1206.0161}{arXiv:1206.0161}]}.
}\end{itemize}
{\phvbn Among the other works by Author, the following two are referred in the text.}
\begin{itemize}{\small
 \item [{\phvbn[F]}]
   S.~Asai, Y.~Azuma, M.~Endo, K.~Hamaguchi, and S.~Iwamoto,
   {\em {Stau Kinks at the LHC}},
   \href{http://dx.doi.org/10.1007/JHEP12(2011)041}{JHEP {\bfseries 1112} (2011) 041}
   {\ttfamily [\href{http://arxiv.org/abs/1103.1881}{arXiv:1103.1881}]}.
 \item [{\phvbn[G]}]
   M.~Endo, K.~Hamaguchi, S.~Iwamoto, K.~Nakayama, and N.~Yokozaki,
   {\em {Higgs mass and muon anomalous magnetic moment in the U(1) extended MSSM}},
   \href{http://dx.doi.org/10.1103/PhysRevD.85.095006}{Phys.\ Rev.\  {\bfseries D85} (2012) 095006}
   {\ttfamily [\href{http://arxiv.org/abs/1112.6412}{arXiv:1112.6412}]}.
}\end{itemize}
}

\newpage
\pdfbookmark[1]{Acknowledgment}{bookmark:ack}
\includepdf[offset=-72 -72,noautoscale=true,pages={3}]{jap/dd.pdf}

\mbox{}\par\vspace{148.8mm}\par
\pdfbookmark[1]{Revision Log}{bookmark:rev}

\clearpage\mbox{}
\clearpage
\subparagraph*{}
\label{sec:abstract-en}
\pdfbookmark[1]{Abstract}{bookmark:abs}
\begin{center}
 \fboxsep=15pt\doublebox{
 \begin{minipage}{.81\textwidth} \vspace{-24pt}
\hfil\fboxsep=0pt\colorbox{white}{\quad{\huge \phvbsc Abstract}\quad}\hfil\par%
 \vspace{10pt}
\begin{minipage}{\textwidth}%
\phvmn

We physicists are on a journey towards the ultimate theory which describes everything in our Universe.
A great milestone achieved in 2012 is the discovery of the Higgs boson by the CMS and the ATLAS collaborations at the Large Hadron Collider (LHC); it completes the Standard Model of particle physics, which was developed through the mid to late twenties century.

\vspace{5pt}

The long twentieth century was over.
Happily making a steppingstone of the Standard Model, we are now heading to more fundamental theories.
Nature has many unsolved features: the Dark Matter, the Dark Energy, and the mechanism which produced current baryon asymmetry of our Universe, etc.
Also we have to build a unified explanation of the three forces embedded in the Standard Model, and to develop a description of the gravitational force in the language of the quantum theories.

\vspace{5pt}

The supersymmetry is one of the most promising candidates for theories beyond the Standard Model, and its tail was expected to be caught at the early stage of the LHC.
However, our expectation was not fulfilled, and we have no footprints observed yet.
What does this mean?

\vspace{5pt}

\starline

\vspace{5pt}

The current status of the supersymmetric theories is described in this dissertation.
First, the simplest model of supersymmetric Standard Model is introduced, which is called the MSSM.
Under this framework, the discrepancy in the muon anomalous magnetic moment between the prediction from the Standard Model and the experimental result suggests the supersymmetric particles are of order 100\,GeV, which is also supported by discussions on the little hierarchy problem.
However, the LHC experiments have found no scalar-quarks or gluinos in such mass range, and moreover, the Higgs boson mass of 126\,GeV requires, within the MSSM framework, the scalar-top mass of order 1--10\,TeV.
This current status forces us to abandon the simplest supersymmetry-breaking frameworks of the CMSSM and the GMSB scenarios.

\vspace{5pt}

Two promising possibilities remain there: the first is that the scalar-quarks and the gluino are much heavier than of order 100\,GeV while the other SUSY particles remain near the order, and the second is to extend the MSSM with extra fields.
The second scenario is investigated in this dissertation; the V-MSSM is proposed as an extension of the MSSM with a ($\TEN+\TENbar$) pair of the SU(5) decuplets.
In the framework the Higgs mass is increased by effect from the extra matters, and thus the 126\,GeV is achieved with the scalar-top having a lighter mass.
This fact resurrects the CMSSM and the GMSB scenarios.
This dissertation examines the GMSB scenario under the V-MSSM; it is called V-GMSB scenario.

\vspace{5pt}

It is shown that the V-GMSB has a potential to realize the 126\,GeV mass of the Higgs boson with holding the explanation of the muon magnetic moment discrepancy, if the masses of the extra quarks are approximately less than 1.2\,TeV.
Constraints on the V-GMSB from the LHC experiments are discussed then; it is concluded that the gluino mass must be approximately heavier than 1.1\,TeV, and that the extra quarks be heavier than 300--650\,GeV depending on the decay branches of them.

\vspace{5pt}

LHC prospects are briefly discussed.
As the extra quarks are expected to be approximately less than 1.2\,TeV, searches for the particles are of great interest at the 14\,TeV LHC;
constraints from the supersymmetry search, especially on the gluino mass, are expected to be much improved there.
Therefore, it is expected that the fate of the V-GMSB is adjudicated at the court of the 14\,TeV LHC.

\end{minipage}\end{minipage}}

\end{center}

\clearpage

\includepdf[offset=72 -72,noautoscale=true,pages={4}]{jap/dd.pdf}
\mbox{}\par{\vskip-3.2em
\pdfbookmark[1]{Abstract (ja)}{bookmark:abs-ja}
}

\newpage
\pdfbookmark[0]{Table of Contents}{bookmark:toc}
\tableofcontents

\newpage
\pdfbookmark[0]{List of Figures}{bookmark:lof}
\listoffigures
\newpage
\pdfbookmark[0]{List of Tables}{bookmark:lot}
\listoftables

\mainmatter
\frontmatterfalse
\chapter{Overture}
\label{cha:overture}

\def\abrace#1{$\langle\!\langle$#1$\rangle\!\rangle$}

\paragraph{A new boson is observed in Higgs hunting.}
The magisterial thesis of Author, submitted in January 2010, begins with the following paragraph:
\begin{quote}
 \begin{minipage}{0.84\textwidth}
 We have the Standard Model, which describes almost all physics below the energy scale 100\,GeV.
 Although it is still under verification, especially the existence of the Higgs boson,
 the experiments held in the Large Hadron Collider (LHC) will work out the answer soon, which will be a declaration of the triumph of our philosophy.
 \end{minipage}
\end{quote}
This prognostication became reality.
On 4th July 2012, the ATLAS and the CMS collaborations claimed that they respectively observed a new boson with a mass approximately $126\GeV$ in searches for the Standard Model Higgs boson~\cite{20120704CMS,20120704ATLAS}.
It is not confirmed that the new particle is the Higgs boson;
we have to measure its property precisely, especially the \YUKAWA{} couplings with the Standard Model particles.
Nonetheless, it allows us to expect that the Standard Model with the Higgs mechanism~\cite{StandardModel,HiggsBoson} will be validated with the LHC and the International Linear Collider (ILC) in the near future.

In this dissertation, we assume that the new particle is the Higgs boson, and discuss the Standard Model with the Higgs mechanism, simply we call it the Standard Model hereafter, and higher-energy theories above it.

\paragraph{Standard Model has been completed, but\dots}
The discovery of the Higgs boson has completed the Standard Model.
This model has the electroweak symmetry breaking as its heart, which is governed by the Higgs mechanism, and explains almost all of Nature.

We physicists, however, expect that more fundamental theories are hidden at higher energy scale $M\s{high}$, and hope that they solve the following prospects or problems:
\begin{itemize}\itemsep=0\baselineskip
 \item[(a)] unification of the electroweak force and the strong force (``grand unification''),
 \item[(b)] description of the gravitational force in harmony with the electroweak and strong forces,
 \item[(c)] the reason why neutrinos have such extremely tiny masses,
 \item[(d)] the mechanism which generated current baryon asymmetry of our Universe,
 \item[(e)] the mechanism which caused the inflation in the early universe,
 \item[(f)] identification of the Dark Matter and the Dark Energy,
\end{itemize}
but these are so fundamental that cannot be solved in a night.
Towards these problems physicists have examined the Standard Model, and have found the following problems, discrepancy, or unnaturalness, which can be used as keys towards more fundamental theories:
\begin{description}\itemindent=0pt
\item[\abrace{problem} Dark Matter problem]
  We know that our familiar matters, e.g., electrons, protons, and neutrons, account for approximately 4\% of the substance of our Universe~\cite{Jarosik:2010iu,Komatsu:2010fb}.
  We consider that approximately 20\% is some other matter, called ``Dark Matter,'' and that the rest is not even matter, which we call ``Dark Energy.''
  We do not know nature of them; as for the Dark Matter, since it is still considered to be ``matter,'' expected to be ascertained more easily.
\item[\abrace{problem} current baryon asymmetry of our Universe]
  Although the baryon number $B$ is slightly violated with the sphaleron effect~\cite{Manton:1983nd,Klinkhamer:1984di,tHooft:1976fv}, the violation is too feeble to explain current baryon asymmetry of our Universe.
\item[\abrace{unnaturalness} fine-tuning problem (hierarchy problem)]
  Based on our current understanding of quantum field theory, the mass of the Higgs boson, $m_h$, should naturally be much above $\Order(100)\GeV$, but we found it is just around $126\GeV$. This means that Nature employs unbelievably unnatural tuning.
  This topic is examined in Sec.~\ref{sec:hierarchy-problem}.
\item[\abrace{discrepancy} muon $g-2$ problem]
  The anomalous magnetic moment of muons, or the muon $g-2$, is precisely measured in experiments, but the measured value has $3\sigma$-level discrepancy compared with the theoretical prediction under the Standard Model, as will be discussed in Sec.~\ref{sec:SMg-2}.
\item[\abrace{discrepancy} slight mismatch on gauge coupling unification]
  Physicists expect the three forces, the electromagnetic force, the weak force, and the strong force, would be unified in more fundamental theories.
  If so, the strength of these forces, expressed in terms of gauge coupling constants, should be common at the energy scale of such theories.
  Actually the coupling constants are dependent on energy scale, and the matching does roughly occur at $\sim10^{16}\GeV$.
  However, there lies a slight mismatch, and we have to modify the energy dependence slightly in order to realize complete unification.
  We will discuss this topic in Sec.~\ref{sec:SMunification}.
\end{description}

\subsubsection{The SUSY and the MSSM}
Amazingly, we have a silver bullet. With examining those hints, physicists invented the supersymmetry (SUSY)~\cite{Haag:1974qh} several decades ago.

The SUSY is a symmetry between bosons and fermions: it transforms bosons into fermions, and vice versa.
The SUSY extends the Standard Model.
The minimal version of the supersymmetric Standard Model, the minimal supersymmetric standard model (MSSM)~\cite{Fayet:1976et,Fayet:1977yc,Farrar:1978xj} has, therefore, scalar quarks (squarks) and scalar leptons (sleptons) as the bosonic partners of the Standard Model fermions.
Also the Higgs boson and the gauge bosons meet their fermionic partners called Higgsino and gauginos.
Chapter \ref{cha:mssm} of this dissertation is devoted to topics around the MSSM.

The MSSM completely solves the fine-tuning problem.
It also has capability to provide a candidate for the Dark Matter, and to solve the muon $g-2$ problem.
Also the slight mismatch of the gauge couplings is resolved.
Moreover, the SUSY is considered as a key to build string theories, which are considered as promising candidates for the ultimate theory.%
\footnote{%
A too philosophical (and not scientific) note:

It is instructive that a symmetry between bosons and fermions solves the hierarchy problem.
The hierarchy problem ultimately originates the fact that a scalar boson, the Higgs field, is appended to the Standard Model, where all matters are fermionic and all forces are governed by vector bosons, in order to realize the electroweak symmetry breaking.
It is somewhat expedient that the electroweak symmetry breaking, the heart of the Standard Model, is realized by a strange, and lastly appended, particle, and we had to worry why we have a sole scalar particle, a muggle, in the theory of fermions.

The history of physics is characterized as the cycle of unifications, from that of motions of apples and the moon to that of the $\gSU(2)\stx{weak}$ and the $\gU(1)_Y$ gauge symmetry (in other words, that of the $W^\pm$ bosons, the $Z$ boson, and the photon).
Therefore, the principal problem of the Standard Model, in this historical and philosophical viewpoint, is the co-existence of bosonic and fermionic matters.
The Standard Model should be, ``historically,'' supersymmetric, and it is one important reason that I, Author of this dissertation, prefer the SUSY as the model beyond the Standard Model.
}

However, there ain't no such thing as a free lunch. The great MSSM has still many problems.
The principal one is the fact that the SUSY is not realized in Nature. We do not have scalar electrons with a mass of $0.511\un{MeV}$.
We thus consider the SUSY is violated so that the masses of the superpartners become much heavier.
This is realized with appending extra terms that do not respect supersymmetry to the (supersymmetric) Lagrangian of the MSSM.
Then this patch causes $CP$- and flavor problems, and another patch, called the $R$-parity, is required to make protons stable.
Another weak point of this model is that no evidence has been found at experiments, although the LHC is expected to observe such signals. It is nothing but a hypothesis, or a daydream, at this stage.
These problems, and solution candidates, are discussed in Chapter~\ref{cha:mssm}.

\subsubsection{About this dissertation}
We in this dissertation will see how the status of this SUSY daydream is altered by the discovery of the Higgs boson.

After discussing the problems related to the Standard Model in the next chapter, we will review the MSSM and our current understanding on it in Chapter~\ref{cha:mssm}.
The Higgs mechanism and the mass of the Higgs boson under the MSSM, and the mechanism to solve the muon $g-2$ problem, are examined there, and current status of LHC SUSY searches is reviewed.
We will then notice that the mass of the Higgs boson, $126\GeV$, is a bit heavier than natural expectations, and that the Higgs boson now requires the squark to have masses of $\Order(1\text{--}10)\TeV$.
This fact forces us to abandon the SUSY solution to the muon $g-2$ problem as long as we adopt the most simple and neat set-up, called the gauge-mediated SUSY-breaking (GMSB) framework.

To revitalize the GMSB framework, in Chapter~\ref{cha:vectorlike}, we extend the MSSM with extra particles.
This model, called the V-MSSM, is the main topic of this dissertation.
In the model the extra quarks yield extra contributions to the Higgs mass, and the $126\GeV$ is realized with the squarks having a mass of $\lesssim1\TeV$.
Such lighter masses of the SUSY particles allow the SUSY contribution to the muon $g-2$ to be large enough to explain the discrepancy even under the GMSB scenario.
We call this the V-GMSB scenario.\footnote{The same discussion can be applied to the CMSSM framework; that is, in the CMSSM framework the muon $g-2$ anomaly cannot be solved under the $126\GeV$ Higgs constraint~\cite{Endo:2011gy}, and extending to the V-MSSM resolves this conflict~\cite{Endo:2011mc,Endo:2012rd}.}

Chapter~\ref{cha:vectorlike} is the main chapter of this dissertation.
There phenomenology of the V-MSSM is discussed.
We start from the discussion on the Higgs boson mass, on the gauge coupling unification, and on the muon $g-2$, under the V-MSSM.
Then the V-GMSB scenario is introduced in Sec.~\ref{sec:vgmsb-model}.
The characteristics of the model are examined, and the vacuum stability bound enters the discussion as a very severe constraint on the V-GMSB scenario.
However, even under the bound, as can be seen quantitatively in Sec.~\ref{sec:vgmsb-numerical-result}, the Higgs mass of $126\GeV$ can be explained with holding the explanation of the  muon $g-2$ discrepancy. Here the GMSB is revitalized.
Finally, interpreting reports from the LHC experiments, we will obtain current collider constraints on the model in Sec.~\ref{sec:vmssm-lhc} and Sec.~\ref{sec:vector}.
Especially, Fig.~\ref{fig:results1} is the conclusive figure of our tour.

Chapter~\ref{cha:coda} is the {\em coda}, where the main theme of this dissertation is recapitulated.


\chapter{Foundation}
\label{cha:foundation}

This chapter is a review of the problems stated in Chapter \ref{cha:overture} of the Standard Model.
We will in Sec.~\ref{sec:SM} start from the Standard Model and review the electroweak symmetry breaking.
The Standard Model is now completed with the discovery of the Higgs boson, but we will see it has a strange problem, ``hierarchy problem,'' once we regard the Standard Model as a low-energy effective theory of more fundamental theories.
This is the main topic of this chapter.
Then in Sec.~\ref{sec:SMg-2} we review the muon $g-2$ problem, a discrepancy in the muon anomalous magnetic moment between the Standard Model prediction and the measured value.
Sec.~\ref{sec:SMunification} is a discussion towards the grand unification theories (GUTs), where we will see a slight mismatch in unification of the gauge coupling constants.
These problems are all solved with the supersymmetry (SUSY) in Chapter~\ref{cha:mssm}.

\section{The Standard Model and Hierarchy Problem}
\label{sec:SM}
\subsection{The Standard Model}
The Standard Model~\cite{StandardModel} is a model of the gauge theory.
It has a gauge symmetry of $G\s{SM}=\gU(1)_Y \times \gSU(2)\s{weak} \times \gSU(3)\s{strong}$.
The symmetries appear in Nature as ``forces,'' which in language of the quantum field theory are governed by gauge bosons: a $B$-boson for $\gU(1)_Y$, three $W$-bosons for $\gSU(2)\s{weak}$, eight gluons for $\gSU(3)\s{strong}$.

However, Nature does not have $\gU(1)_Y\times \gSU(2)\s{weak}$; this symmetry, called the electroweak symmetry, is spontaneously broken with the Higgs mechanism~\cite{HiggsBoson}, discussed below, and  falls into a $\gU(1)$ electromagnetic symmetry, which governs the electromagnetic force with the photon $\gamma$. The remnant yields the weak force with the $W^{\pm}$-bosons and the $Z$-boson.
This is the kernel of the Standard Model, the electroweak symmetry breaking.

Let us start from the Standard Model Lagrangian of our triumph.
\begin{align}
 \Lag &= \Lag\s{gauge} + \Lag\s{Higgs} + \Lag\s{matter} + \Lag\s{\YUKAWA};\\
&\Lag\s{gauge} =
 -\tfrac14B^{\mu\nu}B_{\mu\nu}
 -\tfrac12\Tr(W^{\mu\nu}W_{\mu\nu})
 -\tfrac12\Tr(G^{\mu\nu}G_{\mu\nu}),\\
&\Lag\s{Higgs} =
 \left|\left(\partial_\mu-\ii g_2W_\mu-\tfrac12\ii g_YB_\mu\right)H\right|^2
 - V(H),\\
 \begin{split}
&\Lag\s{matter} =
 \overline Q_i\ii\gamma^\mu\left(\partial_\mu-\ii g_3G_\mu-\ii g_2W_\mu-\tfrac16\ii g_YB_\mu\right)\PL Q_i\\
&\qquad\qquad
 + \overline U_i\ii\gamma^\mu\left(\partial_\mu-\ii g_3G_\mu-\tfrac23\ii g_YB_\mu\right)\PR U_i
 + \overline D_i\ii\gamma^\mu\left(\partial_\mu-\ii g_3G_\mu+\tfrac13\ii g_YB_\mu\right)\PR D_i\\
&\qquad\qquad
 + \overline L_i\ii\gamma^\mu\left(\partial_\mu-\ii g_2W_\mu+\tfrac12\ii g_YB_\mu\right)\PL L_i
 + \overline E_i\ii\gamma^\mu\left(\partial_\mu+\ii g_YB_\mu\right)\PR E_i,
 \end{split}\\
&\Lag\s{\YUKAWA} =
 \overline U_i(y_u)_{ij}H\PL Q_j - \overline D_i (y_d)_{ij}H^\dagger\PL Q_j - \overline E_i(y_e)_{ij}H^\dagger\PL
 L_j + \Hc;\\
&V(H) = \frac12\lambda|H|^4 - \mu^2|H|^2.\label{eq:SMhiggspotential}
\end{align}
$B_\mu$, $W_\mu$ and $G_\mu$ are the gauge fields, and their ``field strengths''
$\partial_\mu X_\nu - \partial_\nu X_\mu - \ii g_X[X_\mu, X_\nu]$ are denoted as $X_{\mu\nu}$.
$H$ is the Standard Model Higgs field, and the Higgs potential $V(H)$ is characterized by the quartic coupling $\lambda$ and the quadratic coupling $\mu$, where $\mu$ is the only parameter with mass dimension in the Standard Model. $y_{u,d,e}$ are matrices of the Yukawa couplings {\em in the Standard Model}\footnote{%
We will later use $Y_{u,d,e}$ for the \YUKAWA\ couplings in the MSSM.}; $i$ and $j$ are generation indices of quarks ($Q$, $U$ and $D$) and leptons ($L$ and $E$).
$\PL$ and $\PR$ are the well-known projection operators.
This Lagrangian describes Nature very well, and exceptions are limited to be, e.g., the tiny neutrino masses and the Dark Matter.

\begin{rightnote}
 Here one should notice that the baryon number $B$ and the lepton number $L$ are respectively conserved in this Lagrangian; this is an accidental symmetry originating in the Standard Model gauge symmetry.
 Actually these symmetries are {\em anomalous}, and quantum effect causes especially in the early universe the so-called ``sphaleron process,'' which violates $B+L$ significantly~\cite{Manton:1983nd,Klinkhamer:1984di,tHooft:1976fv}.
 Meanwhile, the conservation of the number $B-L$ is not anomalous, and thus it is kept even under quantum effect.
\end{rightnote}

Our main concern is the Higgs potential.
For a successful electroweak symmetry breaking $\lambda>0$ and $\mu^2>0$ must hold.
Then the minimum of the Higgs potential falls in $|H|=\mu/\sqrt{\lambda}=:v$, which results in the vacuum expectation value $v$ of the Higgs boson; the value is well-known to be $v\approx 174\GeV$ from the masses of the $W$- and $Z$-bosons.
Finally, the Higgs field is parameterized as
\begin{align}
 H(x)=\pmat{H^+(x)\\H^0(x)}=\pmat{0\\v}+\frac1{\sqrt2}\pmat{\phi_1(x)+\ii\phi_2(x)\\h(x)+\ii\phi_3(x)}.
\end{align}
$h(x)$ is the (Standard Model) Higgs boson, which we first observed in 2012. $\phi_i(x)$ are would-be \NAMBUGOLDSTONE\ bosons, ignored here for simplicity. The potential is now
\begin{equation}
  V(h)
 = \frac{\lambda}{8}h^4+\sqrt{\frac{\lambda}{2}}\mu h^3 +\mu^2 h^2+\text{constant}.
\end{equation}
What we observed in this year 2012 is nothing but $m_h=\sqrt2\mu\approx126\GeV$, and this results in $\lambda\approx 0.26$.
Now we know all the parameters in the Standard Model. The Standard Model is completed.

The electroweak symmetry breaking makes the three $W$-bosons massive, which are observed as $W^\pm$-bosons and a $Z$-boson. This effect emerges from $\Lag\stx{Higgs}$ as, with ignoring the \NAMBUGOLDSTONE\ bosons,
\begin{equation}
 \Lag\stx{Higgs}
= \left|\left(\partial_\mu-\ii g_2W_\mu-\tfrac12\ii g_YB_\mu\right)H\right|^2
\supset \frac12(\partial_\mu h)^2+\frac{v^2}2\left({g_2}^2{W^+}^\mu W^-_\mu + \frac{{g_Z}^2}{2}Z^\mu Z_\mu\right).\label{eq:SMgaugemass}
\end{equation}
Here redefinition of the gauge field is employed as
\begin{align}
 W^\pm_\mu&:=\frac1{\sqrt2}(W^1_\mu\mp\ii W^2_\mu),&
\pmat{Z_\mu \\ A_\mu}
&:= \pmat{%
\cos\theta\s w & -\sin\theta\s w\\
\sin\theta\s w & \cos\theta\s w
}
\pmat{W^3_\mu \\ B_\mu},
\end{align}
where
\begin{align}
 e              &:= -\frac{g_Yg_2}{\sqrt{{g_Y}^2+{g_2}^2}};&
 g_Z            &:= \sqrt{{g_Y}^2+{g_2}^2};&
 g_Y&=\frac{|e|}{\cos\theta\s w}=g_Z\sin\theta\s w,&
 g_2&=\frac{|e|}{\sin\theta\s w}=g_Z\cos\theta\s w.
\end{align}
We can find the mass of the gauge bosons as
\begin{align}
  m_W&=\frac{g_2}{\sqrt2}v,&
  m_Z&=\frac{g_Z}{\sqrt2}v,
\end{align}
where we can obtain the value of $v\approx174\GeV$.

\subsection{Hierarchy problem}
\label{sec:hierarchy-problem}

The Standard Model with the Higgs mechanism is a complete beautiful theory for its own sake.
However, we physicists do not consider the Standard Model as the ultimate theory, but a low-energy effective theory of more fundamental theories hidden at higher energy scales, which we label $M\s{high}$.
Once we take this standpoint, which is the current paradigm of our science, the Higgs particle brings an {\em unnaturalness} into the Standard Model, which has been discussed for several decades, the hierarchy problem~\cite{HierarchyProblem}.
Let us briefly review the unnaturalness through the electroweak symmetry breaking in the Standard Model.

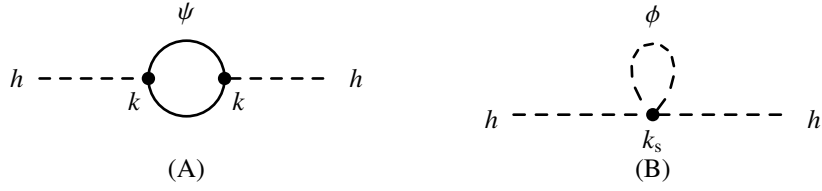
\begin{figure}[t]\begin{center}\begin{fmffile}{feyn/RGdiagramforML}
 \begin{minipage}[t]{0.4\textwidth}\begin{center}\begin{fmfgraph*}(110,50)
 \fmfleft{d1,h1,d2}\fmfright{d3,h2,d4}
 \fmf{phantom}{d1,d3}\fmf{phantom}{d2,d4}
 \fmf{xscalar,tension=0.7}{h1,v1}
 \fmf{xfermion,left,tension=0.5,label=$\psi$}{v1,v2}
 \fmf{xfermion,left,tension=0.5}{v2,v1}
 \fmf{xscalar,tension=0.7}{v2,h2}
 \fmflabel{$h$}{h1}
 \fmflabel{$h$}{h2}
 \fmfv{label=$k$,l.a=-120}{v1}
 \fmfv{label=$k$,l.a=-60}{v2}
 \fmfdot{v1,v2}
 \end{fmfgraph*}\\(A)\end{center}\end{minipage}
 \begin{minipage}[t]{0.4\textwidth}\begin{center}\begin{fmfgraph*}(110,50)
 \fmfleft{d1,h1,d2,d3,d4}\fmfright{d6,h2,d7,d8,d9}
 \fmf{phantom}{d1,d6}\fmf{phantom}{d4,d9}
 \fmf{xscalar,tension=0.7}{h1,v}
 \fmf{xscalar,right,tension=1.3,label=$\phi$}{v,v}
 \fmf{xscalar,tension=0.7}{v,h2}
 \fmflabel{$h$}{h1}
 \fmflabel{$h$}{h2}
 \fmfv{label=$k\s{s}$}{v}
 \fmfdot{v}
 \end{fmfgraph*}\\(B)\end{center}\end{minipage}
\end{fmffile}
\caption[The Feynman diagrams yielding quadratic quantum corrections to the Higgs boson mass.]{
The one-loop level diagrams which give quadratic quantum corrections to the mass of the Higgs boson, provided by (A) a Dirac fermion $\psi$ and (B) a scalar boson $\phi$.
The particles are assumed to be coupled to the Higgs boson $h$ with interactions $kh\overline\psi\psi$ and $k\s{s}|h|^2|\phi|^2$, respectively, where $k$ and $k\s s$ are coupling constants.
}
\label{fig:higgs1loopcorrection}
\end{center}\end{figure}

The mass of the Standard Model Higgs boson receives quadratic quantum corrections from the diagrams in Fig.~\ref{fig:higgs1loopcorrection}.
The contributions are respectively evaluated as, at one-loop level,
\begin{equation}
 \Delta m_h^2 = -\frac{|k|^2}{8\pi^2}M\s{high}^2+\Order(\log M\s{high})\label{eq:1loopfermion}
\end{equation}
from a fermion (Fig.~\ref{fig:higgs1loopcorrection}.A), and
\begin{equation}
 \Delta m_h^2 = \frac{k\s{s}}{16\pi^2}M\s{high}^2+\Order(\log M\s{high})\label{eq:1loopscalar}
\end{equation}
from a scalar boson (Fig.~\ref{fig:higgs1loopcorrection}.B).
The Standard Model has a top quark with $k\sim1$, but no scalars other than the Higgs boson itself. $\Delta m_h^2$ is thus of order $M\s{high}^2$, and we need a finely tuned cancellation in the right hand side of the formula
\begin{equation}
 m_h^2\left(\text{physical}\right) =  m_h^2\left(\text{bare}\right) + \Delta m_h^2
\end{equation}
to realize $m_h^2\sim10^4\GeV^2$.

How should we settle this unnaturalness?
One possibility is to assume that $M\s{high}$ is near 100\GeV, and that our quantum field theory cannot be applied above this scale.
Then the tuning is not so fine, and moreover, we need not care quantum corrections at any higher scales.

A more beautiful solution is provided by the supersymmetry (SUSY)~\cite{Haag:1974qh}.
The SUSY, a symmetry between fermions and bosons, supplies {\em two} scalar bosons for a respective Dirac fermion. In other words, a Weyl fermion is accompanied by a complex scalar field under the SUSY, and vise versa.
Moreover, the SUSY guarantees that the scalar partners $\phi_{1,2}$ for a Dirac fermion $\psi$ that has a coupling $k$ to the Higgs boson do couple to the Higgs boson with the exact coupling $k\s s=|k|^2$.
Therefore, the ``superpartners'' cancel the quadratic divergence as
\begin{equation}
 \Delta m_h^2 = -\frac{|k|^2}{8\pi^2}M\s{high}^2+2\times \frac{k\s{s}}{16\pi^2}M\s{high} + \Order(\log M\s{high}) \leadsto \Order(\log M\s{high}).
\end{equation}
We will discuss this powerful hypothesis in Chapter \ref{cha:mssm}.

\newpage

\section[The Muon \texorpdfstring{$g-2$}{g-2}]{The Muon $\boldsymbol{g-2}$}
\label{sec:SMg-2}
\subsection{Foundation}
The $g$-value of the magnetic moment is one of the most famous and long investigated quantities in frontier of physics. It is defined as a ratio between the spin magnetic dipole moment $\vc{\mu}\s{spin}$ and the spin vector $\vc S$ of a particle as
\begin{equation}
\vc{\mu}\s{spin} = qg\mu\s B\vc S \where\mu\s B:=\frac{|e|}{2m},
\end{equation}
and expressed as a term in the Hamiltonian $H$ as
\begin{equation}
 H\supset -\vc\mu\s{spin}\cdot\vc{B} = -(qg\mu\s B)\vipro SB,
\end{equation}
where $m$ is the mass, and $q|e|$ is the electric charge, of the particle.

The $g$-value for electrons and muons, $g_e$ and $g_\mu$, are predicted as $g_e=g_\mu=2$ under the quantum mechanics.
However, radiative corrections shift the values slightly. The shift is known as the anomalous magnetic moment $g-2$, or $a:=(g-2)/2$.

Historically, the discrepancy was known through the observations of Land\'e $g$-factor $g_J$ of atoms.
In 1947, Schwinger calculated the one-loop level QED contribution to the electron magnetic moment as $a_e=\alpha/2\pi=0.00116$~\cite{Schwinger:1948iu}, where $\alpha\approx1/137$ is the fine structure constant  at the low-energy scale.
Then, in 1948, Kusch and Foley precisely measured the difference as $a_e=0.00119$ with comparing the $g_J$ values of gallium, indium and sodium~\cite{PhysRev.74.250}.

Currently the most precise measurements are achieved by the Harvard group for the electron at the {\em sub-ppb} level~\cite{PhysRevA.83.052122}, and by the muon $g-2$ collaboration at Brookhaven National Laboratory for the muon at the {\em sub-ppm} level. 
Theoretical calculations at similar precision are achieved, as exemplified by the five-loop level calculation for the QED contributions.
The values are summarized as
\begin{align}
 a_e(\text{exp})=&\,\left(11\,596\,521.8073\pm0.0028\right)\EE{-10}\quad\text{(0.24\,ppb)}\quad\text{\cite{PhysRevA.83.052122}},\\
 a_e(\text{SM}) =&\,\left(11\,596\,521.8178\pm0.0077\right)\EE{-10}\quad\text{(0.67\,ppb)}\quad\text{\cite{Aoyama:2012wj}},\\
 a_\mu(\text{exp})=&\,\left(11\,659\,208.9\pm6.3\right)\EE{-10}\quad\text{(0.54\,ppm)}\quad\text{\cite{Bennett:2006fi, Roberts:2010cj}},\\
 a_\mu(\text{SM}) =&\,\left(11\,659\,182.8\pm4.9\right)\EE{-10}\quad\text{(0.42\,ppm)}\quad\text{\cite{Aoyama:2012wk,g-2_EWderafael2002,g-2_EWmarciano2002,g-2_Hagiwara2011,Prades:2009tw}},
\end{align}
and the discrepancies between experiment and theory are
\begin{align}
 a_e(\text{exp}-\text{SM}) =&\,(-1.06\pm0.82)\EE{-12}\quad\text{($1.3\sigma$-level)},\\
 a_\mu(\text{exp}-\text{SM}) =&\,(26.1\pm8.0)\EE{-10}\quad\text{($3.3\sigma$-level)}.\label{eq:muon_g-2}
\end{align}
The Standard Model predicts the experimental values very well, but there exists $3\sigma$-level discrepancy in the muon $g-2$. This is the muon $g-2$ problem.

Now let us see the muon $g-2$ prediction under the Standard Model in detail.
The $g$-factor of the muon magnetic moment is expressed in the QFT language as
\begin{equation}
 g=2\left[F_1(0)+F_2(0)\right],
\end{equation}
where $F_i(q^2)$ are the form factors of the $\mu$--$\mu$--$\gamma$ vertex function $\Gamma^\mu$, which is described in Fig.~\ref{fig:g-2_SM}.A:
\begin{equation}
\Gamma^\mu(p',p) = \gamma^\mu F_1(q^2)+\frac{\ii\sigma^{\mu\nu}q_\nu}{2m}F_2(q^2).
\end{equation}
(See, e.g., Ref.~\cite{PeskinQFT} for a review.)

At the tree level $F_1(0)=1$ and $F_2(0)=0$, and thus $g=2$.
Quantum corrections modify these factors, but since the correction to $F_1(q^2)$ is sunk into the renormalization of $e$, the muon anomalous magnetic moment can be expressed as
\begin{equation}
 a_\mu = F_2(0).
\end{equation}

\subsection[Standard Model evaluation on the muon \texorpdfstring{$g-2$}{g-2}]{Standard Model evaluation on the muon $\boldsymbol{g-2}$}
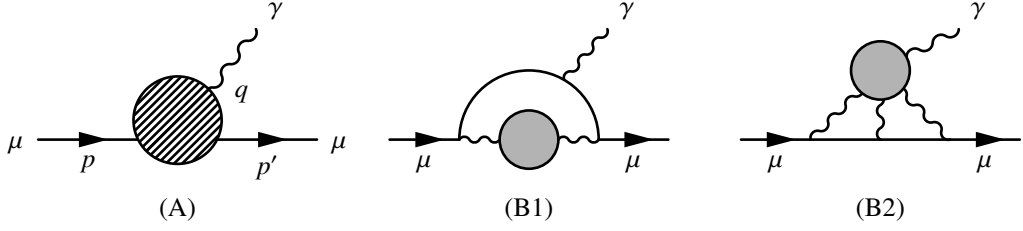
\begin{figure}[t]\begin{center}
 \begin{fmffile}{feyn/g_2-SM}
  \begin{minipage}[t]{0.3\textwidth}
   \begin{center}
    \begin{fmfgraph*}(110,70)
\fmfleft{d1,p1,d2,d3,d4}\fmfright{d5,p2,d6,d7,gc}
\fmf{fermion,label=$p$,l.s=right}{p1,x1}
\fmf{vanilla,tension=1}{x2,x1}
\fmf{phantom,tension=1}{x2,x4,xx,x3,x1}
\fmf{fermion,label=$p'$,l.s=right}{x2,p2}
\fmf{phantom,tension=5}{gc,ge}
\fmf{phantom,tension=1}{ge,d1}
\fmfposition
\fmf{phantom}{gc,yy,d1}
\fmf{phantom}{yy,gb,xx}
\fmf{phantom,left,tag=1,l.s=right}{x3,x4}
\fmfipath{p[]}
\fmfiset{p1}{vpath1(__x3,__x4)}
\fmfi{photon,label=$q$}{point length(p1)/3*2 of p1 -- vloc(__ge)}
\fmflabel{$\gamma$}{ge}
\fmflabel{$\mu$}{p1}
\fmflabel{$\mu$}{p2}
\fmfblob{0.3w}{gb}
    \end{fmfgraph*}
    \\(A)
   \end{center} 
  \end{minipage}
  \begin{minipage}[t]{0.3\textwidth}
   \begin{center}
    \begin{fmfgraph*}(110,70)
\fmfleft{d1,p1,d2,d3,d4}\fmfright{d5,p2,d6,d7,gc}
\fmf{fermion,label=$\mu$,l.s=right}{p1,x1}
\fmf{photon}{x2,xx,x1}
\fmf{fermion,label=$\mu$,l.s=right}{x2,p2}
\fmf{phantom,tension=5}{gc,ge}
\fmf{phantom,tension=1}{ge,d1}
\fmfposition
\fmf{vanilla,left,tag=1,l.s=right}{x1,x2}
\fmfipath{p[]}
\fmfiset{p1}{vpath1(__x1,__x2)}
\fmfi{photon}{point length(p1)/3*2 of p1 -- vloc(__ge)}
\fmflabel{$\gamma$}{ge}
\fmfv{d.size=0.2w,d.shape=circle,d.filled=30}{xx}
    \end{fmfgraph*}
    \\(B1)
   \end{center} 
  \end{minipage}
  \begin{minipage}[t]{0.3\textwidth}
   \begin{center}
    \begin{fmfgraph*}(110,70)
\fmfleft{d1,p1,d2,d3,d4}\fmfright{d5,p2,d6,d7,gc}
\fmf{fermion,label=$\mu$,l.s=right}{p1,x1}
\fmf{vanilla}{x2,xx,x1}
\fmf{fermion,label=$\mu$,l.s=right}{x2,p2}
\fmf{phantom,tension=5}{gc,ge}
\fmf{phantom,tension=1}{ge,d1}
\fmfposition
\fmf{phantom,left,tag=1,l.s=right}{x1,x2}
\fmfipath{p[]}
\fmfiset{p1}{vpath1(__x1,__x2)}
\fmfipair{mid}
\fmfiset{mid}{point length(p1)/2 of p1}
\fmfi{photon}{mid -- vloc(__ge)}
\fmfi{photon}{mid -- vloc(__xx)}
\fmflabel{$\gamma$}{ge}
\fmfi{photon,left}{mid -- vloc(__x1)}
\fmfi{photon,left}{mid -- vloc(__x2)}
\fmfiv{d.shape=circle,d.filled=30,d.size=0.2w}{mid}
    \end{fmfgraph*}
    \\(B2)
   \end{center} 
  \end{minipage}
  \caption[The Feynman diagram which contributes to the $g$-factor of the muon magnetic moment.]
{(A) The Feynman diagram which contributes to the $g$-factor of the muon magnetic moment. The tree level diagram corresponds to the classical result $g_\mu=2$, and quantum corrections, i.e., loop diagrams, yield deviation from 2. $p$, $p'$ and $q=p'-p$ are momenta of the particles. (B1) The hadronic vacuum-polarization contributions to the muon $g-2$. (B2) The hadronic light-by-light contributions to the muon $g-2$. The gray circles denote hadronic loop diagrams.}
  \label{fig:g-2_SM}
 \end{fmffile}
\end{center}
\end{figure}

The Standard Model prediction of the muon $g-2$ is summarized as:
{\catcode`?=\active \def?{\phantom{0}} \catcode`@=\active \def@{\mathbin{\phantom{-}}}
\begin{align}
 a_\mu(\text{QED})    =&\,@\left(11\,658\,471.8951\pm0.0080\right)\EE{-10}\quad\text{\cite{Aoyama:2012wk}},\\
 a_\mu(\text{EW})     =&\,@\left(??\,???\,?15.4???\pm0.2???\right)\EE{-10}\quad\text{\cite{g-2_EWderafael2002,g-2_EWmarciano2002}},\\
 a_\mu(\text{HVP-LO}) =&\,@\left(??\,???\,692.3???\pm4.2???\right)\EE{-10}\quad\text{\cite{g-2_Davier2010}},\\
                       &\,@\left(??\,???\,694.91??\pm4.27??\right)\EE{-10}\quad\text{\cite{g-2_Hagiwara2011}},\\
 a_\mu(\text{HVP-HO}) =&\,-\left(??\,???\,??9.84??\pm0.07??\right)\EE{-10}\quad\text{\cite{g-2_Hagiwara2011}},\\
 a_\mu(\text{HLbL})   =&\,@\left(??\,???\,?10.5???\pm2.6???\right)\EE{-10}\quad\text{\cite{Prades:2009tw}},\\
                       &\,@\left(??\,???\,?11.6???\pm4.0???\right)\EE{-10}\quad\text{\cite{Nyffeler:2009tw,Jegerlehner:2009ry}}.
\end{align}
}
For the ``HVP-LO'' and the ``HLbL'' contributions two values are cited as a reference.

The respective categories are defined as follows:
\begin{description}
 \item[QED contribution]
  comes from the diagrams only with leptons and photons.
  It is calculated analytically up to the three-loop level (of order $\alpha^{3}$), and recently an automated computation of the Feynman diagrams finished its five-loop level (of order $\alpha^{5}$) calculation~\cite{Aoyama:2012wj,Aoyama:2012wk}.
  The uncertainty of the QED contribution, dominated by the uncertainty in measurement of $\alpha$, is much smaller than those of the other contributions.
 \item[Electroweak contribution] is from the diagrams with Higgs, $W$ and/or $Z$ boson but without gluons.
  The contribution is evaluated at the two-loop level with including leading log three-loop effects~\cite{g-2_EWderafael2002,g-2_EWmarciano2002}.
  A hadronic loop uncertainty of $\pm0.1\EE{-10}$, and an uncertainty of $\pm0.2\EE{-10}$ from the ``then-unknown'' Higgs boson mass are considered, but the latter corresponds to the mass range of $114\GeV\lesssim m_h\lesssim 250\GeV$ allowed in those days, and thus is considered to be improved.
 \item[Hadronic contribution] is that including QCD interaction.
  This contribution is separated into two types: the diagrams of Fig.~\ref{fig:g-2_SM}.B1, called the hadronic vacuum polarization (HVP) contribution, and those of Figs.~\ref{fig:g-2_SM}.B2, the hadronic light-by-light (HLbL) contribution.
  The lowest order contribution of the HVP, or HVP-LO, enters at the order $\alpha^2$, while higher order contributions (HVP-HO) are of order $\alpha^3$ as well as the HLbL.

  The HVP contributions cannot be calculated directly, and are evaluated through the dispersion relation using experimental data of the cross sections $\sigma\s{tot}^0\left(e^+e^-\to\gamma^*\to\text{hadrons}\right)$.
  Due to limited accuracy of those data, the hadronic contributions are the dominant source of the uncertainty of the muon $g-2$ prediction;
  especially there lies a disagreement among experimental data for the $2\pi$-channel region.
  Several results from different collaborations are currently available~\cite{g-2_Davier2010,g-2_Hagiwara2011,Benayoun:2012wc}, two among which~\cite{g-2_Davier2010,g-2_Hagiwara2011} are quoted as a reference.
   \begin{rightnote}
    The evaluation via the dispersion relation is simply summarized as the following equation:
    \begin{equation}
     a_\mu(\text{HVP-LO})=\frac{1}{4\pi^3}\int_{m_\pi^2}^\infty \dd s\ \sigma\s{tot}^0\left(e^+e^-\to\gamma^*\to\text{hadrons}\right)K(s),
    \end{equation}
   where $K(s)$ is a kernel function (See, e.g., Ref.~\cite{Hagiwara:2003da}).
   Note that $\sigma\s{tot}^0$ is evaluated with final state radiations but without initial state radiations and vacuum polarization corrections.
   \end{rightnote}
  The HLbL contributions cannot be calculated directly, nor be evaluated with experimental data.
  Lattice calculations~\cite{g-2_lattice} are expected, but currently low-energy effective theories~\cite{Prades:2009tw,Nyffeler:2009tw,Jegerlehner:2009ry} are exploited.
\end{description}
Summing up all the above contributions, we obtain the Standard Model expectation. For example, if we combine the values from Refs.~\cite{Aoyama:2012wk,g-2_EWderafael2002,g-2_EWmarciano2002,g-2_Hagiwara2011,Prades:2009tw} with errors in quadrature, the muon $g-2$ is predicted as
\begin{equation}
  a_\mu(\text{SM}) =\left(11\,659\,182.8\pm4.9\right)\EE{-10},
\end{equation}
and we face a $3\sigma$-level discrepancy.

\starline

Several models have been invented to solve the muon $g-2$ anomaly.
The SUSY can be a solution again, where the superpartners, heavy but having ordinary couplings to muons, yield loop level contributions to shift the predicted value.
Another candidate is a hidden photon~\cite{Fayet:2007ua,Pospelov:2008zw}, an extra $\gU(1)$ gauge boson with a feeble mixing with the photon.
In this case, the mixing is as small as $10^{-4}$, but if the mass of the hidden photon is of order $100\MeV$, the theoretical value of the muon $g-2$ is shifted enough to match the experimental result.
This model is out of scope of this dissertation; we will concentrate on the SUSY, and see how the SUSY solves the muon $g-2$ problem in Sec.~\ref{sec:muon-g-2-mssm}.

\section{Slight Mismatch of the Gauge Coupling Constants}
\label{sec:SMunification}

As is mentioned when we discussed the hierarchy problem in Sec.~\ref{sec:SM}, the Standard Model is considered as a low-energy effective theory of more fundamental theories, where we expect that some features of the Standard Model are ``unified'' with beautiful and sophisticated manners.
An example is the grand unification, the unification of the three forces.

Several models are proposed to realize the grand unification, and the most famous ones are the $\gSU(5)$ grand unification theories ($\gSU(5)$-GUTs)~\cite{Georgi:1974sy}.
Since the $\gSU(5)$ gauge group includes $G\s{SM}$ as a subgroup, this scenario is very promising, and has been studied for several decades.
$\gSU(5)$-GUTs have only one force of $\gSU(5)$. Let us call the gauge coupling $g_5$. The gauge group is expected (and assumed) to confront spontaneous symmetry breaking and to break down into three forces at a scale higher than $\Order(100)\GeV$: $\gSU(3)$ with a gauge coupling $g_{\gSU(3)}=g_5$, $\gSU(2)$ with $g_{\gSU(2)}=g_5$, and $\gU(1)$ with $g_{\gU(1)}=\sqrt{3/5}g_5$.

Here we face one problem. Our three forces do not have the strengths of $(g\s s, g_2, g_Y)=(g_5, g_5, \sqrt{3/5}g_5)$, but instead $\sim(1.2, 0.65, 0.36)$ at the electroweak symmetry breaking scale $m_Z$.
Fortunately, however, the gauge couplings depend on the scale.
The dependence is embedded in the renormalization group equations (RGEs) of the gauge coupling.
In the $\MSbar$ scheme~\cite{Bardeen:1978yd}, the RGEs are given as~\cite{Machacek:1983tz}
\begin{align}
\diff{g_1}{\log Q} &= \frac{1}{16\pi^2}\biggl( \frac{41}{10}g_1^3 \biggr)
+\frac{1}{(16\pi^2)^2}\left(\frac{199}{50}g_1^2+\frac{27}{10}g_2^2+\frac{44}{5}g_3^2-\frac{17}{10}y_t^2-\frac{1}{2}y_b^2-\frac{3 }{2}y_\tau^2\right)g_1^3,\\
\diff{g_2}{\log Q} &= \frac{1}{16\pi^2}\biggl( -\frac{19}{6}g_2^3 \biggr)
+\frac{1}{(16\pi^2)^2}\left(\frac{9}{10}g_1^2+\frac{35}{6}g_2^2+12 g_3^2-\frac{3}{2}y_t^2-\frac{3}{2}y_b^2-\frac{1}{2}y_{\tau}^2\right)g_2^3,\\
\diff{g_3}{\log Q} &= \frac{1}{16\pi^2}\biggl( -7g_3^3            \biggr)
+\frac{1}{(16\pi^2)^2}\left(\frac{11}{10}g_1^2+\frac{9}{2}g_2^2-26 g_3^2-2 y_t^2-2 y_b^2\right)g_3^3,
\end{align}
at the two-loop level, where $Q$ is an energy scale of evaluation and we have defined
\begin{equation}
 \left(g_3, g_2, g_1\right):=\left(g\s s, g_2, \sqrt{\frac53}g_Y\right).
\end{equation}
Using these RGEs together with one-loop level RGEs for the \YUKAWA\ couplings, which are given as
\begin{align}
 \diff{y_t}{\log Q} &=   \frac{1}{16\pi^2} \left(-\frac{17}{20}g_1^2-\frac{9}{4}g_2^2-8 g_3^2+\frac{9}{2}y_t^2+\frac{3}{2}y_b^2+y_{\tau }^2\right)y_t,\\
 \diff{y_b}{\log Q} &=   \frac{1}{16\pi^2} \left(-\frac{1}{4}g_1^2-\frac{9}{4}g_2^2-8 g_3^2+\frac{3}{2}y_t^2+\frac{9}{2}y_b^2+y_{\tau }^2\right)y_b,\\
 \diff{y_\tau}{\log Q}&= \frac{1}{16\pi^2} \left(-\frac{9}{4}g_1^2-\frac{9}{4}g_2^2+3 y_t^2+3 y_b^2+\frac{5}{2}y_\tau^2\right) y_\tau,
\end{align}
we can evaluate the values of the gauge couplings at any energy scales.

\begin{figure}[t]
\begin{center}
   \includegraphics[width=0.6\textwidth]{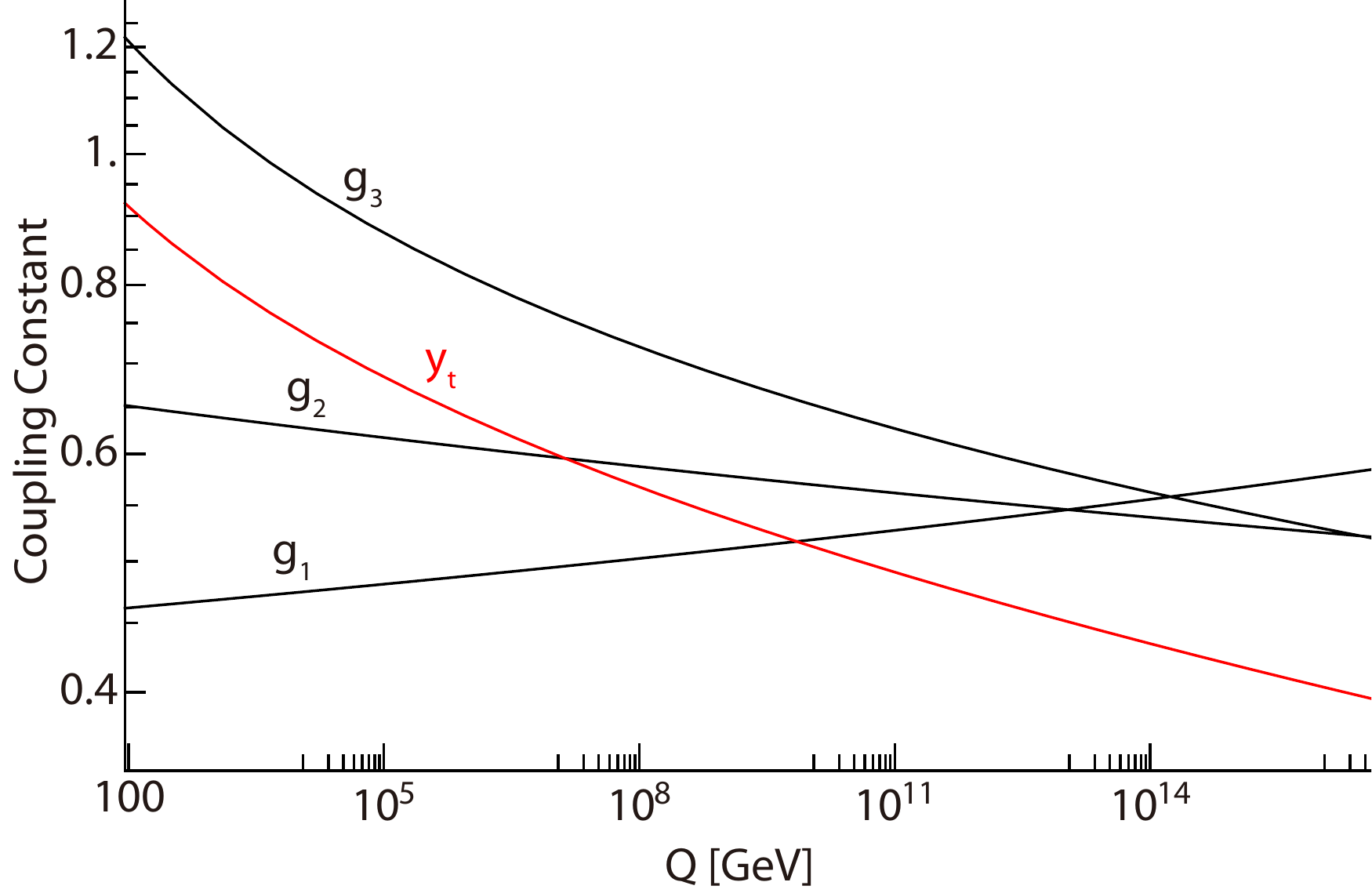}
  \caption[The gauge coupling running under the Standard Model.]{%
 The renormalization group evolution of the gauge coupling constants $g_3$, $g_2$ and $g_1$ (black lines, from top to bottom) under the Standard Model, together with the evolution of $y_t$ (red line).
 Evaluation is done at the two-loop level for the gauge couplings. The \YUKAWA\ couplings entering the evaluation are evolved with the one-loop level RGEs.
  Several remarks on this figure are documented in Appendix~\ref{app:rge_figures}.
   }
  \label{fig:SM_gcu}
\end{center} 
\end{figure}

The result is shown in Fig.~\ref{fig:SM_gcu}. The couplings approach to each other, and gather closely at $Q\sim 10^{15}\GeV$, but complete matching is not achieved; there is a slight mismatch of the gauge coupling constants.

How can we settle this situation?
First we must keep in mind that $\gSU(5)$-breaking effect might change the runnings near $10^{15}\GeV$.
Since all what we need is just a slight shift, we can optimistically hope that the gauge coupling unification is realized with help from the effect, called ``threshold corrections.''

Another solution is, again, the SUSY.
Since the SUSY introduces many fields, the renormalization group running is modified.
We here postpone the discussion in Sec.~\ref{sec:mssm-gcu}, where we will see the SUSY performs fabulous miracle.

\section{Concluding Remarks}
In this chapter, the Standard Model and related issues are introduced. We saw the hierarchy problem in Sec.~\ref{sec:SM}, the muon $g-2$ problem in Sec.~\ref{sec:SMg-2}, and the unification problem in Sec.~\ref{sec:SMunification}.

Miraculously the SUSY can solve these three problems.
We have now enough reasons to believe Nature adopts the SUSY.
However, on the contrary, we have no signature of the SUSY although it was optimistically expected to be found at the Large Hadron Collider.
We have indirect evidences but no direct evidences.
How should we understand this, somewhat strange, situation?

Jurists have the principle of {\sc ``in dubio pro reo''}.
We scientists respect a similar principle; experimental evidence is the most important for the theory, and any beautiful theories could not win their validity without signatures in experiments.
The SUSY is, therefore, just a hypothesis yet.

Then, what we can do now is to fully utilize the LHC to determine whether the SUSY exists or not, and therefore what is important is to examine LHC SUSY searches without optimism and to invent more efficient way to detect the SUSY.
From this viewpoint, the next chapter is dedicated to an overview of the ATLAS detector, one of the experiments at the LHC.
The SUSY comes on the discussion after this excursion.


\chapter{The ATLAS Experiment}
\label{cha:atlas}
\newcommand{\vPT}{\vc p\s T}
\newcommand{\PT}{p\s T}
\newcommand{\ET}{E\s T}

\newcommand{\MET}{\cancel{E}\s{T}}
\newcommand{\vMET}{\cancel{\vc{E}}\s T}
\newcommand{\CL}[1]{\text{CL}_{\rm #1}}
\newcommand{\CLexp}[1]{\CL{#1}\suprm{exp}}
\newcommand{\stauL}{\tilde \tauL}
\newcommand{\stauR}{\tilde \tauR}
\newcommand{\mstauL}{m_{L_3}}
\newcommand{\mstauR}{m_{\bE_3}}

The ATLAS experiment~\cite{aad:2008zzm,Aad:2009wy} is a general-purpose detector for the Large Hadron Collider (LHC).
With utilizing the detector, the ATLAS collaboration recorded a great milestone of the Higgs discovery in July 2012~\cite{20120704ATLAS} together with the CMS collaboration~\cite{20120704CMS}.
They have also searched for signatures from models beyond the Standard Model such as the supersymmetry (SUSY).

In Chapter~\ref{cha:vectorlike} we will apply a result of a SUSY search from the ATLAS experiment with detailed discussion on the ATLAS detector, such as consideration of lepton detection efficiency.
In preparation for the discussion we here briefly review the detector instrument and how physical objects (such as electrons and muons) are detected.

The CMS detector~\cite{Chatrchyan:2008aa} is the other general-purpose detector for the LHC.
Its apparatuses are slightly different, but basic ideas of design and general strategy for the object detection are similar to those of the ATLAS detector.
Therefore, although we simply concentrate on the ATLAS detector in this dissertation, the discussions can to a considerable degree be applied to the CMS detector.

\section{Overview}
\label{sec:atlas-overview}

 \begin{figure}[p]
   \begin{center}
   \includegraphics[width=0.9\textwidth]{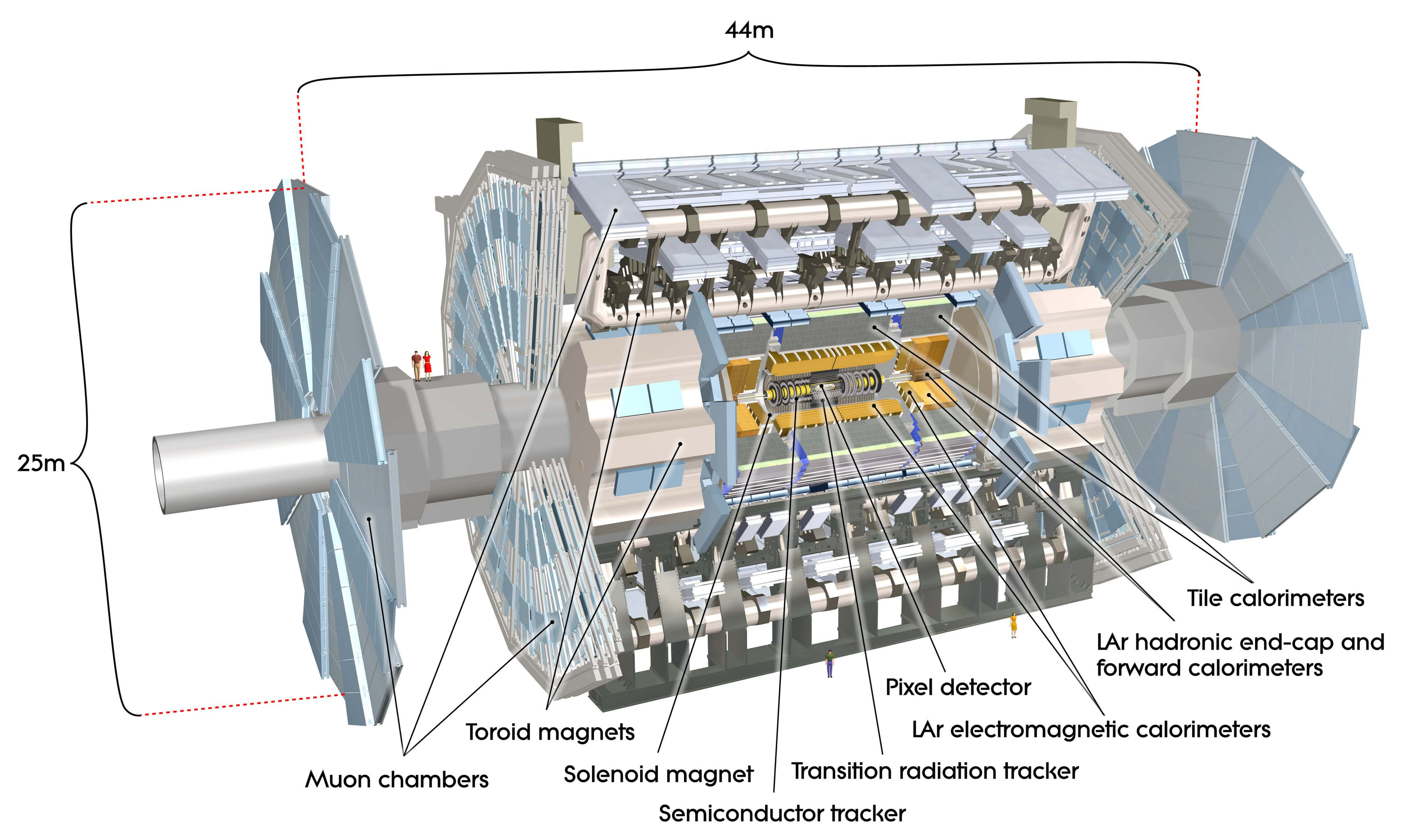}
    \caption[An overview of the ATLAS detector and its subdetectors.]{An overview of the ATLAS detector and its subdetectors. The innermost layer is the tracker, which is composed by the pixel detector, the semiconductor tracker (SCT), and the transition radiation tracker (TRT). The calorimeters surround the tracker, and the muon chambers (MS) is installed in the outermost sector. {\bf This figure is cited from Ref.~\cite{aad:2008zzm}. (ATLAS Experiment \copyright 2008 CERN)}}
    \label{fig:ATLAS_fulldetector}
   \end{center}
 \end{figure}

 \begin{figure}[p]
\begin{center}
   \includegraphics[width=0.7\textwidth]{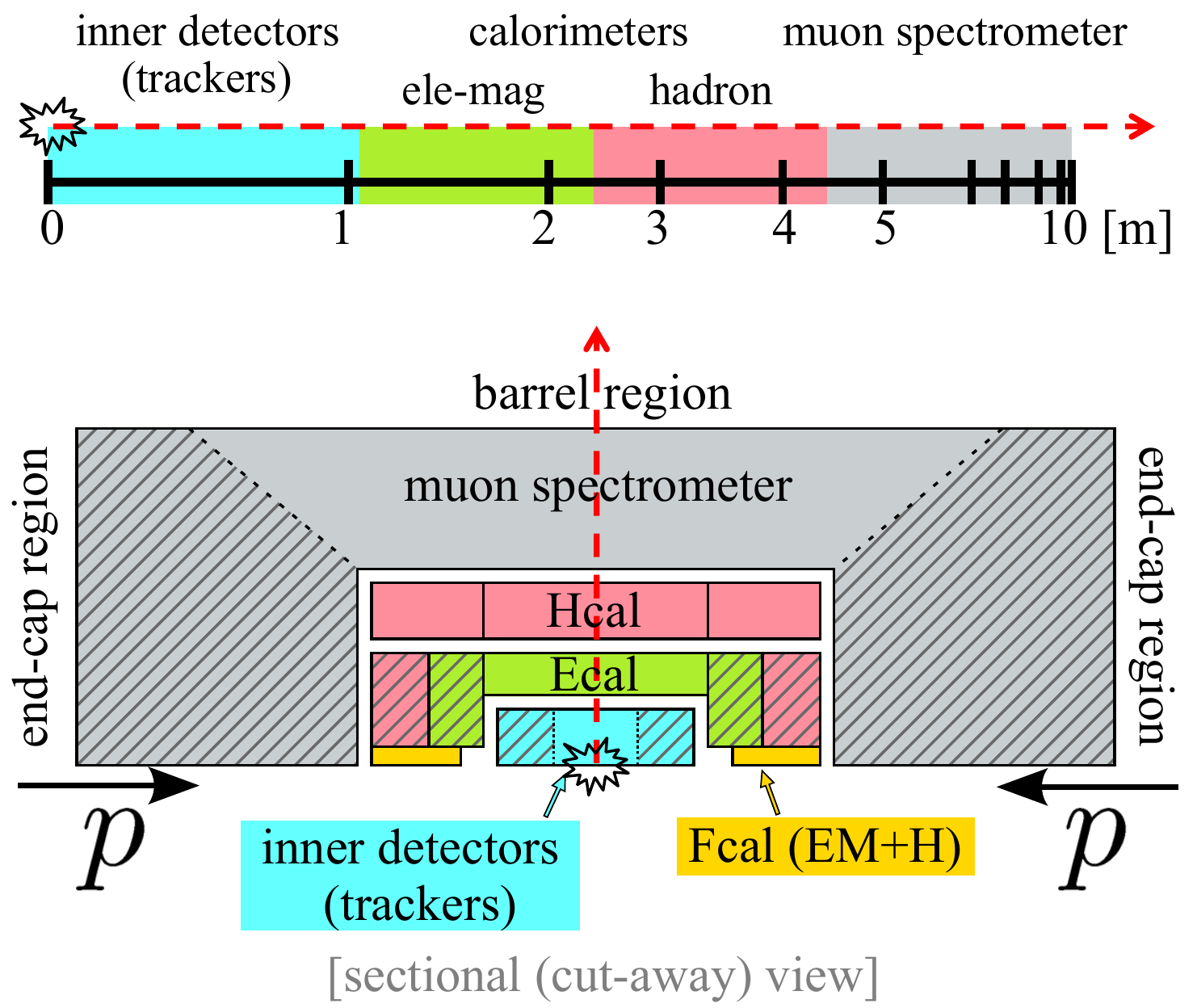}
  \caption[Much simplified overview of the ATLAS detector.]{An extremely-simplified (and thus inaccurate) version of Fig.~\ref{fig:ATLAS_fulldetector}. The red line in each figure illustrates an imaginary trajectory of a particle produced at the collision point.
 The upper figure shows the approximate scale of the detector, and the lower is a cartoon describing the detector system. The detectors are roughly separated to the ``barrel'' region (not shaded) and the ``end-cap'' region (shaded).}
  \label{fig:ATLAS_mydetector}
\end{center} 
\end{figure}

 \begin{figure}[p]
   \begin{center}
    \includegraphics[width=0.8\textwidth]{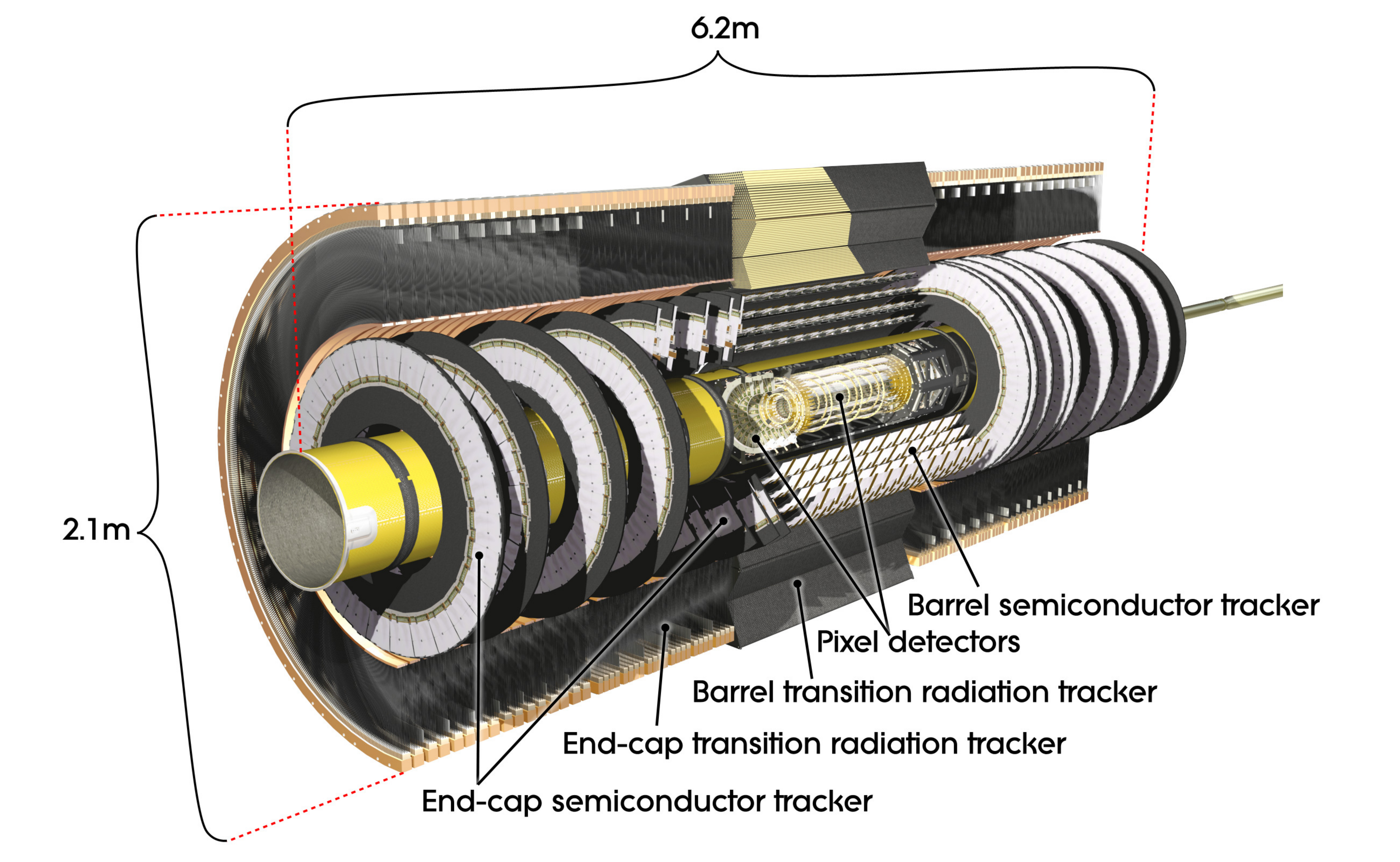}
    \caption[An overview of the ATLAS inner detectors.]{An overview of the ATLAS inner detectors (ID), which can be found at the center of Fig.~\ref{fig:ATLAS_fulldetector}. The pixel detector, the semiconductor tracker (SCT), the transition radiation tracker (TRT) are respectively indicated. See Fig.~\ref{fig:ATLAS_innerdetector} for the actual installation of the detectors. {\bf This figure is cited from Ref.~\cite{aad:2008zzm}. (ATLAS Experiment \copyright 2008 CERN)}}
    \label{fig:ATLAS_ID}
   \end{center}
 \end{figure}

 \begin{figure}[p]
   \begin{center}
   \includegraphics[width=0.8\textwidth]{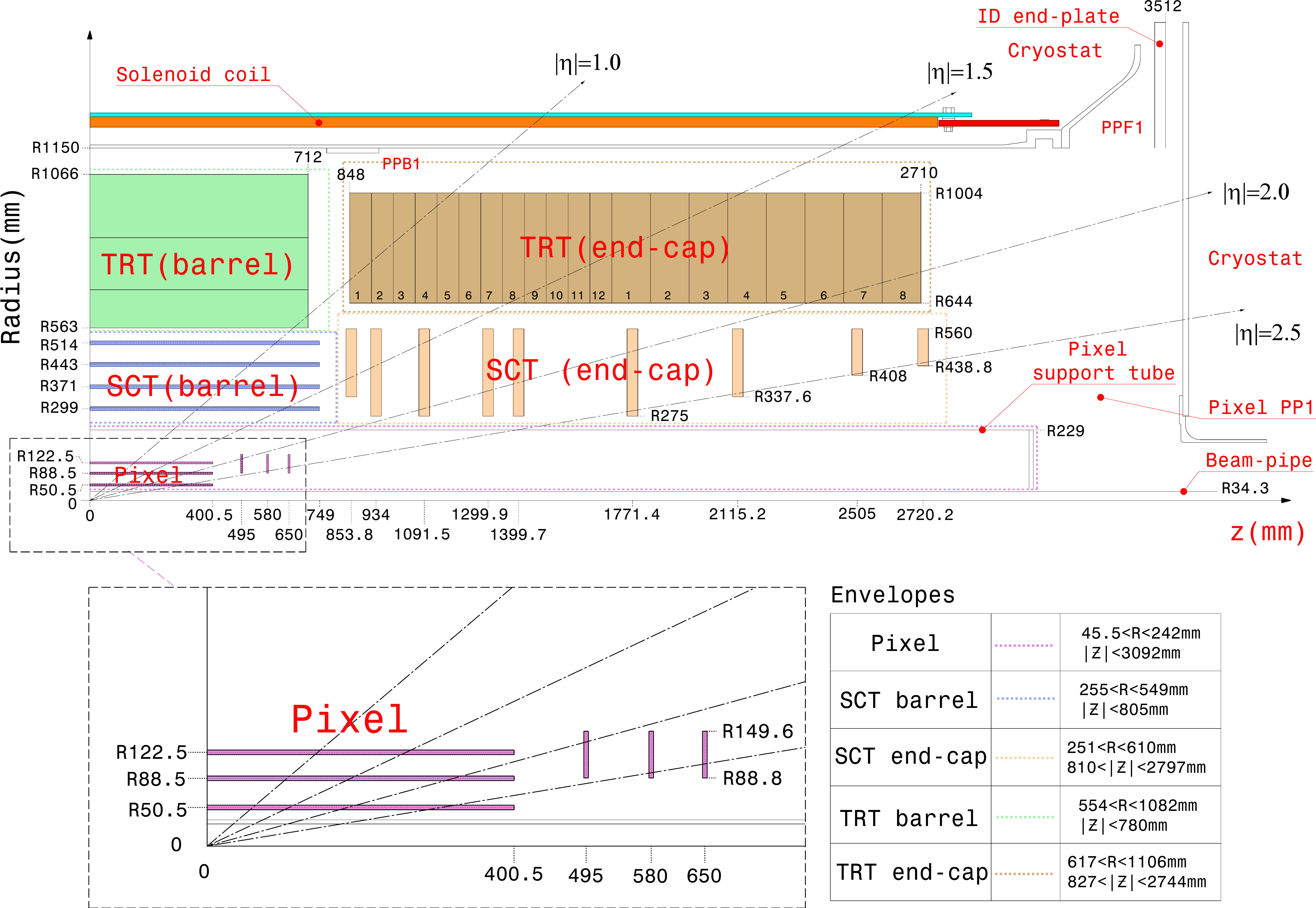}
    \caption[A cut away view of the ATLAS inner detectors.]{A cut-away view of the ATLAS inner detectors (ID). The collision point locates in the left--bottom corner of this figure. We can see that each detector is separated to the barrel part and the end-cap part.
    The trackers are installed within $|\eta|<2.5$, but the TRT coverage is restricted to $|\eta|<1.9$.
    {\bf This figure is cited from Ref.~\cite{aad:2008zzm}. (ATLAS Experiment \copyright 2008 CERN)}}
    \label{fig:ATLAS_innerdetector}
   \end{center}
 \end{figure}

\subsection{Coordinates}
For later convenience the definition of coordinates in the ATLAS detector is documented at first.

The ATLAS collaboration uses a right-handed coordinate system with the origin at the nominal collision point. From the collision point the $x$-axis directs the center of the LHC ring, $y$-axis points upwards, and $z$-axis is defined along the beam axis.
Spherical coordinates $(r,\theta,\phi)$ is defined as usual: $r:=\sqrt{x^2+y^2+z^2}$, $\cos\theta:=z/r$, and $\tan\phi:=y/x$. The pseudorapidity $\eta$ is defined as $\eta:=-\log\tan(\theta/2)$. The transverse direction corresponds $\eta=0$, while $\eta=\pm\infty$ point the beam direction.

Momenta and energies in the detector are usually expressed as ``transverse momenta'' $\vPT$ and ``transverse energies'' $\ET$, which is defined as $\vPT=(p_x, p_y, 0)$, $\PT=\norm{\vPT}=p\sin\theta$ and $\ET=E\sin\theta$.
Angular distances are often expressed in terms of $\Delta R:=\sqrt{(\Delta\eta)^2+(\Delta\phi)^2}$.

\subsection{Detector}
The ATLAS detector~\cite{aad:2008zzm} consists of three apparatuses: trackers, calorimeters, and muon spectrometer (MS).
Fig.~\ref{fig:ATLAS_fulldetector} provides an overview of the detector system, where we can find the trackers at the very center, the calorimeters as the brown apparatus surrounding the trackers and the gray surrounding the brown, and the MS as the outermost part.
In the figure magnets are also drawn, which provide strong magnetic field in the ATLAS detector.
A simplified cartoon is shown in Fig.~\ref{fig:ATLAS_mydetector} for a better understanding.

The trackers, or the inner detectors (ID), are the innermost apparatus surrounding the LHC beam pipe cylindrically.
Charged particles flying through the detectors provide many hits, where actually they ionize detector materials, on the layers of the trackers, and the hits are reconstructed as charged tracks.
The ID itself has also layered structure; from inner to outer, the pixel detector, the semiconductor tracker (SCT), and the transition radiation tracker (TRT) are installed.
In Fig.~\ref{fig:ATLAS_fulldetector}, the pixel is installed inside the very central yellow tube, and the SCT surrounds it.
They as well as the TRT are clearly observed in Fig.~\ref{fig:ATLAS_ID}, a detailed view of the ID.
In addition, to see the exact size of each inner detector, a cut-away view of the ID is provided in Fig.~\ref{fig:ATLAS_innerdetector}.

 \begin{figure}[t]
   \begin{center}
   \includegraphics[width=0.8\textwidth]{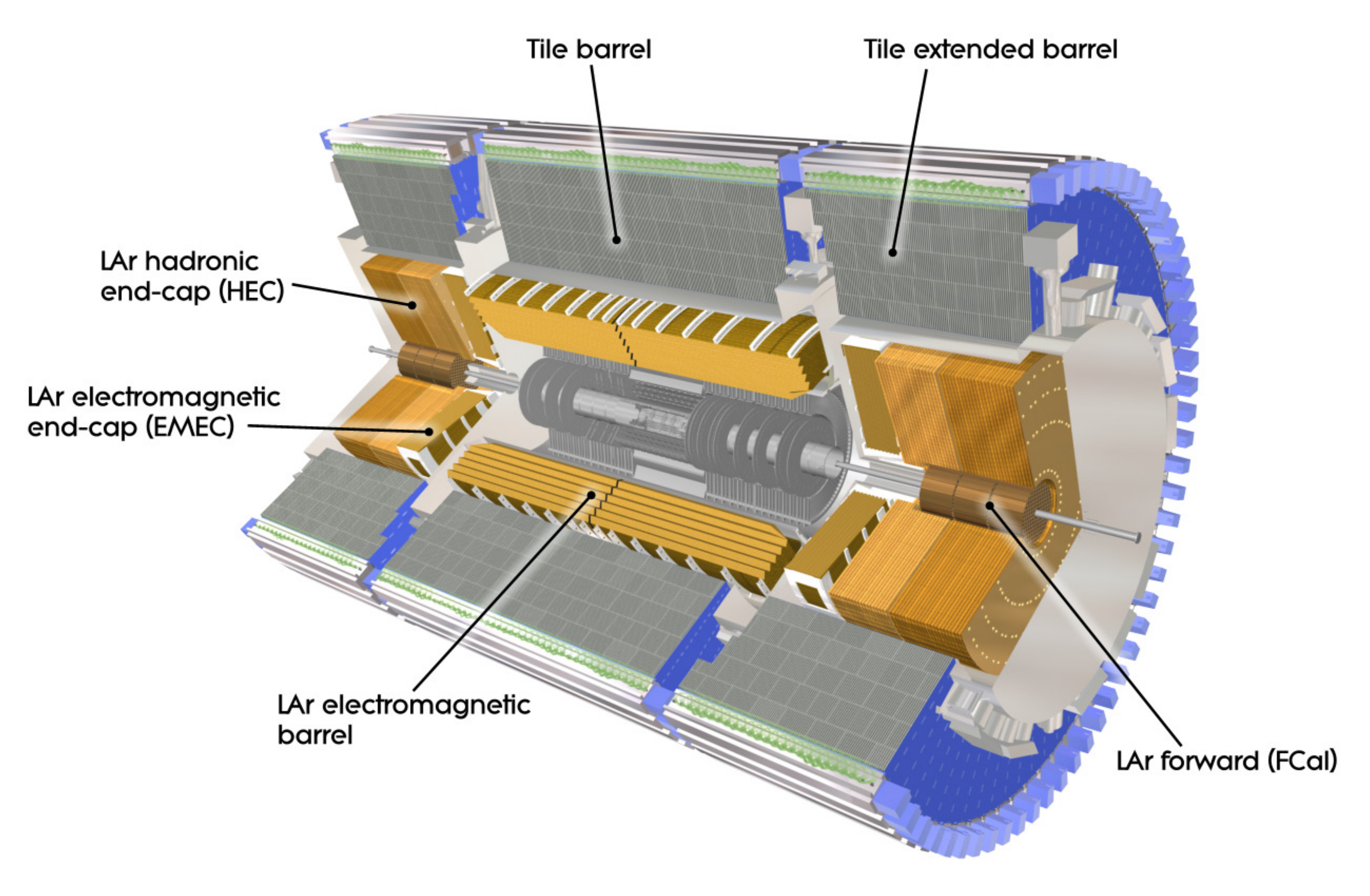}
    \caption[A cut away view of the ATLAS calorimeters.]{A cut-away view of the ATLAS calorimeters. Detailed explanation is found in the text. {\bf This figure is cited from Ref.~\cite{aad:2008zzm}. (ATLAS Experiment \copyright 2008 CERN)}}
    \label{fig:ATLAS_Calo}
   \end{center}
 \end{figure}

The calorimeters, surrounding the ID, are composed by the electromagnetic calorimeter (ECAL), the hadronic calorimeter (HCAL), and the forward calorimeter (FCAL).
A detailed cut-away view is provided in Fig.~\ref{fig:ATLAS_Calo}, where we can clearly find the ECAL as ``LAr electromagnetic barrel'' and ``LAr electromagnetic end-cap (EMEC),'' the HCAL as ``Tile (extended) barrel'' and ``LAr hadronic end-cap (HEC)'', and the FCAL as ``LAr forward (FCal).''

The ECAL is the inner part of the calorimeter, which consists of a barrel part (``LAr electromagnetic barrel'') for $|\eta|<1.475$ and two end-cap parts (``LAr electromagnetic end-caps'') for $1.375<|\eta|<3.2$.
They all are sampling calorimeter with liquid argon (LAr) and lead.
Electrons and photons mainly interact with the ECAL, and lose almost all of their energy to be observed as electromagnetic jets in the ECAL.
The energy loss of electrons is through Bremsstrahlung, while photons have various interactions such as $e^+e^-$ pair-production, photo-electric effect, and Compton scattering.
Note that the muon does not stop here because its energy is too small to cause significant Bremsstrahlung; the energy loss $\dd E/\dd x$ with Bremsstrahlung of a particle whose mass is $m$ is approximately proportional to $E/m^2$, and thus, $E\gtrsim 1\TeV$ for muon is required to trigger Bremsstrahlung.

The HCAL of the ATLAS detector is a sampling calorimeter for hadronic particles.
It is with large dense, and hadronic particles interact with the detector material to deposit their all energy and stops; finally they are observed as hadronic jets.
Iron--scintillating-tile system is used to cover the barrel region: the apparatus for $|\eta|<1.0$ is called ``tile barrel'', and for $0.8<|\eta|<1.7$ are ``tile extended barrel'' as shown in Fig.~\ref{fig:ATLAS_Calo}.
For the end-cap region copper--LAr system is used, which is displayed as ``LAr hadronic end-cap'' in the figure.

The very forward regions of $3.1<|\eta|<4.9$ are covered with the FCAL.
Copper--LAr system for electromagnetic objects and Wolfram--LAr system for hadrons are utilized.

The MS, the outermost apparatus of the ATLAS detector, is another tracker for muons.
All the other (known) particles but muons and neutrinos are already trapped in the previous detectors, and here momenta of muons are measured.

In addition to the above detectors, the ATLAS experiment has three smaller detector systems in a very forward (and backward) region: LUCID (luminosity measurement using Cerenkov integrating detector), ZDC (zero-degree calorimeter), and ALFA (absolute luminosity for ATLAS).
They lie $\pm17\un{m}$, $\pm{140}\un{m}$, and $\pm{240}\un{m}$ from the collision point, respectively.
The main purposes of the LUCID and the ALFA are to measure the integrated luminosity and to provide online monitoring of the instantaneous luminosity and beam conditions~\cite{aad:2008zzm}.
The ZDC is mainly utilized to detect very forward neutrons in heavy-ion collisions~\cite{aad:2008zzm}.

\starline

The performance of the ATLAS detector is fully exploited within $|\eta|<2.5$, or $9^\circ<\theta<171^\circ$.
The coverage of the detectors is approximately summarized as: $|\eta|<2.5$ for the trackers, $|\eta|<3.2$ for the ECAL and the HCAL, and $|\eta|<2.7$ for the MS. The FCAL is installed at $3.1<|\eta|<4.9$.
Only the central region is utilized for usual searches for models beyond the Standard Model.

\subsection{Triggering}
The nominal bunch-crossing rate of the ATLAS detector is $40\un{MHz}$, which means that enormously vast numbers of collisions occur at the center of the ATLAS detector.
Not all of them can be recorded to the storage due to limitations of processing speed and storage capacity.
Therefore, a system of event selections, called triggering system, is installed in the data-taking system. The selections are designed to pickup events of our interest to reduce the output rate to $\sim200\un{Hz}$ ($\approx 300\un{MB/s}$), which is acceptable by the data-taking system~\cite{Aad:2009wy}.

Several selections are installed inclusively; i.e., an event passes the trigger and is stored for the physics analysis if it fulfills at least one of the trigger requirements.
A famous selection is that requiring hard jets plus missing energy; for example, in the analysis which we will examine in Sec.~\ref{sec:neutralinoNLSP}, events which passed the trigger requiring missing energy of $\MET>100\GeV$ and at least one jet with $\PT>80\GeV$ are utilized.

Note that the criteria depend on the instantaneous luminosity of the LHC; in the run of 2012, the bunch-crossing rate was $<20\un{MHz}$, and thus the criteria were looser than those designed for the nominal rate.

\section{Object Identification and Reconstruction}
\label{sec:atlas-object}

The detectors are installed to detect particles produced in collisions, that is, in fact, to identify their species and to reconstruct their momenta and energy.
Detection is always accompanied by efficiency, resolution, and fake rate.
These are studied at first in silico, i.e., in Monte Carlo simulations, and now are measured in situ.

We will briefly review those issues on identifications, especially for jets, electrons, and muons in detail, which are important in our numerical evaluations of the SUSY model discussed later.

\subsubsection{Hadron}
Hadronic particles are captured in the HCAL and produce hadronic showers to be observed as ``jets.''
In the calorimeter, which consists of many cells, measured is how much energy is deposited in the respective cells.
Thus we have to reconstruct jets from such information.
Many algorithms, called jet algorithms, are invented to this end.
Recent ATLAS SUSY searches usually utilized the anti-$k_t$ algorithm~\cite{anti-kt} with a distance parameter of $0.4$.

In the actual experiment the energy of the jet is calibrated to match the result from their full detector simulations, but it is beyond the scope of this dissertation.

\subsubsection[\texorpdfstring{$b$}{b}-tagging]{$\vc b$-tagging}
The $b$-quark always appears in new physics searches at the collider experiment in these days; the top quark always produces a $b$-quark, and the dominant decay branch of the Higgs boson is expected as $h\to b\bar b$.
Therefore, it is very important to identify hadronic particles which originate in $b$-quarks.
This is called ``$b$-tagging,'' and several algorithms are invented to this end.

One famous method for $b$-tagging is the secondary-vertex method, which exploits a characteristic feature of the  $b$-mesons that their lifetime is slightly longer than that of other mesons.
It is about $1.5\un{ps}$, and thus the flight is a few millimeters, which is observable with the ID~\cite{ATLAS2010099}.

$b$-tagging algorithms are applied to the reconstructed jets, and decisions are made that they are $b$-jets or not.
Here the efficiency and the fake rate appear.
It is important that the fake rate for jets containing (or, originating in) $c$-quark is remarkably worse than that for the light-jets (jets without $c$- and $b$-quarks).

Recent analyses by the ATLAS collaboration utilize the {\bf MV1} algorithm, a neural-network-based algorithm whose input is taken from several stand-alone $b$-tagging algorithms.
As is expected, this neural-network-based algorithm generally yields better efficiency~\cite{ATLAS2012043} and better jet rejection~\cite{ATLAS2012040,ATLAS2012043}.

\subsubsection{Electron}
Electrons produced at the collision point leave tracks in the trackers, and provide shower to stop in the ECAL.
Thus identification and reconstruction of electrons utilizes the trackers and ECAL.
Since similar signatures are provided by charged hadrons, it is important to reject jets faking electrons, but then we face a difficult trade-off between identification efficiency and rejection power.
As a solution several criteria are defined for electron identification.

The recent techniques for electron reconstruction are summarized in Ref.~\cite{Aad:2011mk}.
In the document three criteria for the central region and two for the forward region are defined.

For electron identification in the central region with the tracker coverage, $|\eta|<2.47$, ``loose'', ``medium'' and ``tight'' selections are provided with an {\em expected} jet rejection of 500, 5000 and 50000, respectively.
The ``loose'' selection employs only the calorimeter information. The ``medium'' selection further requires a track--cluster matching, and track quality provided by the pixel and the SCT.
In the ``tight'' selection the criteria of the ``medium'' are tightened, the information from the TRT detector is used, and discrimination against photon conversion is employed.

For the forward region of $2.5<|\eta|<4.9$, the region without trackers, ``forward loose'' and ``forward tight'' selections are defined, which utilize only the calorimeter information
Electrons in this region are usually not used in searches for models beyond the Standard model.

The electron reconstruction efficiency and the momentum resolution are publicly reported in Ref.~\cite{Aad:2011mk}\footnote{%
This is the latest public report (conference note) at the date of November 2012.%
} with the tag-and-probe method based on $\sqrt s=7\TeV$ collision data corresponding to the integrated luminosity of $40\invpb$, which were taken in the year 2010.
Here one should note that the efficiency for an electron $\epsilon\suprm{elec}$ is decomposed into two factors as
\begin{equation}
\epsilon\suprm{elec}=\alpha\s{reco}\cdot\epsilon\s{ID}.
\end{equation}
The first term $\alpha\s{reco}$ is the efficiency for the cluster in the ECAL from an electron to be reconstructed well, and the second term $\epsilon\s{ID}$ is that for an electron to pass the selection criteria discussed above.
$\alpha\s{reco}$ is measured with $Z\to ee$ events, and $\epsilon\s{ID}$ is with $Z\to ee$ for $\PT\in[20,50]\GeV$, $W\to e\nu$ for $\PT\in[15,50]\GeV$ and $J/\psi\to ee$ for $\PT\in[4,20]\GeV$.

The Bremsstrahlung effect with materials {\em before} the ECAL is not negligible for electrons.
To correct the energy loss, a technology called the Gaussian sum filter is utilized in some of recent analyses, e.g. in Ref.~\cite{ATLAS2012151}, which provides better performance in the electron detection~\cite{ATLAS2012047}.

\begin{rightnote}
 The ATLAS collaboration seems to have re-optimized the selections and defined ``loose++'', ``medium++'', and ''tight++''~\cite{ATLASCOMPHYS2012783}, whose definitions are hardly found in public documents.
 We do not discriminate the re-optimized selection with the original ones for simplicity.
\end{rightnote}

These reports are summarized on an web page\footnote{%
\url{https://twiki.cern.ch/twiki/bin/view/AtlasPublic/ElectronGammaPublicCollisionResults} for electrons, and \url{https://twiki.cern.ch/twiki/bin/view/AtlasPublic/MuonPerformancePublicPlots} for muons, as of November 2012.}%
, where preliminary plots of the efficiency and the resolution measurements are also available.

\makeatletter
\newcounter{tmpfootnote}{\setcounter{tmpfootnote}{\@arabic\c@footnote}}
\makeatother

\subsubsection{Muon}
Muons in the ATLAS detector, produced at the center, fly through the ID, pass through the calorimeters with a little energy deposits, and reach the muon spectrometer (MS).
Muons are reconstructed with the MS, which is essential for muon identification and reconstruction.
The ID are also used to increase resolution and efficiency.

Recently the ATLAS collaboration reported three methods of the muon identification~\cite{ATLAS2011046,ATLAS2011063}.
{\bf Stand-alone (SA) muon} is the muon which is reconstructed with utilizing only the MS, i.e., without using the ID.
Their direction and momenta are determined with the trajectory in the MS, taking energy losses in the calorimeters into account.
{\bf Combined (CB) muon} and {\bf segment tagged (ST) muon} are, meanwhile, reconstructed with utilizing both the ID and the MS.
In the former case the identification is respectively done in the MS and the ID, and a combination of an MS track and an ID track is employed. A successful combination results in a CB muon.
In the latter case the identification is done in the ID. The ID track is extrapolated to the MS, and a successful association with MS track segments is reconstructed as a ST muon.

CB muons are with the highest purity, and thus usually utilized.
ST muons are sometimes used supplementarily to compensate the inefficiency of that for CB muons.

The muon reconstruction efficiency is publicly reported in Ref.~\cite{ATLAS2011063} and Ref.~\cite{ATLAS2012125}\footnote{%
These are the latest public reports (conference notes) at the date of November 2012.%
}.
In Ref.~\cite{ATLAS2011063} the efficiency for $\PT>20\GeV$ is measured with $Z\to\mu\mu$ events.
The efficiency for $\PT\in [2,10]\GeV$ is reported in Ref.~\cite{ATLAS2012125}, where $J/\psi\to\mu\mu$ events are used.
Both measurements utilize the tag-and-probe method, and are based on the $\sqrt s=7\TeV$ collision data corresponding to the integrated luminosity of $40\invpb$, which were taken in the year 2010.

The muon momentum resolution is reported in Ref.~\cite{ATLAS2011046}\footnote{%
This is the latest public report (conference note) at the date of November 2012.%
}, which is based on $Z\to\mu\mu$ and$W\to\mu\nu$ events in the 2010 data of $7\TeV$ and $40\invpb$.

These reports are summarized on an web page$^{*\thetmpfootnote}$, where preliminary plots of the efficiency and the resolution measurements are also available.
Especially several plots for the efficiency measurement at the $8\TeV$ LHC can be found as a rapid communication~\cite{ATLASCOMPHYS2012716}.

\section{Concluding Remark}
\label{sec:atlas-concluding-remark}
With the detector and the strategy we have seen in this chapter, the discovery of the Higgs boson was achieved.
However, here another factor which helped the discovery should be emphasized.
It is provided by the effort of the collaboration for the LHC accelerator.

The LHC ran at the energy of $E\s{CM}=7\TeV$ in 2010 and 2011, and data corresponding to an integrated luminosity of $5.3\invfb$ were recorded at the ATLAS experiment.
Then in 2012 the energy was increased to $E\s{CM}=8\TeV$, and data of $21.7\invfb$ were obtained.
Especially in 2012 run, the collision rate was increased to $20\un{MHz}$, and the peak luminosity of $7.73\times10^{33}\un{cm^{-2}s^{-1}}$ is recorded with fully utilizing the accelerator.
With these great utilizations of the LHC, the discovery of the Higgs boson was achieved.

On 17 December 2012, the LHC was shut down to prepare for collisions with $E\s{CM}=13\TeV$ and $14\TeV$.
In the $13\TeV$ run expected to start in 2015, the collision rate will be increased to the nominal $40\un{MHz}$, which will provide larger instantaneous luminosity of $\sim1\times10^{34}\un{cm^{-2}s^{-1}}$ to allow us to expect data corresponding to $\Order(100)\invfb$ in the $13\text{--}14\TeV$ runs.\footnote{Also in order to obtain much more data the HL-LHC (High Luminosity Large Hadron Collider) is proposed, where an instantaneous luminosity of $\sim3\EE{34}\un{cm^{-2}s^{-1}}$ and an integrated luminosity of $\sim3000\invfb$ are expected.}

Now we have to wait for three years, but now, contrary to that after the accident in 2009, we have data, which are enough to allow us to find the ``tail'' of the physics beyond the Standard Model buried inside them.

In the next chapter, we will review the MSSM as a promising candidate for such theories, and discuss its current status with examining the data.


\chapter{The MSSM and Its Current Status}
\label{cha:mssm}

\def\DRbar{\overline{\rm DR}}
\def\DRbarPrime{\overline{\rm DR}{}'}
\def\dd{{\mathrm d}}

\begin{table}[p]
\begin{center}
\caption[The field content of the MSSM.]{The field content of the MSSM. In the leftmost column the superfield notation is used, and the rightmost two columns describe the included fields of the superfield: complex scalar fields for the spin $0$ particles, Weyl spinors for the spin $1/2$ ones, and vector fields for the spin $1$'s.
The gauge indices are omitted, while indices for the three generations are denoted as subscripts $i$.}

\vspace{\baselineskip}

\renewcommand{\arraystretch}{1.4}
 \begin{tabular}[b]{@{\Vrule\ }c|ccc|cc@{\ \Vrule}}\Hrule
\multicolumn{6}{@{\Vrule\ }c@{\ \Vrule}}{{\bf Matter and Higgs fields} (chiral multiplet)}\\\Hrule
      & $\gSU(3)$& $\gSU(2)$& $\gU(1)$& spin 0 & spin $1/2$\\\hline
 $Q_i$   & $\vc 3$ & $\vc 2$ & $1/6$  & $(\tilde\uL,\tilde \dL)$   & $(\uL,\dL)$ \\\hline
 $\bU_i$ & $\bar{\vc 3}$ & $\vc 1$ & $-2/3$ & $\tilde \uR^*$             & $\uR^\dagger$ \\\hline
 $\bD_i$ & $\bar{\vc 3}$ & $\vc 1$ & $1/3$  & $\tilde \dR^*$             & $\dR^\dagger$ \\\hline
 $L_i$   & $\vc 1$ & $\vc 2$ & $-1/2$ & $(\tilde \nu, \tilde \eL)$ & $(\nu,\eL)$ \\\hline
 $\bE_i$ & $\vc 1$ & $\vc 1$ & $1$    & $\tilde \eR^*$             & $\eR^\dagger$ \\\hline
 $\Hu$   & $\vc 1$ & $\vc 2$ & $1/2$  & $(\HuP,\HuZ)$ & $(\tilde\HuP,\tilde\HuZ)$ \\\hline
 $\Hd$   & $\vc 1$ & $\vc 2$ & $-1/2$ & $(\HdZ,\HdM)$ & $(\tilde\HdZ,\tilde\HdM)$ \\\Hrule
\multicolumn{6}{@{\Vrule\ }c@{\ \Vrule}}{{\bf Gauge fields} (vector multiplet)}\\\Hrule
   & $\gSU(3)$ & $\gSU(2) $ & $\gU(1)$ & spin $1/2$ & spin 1 \\\hline
 $G$     & $\vc 8$ & $\vc 1$ & $0$ & $\tilde g$ & $g$ \\\hline
 $W$     & $\vc 1$ & $\vc 3$ & $0$ & $\tilde W$ & $W$ \\\hline
 $B$     & $\vc 1$ & $\vc 1$ & $0$ & $\tilde B$ & $B$ \\\Hrule
\end{tabular}
\renewcommand{\arraystretch}{1}
\label{tab:FieldContentOfMSSM}
\end{center}
\end{table}

\begin{figure}[p]\begin{center}
\begin{fmffile}{feyn/ProtonDecay}
\begin{fmfgraph*}(130,80)
\fmfstraight
\fmfleft{d1,x2,u1,u2,x3}
\fmfright{ep,x1,us,u3,x4}
\fmf{xquark}{u2,u3}
\fmf{xquark}{v1,u1}
\fmf{xquark}{v1,d1}
\fmf{squark,label=$\tilde s\s R^*$}{v1,v2}
\fmfdot{v1,v2}
\fmf{xlepton}{ep,v2}
\fmf{xquark}{us,v2}
\fmflabel{$u$}{u1}
\fmflabel{$u$}{u2}
\fmflabel{$d$}{d1}
\fmflabel{$e^+$}{ep}
\fmflabel{$u^\dagger$}{us}
\fmflabel{$u$}{u3}
\end{fmfgraph*}
\end{fmffile}\vspace{18.5pt}
\caption[A Feynman diagram of the proton decay under the MSSM with the $R$-parity violation.]{Feynman diagram of the proton decay caused by the interactions $\bU_1\bD_1\bD_2$ and $L_1Q_1\bD_2$. To forbid this interaction we have to introduce the $R$-parity.}
\label{fig:ProtonDecayDim4app}
\end{center}\end{figure}
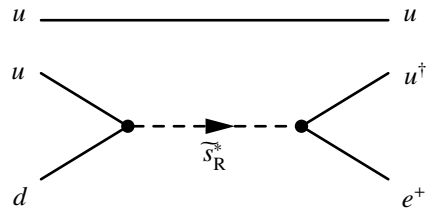

\section{The MSSM}
\label{sec:mssm}
The minimal supersymmetric standard model (MSSM)~\cite{Fayet:1976et,Fayet:1977yc,Farrar:1978xj} is the minimal supersymmetric extension of the Standard Model.
It is characterized by imposed gauge symmetries and field content, as well as the Standard Model.
The gauge symmetry of the Standard Model is the same as the Standard Model:
\begin{equation}
 G\s{MSSM}=\gSU(3)\s{strong}\times\gSU(2)\s{weak}\times\gU(1)_{Y}.
\end{equation}
The field content is shown in Table~\ref{tab:FieldContentOfMSSM}.
Here, unlike the Standard Model, we need two Higgs doublets $\Hu$ and $\Hd$.
This is because the SUSY severely constrains the Lagrangian to forbid a single Higgs doublet to have \YUKAWA{} interactions both with the up-type quarks ($Q$ and $\bU$) and down-type quarks ($Q$ and $\bD$).
Note that the anomaly cancellation condition, in the consideration of which fermionic partners of the Higgs boson must be taken into account, is satisfied with these two doublets.

The symmetry and the field content lead us to the following superpotential of the MSSM:
\begin{equation}
\begin{split}
 W\s{full}&= \mu\Hu\Hd
      - \left(Y_u\right)_{ij} \Hu Q_i \bU_j
      + \left(Y_d\right)_{ij} \Hd Q_i \bD_j
      + \left(Y_e\right)_{ij} \Hd L_i \bE_j\\
 &\quad + \kappa_i\Hu L_i
        + \frac12{\lambda  }_{ijk} L_i L_j \bE_k
        + {\lambda' }_{ijk} L_i Q_j \bD_k
        + \frac12{\lambda''}_{ijk} \bU_i \bD_j \bD_k.
\end{split}
\end{equation}
In the terms in the second line of the expression, however, the baryon number $B$ and the lepton number $L$ are not conserved.
This is not only inconsistent with the fact that $B$- and $L$-violations are not observed yet in experiments, but also makes protons decay in a very rapid rate via, for example, the Feynman diagram described in Fig.~\ref{fig:ProtonDecayDim4app}.
The decay rate of the proton is severely constrained by Super-Kamiokande experiments~\cite{Nishino:2009aa}. For this channel, the constraint is
\begin{equation}
   \Gamma(p\to\pi e^+)
\sim \left|\lambda'_{112}\lambda''_{112}\right|^2\frac{m\s{proton}^5}{m_{\tilde \sR}^4}
 =   \frac{\left|\lambda'_{112}\lambda''_{112}\right|^2}{2.9\EE{-20}\un{yr}}
\left(\frac{1\TeV}{m_{\tilde\sR}}\right)^4 < \left(8.2\EE{33}\un{yr}\right)^{-1},
\end{equation}
which results in the constraint on the coupling constants of $\left|\lambda'_{112}\lambda''_{112}\right|\lesssim 10^{-27}$.

In order to solve this unnaturalness, we usually install the conservation of the $R$-parity~\cite{Farrar:1978xj} into the MSSM, which is a discrete $\gZ_2$ symmetry defined as $P_R:=(-1)^{3B-L+2s}$. Here $s$ is the spin of the particle.
The exact conservation of the $R$-parity restricts the superpotential as
\begin{equation}
\label{eq:MSSMsuperpotential}
W= \mu\Hu\Hd
      - \left(Y_u\right)_{ij} \Hu Q_i \bU_j
      + \left(Y_d\right)_{ij} \Hd Q_i \bD_j
      + \left(Y_e\right)_{ij} \Hd L_i \bE_j,
\end{equation}
and now the baryon number $B$ and the lepton number $L$ are conserved at the classical level. (Note that the sphaleron process, mentioned in Sec.~\ref{sec:SM}.)
Conservation of $B$ and $L$ at the tree level is achieved accidentally in the Standard Model, but in the MSSM we have to impose it by hand.
We in this dissertation assume the $R$-parity conservation, and use this superpotential \eqref{eq:MSSMsuperpotential}.

This ``MSSM with $R$-parity'' provides a very nice explanation of the Dark Matter problem.
Under the $R$-parity conservation the lightest particle among those having odd parity, actually which is the lightest supersymmetric particle (LSP), becomes stable.
The LSP is considered as a promising candidate for the Dark Matter~\cite{LSPasDarkMatter}.

\begin{rightnote}
Strictly speaking, the $R$-parity does not make protons stable.
Besides the sphaleron process, the $B$ and $L$ are violated in higher dimensional interactions even in the presence of the $R$-parity conservation.
Such terms can be forbidden by imposing, e.g., the baryon triality $B_3$~\cite{Ibanez:1991pr} or the proton hexality~\cite{Dreiner:2005rd} instead of the $R$-parity, where these symmetries conserve $B$ and $L$ up to dimension five operators with holding the stability of the LSP.

On the other hand, just for avoiding the proton decay problem, we do not have to forbid both the $B$- and $L$-violations.
Protons remain stable if the baryon number is conserved, or the lepton number is conserved and the LSP is heavier than protons.
In both choices the LSP becomes unstable, and difficult to be served as a Dark Matter candidate.
\end{rightnote}

Now we can write down the full Lagrangian of the MSSM, but it is still insufficient because we know our Universe is not supersymmetric.
We have to introduce additional ``SUSY-breaking (\SUSYbreaking)'' terms, which do not respect the SUSY, into the Lagrangian.
However, we must be careful not to disgrace the MSSM.
The hierarchy problem is solved in the MSSM because the SUSY guarantees the condition $|k|^2=k\s{s}$ in Eqs.~\eqref{eq:1loopfermion} and \eqref{eq:1loopscalar}.
For this reason we usually introduce only the ``soft'' SUSY-breaking terms, or terms whose coupling constants have positive mass dimension.
Then the condition $|k|^2=k\s{s}$ does not violated even in quantum corrections because soft couplings do not appear in the renormalization group equations of the dimensionless couplings.

The soft SUSY-breaking Lagrangian of the MSSM is given as
\begin{equation}\label{eq:MSSMsusybreaking}\begin{split}
 -\Lag\s{\SUSYbreaking}
&= \frac12\left(M_3\tilde g\tilde g+M_2\tilde W\tilde W+M_1\tilde B\tilde B+\Hc\right)\\
&  \quad+\left[
    - (a_u)_{ij}\Hu\tilde Q_i\tilde{\bar u}_j
    + (a_d)_{ij}\Hd\tilde Q_i\tilde{\bar d}_j
    + (a_e)_{ij}\Hd\tilde L_i\tilde{\bar e}_j + \Hc
   \right]\\
&  \quad+\left[
      \left(m^2_Q\right)_{ij}  \tilde {Q}^*_i\tilde Q_j
     +\left(m^2_L\right)_{ij}  \tilde {L}^*_i\tilde L_j
     +\left(m^2_\bU\right)_{ij}\tilde {\bar u}^*_i\tilde{\bar u_j}
     +\left(m^2_\bD\right)_{ij}\tilde {\bar d}^*_i\tilde{\bar d_j}
     +\left(m^2_\bE\right)_{ij}\tilde {\bar e}^*_i\tilde{\bar e_j}
   \right]\\
&  \quad+\left[
     m^2_{\Hu}\Hu^*\Hu + m^2_{\Hd}\Hd^*\Hd + \left(b\Hu\Hd+\Hc\right)
   \right],
\end{split}\end{equation}
where $M_a$ and $m^2_X$ are the (quadratic) masses for the gauginos and the scalar bosons (Higgs bosons, squarks and sleptons), $a_{u,d,e}$ are called trilinear scalar couplings, and $b$ is the off-diagonal mass term of the Higgs bosons.
For later convenience, we here define the following parameters as usual:
\begin{align}
 (A_u)_{ij}& := \frac{(a_u)_{ij}}{(Y_u)_{ij}},&
 (A_d)_{ij}& := \frac{(a_d)_{ij}}{(Y_d)_{ij}},&
 (A_e)_{ij}& := \frac{(a_e)_{ij}}{(Y_e)_{ij}},&
 B &:= \frac{b}{\mu}.\label{eq:aaab-definition}
\end{align}

In the MSSM we have redefined the phases so that the coefficient $b$, the vacuum expectation values of the Higgs bosons, and fermion masses are to be positive. Then the CKM matrix has one phase as is in the case of the Standard Model, and the phases are thrust to the SUSY-breaking parameters $M_a$ and $a_i$.
Nevertheless, we ignore these phases in this dissertation;
in particular, $M_a$ are assumed positive throughout our discussions.

\starline

Breaking a symmetry by hand is always accompanied by a new problem, and this is the case, too.
Not only $M_a$ but also $m^2$ and $a$ in Eq.~\eqref{eq:MSSMsusybreaking} have complex phases in general and induces $CP$-violation.
Also those for quarks and leptons can have off-diagonal component, and large flavor violation may be yielded.
These $CP$- and flavor violation may easily contradict with current observations; these are called the SUSY $CP$- and flavor problems.
These problems originate in a fundamental question that what kind of mechanism generates the SUSY-breaking\ effects.
Various models have been built, or are being built, to achieve a realistic SUSY-breaking.

A promising model for SUSY-breaking is the gauge-mediated SUSY-breaking (GMSB) scenario~\cite{GMSB1oraifeartaigh,GMSB2dynamical}.
In this scenario the source of SUSY-breaking is ``hidden'' from the MSSM particles, i.e. there are no direct interactions between the MSSM fields and the source of SUSY-breaking.
Instead the existence of the source, or the hidden sector, is mediated with extra fields, called messengers, which interact with the MSSM particles via the Standard Model gauge interactions.
Since the gauge interactions do not ``know'' $CP$-violating phases or the flavor indices, we can circumvent the SUSY $CP$- and flavor problems.
This GMSB framework is adopted to the V-MSSM, introduced in Sec.~\ref{cha:vectorlike}, which we call ``V-GMSB scenario.''

\section{Higgs Mass in the MSSM}
\label{sec:higgs-mass-mssm}
The Higgs sector of the MSSM is completely different from that of the Standard Model.
We thus begin with the discussion on the SUSY from reviewing the Higgs sector and the electroweak symmetry breaking inside it.

\subsection{A brief review of tree-level result}
\label{sec:MSSM-higgs-tree}
We start from the tree-level discussion\footnote{Here we just incorporate a brief review. More detailed discussions can be found in, e.g., Ref.~\cite{Martin:1997ns}.}.
The tree-level scalar potential related to the Higgs fields is given directly from the superpotential \eqref{eq:MSSMsuperpotential} and the SUSY-breaking\ Lagrangian~\eqref{eq:MSSMsusybreaking} as
\begin{equation}\begin{split}
 V\stx{Higgs}^{(0)}&=
   \left(\abssq\mu+m_{\Hu}^2\right)\left(\abssq\HuP+\abssq\HuZ\right)
 + \left(\abssq\mu+m_{\Hd}^2\right)\left(\abssq\HdZ+\abssq\HdM\right)
 + \left[b\left(\HuP\HdM-\HuZ\HdZ\right)+\Hc\right]
\\&\qquad
 + \frac{g_Y^2+g_2^2}{8}\left(\abssq\HuP+\abssq\HuZ-\abssq\HdZ-\abssq\Hd\right)^2
 + \frac{g_2^2}{2}\abssq{\HuP\HdZ^*+\HuZ\HdM^*}.
\end{split}\end{equation}
We choose the $\gSU(2)$ basis not to break the electromagnetic symmetry, which leads us to
\begin{equation}\begin{split}
 V_{\text{Higgs}}^{(0)}
&=
   \left(\abssq\mu+m_{\Hu}^2\right) \abssq\HuZ
 + \left(\abssq\mu+m_{\Hd}^2\right) \abssq\HdZ -\left(b\HuZ\HdZ+\Hc\right)
 + \frac{g_Y^2+g_2^2}{8}\left(\abssq\HuZ-\abssq\HdZ\right)^2.\label{eq:MSSMhiggspotTree}
\end{split}\end{equation}
As discussed below Eq.~\eqref{eq:aaab-definition}, we take as a convention $b>0$ by redefining the phases of $\Hd$. Then at minima of $V^{(0)}\s{Higgs}$ $\HuZ\HdZ>0$, and thus we can take the convention under which the vacuum expectation values of $\HuZ$ and $\HdZ$ are real and positive.

We would like to derive the vacuum expectation values
\begin{align}
 \vev{\Hu}&=\vu=v\sin\beta,& \vev{\Hd}&=\vu=v\cos\beta,& v&\approx 174\GeV,
\end{align}
from this potential. Ideologically $(v,\beta)$ are extracted from the condition
\begin{align}
 \vev{\pdiff{}{\HuZ}V^{(0)}\stx{Higgs}} =  \vev{\pdiff{}{\HdZ}V^{(0)}\stx{Higgs}} &= 0,
\label{eq:higgsvevcondition}
\end{align}
but, since we actually know the value of $v$ as $174\GeV$, we use these conditions not to determine $(v,\tan\beta)$ but to constrain the parameters of the MSSM $\left(m_{\Hu}^2,m_{\Hd}^2,\mu,b\right)$ with treating $(v,\tan\beta)$ as input values.
Then we obtain
\begin{align}
 m_{\Hu}^2+\abssq\mu-b\cot\beta-\frac12 m_Z^2\cos2\beta&=0,&
 m_{\Hd}^2+\abssq\mu-b\tan\beta+\frac12 m_Z^2\cos2\beta&=0,\label{eq:higgsvevcondition2}
\end{align}
where the $Z$-boson mass $m_Z^2=\left(g_Y^2+g_2^2\right)v^2/2$ is used, or equivalently
\begin{align}
 \sin2\beta&=\frac{2b}{m^2_{\Hu}+m^2_{\Hd}+2\abssq\mu},&
 m_Z^2&=\frac{\left|m_{\Hd}^2-m_{\Hu}^2\right|}{\sqrt{1-\sin^22\beta}}-m^2_{\Hu}-m^2_{\Hd}-2\abssq\mu,\label{eq:HiggsPotentialConstraints}
\end{align}
as the constraints.

With the determined values of $\left(m_{\Hu}^2,m_{\Hd}^2,\mu,b\right)$, we can evaluate the mass terms of the Higgs sector.
The mass matrix for the $CP$-even Higgs bosons, $(h,H)$, is obtained to be
\begin{align}
 V&\ni\frac12\pmat{h&H}\trans{R_\alpha}\pmat{
\mathcal M^{(0)}\s{uu}&\mathcal M^{(0)}\s{ud}\\\mathcal M^{(0)}\s{ud}&\mathcal M^{(0)}\s{dd}}
R_\alpha\pmat{h\\H},
\end{align}
where
\begin{align}
 \begin{split}
  \mathcal M^{(0)}\s{uu}&=\abssq\mu+m_{\Hu}^2+\frac12{m_Z^2}(1-2\cos2\beta),\qquad
  \mathcal M^{(0)}\s{ud}=-b-\frac12{m_Z^2}\sin2\beta,\\
  \mathcal M^{(0)}\s{dd}&=\abssq\mu+m_{\Hd}^2+\frac12{m_Z^2}(1+2\cos2\beta),
 \end{split}
\end{align}
and the tree-level masses are as
\begin{align}
 m^2_A&=\frac{2b}{\sin2\beta}=2\abssq\mu+m_{\Hu}^2+m_{\Hd}^2
\label{eq:mA_tree}\\
 m^2_{h,H}&=\frac12\left(m_Z^2+m_A^2
\mp\sqrt{\left(m^2_A-m^2_Z\right)^2+4m_Z^2m_A^2\sin^22\beta}
\right),
\label{eq:hH_tree}\\
 m^2_{H^\pm}&=m^2_A+m_W^2.
\label{eq:Hpm_tree}
\end{align}
Here we have defined the well-known Higgs bosons\footnote{%
Throughout this dissertation, the term ``Higgs boson'' in the context of the MSSM (and the V-MSSM) refers to the lighter $CP$-even Higgs boson $h$.}, $h$, $H$, $A$, $H^\pm$, and the \NAMBUGOLDSTONE\ bosons, $G^0$, $G^\pm$, as
\begin{align}
 \pmat{\HuZ\\\HdZ}&=\pmat{\vu\\\vd}+\frac{1}{\sqrt2}R_\alpha\pmat{h\\H}+\frac{\ii}{\sqrt2}R_{\beta_0}\pmat{A\\G^0},&
 \pmat{\HuP\\\HdM^*}&=R_{\beta_+}\pmat{G^+\\H^+},&
\end{align}
with the four rotation matrices
\begin{align}
 R_\alpha     &=\pmat{\cos\alpha&\sin\alpha\\-\sin\alpha&\cos\alpha},&
 R_{\beta_0}  &=\pmat{\sin\beta_0&\cos\beta_0\\-\cos\beta_0&\sin\beta_0},&
 R_{\beta_\pm}  &=\pmat{\sin\beta_\pm&\cos\beta_\pm\\-\cos\beta_\pm&\sin\beta_\pm}.
\end{align}
The mixing angle $\alpha$, which is traditionally chosen to be negative, is related to $\tan\beta$ at the tree level as
\begin{equation}
 \frac{\tan2\alpha}{\tan2\beta}=\frac{m_A^2+m_Z^2}{m_A^2-m_Z^2}.
\end{equation}

The case where $m_A^2\gg m_Z^2$ is called ``decoupling limit''. In this limit the angles $\alpha$ and $\beta$ are related as
\begin{equation}
 \frac{\tan2\alpha}{\tan2\beta}\to 1 \qquad\left(\alpha\to\beta-\frac{\pi}2\right).
\end{equation}

\starline

Let us check resulting features of the electroweak symmetry breaking.
It is straightforward to check that the $\gSU(2)$ gauge bosons successfully obtain masses:
\begin{equation}
\begin{split}
  \Lag
 &\supset
 \left|\left(\partial_\mu-\ii g_2W_\mu-\tfrac12\ii g_YB_\mu\right)\Hu\right|^2
 +
 \left|\left(\partial_\mu-\ii g_2W_\mu+\tfrac12\ii g_YB_\mu\right)\Hd\right|^2\\
 &
 \supset \frac12(\partial_\mu h)^2+\frac{v^2}2\left({g_2}^2{W^+}^\mu W^-_\mu + \frac{{g_Z}^2}{2}Z^\mu Z_\mu\right),
\end{split}
\end{equation}
which is completely the same as Eq.~\eqref{eq:SMgaugemass}, the Standard Model version.
Meanwhile the fermion terms are slightly changed as
\begin{equation}
\begin{split}
 -\Lag &\supset
        \left(Y_u\right)_{ij} \HuZ \uL{}_i \uR^\dagger{}_j
      + \left(Y_d\right)_{ij} \HdZ \dL{}_i \dR^\dagger{}_j
      + \left(Y_e\right)_{ij} \HdZ \eL{}_i \eR^\dagger{}_j\\
      &\leadsto
        \left(vY_u{}_{ij}\sin\beta\right)\uL{}_i \uR^\dagger{}_j
      + \left(vY_d{}_{ij}\cos\beta\right)\dL{}_i \dR^\dagger{}_j
      + \left(vY_e{}_{ij}\cos\beta\right)\eL{}_i \eR^\dagger{}_j.
\end{split}
\end{equation}

\subsection{With one-loop level effective potential}
\label{sec:MSSM-higgs-1loop}
The above discussion is based on the tree level result. As will be mentioned in the discussion section (Sec.~\ref{sec:mssm-h-discussion}), quantum corrections are crucial to the mass of the Higgs boson in the MSSM, which currently is calculated up to the three-loop level accuracy~\cite{Martin:2007pg,Kant:2010tf}.
Here, to examine the corrections analytically, we discuss the one-loop level correction with utilizing the effective potential method.

We begin with the discussion how we should treat the effective potential in the calculation.
What we should consider carefully is that the vacuum expectation value $v\approx174\GeV$ should not be changed even if the potential receives higher order corrections.
That is, even the potential is modified as
\begin{equation}
 V^{(0)}\stx{Higgs}\longrightarrow
 V^{\rm eff}\stx{Higgs}=V^{(0)}\stx{Higgs}+\Delta V,
\end{equation}
the conditions
\begin{align}
 \vev{\pdiff{}{\HuZ}V^{\rm eff}\stx{Higgs}} =  \vev{\pdiff{}{\HdZ}V^{\rm eff}\stx{Higgs}} &= 0
\end{align}
must hold {\em with the same value of $v\approx174\GeV$}.
Thus the constraints \eqref{eq:higgsvevcondition2} are modified to be
\begin{align}
 \begin{split}
 m_{\Hu}^2+\frac{1}{2\vu}\vev{\pdiff{\Delta V}{\HuZ}}+\abssq\mu-b\cot\beta-\frac12 m_Z^2\cos2\beta&=0,\\
 m_{\Hd}^2+\frac{1}{2\vd}\vev{\pdiff{\Delta V}{\HdZ}}+\abssq\mu-b\tan\beta+\frac12 m_Z^2\cos2\beta&=0,
 \end{split}
\end{align}
and the resulting parameters $\left(m_{\Hu}^2,m_{\Hd}^2,\mu,b\right)$ are different from those at the tree level. Here we have already imposed that $\tan\beta$ is not modified, or in other words, treated it as an input value.

Here {\em let us assume that $(\mu,b)$ are kept unchanged, and the change of $\alpha$, the mixing angle of $h$ and $H$, can be neglected.} Then we can write down the parameters as
\begin{align}
 \left(m_{\Hu}^2,m_{\Hd}^2,\mu,b\right)\stx{tree}
&=\left(m_{\Hu}^2+\frac{1}{2\vu}\vev{\pdiff{\Delta V}{\HuZ}},\quad
m_{\Hd}^2+\frac{1}{2\vd}\vev{\pdiff{\Delta V}{\HdZ}},\quad
\mu,\quad
b\right)\stx{under $\Delta V$},
\end{align}
and the components of the mass matrix for the $CP$-even Higgs bosons are obtained as
\begin{align}
\begin{split}
  \mathcal M\s{uu}\suprm{eff}&=
 \vev{\frac12
 \pdiffn2{V^{(0)}\stx{Higgs}}{(\HuZ)} + \frac12\pdiffn2{\Delta V}{(\HuZ)}
 }\stx{under $\Delta V$}\\
 &=
 \vev{
 \abssq\mu+m_{\Hu}^2+\frac12{m_Z^2}(1-2\cos2\beta)
 }\stx{under $\Delta V$}
 +\frac12\vev{\pdiffn2{\Delta V}{(\HuZ)}}\stx{under $\Delta V$}
 \\&=
 \mathcal M\s{uu}^{(0)} - \frac{1}{2\vu}\vev{\pdiff{\Delta V}{\HuZ}}\stx{under $\Delta V$}
 +\frac12\vev{\pdiffn2{\Delta V}{(\HuZ)}}\stx{under $\Delta V$},
\end{split}
\end{align}
et cetera, or simply evaluated as
\begin{align}
 \Delta\mathcal M\s{uu}&=\frac12\left(\pdiffn2{}{\vu}-\frac{1}{\vu}\pdiff{}{\vu}\right)\vev{\Delta V},&
 \Delta\mathcal M\s{ud}&=\frac12\pdiff{^2}{\vu\partial\vd}\vev{\Delta V},&
 \Delta\mathcal M\s{dd}&=\frac12\left(\pdiffn2{}{\vd}-\frac{1}{\vd}\pdiff{}{\vd}\right)\vev{\Delta V}.
\end{align}
Especially the mass of the lighter Higgs boson is modified to be
\begin{align}
 m^2_h&=\left[m^2_h\right]\s{tree}+
\left[
 \frac{\cos^2\alpha}{2}\left(\pdiffn2{}{\vu}-\frac{1}{\vu}\pdiff{}{\vu}\right)
 +\frac{\sin^2\alpha}{2}\left(\pdiffn2{}{\vd}-\frac{1}{\vd}\pdiff{}{\vd}\right)
 -\sin\alpha\cos\alpha\pdiff{^2}{\vu\partial\vd}
\right]\vev{\Delta V},\label{eq:effectivehiggs}
\end{align}
and if we take the decoupling limit, we obtain
\begin{align}
 m^2_h&=\left[m^2_h\right]\s{tree}+
\left[
 \frac{\sin^2\beta}{2}\left(\pdiffn2{}{\vu}-\frac{1}{\vu}\pdiff{}{\vu}\right)
 +\frac{\cos^2\beta}{2}\left(\pdiffn2{}{\vd}-\frac{1}{\vd}\pdiff{}{\vd}\right)
 +\sin\beta\cos\beta\pdiff{^2}{\vu\partial\vd}
\right]\vev{\Delta V}.\label{eq:effectivehiggsDecoup}
\end{align}
Note that the tree level mass $[m^2_h]\s{tree}$ should be evaluated with absence of,  and the correction terms should be with presence of, the correction $\Delta V$.

 \begin{rightnote}
  Here we set $(\mu,b)$ unchanged and neglected the change of $\alpha$ to obtain the equation \eqref{eq:effectivehiggs}, and took the decoupling limit to Eq.~\eqref{eq:effectivehiggsDecoup}.
  However, even if we instead set $(m_{\Hu}, m_{\Hd})$ unchanged, Eq.~\eqref{eq:effectivehiggsDecoup} can still be obtained at the decoupling limit.
  We should note that there lie several approximations, but since, anyway, corrections at higher levels are more important, we do not care these issues.
 \end{rightnote}

\subsection{MSSM Higgs mass at the one-loop level}
\label{sec:MSSM-higgs-1loopmass}
The effective potential of the MSSM at the one-loop level is known as, in the $\DRbarPrime$ scheme~\cite{Jack:1994rk},\footnote{%
The $\DRbarPrime$ scheme is a modified version of the $\DRbar$ scheme~\cite{DRbarSCHEME} in which effects of the mass of the $\epsilon$-scalars, appearing as the extra components of the vector bosons corresponding to the extra $2\epsilon$ dimension in the dimensional reduction methodology, are properly removed. More detailed description can be found in Ref.~\cite{Martin:1993zk}.}
\begin{equation}
\Delta V^{(1)} = \frac{1}{16\pi^2}
\left[
 \sum_{X=\text{spin}\,0}F\left(m_X^2\right)
-2\sum_{X=\text{spin}\,1/2}F\left(m_X^2\right)
+3\sum_{X=\text{spin}\,1}F\left(m_X^2\right)
\right],
\label{eq:MSSMEffPot}
\end{equation}
where
\begin{equation}
 F(x)=\frac{x^2}4\left(\log\frac{x}{Q^2}-\frac32\right)
\end{equation}
with renormalization scale $Q$, which is irrelevant in calculation of the Higgs mass.
Note that the summation should be taken for all particles; i.e., one must not forget the color factor $N\s{c}=3$.

The dominant contributions come from the top--stop sector.
The masses expressed as functions of $(\vu, \vd)$ are obtained from the mass matrices
\begin{align}
  M\s{stop}&=\pmat{%
  m^2_{Q_{33}}+Y_t^2\vu^2+\frac12\left(\frac12g_2^2-\frac16g_Y^2\right)\left(\vd^2-\vu^2\right)&
  Y_t(\vu A_t^* - \mu\vd)\\
  Y_t(\vu A_t^* - \mu\vd)&
  m^2_{\bU_{33}}+Y_t^2\vu^2+\frac12\cdot\frac23g_Y^2\left(\vd^2-\vu^2\right)
},
& M\s{top}&=Y_t\vu.
\end{align}
Then through a bit tough calculation we can obtain
\begin{align}
  \mathcal M^{(1)}\s{uu}\Big|\stx{top--stop}&=\frac{3m_t^4}{2\pi^2\vu^2}\left[
 \log\frac{m_S^2}{m_t^2}+\frac{A_tX_t}{m_S^2}\left(1-\frac{A_tX_t}{12m_S^2}\right)
\right],\\
  \mathcal M^{(1)}\s{ud}\Big|\stx{top--stop}&=\frac{3m_t^4}{2\pi^2\vu^2}
\frac{(A_t-X_t)X_t(A_tX_t-6m_S^2)}{12m_S^4\cot\beta}
,\\
 \mathcal M^{(1)}\s{dd}\Big|\stx{top--stop}&=\frac{3m_t^4}{2\pi^2\vu^2}
\frac{-(A_t-X_t)^2X_t^2}{12m_S^4\cot^2\beta},
\end{align}
and, according to the previous discussion,
\begin{align}
 \Delta m^2_h{}^{(1)}\Big|\stx{top--stop}
&\approx
\left(
 \frac{\sin^2\beta}{2}\mathcal M^{(1)}\s{uu}\Big|\stx{top--stop}
 +\frac{\cos^2\beta}{2}\mathcal M^{(1)}\s{dd}\Big|\stx{top--stop}
 +\sin\beta\cos\beta\mathcal M^{(1)}\s{ud}\Big|\stx{top--stop}
\right)\\
&=
 \frac{3Y_t^4\vu^2\sin^2\beta}{4\pi^2}\left[
\log\frac{\sqrt{m_S^2-\Delta^2}}{m_t^2}
+\left(\frac{X_t^2}{2\Delta^2}-\frac {m_S^2 X_t^4}{8\Delta^6}\right)
\log\frac{m_S^2+\Delta^2}{m_S^2-\Delta^2}
+\frac{X_t^4}{4\Delta^4}
\right]
+\Order\left(g_2^2,g_Y^2\right)
\\&\approx
 \frac{3m_t^4}{4\pi^2 v^2}\left[
\log\frac{m_S^2}{m_t^2}+\frac{X_t^2}{m_S^2}\left(1-\frac{X_t^2}{12m_S^2}\right)
-\frac{1}{2}\left(\frac{\Delta^2}{m_S^2}\right)^2
+\cdots
\right]\label{eq:MSSM1loopHiggsBound}
\end{align}
where
\begin{align}
 m_S^2&=\frac{m_{\tilde t_1}^2+m_{\tilde t_2}^2}{2},&
 \Delta^2&=\frac{m_{\tilde t_2}^2-m_{\tilde t_1}^2}{2},&
 X_t &=A_t-\mu\cot\beta,&v\approx174\GeV.
\end{align}

\subsection{Discussion --- \texorpdfstring{The 126\,GeV Higgs}{The 126GeV Higgs}}
\label{sec:mssm-h-discussion}

The most important difference on the Higgs mass between the Standard Model and the MSSM is the Higgs quartic coupling.
In the Standard Model the coupling, denoted by $\lambda$ in Eq.~\eqref{eq:SMhiggspotential}, is an unknown parameter, and we can tune it to obtain the Higgs mass of $\sim126\GeV$.
On the contrary, under the MSSM, the quartic coupling is fixed by the SUSY, as is shown in Eq.~\eqref{eq:MSSMhiggspotTree}.
Therefore, the Higgs boson mass is restricted, and actually the mass has an upper bound at the tree level of~\cite{MSSMHiggsTreeBound}
\begin{equation}
  m^2_h\approx m_Z^2\cos^22\beta<m_Z^2.
\end{equation}

As a result, in 1980's, the Higgs boson under the MSSM was considered lighter than $Z$-boson.
Therefore, people considered that the Large Electron--Positron Collider (LEP), which was designed in early 1980's with $E\s{CM}=100\GeV$ and started operation in 1989, would discover the Higgs boson or reject the MSSM.
In those days, the most promising channel for the Higgs search was therefore $e^+e^-\to Z\to he^+e^-$ for the early stage and $e^+e^-\to Zh$ for $E\s{CM}\gtrsim180\GeV$, and this upper bound was referred in many reports from the LEP experiments.
Ref.~\cite{Abreu:1990by} from the DELPHI collaboration is an example, which was published in August 1990 to report that the MSSM Higgs boson must be heavier than $28\GeV$ for all values of $\tan\beta$.

However, Nature was not so simple. In the year 1990, several groups found that radiative corrections could raise the Higgs boson mass~\cite{Okada:1990gg, Ellis:1990nz}.
In 1991, before the exclusion limit reached to the $Z$-boson mass, those papers were published and the upper bound on the MSSM Higgs boson was loosened~\cite{Okada:1990vk,Ellis:1990nz,Haber:1990aw,Okada:1990gg,Ellis:1991zd}.
For example, if one substitutes $\tan\beta=10$, $M_A=M_S=1\TeV$, $X_t=0$ and $\Delta^2=0$ in Eqs.~\eqref{eq:hH_tree} and \eqref{eq:MSSM1loopHiggsBound}, the Higgs mass is evaluated to be
\begin{equation}
m_h\sim \sqrt{(89.4\GeV)^2+(88.8\GeV)^2}=126\GeV \qquad\text{($X_t=0$ : no-mixing)},
\end{equation}
which could not be reached even after the upgrade of the LEP collider, i.e. with the LEP-II experiment.
Higgs mass increase is enhanced at $X_t=\pm\sqrt{6}M_S$ due to the stop mixing, which is called $m_h$-max scenario or maximal mixing scenario. The Higgs mass becomes, with the same parameters and simply assuming $\Delta=0$,
\begin{equation}
m_h\sim \sqrt{(126\GeV)^2+(82.1\GeV)^2}=150\GeV \qquad\text{($X_t=\pm\sqrt6 M_S$ : maximal-mixing)}.
\end{equation}

The above calculation was based on the approximation that only the top--stop sector was included.
Other one-loop level contributions, such as threshold effects on the top-quark \YUKAWA\ coupling and corrections from bottom--sbottom and tau--stau sector, are important as well as higher-level loop corrections.
The bottom--sbottom and the tau--stau one-loop level contributions are approximately given as~\cite{Carena:2011aa,Carena:1995wu}
\begin{equation}
 \Delta m_h^2\simeq  - \frac{m_b^4}{16\pi^2 v^2\cos^4\beta}\frac{\mu^4}{M\s{SUSY}^4}
 \left[1+\frac{1}{16\pi^2}\left(
9hb^2-\frac{5m_t^2}{v^2}-64\pi\alpha_3
\right)
\log\frac{M\s{SUSY}^2}{m_t^2}\right]
\end{equation}
and
\begin{equation}
 \Delta m_h^2\simeq -\frac{m_{\tau}^4}{48\pi^2v^2\cos^4\beta}\frac{\mu^4}{M_{\tilde\tau}^4},
\end{equation}
respectively.
They give negative contributions of $\Delta m_h\sim -10\GeV$ to the Higgs mass.

Conclusively, to realize the $126\GeV$ Higgs boson within the MSSM, the stop mass $M_S$ must be $\sim \Order(10)\TeV$ in the no-mixing case and $\sim 1\text{--}2\TeV$ in the maximal-mixing scenario (See, e.g., Ref.~\cite{Draper:2011aa}).

\subsection{Discussion --- The little hierarchy problem}
\label{sec:mssm-little-discussion}
Another interesting feature of the Higgs sector in the MSSM is the so-called little hierarchy problem~\cite{LittleHierarchy}.
This problem lies in Eq.~\eqref{eq:higgsvevcondition2}.
The equation must be hold for the Higgs fields to have vacuum expectation values of $v=\sqrt{\vu^2+\vd^2}=174\GeV$, but it includes a cancellation among $m_{\Hu}^2$, $m_{\Hd}^2$ and $\mu^2$.
There we actually need a ``little'' tuning of order
\begin{equation}
 \text{tuning} \sim
\frac{m_Z^2}{m_{\Hu}^2}\sim
\frac{m_Z^2}{m_{\Hd}^2}\sim
\frac{m_Z^2}{\mu^2},\label{eq:little_hierarchy}
\end{equation}
where the MSSM parameters should be evaluated at the EWSB scale $m_Z$.

This tuning is related to the mass of the stop.
The parameter $m^2_{\Hu}$ receives the radiative correction of
\begin{equation}
 \Delta m_{\Hu}^2\sim -\frac{3Y_t^2}{4\pi^2}m_{\tilde t}^2 \log\frac{\Lambda}{m_{\tilde t}},\label{eq:little_hierarchy_correction}
\end{equation}
which can be seen from the renormalization group equations (RGEs) summarized in Appendix \ref{sec:MSSM_RGE}, or more simply, from the effective potential approach as
$\Delta m_{\Hu}^2 \approx \expval{\dd\Delta V/\dd\HuZ}/(2\vu)$.
Thus the tuning can be interpreted as
\begin{equation}
 \text{tuning} \approx 0.3\%\times \left(\frac{m_{\tilde t}}{1\TeV}\right)^{-2}.
\end{equation}
We hope a lighter stop, or a lighter SUSY scale $M\s{SUSY}$, to eliminate this little tuning.

However, as we saw in the last section, the SUSY scale seems to be heavier than $1\TeV$.
How can we settle this collision?
Is this inevitable price to pay for eliminating the fine-tuning?
This is one of the hottest topics after the Higgs discovery, and currently under discussions.

\section[Muon \texorpdfstring{$g-2$}{g-2} in the MSSM]{Muon $\boldsymbol{g-2}$ in the MSSM}
\label{sec:muon-g-2-mssm}

As we saw in Sec.~\ref{sec:SMg-2}, the muon $g-2$ has $3\sigma$ level discrepancy between its experimental result and the Standard Model theoretical value.
The MSSM has extra contributions to the $g-2$, and might be a solution to the $g-2$ problem if the sign of the correction falls as the desired one, i.e., positive~\cite{SUSY_gminus2}.

As we will saw later in Eqs.~\eqref{eq:MSSMg-2chargino}--\eqref{eq:MSSMg-2right}, all the extra contributions for a lepton are proportional to the mass squared of the lepton. Therefore, the SUSY contribution to the electron $g-2$ can be neglected, which, thus, we do not consider afterwards.

The dominant contributions come from the neutralino--smuon and the chargino--muon-sneutrino loop diagrams shown in Fig.~\ref{fig:MSSM_g-2}.
The calculation is straightforward, but summarized here as a reference.

\subsection{Formulae in mass eigenstates}
\label{sec:muon-g-2-mssm-mass}
Relevant interactions of the MSSM are included in the Lagrangian as
\begin{equation}
\begin{split}
  \Lag \supset&
 -\sqrt2 g_Y\left(-\frac12\smuL^*\muL\bino + \smuRc^*\muRc\bino\right)
 -\sqrt2 g_2\pmat{\snumu^*&\smuL^*}\frac12\pmat{\wino^0&\sqrt2\wino^+\\\sqrt2\wino^-&-\wino^0}\pmat{\nu\\\muL}
 \\&
 -Y_\mu\left[\left(\HnodZ\smuL-\HnodM\snumu\right)\muRc+\left(\HnodZ\muL-\HnodM\nu\right)\smuRc\right]+\Hc
\end{split}
\end{equation}
in the gauge eigenstates. Defining the mass eigenstates $\neut[i]$, $\chgPM[i]$ and $\smu_{a}$ as
\begin{align}\begin{cases}
 \bino   \!\!\!\!&=N_{1i}\, \neut[i]\\
 \wino^0 \!\!\!\!&=N_{2i}\, \neut[i]\\
 \HnodZ  \!\!\!\!&=N_{3i}\, \neut[i]\\
 \HnouZ  \!\!\!\!&=N_{4i}\, \neut[i]\ \ ,
\end{cases}
&&
\begin{cases}
 \wino^- \!\!\!\!&=C_{1i}\, \chgM[i]\\
 \HnodM  \!\!\!\!&=C_{2i}\, \chgM[i]
\ \ ,\end{cases}
&&
\begin{cases}
 \wino^+ \!\!\!\!&=D_{1i}\, \chgP[i]\\
 \HnouP  \!\!\!\!&=D_{1i}\, \chgP[i]
\ \ ,\end{cases}
&&
\begin{cases}
 \smuL \!\!\!\!&=E_{1a}\, \smu_{a}\\
 \smuR \!\!\!\!&=E_{2a}\, \smu_a
\ \ ,\end{cases}
\end{align}
the terms are rewritten as
\begin{align*}
 \Lag \supset&
-\sqrt2 g_Y N_{1i}\left(-\frac12\smuL^*\muL\neut[i] + \smuRc^*\muRc\neut[i]\right)
-g_2\left(D_{1i}\ \snumu^*\chgP[i]\muL-\frac{N_{2i}}{\sqrt2}\ \smuL^* \neut[i]\muL\right)
\\&
-Y_\mu\left(N_{3i}\ \neut[i]\smuL\muRc-C_{2i}\ \chgM[i]\snumu\muRc+N_{3i}\ \neut[i]\muL\smuRc\right)+\Hc
\\=&
\smu_a^*\overline{\neut[i]}\left[
\left(\frac{g_YN_{1i}^*+g_2 N_{2i}^*}{\sqrt2}E_{1a}^*  - Y_\mu N_{3i}^*E_{2a}^*\right) \PL
+ \left(-Y_\mu N_{3i}E_{1a}^* - \sqrt2 g_YN_{1i}E_{2a}^*\right)\PR
\right]\mu
\\&
+\snumu^*\overline{\chgM[i]}\left[Y_\mu C_{2i}\PR-g_2 D_{1i}^*\PL\right]\mu+\Hc
\end{align*}
Here one should impose proper phase-shifts to the fields so that the masses become positive; thus $N_{xi}$'s etc. are complex in general.
Also the small mixing between $\smuL$ and $\smuR$ must not be neglected, which actually does contribute to the final result.

With these obtained couplings, employing the usual calculation of Feynman diagrams, we can obtain the MSSM contribution $\Delta a_\mu$ to the muon anomalous magnetic moment $a_\mu:=(g_\mu-2)/2$ as
\begin{align}
 \begin{split}
 &\Delta a_\mu= \sum_{a}\sum_{i}
f\s N\left(m_{\smu_a}, m_{\neut[i]},
\frac{g_YN_{1i}^*+g_2 N_{2i}^*}{\sqrt2}E_{1a}^*  - Y_\mu N_{3i}^*E_{2a}^*,
-Y_\mu N_{3i}E_{1a}^* - \sqrt2 g_YN_{1i}E_{2a}^*\right)\\
&\qquad
+\sum_i
f\s C\left(m_{\snumu},m_{\chgPM[i]}, -g_2D_{1i}, Y_\mu C_{2i}\right);\label{eq:MSSMg-2}
 \end{split}
\\
&f\s N(M, M_{\tilde\chi},g\s L, g\s R)=
\frac{1}{16\pi^2}
\left[
- \Re\left(g\s L^*g\s R\right) \frac{m_\mu M_{\tilde\chi}}{M^2}N_1\left(\frac{M_{\tilde\chi}^2}{M^2}\right)
-\frac{|g\s L|^2+|g\s R|^2}{6} \frac{m_\mu^2}{M^2}N_2\left(\frac{M_{\tilde\chi}^2}{M^2}\right)
\right]
\notag,\\
&
f\s C(M, M_{\tilde\chi},g\s L, g\s R)=
\frac{1}{16\pi^2}
\left[
-\Re\left(g\s L^*g\s R\right)\frac{m_\mu M_{\tilde\chi}}{M^2}C_1\left(\frac{M_{\tilde\chi}^2}{M^2}\right)
+\frac{|g\s L|^2+|g\s R|^2}{6} \frac{m_\mu^2}{M^2}C_2\left(\frac{M_{\tilde\chi}^2}{M^2}\right)
\right]
\notag,
\end{align}
\begin{align*}
 \text{where}\quad
N_1(x)&:=\frac{1-6x+3x^2+2x^3-6x^2\log x}{(1-x)^4},&
N_2(x)&:=\frac{1-x^2+2x\log x}{(1-x)^3},\\
C_1(x)&:=\frac{3-4x+x^2+2\log x}{(1-x)^3},&
C_2(x)&:=\frac{2+3x-6x^2+x^3+6x\log x}{(1-x)^4}.
\end{align*}

\subsection{Formulae in gauge eigenstates}
\label{sec:muon-g-2-mssm-gauge}
The above formula \eqref{eq:MSSMg-2} is accurate and stiff. However, from a theoretical viewpoint formulae in the gauge eigenstates are more interesting and convenient even at the expense of accuracy.
The formulae can be obtained from the Feynman diagrams shown in Fig.~\ref{fig:MSSM_g-2_gauge} with the mass insertion method, which results in~\cite{Cho:2011rk}
\begin{align}
&& \text{(a)} &&
\Delta a_\mu\left(\text{``chargino''}\right)&={\phantom-}
\frac{g_2^2}{8\pi^2}m_\mu^2\cdot\frac{M_2\cdot\mu\tan\beta}{m_{\snumu}^4}
\cdot F_a\left(M_2, \mu;\ m_{\snumu}\right),\label{eq:MSSMg-2chargino}
&&\\
&& \text{(b)} &&
\Delta a_\mu\left(\text{``pure-bino''}\right)&={\phantom-}
\frac{g_Y^2}{8\pi^2}m_\mu^2\cdot\frac{\mu\tan\beta}{M_1^3}
\cdot F_b\left(m_{\smuL}, m_{\smuR};\ M_1\right),\label{eq:MSSMg-2purebino}
&&\\
&& \text{(c)} &&
\Delta a_\mu\left(\text{``$\smuL$ (bino)''}\right)&={\phantom-}
\frac{g_Y^2}{16\pi^2}m_\mu^2\cdot\frac{M_1\cdot\mu\tan\beta}{m_{\smuL}^4}
\cdot F_b\left(M_1,\mu;\ m_{\smuL}\right),\label{eq:MSSMg-2leftbino}
&&\\
&& \text{(d)} &&
\Delta a_\mu\left(\text{``$\smuL$ (wino)''}\right)&=
-\frac{g_2^2}{16\pi^2}m_\mu^2\cdot\frac{M_2\cdot\mu\tan\beta}{m_{\smuL}^4}
\cdot F_b\left(M_2,\mu;\ m_{\smuL}\right),\label{eq:MSSMg-2leftwino}
&&\\
&& \text{(e)} &&
\Delta a_\mu\left(\text{``$\smuR$ (bino)''}\right)&=
-\frac{g_Y^2}{8\pi^2}m_\mu^2\cdot\frac{M_1\cdot\mu\tan\beta}{m_{\smuR}^4}
\cdot F_b\left(M_1,\mu;\ m_{\smuR}\right),\label{eq:MSSMg-2right}
&&
\end{align}
where
\begin{align}
 F_a(x,y; z)&:=+\frac12\frac{C_1(x^2/z^2)-C_1(y^2/z^2)}{x^2/z^2-y^2/z^2},&
 F_b(x,y; z)&:=-\frac12\frac{N_2(x^2/z^2)-N_2(y^2/z^2)}{x^2/z^2-y^2/z^2}.
\end{align}
Since $\dd N_2(x)/\dd x<0$ and $\dd C_1(x)/\dd x>0$ for $x>0$, these functions $F_a$ and $F_b$ return positive values for any mass parameters.

\newpage

\begin{figure}[t]
 \begin{center}\vspace{\baselineskip}
 \begin{fmffile}{feyn/neut_mul_g-2}
\begin{fmfgraph*}(125,70)
\fmfleft{d1,p1,d2,d3,d4}\fmfright{d5,p2,d6,d7,gc}
\fmf{fermion,label=$\mu$,l.s=right}{p1,x1}
\fmf{fermion,label=$\mu$,l.s=right}{x2,p2}
\fmf{xgaugino,tension=0.5,lab=$\neut[a]$,l.s=left}{x2,x1}
\fmf{phantom,tension=5}{gc,gb}
\fmf{phantom,tension=1}{gb,d1}
\fmfposition
\fmf{phantom,left,tag=1,lab=$\smu_i$,l.s=right}{x1,x2}
\fmfipath{p[]}
\fmfiset{p1}{vpath1(__x1,__x2)}
\fmfi{photon}{point length(p1)/3*2 of p1 -- vloc(__gb)}
\fmfv{label=$\gamma$}{gb}
\fmfi{dashes}{p1}
\end{fmfgraph*}
\end{fmffile}
\hspace{10pt}
 \begin{fmffile}{feyn/neut_nu_g-2}
\begin{fmfgraph*}(125,70)
\fmfleft{d1,p1,d2,d3,d4}\fmfright{d5,p2,d6,d7,gc}
\fmf{fermion,label=$\mu$,l.s=right}{p1,x1}
\fmf{fermion,label=$\mu$,l.s=right}{x2,p2}
\fmf{dashes,tension=0.5,lab=$\snumu$,l.s=left}{x2,x1}
\fmf{phantom,tension=5}{gc,gb}
\fmf{phantom,tension=1}{gb,d1}
\fmfposition
\fmf{phantom,left,tag=1,lab=$\chgM[a]$,l.s=right}{x1,x2}
\fmfipath{p[]}
\fmfiset{p1}{vpath1(__x1,__x2)}
\fmfi{photon}{point length(p1)/3*2 of p1 -- vloc(__gb)}
\fmfv{label=$\gamma$}{gb}
\fmfi{xgaugino}{p1}
\end{fmfgraph*}
 \end{fmffile}
\caption[The MSSM dominant contributions to the muon $g-2$ (in the mass eigenstates).]{The diagrams of the MSSM dominant contributions to the muon $g-2$.}
\label{fig:MSSM_g-2}
\end{center}

\vspace{3\baselineskip
}
 \begin{center}
 \begin{fmffile}{feyn/g_2-gauge}
 \begin{minipage}[t]{0.3\textwidth}
\begin{center}
   \begin{fmfgraph*}(125,70)
 \fmfleft{d1,p1,d2,d3,d4}\fmfright{d5,p2,d6,d7,gc}
 \fmf{fermion,label=$\muL$,l.s=right}{p1,x1}
 \fmf{fermion,label=$\muR$,l.s=right}{x2,p2}
 \fmf{dashes,tension=0.5,lab=$\snumu$,l.s=left}{x2,x1}
 \fmf{phantom,tension=5}{gc,gb}
 \fmf{phantom,tension=5}{gb,ga}
 \fmf{phantom,tension=1}{ga,d1}
 \fmfposition
 \fmf{phantom,left,tag=1,lab=$\wino^-$--$\HnodM$,l.s=left}{x1,x2}
 \fmfipath{p[]}
 \fmfiset{p1}{vpath1(__x1,__x2)}
 \fmf{photon}{gb,ga}
 \fmfv{label=$\gamma$}{gb}
 \fmfiv{d.sh=cross,d.size=5thick}{point length(p1)/2 of p1}
 \fmfi{xgaugino}{p1}
   \end{fmfgraph*}
 \\(a)
\end{center} 
\end{minipage}
 \begin{minipage}[t]{0.3\textwidth}
\begin{center}
   \begin{fmfgraph*}(125,70)
 \fmfleft{d1,p1,d2,d3,d4}\fmfright{d5,p2,d6,d7,gc}
 \fmf{fermion,label=$\muL$,l.s=right}{p1,x1}
 \fmf{fermion,label=$\muR$,l.s=right}{x2,p2}
 \fmf{xgaugino,tension=0.5,lab=$\bino$,l.s=left}{x2,x1}
 \fmf{phantom,tension=5}{gc,gb}
 \fmf{phantom,tension=5}{gb,ga}
 \fmf{phantom,tension=1}{ga,d1}
 \fmfposition
 \fmf{phantom,left,tag=1,lab=$\smuL$--$\smuR$,l.s=left}{x1,x2}
 \fmfipath{p[]}
 \fmfiset{p1}{vpath1(__x1,__x2)}
 \fmf{photon}{gb,ga}
 \fmfv{label=$\gamma$}{gb}
 \fmfiv{d.sh=cross,d.size=5thick}{point length(p1)/2 of p1}
 \fmfi{dashes}{p1}
  \end{fmfgraph*}
 \\(b)
\end{center} 
\end{minipage}
 \begin{minipage}[t]{0.3\textwidth}
\begin{center}
   \begin{fmfgraph*}(125,70)
 \fmfleft{d1,p1,d2,d3,d4}\fmfright{d5,p2,d6,d7,gc}
 \fmf{fermion,label=$\muL$,l.s=right}{p1,x1}
 \fmf{fermion,label=$\muR$,l.s=right}{x2,p2}
 \fmf{xgaugino,tension=0.5,lab=$\bino$--$\HnodZ$,l.s=left}{x2,x1}
 \fmf{phantom,tension=5}{gc,gb}
 \fmf{phantom,tension=1}{gb,d1}
 \fmfposition
 \fmf{phantom}{x1,xx,x2}
 \fmf{phantom,left,tag=1,lab=$\smuL$,l.s=left}{x1,x2}
 \fmfipath{p[]}
 \fmfiset{p1}{vpath1(__x1,__x2)}
 \fmfi{photon}{point length(p1)/3*2 of p1 -- vloc(__gb)}
 \fmfv{label=$\gamma$}{gb}
 \fmfv{d.sh=cross,d.size=5thick}{xx}
 \fmfi{dashes}{p1}
  \end{fmfgraph*}
\\(c)
\end{center}
\end{minipage}

\vspace{10pt}

 \begin{minipage}[t]{0.3\textwidth}
\begin{center}
   \begin{fmfgraph*}(125,70)
 \fmfleft{d1,p1,d2,d3,d4}\fmfright{d5,p2,d6,d7,gc}
 \fmf{fermion,label=$\muL$,l.s=right}{p1,x1}
 \fmf{fermion,label=$\muR$,l.s=right}{x2,p2}
 \fmf{xgaugino,tension=0.5,lab=$\wino^0$--$\HnodZ$,l.s=left}{x2,x1}
 \fmf{phantom,tension=5}{gc,gb}
 \fmf{phantom,tension=1}{gb,d1}
 \fmfposition
 \fmf{phantom}{x1,xx,x2}
 \fmf{phantom,left,tag=1,lab=$\smuL$,l.s=left}{x1,x2}
 \fmfipath{p[]}
 \fmfiset{p1}{vpath1(__x1,__x2)}
 \fmfi{photon}{point length(p1)/3*2 of p1 -- vloc(__gb)}
 \fmfv{label=$\gamma$}{gb}
 \fmfv{d.sh=cross,d.size=5thick}{xx}
 \fmfi{dashes}{p1}
  \end{fmfgraph*}
\\(d)
\end{center} 
\end{minipage}
 \begin{minipage}[t]{0.3\textwidth}
\begin{center}
   \begin{fmfgraph*}(125,70)
 \fmfleft{d1,p1,d2,d3,d4}\fmfright{d5,p2,d6,d7,gc}
 \fmf{fermion,label=$\muL$,l.s=right}{p1,x1}
 \fmf{fermion,label=$\muR$,l.s=right}{x2,p2}
 \fmf{xgaugino,tension=0.5,lab=$\HnodZ$--$\bino$,l.s=left}{x2,x1}
 \fmf{phantom,tension=5}{gc,gb}
 \fmf{phantom,tension=1}{gb,d1}
 \fmfposition
 \fmf{phantom}{x1,xx,x2}
 \fmf{phantom,left,tag=1,lab=$\smuR$,l.s=left}{x1,x2}
 \fmfipath{p[]}
 \fmfiset{p1}{vpath1(__x1,__x2)}
 \fmfi{photon}{point length(p1)/3*2 of p1 -- vloc(__gb)}
 \fmfv{label=$\gamma$}{gb}
 \fmfv{d.sh=cross,d.size=5thick}{xx}
 \fmfi{dashes}{p1}
  \end{fmfgraph*}
\\(e)
\end{center} 
\end{minipage}
\end{fmffile}
\caption[The MSSM dominant contributions to the muon $g-2$ (in the gauge eigenstates).]{The same as Fig.~\ref{fig:MSSM_g-2}, but expressed in terms of the gauge eigenstates.}
\label{fig:MSSM_g-2_gauge}
\end{center}
\end{figure}

\subsection{Interpretation}
\label{sec:muon-g-2-mssm-interp}
As is discussed in Sec.~\ref{sec:SMg-2}, we need a {\em positive} shift of the muon $g-2$ to explain the observed value, and are happy if the shift is $\Delta a_\mu\sim+10^{-9}$.
From this viewpoint, here we will briefly interpret the obtained result \eqref{eq:MSSMg-2chargino}--\eqref{eq:MSSMg-2right}.

First let us take the limit where all the relevant mass parameters are common:  $M_1=M_2=|\mu|=m_{\smuL}=m_{\smuR}=m_{\snumu}=:M\s{SUSY}$. Then, as the functions return $F_a\to1/4$ and $F_b\to 1/12$,
the results are approximated to be
\begin{align}
  \text{[degenerated]}\quad
 \Delta a_\mu(\text{``chargino''})&\sim \dfrac{g_2^2}{32\pi^2}\dfrac{m_\mu^2\cdot \mu\tan\beta}{M\s{SUSY}^3}
\approx 1.5\EE-9\cdot\left[\frac{\mu\tan\beta}{M^3\s{SUSY}}\right]\Big/(100\GeV)^2
,\\
 \Delta a_\mu(\text{``neutralino''})&\sim -\dfrac{g_2^2-g_Y^2}{192\pi^2}\dfrac{m_\mu^2\cdot \mu\tan\beta}{M\s{SUSY}^3}
\approx -0.17\EE-9\cdot\left[\frac{\mu\tan\beta}{M^3\s{SUSY}}\right]\Big/(100\GeV)^2
.
\end{align}
In this case the chargino diagram (a) gives the dominant contribution, and from the expression we realize  $M\s{SUSY}\sim\Order(100)\GeV$ is required to explain the discrepancy in this degenerated scenario.

What will happen if one of the smuons is extremely heavy?
The answer is simple: corresponding neutralino contributions just vanish.
In particular, the case where $m_{\smuL}\gg m_{\smuR}$ is very interesting. The neutralino contribution becomes negative with $\mu>0$, and moreover, since $m_{\snumu}\sim m_{\smuL}$ is expected, the total contribution is given as
\begin{equation}
\text{[$m_{\snumu}\sim m_{\smuL}\gg m_{\smuR}$]}\quad
\Delta a_\mu=
-\frac{g_Y^2}{8\pi^2}m_\mu^2\cdot\frac{M_1\cdot\mu\tan\beta}{m_{\smuR}^4}
\cdot F_b\left(M_1,\mu;\ m_{\smuR}\right).
\end{equation}
Thus $\mu$ must be, different from the usual scenario, negative to explain the discrepancy.

The other extreme cases are $M_1\sim M_2\ll \mu$ and $M_1\sim M_2\gg \mu$.
In the former only the diagram (a) contributes. Both of the smuons must be light, and $\mu>0$ should be hold as usual.
The latter case has no diagrams to contribute: $\Delta a_\mu\simeq0$.

The above results are summarized in the next table.
 \begin{table}[h]
\begin{center}\renewcommand{\arraystretch}{1.2}
  \caption{Summary of the SUSY explanation of the muon $g-2$ in several extreme scenarios.}
 \label{tab:SUSYg-2summary}
 \begin{tabular}[t]{|c|c|c|c|c|c|c|}\hline
 & (a) & (b) & (c) & (d)  & (e) & \multirow{2}{*}{Note (to solve the discrepancy)}\\
 & $\chgPM$ & pure-$\bino$ & $\smuL$ ($\bino$) & $\smuL$ ($\wino$) & $\smuR$ ($\bino$) & \\\hline
 degenerated case & $\checkmark$ & ($\checkmark$)& ($\checkmark$) & ($\checkmark$)& ($\checkmark$)&\\\hline
 $\smuL,\snumu$-decoupled &&&&& $\checkmark$ & $\mu<0$ for a positive shift. \\\hline
 $\smuR$-decoupled        & $\checkmark$ &&($\checkmark$)&($\checkmark$) && \\\hline
 Higgsino-decoupled       & $\checkmark$ &&&&& Both $\smuL$ and $\smuR$ must be light. \\\hline
 gaugino-decoupled        &&&&&& Impossible.\\\hline
 \end{tabular}
\end{center} 
\end{table}

 \begin{figure}[p]
\begin{center}
   \includegraphics[width=0.6\textwidth]{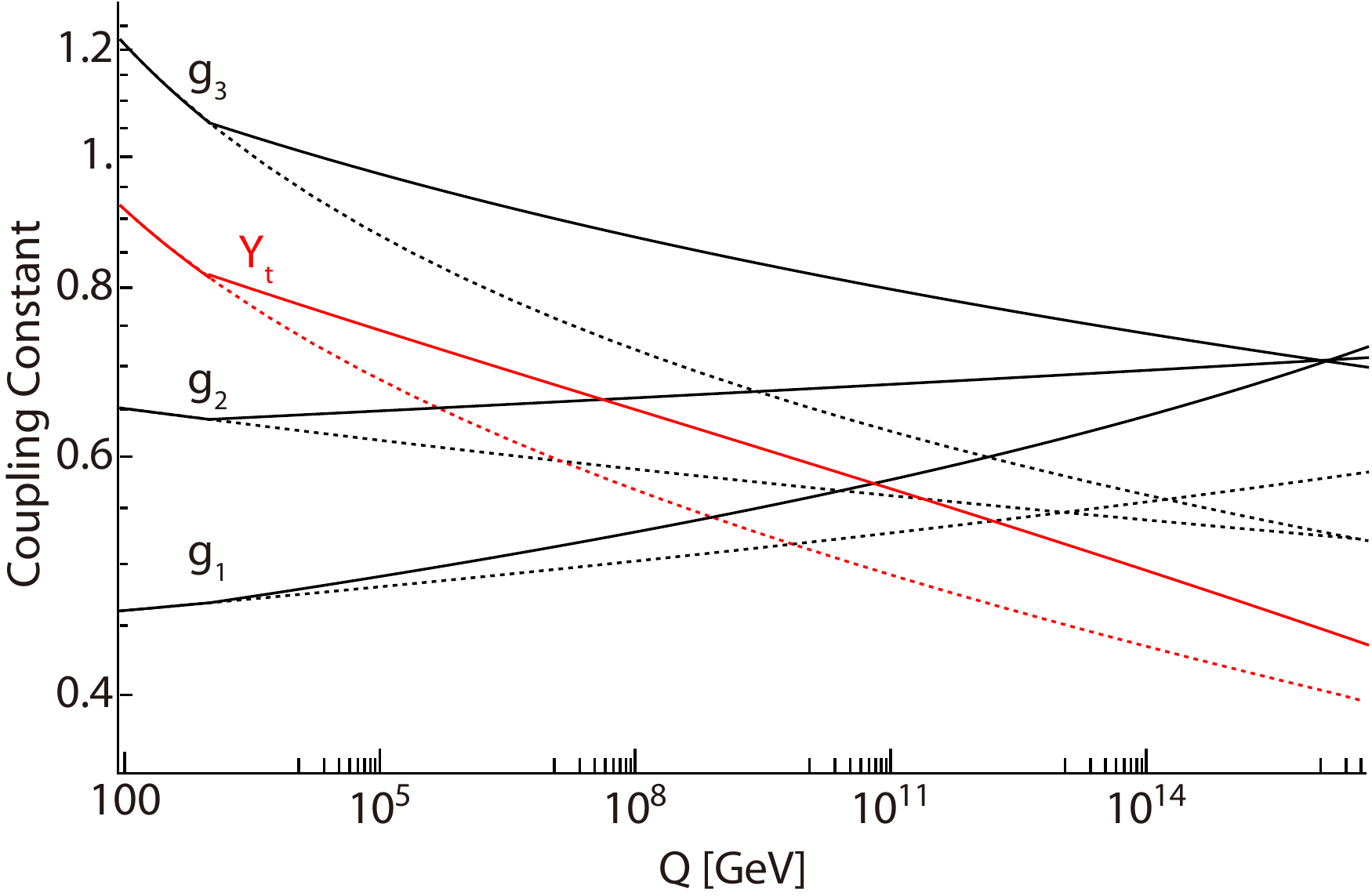}

   \includegraphics[width=0.6\textwidth]{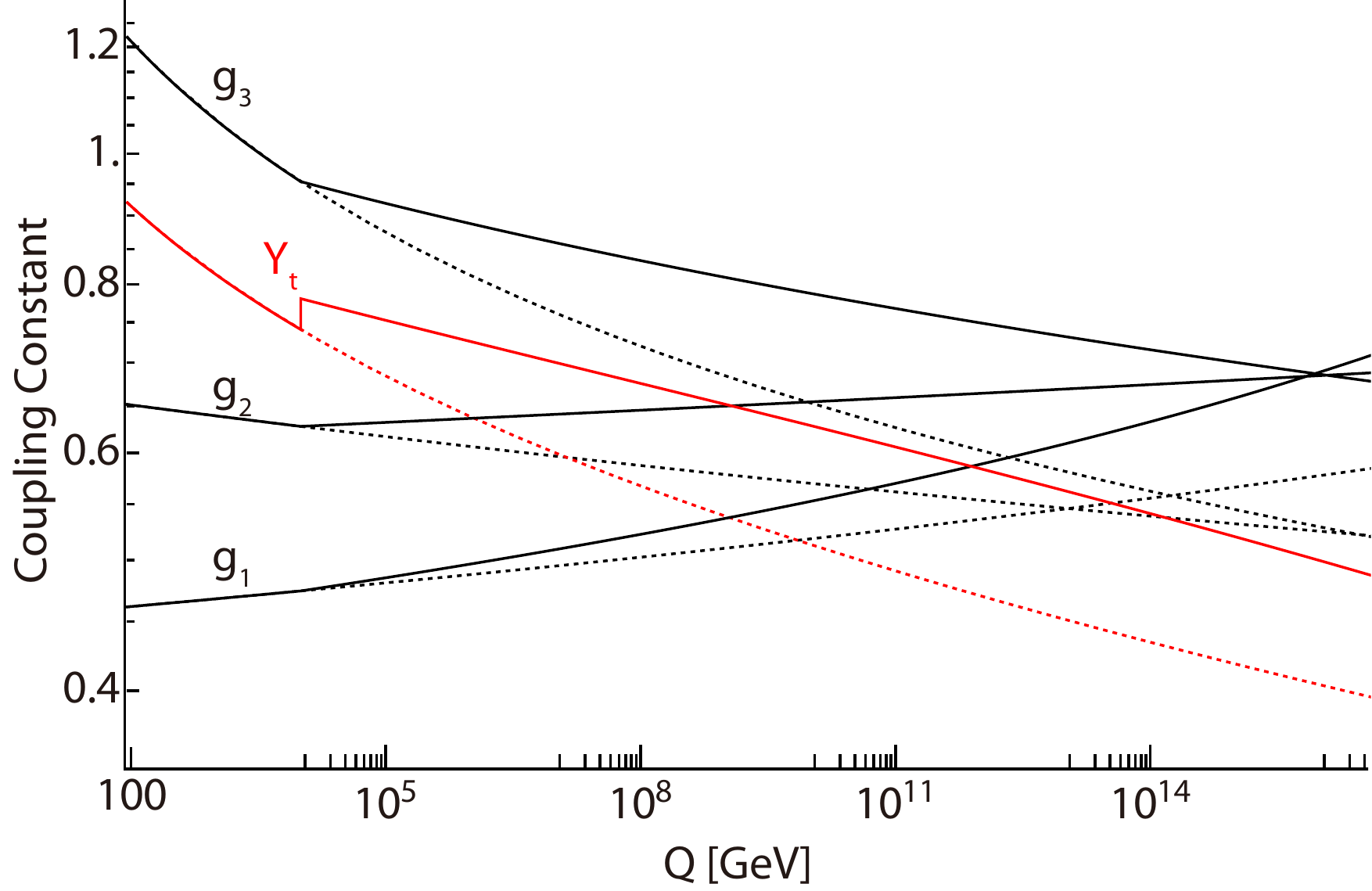}

   \includegraphics[width=0.6\textwidth]{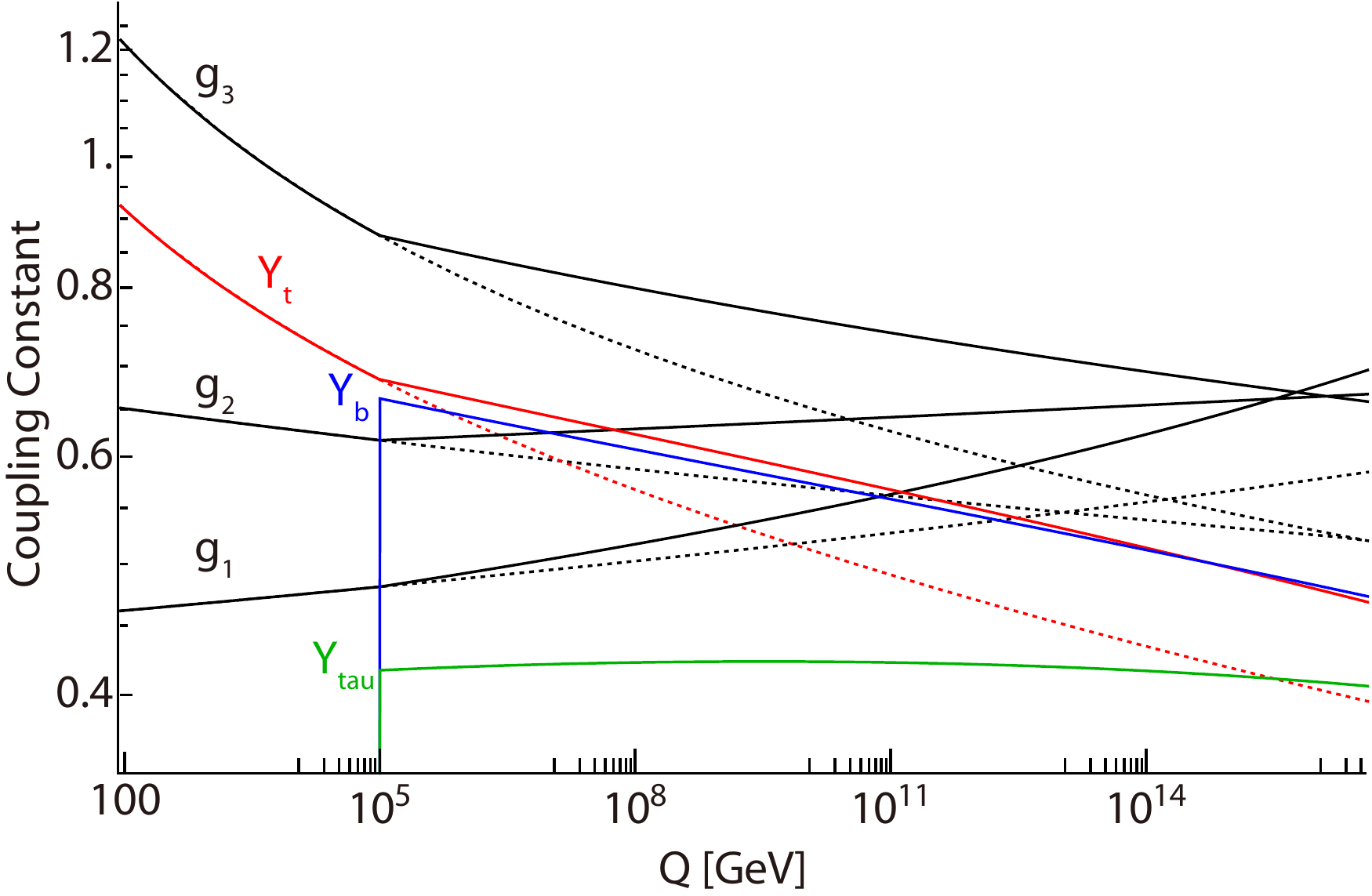}
  \caption[The gauge coupling running under the MSSM.]{%
 The renormalization group evolution of the gauge couplings $g_3$, $g_2$ and $g_1$ (black lines, from top to bottom) under the MSSM, together with the evolution of $Y_t$ (a red line), $Y_b$ (a blue line), and $Y_\tau$ (a green line).
 For comparison the evolutions under the Standard Model are drawn with dotted lines.
 In the three figures $\tan\beta$ and the SUSY scale $M\s{SUSY}$ are differently set: $(M\s{SUSY},\tan\beta)=(1\,{\rm TeV}, 10)$, $(10\,{\rm TeV},3)$ and $(100\,{\rm TeV}, 40)$ from top to bottom.
 $Y_b$ and $Y_\tau$ appear only in the bottom figure: they are below the plotted range in the other figures.
 The input values and the schemes of the RGEs are summarized in Appendix~\ref{app:rge_figures}, together with a note on approximations employed in drawing these figures.
}
  \label{fig:MSSM_gcu}
\end{center} 
\end{figure}

\section{The Gauge Coupling Unification}
\label{sec:mssm-gcu}
Remember the discussion in Sec.~\ref{sec:SMunification}: physicists desire the gauge coupling unification to realize the grand unification, a unified explanation of the three forces; under the renormalization group running they approach to each other at a high-energy scale of $\sim10^{15}\GeV$, but the trajectories have a slight mismatch, which is summarized in Fig.~\ref{fig:SM_gcu}.

The unification of the gauge coupling is a virtue of the SUSY theories.
In the MSSM the RGEs are modified with the presence of the SUSY particles as well as two Higgs doublets; they are summarized in the appendix of this chapter, Appendix~\ref{sec:MSSM_RGE}.

In particular, for the gauge couplings, $(g_1, g_2, g_3):=(\sqrt{5/3}g_Y, g_2, g\s s)$, the one-loop level RGEs are well-known to be
\begin{equation}
 \diff{g_a}{\ln Q} = \frac{g_a^3}{16\pi^2}\left[-3C_a(G)+\sum_{i=\text{matters}}I_a(i)\right],
\end{equation}
where $C_a(G)$ is the quadratic Casimir invariant for the adjoint representation of the group, and $I_a(i)$ is the Dynkin index of the chiral supermultiplets appearing in the model.
For the MSSM these equations are evaluated as
\begin{align}
 \diff{g_3}{\ln Q} &= \frac{g_3^3}{16\pi^2}\left[-3\times3+\frac{1}{2}\times12\right],\\
 \diff{g_2}{\ln Q} &= \frac{g_2^3}{16\pi^2}\left[-3\times2+\frac{1}{2}\times14\right],\\
 \diff{g_1}{\ln Q} &= \frac{g_1^3}{16\pi^2}\left[-3\times0+
\left(\frac{1}{6}\right)^2\times18+
\left(\frac{-2}{3}\right)^2\times9+
\left(\frac{1}{3}\right)^2\times9+
\left(\frac{1}{2}\right)^2\times6+
3+
\left(\frac{1}{2}\right)^2\times4\right]\times\frac{3}{5}.
\end{align}

These RGEs result in the well-known result displayed in Fig.~\ref{fig:MSSM_gcu}, where the unification is much improved.
These figures are drawn with the two-loop level RGEs, and three sets of $\tan\beta$ and the SUSY scale $M\s{SUSY}$ are taken for illustration: the top figure is with $(M\s{SUSY},\tan\beta) = (1\TeV,10)$, the middle one is $(10\TeV,3)$, and for the bottom $(100\TeV,40)$. In any choices the unification is much improved from the Standard Model case.

\newcommand{\glu}{\tilde g}
\newcommand{\squ}{\tilde q}
\newcommand{\squs}{\tilde q^{(*)}}

\section[Current Status of the MSSM --- Higgs, \texorpdfstring{$g-2$}{g-2}, and LHC SUSY Searches]%
        {Current Status of the MSSM \texorpdfstring{\\\mbox{\kern110pt}}{}--- Higgs, $\boldsymbol{g-2}$, and LHC SUSY Searches}
\label{sec:mssm-discussion}
Let us summarize what we discussed in this chapter.

First we saw the mass of the Higgs boson under the MSSM greatly depends on the stop sector, especially on the stop mass $m\s{\tilde t}$ and the stop mixing parameter $X_t=A_t-\mu\cot\beta$.
To realize the mass $126\GeV$ under the MSSM, the stop mass is required to be $\Order(10)\TeV$ without large stop mixing, and even with the maximal-mixing of $X_t\sim\sqrt\pm6m_{\tilde t}$, it should be $\sim1$--$2\TeV$. This fact prefers {\em heavy} SUSY.

On the other hand, the argument of the little hierarchy, embedded in the Higgs potential, prefers {\em light} SUSY; in particular, light $m\s{\tilde t}$.
This is obviously in collision with the argument of the $126\GeV$ Higgs mass.

We then, in Sec.~\ref{sec:muon-g-2-mssm}, saw that the  muon $g-2$ discrepancy is explained with the SUSY contributions if neutralinos,  charginos, and smuons are of order $100\GeV$.
Here {\em light} SUSY is preferred; note that this requirement is not for the {\em colored} sector as the previous two, but for the {\em non-colored} sector of the MSSM.

Now let us move on to the discussion on results from the LHC experiments.
In this dissertation we concentrate on the $R$-parity conserving SUSY; 
the other possibility, $R$-parity violating SUSY, is beyond the scope of this dissertation and not discussed.

The most promising channel in LHC SUSY searches is pair-production of colored superparticles, that is, squarks $\squ$ and gluinos $\glu$.
Especially, the events from $pp\to \glu\glu$, $\glu\squs$ and $\squs\squs$ channels can be detected well in the detectors; multiple hard jets plus large missing energy signature, where jets come from cascade decays of superparticles, and large missing energy is provided by the LSP.
However, we have not detected such signatures.
Now colored superparticles $\glu$ and $\squs$ are approximately constrained as $m_{\glu}\gtrsim900\GeV$ and $m_{\squ}\gtrsim1400\GeV$, respectively,  which are relatively heavier than we optimists expected.
These constraints were obtained by the ATLAS collaboration~\cite{ATLAS2012109}, where they analyzed their data corresponding to an integrated luminosity of $5.8\invfb$ collected at the $8\TeV$ LHC.
The CMS collaboration also analyzed their $\sim5\invfb$ data obtained at $E\s{CM}=7\TeV$, and obtained similar results~\cite{chatrchyan:2012jx,chatrchyan:2012mfa}.

Searches focusing on the third generation squarks are also employed.
These searches, especially direct stop searches, are important for the following two reasons.
First, the lighter stop is expected to be lighter due to the mixing, and thus be produced more, than the other squarks.
In addition, stop mass is crucial for the Higgs boson mass and the discussion of the little hierarchy.
Such searches are performed with requiring leptons and/or $b$-jets (cf.~Sec.~\ref{sec:atlas-object}), which are expected from the decay of top quarks.
However, we have observed no excess, and the stop mass is constrained as $m\s{\tilde t_1}\gtrsim500\GeV$ as long as the separation between the stop mass and the LSP mass is larger than $\sim 100\GeV$~\cite{ATLAS2012166,ATLAS2012167}.

These two LHC results indicate that {\em colored} sector of the MSSM is {\em heavy}, which is also supported by the $126\GeV$ Higgs mass.

Here one possibility arises: colored superparticles are heavy, but non-colored are light to keep the SUSY explanation of the muon $g-2$ anomaly.
The SUSY with such a scenario can be searched with focusing on pair-production of charginos, neutralinos, and sleptons via electroweak interactions, i.e., $pp\to\tilde\chi^0_2\tilde\chi^0_2, \tilde\chi^0_2\tilde\chi^+_1, \tilde\mu\tilde\mu^*$, etc.
Generally it is difficult to be searched for at the LHC, a hadron collider, because hard jets which are useful to distinguish signal events from background ones are not expected; generally multiple leptons are instead required in relevant searches.
Obtained bounds are $\sim300\text{--}500\GeV$ on masses of $\neut[2]$ and $\chgP[1]$, but these bounds depend to a large extent on the mass spectrum in non-colored sector, and thus further searches are required.
Here it should be noted that these searches do not need higher energy, but a higher luminosity.
As the instantaneous luminosity is to be increased and data of $\Order(100)\invfb$ are expected in the 13--14\,TeV runs of the LHC, the runs are of great importance for this scenario.

The following prognostication should also be emphasized: if the three forces are unified under the $\gSU(5)$-GUTs, gaugino masses $M_a$ are expected to be yielded from a common mechanism; then, if charginos and neutralinos have masses of order $100\GeV$, gluino is expected to have a mass of the same order.
For example, if they have the same masses $M_1=M_2=M_3$ at the GUT scale, their masses are expected to be
\begin{equation}
 M_1:M_2:M_3 \approx g_1^2:g_2^2:g_3^2 \approx 1:2:7
\end{equation}
at the low-energy scale.
Therefore, if masses of charginos and neutralinos are $\lesssim1\TeV$, the gluino mass is expected to be $\lesssim3.5\TeV$.
This possibility, i.e., charginos, neutralinos, and sleptons have $\Order(100)\GeV$, gaugino mass is a few TeV, and the squarks are too heavy to be produced at the LHC, is also expected to be focused on at the $14\TeV$ LHC.

\starline

However, from a theoretical viewpoint, this scenario with heavy-colored light-non-colored particles is not preferred, because it does not respect the $\gSU(5)$-GUTs.
The masses of the squarks and the sleptons should be close, because they are considered to have a common origin, i.e. to originate from the same multiplets, under the $\gSU(5)$-GUTs.
Actually, this $\gSU(5)$-GUTs hypothesis is fully respected in the simplest models for the SUSY-breaking, such as the CMSSM (constrained MSSM) scenario, which is sometimes called the mSUGRA (minimal supergravity), and the GMSB (gauge-mediated SUSY-breaking) scenario.

Unfortunately, experimental results do not support this anticipation.
Under the assumption that the masses of squarks and sleptons are at the same order, it is not so easy to realize the $126\GeV$ Higgs mass with explaining the muon $g-2$ anomaly.
Especially, it is known that the CMSSM scenario does not realize these two features of the SUSY~\cite{Ghilencea:2012gz,Endo:2011gy}, and the GMSB neither.
Nature seems to force us to leave this dream.

The GMSB scenario is, however, too beautiful to be abandoned; the SUSY $CP$- and flavor problems do not arise in this framework.
Therefore, we consider an extension of the MSSM: the V-MSSM, in which a vector-like pair of the decuplets of the $\gSU(5)$, $\TEN+\TENbar$, is appended as extra matters to the MSSM.
In the model, which respects the $\gSU(5)$-GUTs, the extra matters contribute to raise the Higgs boson mass, and therefore the squarks do not need to be so heavy as in the MSSM case.
Since the muon $g-2$ is scarcely affected by the extra matters, both features, the $126\GeV$ Higgs boson and the SUSY explanation of the muon $g-2$, can be simultaneously realized within the simplest SUSY-breaking frameworks, the GMSB and the CMSSM.

In the next chapter, we will discuss the V-MSSM.
Especially, we will examine the V-GMSB scenario, i.e.,  the V-MSSM with the GMSB framework, and to see that the $126\GeV$ mass and the explanation of the muon $g-2$ problem can be simultaneously realized, even under the current LHC constraints.
This is the main topic of this dissertation.

\subappendix
\section{Renormalization Group Equations for the MSSM}\label{sec:MSSM_RGE}
In Chapter~\ref{cha:vectorlike}, we will discuss an extension of the MSSM, called the V-MSSM, and introduce the renormalization group equations (RGEs) for the model.
Here, paying the cost of verbosity for completeness, the RGEs for the MSSM parameters are displayed up to two-loop level.

The $\beta$-functions are calculated with \withpackage[1.1]{Susyno}~\cite{Susyno} and checked by comparing with Refs.~\cite{Martin:1993zk,Yamada:1994id}.\footnote{%
The chosen scheme of Ref.~\cite{Martin:1993zk} was originally noted as the $\DRbar$ scheme, but actually was the $\DRbarPrime$ scheme, as is mentioned in {\em note added} in the paper or, e.g., Ref.~\cite{Martin:2002iu}.}

\subsection{Restriction and notation}
Here we simply employ the following assumptions, that is, we ignore $CP$- and flavor violations thoroughly.
\begin{itemize}
 \item The $R$-parity is conserved.
 \item The scalar soft mass terms $m^2_X$ are diagonal,
 \item For the $A$-terms $a_X$ and the \YUKAWA~coupling $Y_X$, all the components but the $(3,3)$ are neglected.
 \item The gaugino masses $M_a$ are real.
\end{itemize}

The $\beta$-function is defined as
\begin{equation}
 \diff{X(Q)}{\log Q} = \frac{1}{16\pi^2} \beta^{(1)}\left[X\right] + \frac{1}{\left(16\pi^2\right)^2} \beta^{(2)}\left[X\right],
\end{equation}
where $X$ are one of the MSSM parameters. Their definitions are found in Eqs.~\eqref{eq:MSSMsuperpotential} and \eqref{eq:MSSMsusybreaking}.
and $Q$ is the renormalization scale.
The $\overline{\rm DR}'$ scheme~\cite{Jack:1994rk} is chosen as the renormalization scheme.

The following variables are used in the expressions of the $\beta$-functions.
\begin{align*}
 X_t    &:= 2a_t^2    + 2Y_t^2    \left(m^2_{\Hu}+(m^2_{Q})_{33}+(m^2_{\bU})_{33}\right),\\
 X_b    &:= 2a_b^2    + 2Y_b^2    \left(m^2_{\Hd}+(m^2_{Q})_{33}+(m^2_{\bD})_{33}\right),\\
 X_\tau &:= 2a_\tau^2 + 2Y_\tau^2 \left(m^2_{\Hd}+(m^2_{L})_{33}+(m^2_{\bE})_{33}\right),\\
\tilde a_{(t,b,\tau)} &:= Y_{(t,b,\tau)}a_{(t,b,\tau)},\\
 S      &:= m^2_{\Hu}-m^2_{\Hd} +\sum_{i=1}^3\left[ (m^2_{Q})_{ii} -2 (m^2_{\bU})_{ii}+ (m^2_{\bD})_{ii}- (m^2_{L})_{ii}+ (m^2_{\bE})_{ii}\right],\\
 S^{(2)}&:=
-Y_t^2\left(3m^2_{\Hu}+(m^2_Q)_{33}-4(m^2_{\bU})_{33}\right)
+Y_b^2\left(3m^2_{\Hd}-(m^2_Q)_{33}-2(m^2_{\bD})_{33}\right)
+Y_\tau^2\left(m^2_{\Hd}+(m^2_L)_{33}-2(m^2_{\bE})_{33}\right)
 \\&\qquad
+\left(\frac3{10}{g_1^2}+\frac32g_2^2\right)\left(m^2_{\Hu}-m^2_{\Hd}-\sum_{i=1}^3 (m^2_{L})_{ii}\right)
+\left(\frac1{30}g_1^2+\frac32g_2^2+\frac83g_3^2\right)
\sum_{i=1}^3 (m^2_{Q})_{ii}
 \\&\qquad
-\left(\frac{16}{15}g_1^2+\frac{16}3g_3^2\right)\sum_{i=1}^3 (m^2_{\bU})_{ii}
+\left(\frac{2}{15}g_1^2+\frac83g_3^2\right)\sum_{i=1}^3 (m^2_{\bD})_{ii}
+\frac65g_1^2\sum_{i=1}^3 (m^2_{\bE})_{ii},\\
\sigma_1&:=\frac15 g_1^2\left[ 3m^2_{\Hu}+3m^2_{\Hd}+\sum_{i=1}^3\left((m^2_Q)_{ii}+8(m^2_{\bU})_{ii}+2(m^2_{\bD})_{ii}+3(m^2_L)_{ii}+6(m^2_{\bE})_{ii}\right)\right],\\
\sigma_2&:=g_2^2\left[m^2_{\Hu}+m^2_{\Hd}+\sum_{i=1}^3\left((3m^2_Q)_{ii}+(m^2_L)_{ii}\right)\right],\\
\sigma_3&:=g_3^2\sum_{i=1}^3\left(2(m^2_Q)_{ii}+(m^2_{\bU})_{ii}+(m^2_{\bD})_{ii}\right).
\end{align*}
\allowdisplaybreaks[4]
\subsection[One-loop level \texorpdfstring{$\beta$}{beta}-functions]{One-loop level $\boldsymbol \beta$-functions}

\begin{align}
 \begin{split}\label{eq:MSSM_RGE_1_begin}
\beta^{(1)}\left[g_a\right] &= B_a^{(1)}g_a^3\\
\beta^{(1)}\left[M_a\right] &= 2B_a^{(1)} g_a^2M_a
 \end{split}
\qquad \where \left(B_{1}^{(1)},B_{2}^{(1)},B_{3}^{(1)}\right)=\left(\frac{33}5,1,-3\right)\\
\beta^{(1)}\left[Y_{t}\right] &=  \left(Y_{b}^2-\frac{13}{15} g_1^2-3g_2^2-\frac{16}{3} g_3^2+6 Y_{t}^2\right)Y_t\\
\beta^{(1)}\left[Y_{b}\right] &=  \left(6 Y_{b}^2-\frac{7}{15} g_1^2-3g_2^2-\frac{16}{3} g_3^2+Y_{t}^2+Y_{\tau}^2\right)Y_b\\
\beta^{(1)}\left[Y_{\tau}\right] &= \left(3 Y_{b}^2-\frac{9}{5} g_1^2-3g_2^2+4 Y_{\tau}^2\right)Y_\tau\\
\beta^{(1)}\left[a_{t}\right] &=\left(18 Y_{t}^2+Y_{b}^2-\frac{13}{15} g_1^2-3 g_2^2-\frac{16}{3} g_3^2\right)a_t
+\left(2 \tilde a_{b}+\frac{26}{15}g_1^2 M_1+6 g_2^2 M_2+\frac{32}{3} g_3^2 M_3\right) Y_{t}
\\
 \begin{split}
\beta^{(1)}\left[a_{b}\right] &=\left(Y_{t}^2+18 Y_{b}^2+ Y_{\tau}^2-\frac{7}{15} g_1^2-3 g_2^2 -\frac{16}{3}g_3^2\right) a_{b}\\
&\qquad+\left(2 \tilde a_{t}+2 \tilde a_{\tau}+\frac{14}{15} g_1^2 M_1+6g_2^2 M_2+\frac{32}{3} g_3^2 M_3 \right)Y_{b}
 \end{split}
\\
\beta^{(1)}\left[a_{\tau}\right] &=\left(3 Y_{b}^2+12 Y_{\tau}^2-\frac{9}{5} g_1^2-3 g_2^2\right) a_{\tau}
+\left(6 \tilde a_{b}+\frac{18}{5} g_1^2 M_1+6 g_2^2 M_2\right) Y_{\tau}
\\
\beta^{(1)}\left[\mu\right] &= \left(3 Y_{b}^2-\frac{3}{5} g_1^2-3 g_2^2+3Y_{t}^2+Y_{\tau}^2\right)\mu\\
\beta^{(1)}\left[b\right] &= \left(6 \tilde a_{t} +6 \tilde a_{b} +2 \tilde a_{\tau}+\frac{6}{5} g_1^2 M_1+6 g_2^2 M_2\right)\mu+ \left(3 Y_{t}^2+3 Y_{b}^2+Y_{\tau}^2-\frac{3}{5}g_1^2-3 g_2^2\right)b\\
\beta^{(1)}\left[m_{\Hu}^2\right] &= -\frac{6}{5} g_1^2 M_1^2-6 g_2^2 M_2^2+\frac{3}{5}g_1^2 S+3 X_t\\
\beta^{(1)}\left[m_{\Hd}^2\right] &= -\frac{6}{5} g_1^2 M_1^2-6 g_2^2M_2^2-\frac{3}{5} g_1^2 S+3X_b+X_{\tau}\\
\beta^{(1)}\left[(m_{Q}^2){}_{ii}\right] &= -\frac{2}{15} g_1^2 M_1^2-6 g_2^2M_2^2-\frac{32}{3} g_3^2 M_3^2+\frac{1}{5}g_1^2 S+
\oneloopforthree{X_t+X_b}\\
\beta^{(1)}\left[(m_{\bU}^2){}_{ii}\right] &= -\frac{32}{15} g_1^2 M_1^2-\frac{32}{3}g_3^2 M_3^2-\frac{4}{5} g_1^2 S+
\oneloopforthree{2X_t}\\
\beta^{(1)}\left[(m_{\bD}^2){}_{ii}\right] &= -\frac{8}{15} g_1^2M_1^2-\frac{32}{3} g_3^2 M_3^2+\frac{2}{5} g_1^2 S+
\oneloopforthree{2X_b}\\
\beta^{(1)}\left[(m_{L}^2){}_{ii}\right] &= -\frac{6}{5} g_1^2 M_1^2-6 g_2^2M_2^2-\frac{3}{5} g_1^2 S+
\oneloopforthree{X_\tau}\\
\beta^{(1)}\left[(m_{\bE}^2){}_{ii}\right] &= -\frac{24}{5} g_1^2 M_1^2+\frac{6}{5} g_1^2S+
\oneloopforthree{2X_\tau}\\
\label{eq:MSSM_RGE_1_end}
\end{align}


\subsection[Two-loop level \texorpdfstring{$\beta$}{beta}-functions]{Two-loop level $\boldsymbol \beta$-functions}

\begin{align}
\label{eq:MSSM_RGE_2_begin}
\beta^{(2)}\left[g_1\right] &= \left(-\frac{26}{5} Y_t^2-\frac{14}{5}Y_b^2-\frac{18}{5} Y_\tau^2+\frac{199}{25} g_1^2+\frac{27}{5}g_2^2+\frac{88}{5} g_3^2\right)g_1^3\\
\beta^{(2)}\left[g_2\right] &= \left(-6 Y_t^2-6 Y_b^2-2 Y_\tau^2+\frac{9}{5} g_1^2+25 g_2^2+24 g_3^2\right)g_2^3\\
\beta^{(2)}\left[g_3\right] &=  \left(-4 Y_t^2-4 Y_b^2+\frac{11}{5} g_1^2+9g_2^2+14 g_3^2\right)g_3^3\\
 \begin{split}
\beta^{(2)}\left[M_1\right] &= 
\frac15\Bigg[52 \left(\tilde a_t-Y_t^2M_1\right)+{28} \left(\tilde a_b-Y_b^2M_1\right)+{36} \left(\tilde a_\tau-Y_\tau^2 M_1\right)
\\&\hspace{100pt}
+\frac{796}{5} g_1^2M_1 +{54} g_2^2\left(M_1+M_2\right)+{176}g_3^2\left(M_1+M_3\right)\Bigg]
g_1^2 \end{split}\\
 \begin{split}
\beta^{(2)}\left[M_2\right] &= \Bigg[12 \left(\tilde a_t- Y_t^2M_2\right)+12 \left(\tilde a_b- Y_b^2M_2\right)+4 \left(\tilde a_\tau-Y_\tau^2 M_2\right)
\\&\hspace{100pt}
+\frac{18}{5} g_1^2 \left(M_1+M_2\right)+100 g_2^2M_2+48 g_3^2 \left(M_2+M_3\right)\Bigg]
g_2^2 \end{split}\\
 \beta^{(2)}\left[M_3\right] &= \Bigg[8 \left(\tilde a_t-Y_t^2M_3\right)+8 \left(\tilde a_b-Y_b^2M_3\right)+\frac{22}{5} g_1^2 \left(M_1+M_3\right)+18 g_2^2\left(M_2+M_3\right) +56 g_3^2 M_3\Bigg]g_3^2\\
\begin{split}
 \beta^{(2)}\left[Y_t\right] &= \Bigg[
\left(\frac{6}{5}Y_t^2+\frac{2}{5}Y_b^2 \right)g_1^2
+6 Y_t^2g_2^2+16Y_t^2 g_3^2-22Y_t^4 -5 Y_b^4-5Y_t^2Y_b^2-Y_b^2Y_\tau ^2
 \\&\hspace{100pt}
 +\frac{2743}{450} g_1^4+\frac{15}{2}g_2^4-\frac{16}{9} g_3^4+g_1^2 g_2^2+\frac{136}{45} g_1^2 g_3^2+8 g_2^2g_3^2\Bigg]Y_t
\end{split}\\
\begin{split}
 \beta^{(2)}\left[Y_b\right] &= \Bigg[
 \left(\frac{4}{5}  Y_t^2-\frac{2}{5} Y_b^2+\frac{6}{5} Y_\tau ^2\right)g_1^2
+6 Y_b^2g_2^2+16 Y_b^2g_3^2-5Y_t^4-22Y_b^4-5 Y_t^2Y_b^2-3Y_b^2 Y_\tau ^2-3 Y_\tau ^4
 \\&\hspace{100pt}
+\frac{287}{90} g_1^4+\frac{15}{2} g_2^4-\frac{16}{9} g_3^4+g_1^2g_2^2+\frac{8}{9} g_1^2 g_3^2+8 g_2^2 g_3^2\Bigg]Y_b
\end{split}\\
\begin{split}
 \beta^{(2)}\left[Y_\tau \right] &= \Bigg[
 \left(-\frac{2}{5}Y_b^2+\frac{6}{5} Y_\tau^2\right)g_1^2
 +6Y_\tau^2g_2^2 +16g_3^2Y_b^2 
 \\&\hspace{100pt}
-9 Y_b^4-10Y_\tau^4 -3 Y_t^2Y_b^2-9 Y_b^2Y_\tau^2
 +\frac{27}{2} g_1^4+\frac{15}{2} g_2^4+\frac{9}{5} g_1^2g_2^2\Bigg]Y_\tau
\end{split}\\
\begin{split}
 \beta^{(2)}\left[a_t\right] &=
 \frac{2743}{450}g_1^4\left( a_t-4Y_tM_1\right)+\frac{15}{2}g_2^4\left(a_t-4Y_t M_2\right)+\frac{16}{9}g_3^4 \left(4Y_tM_3- a_t\right)
 \\&\hspace{20pt}
 +g_1^2g_2^2\left(a_t-2 Y_tM_1-2 Y_tM_2\right)
 +\frac{136}{45}g_1^2g_3^2 \left(a_t-2 Y_tM_1-2Y_tM_3\right)
 \\&\hspace{20pt}
 +8g_2^2g_3^2 \left(a_t-2 Y_tM_2-2Y_tM_3\right)
 +g_1^2\left(\frac{2}{5} a_t Y_b^2+\frac{4}{5} Y_t \tilde{a}_b+\frac{18}{5} \tilde{a}_t Y_t-\frac{4}{5} Y_tY_b^2M_1-\frac{12}{5} Y_t^3M_1\right)
 \\&\hspace{20pt}
 +g_2^2 \left(-12 M_2 Y_t^3+18 \tilde{a}_t Y_t\right)
 +g_3^2 \left(48 \tilde{a}_t Y_t-32M_3 Y_t^3\right)
 \\&\hspace{20pt}
-110 \tilde{a}_t Y_t^3 -a_t Y_b^2 Y_\tau ^2-2 Y_t Y_b^2\tilde{a}_\tau -2 Y_t \tilde{a}_b Y_\tau ^2
 -5 a_tY_b^4-15 \tilde{a}_t Y_t Y_b^2-20 Y_t \tilde{a}_b Y_b^2-10Y_t^3 \tilde{a}_b
\end{split}\\
\begin{split}
\beta^{(2)}\left[a_b\right]&=
\frac{287}{90}g_1^4 \left( a_b-4 Y_bM_1\right)
+\frac{15}{2}g_2^4 \left( a_b-4 Y_bM_2\right)
-\frac{16}{9}g_3^4 \left(a_b-4 Y_bM_3\right)
 \\&\hspace{20pt}
+g_1^2g_2^2 \left(a_b-2 M_1 Y_b-2 M_2Y_b\right)
+\frac89g_1^2g_3^2 \left(a_b-2Y_b M_1-2Y_bM_3\right)
 \\&\hspace{20pt}
+ g_2^2g_3^2 \left(8a_b-16 M_2 Y_b-16 M_3 Y_b\right)
 \\&\hspace{20pt}
+g_1^2 \left[\frac{4}{5} Y_t^2 a_b-\frac{8}{5}M_1 Y_t^2 Y_b+\frac{8}{5} \tilde{a}_t Y_b+\frac{6}{5} a_bY_\tau ^2-\frac{12}{5} M_1 Y_b Y_\tau ^2+\frac{12}{5} Y_b\tilde{a}_\tau
-\frac{4}{5} M_1 Y_b^3+\frac{6}{5}\tilde{a}_b Y_b
\right]
 \\&\hspace{20pt}
+g_2^2\left(18 \tilde{a}_bY_b-12 M_2 Y_b^3\right)
 +g_3^2 \left(48 \tilde{a}_b Y_b-32 M_3Y_b^3\right)-110 \tilde{a}_b Y_b^3
 \\&\hspace{20pt}
-5 Y_t^4 a_b-20 \tilde{a}_t Y_t^2Y_b-15 Y_t^2 \tilde{a}_b Y_b-10 \tilde{a}_t Y_b^3-3 a_bY_\tau ^4-9 \tilde{a}_b Y_b Y_\tau ^2-6 Y_b^3 \tilde{a}_\tau -12 Y_b \tilde{a}_\tau  Y_\tau ^2
\end{split}\\
\begin{split}
\beta^{(2)}\left[a_\tau \right] &=
\frac{27}{2}g_1^4 \left(a_\tau-4M_1 Y_\tau \right)+\frac{15}{2}g_2^4 \left(a_\tau-4 M_2Y_\tau \right)
+\frac95 g_1^2g_2^2\left(a_\tau-2 Y_\tau M_1-2Y_\tau M_2\right)
\\&\hspace{20pt}
+g_1^2 \left(-\frac{2}{5} Y_b^2 a_\tau +\frac{4}{5} M_1Y_b^2 Y_\tau -\frac{4}{5} \tilde{a}_b Y_\tau-\frac{12}{5} M_1 Y_\tau ^3+\frac{18}{5}\tilde{a}_\tau  Y_\tau\right)
\\&\hspace{20pt}
+g_2^2 \left(18 \tilde{a}_\tau  Y_\tau -12 M_2Y_\tau ^3\right)
+g_3^2 \left(16Y_b^2 a_\tau -32 M_3 Y_b^2 Y_\tau +32 \tilde{a}_b Y_\tau \right) -3 Y_t^2 Y_b^2 a_\tau 
\\&\hspace{20pt}
-6 \tilde{a}_tY_b^2 Y_\tau-6 Y_t^2 \tilde{a}_b Y_\tau
-9 Y_b^4 a_\tau -36\tilde{a}_b Y_b^2 Y_\tau -27 Y_b^2 \tilde{a}_\tau  Y_\tau -18 \tilde{a}_b Y_\tau ^3
-50 \tilde{a}_\tau  Y_\tau ^3
\end{split}\\
\begin{split}
 \beta^{(2)}\left[\mu\right] &= \Bigg[
\left(\frac45Y_t^2-\frac25Y_b^2+\frac65Y_\tau^2\right)g_1^2
+16\left(Y_t^2+Y_b^2\right)g_3^2
 \\&\hspace{100pt}
-9Y_t^4-9Y_b^4-3Y_\tau^4-6Y_t^2Y_b^2
+\frac{207}{50} g_1^4+\frac{15}{2} g_2^4+\frac{9}{5} g_1^2g_2^2
 \Bigg]\mu
\end{split}\\
\begin{split}
 \beta^{(2)}\left[b\right] &= \Bigg[
 \frac85 \left(\tilde{a}_t-Y_t^2M_1\right)g_1^2-\frac45 \left(\tilde{a}_b-Y_b^2M_1\right)g_1^2+\frac{12}5 \left(\tilde{a}_\tau-Y_\tau^2M_1\right)g_1^2
 \\&\hspace{20pt}
+32\left(\tilde{a}_t-M_3 Y_t^2+\tilde{a}_b-M_3 Y_b^2\right)g_3^2
-12\left(
3\tilde{a}_t Y_t^2+3\tilde{a}_b Y_b^2+ \tilde{a}_\tau  Y_\tau ^2+ \tilde{a}_t Y_b^2+ Y_t^2 \tilde{a}_b
\right)
 \\&\hspace{20pt}
-\frac{414}{25} g_1^4 M_1-30 g_2^4 M_2-\frac{18}5g_1^2g_2^2 \left(M_1+M_2\right)
\Bigg]\mu
 \\&\quad
+\Bigg[
\left(\frac45Y_t^2-\frac25Y_b^2+\frac65Y_\tau^2\right)g_1^2
+16\left(Y_t^2+Y_b^2\right)g_3^2
 \\&\hspace{100pt}
-9Y_t^4-9Y_b^4-3Y_\tau^4-6Y_t^2Y_b^2
+\frac{207}{50} g_1^4+\frac{15}{2} g_2^4+\frac{9}{5} g_1^2g_2^2
\Bigg]b
 \end{split}\\
\begin{split}
 \beta^{(2)}\left[m_{\Hu}^2\right] &=
 \frac{621}{25}g_1^4 M_1^2+33 g_2^4 M_2^2
 +\frac{18}{5}g_1^2g_2^2\left(M_1^2+M_2^2+ M_1M_2\right)
+\frac{6}{5}g_1^2 {{S^{(2)}}} +\frac{3}{5}g_1^2 \sigma _1+3 g_2^2 \sigma _2
 \\&\hspace{50pt}
 +\frac45g_1^2\left(X_t-4 M_1 \tilde{a}_t+4 M_1^2 Y_t^2\right)
 +16g_3^2 \left(X_t-4 M_3 \tilde{a}_t+4 M_3^2 Y_t^2\right)
 \\&\hspace{50pt}
 -18 Y_t^2 X_t-36 \tilde{a}_t^2-3 Y_b^2 X_t
 -12 \tilde{a}_t \tilde{a}_b-3 Y_t^2 X_b
 \end{split}\\
\begin{split}
 \beta^{(2)}\left[m_{\Hd}^2\right] &= 
\frac{621}{25} g_1^4 M_1^2+33 g_2^4 M_2^2
 +\frac{18}{5}g_1^2g_2^2\left(M_1^2+M_2^2+ M_1M_2\right)
-\frac{6}{5}g_1^2{{S^{(2)}}}+\frac{3}{5}g_1^2 \sigma _1+3g_2^2 \sigma _2
 \\&\hspace{50pt}
+\frac15{g_1^2}\left(8M_1 \tilde{a}_b-8 M_1^2 Y_b^2-2X_b-24M_1\tilde{a}_\tau+24M_1^2 Y_\tau ^2+6X_{\tau}\right)
 \\&\hspace{50pt}
+16g_3^2 \left(X_b-4 M_3 \tilde{a}_b+4 M_3^2 Y_b^2\right)
 \\&\hspace{50pt}
-12 \tilde{a}_t \tilde{a}_b-3 Y_t^2 X_b
-18 Y_b^2 X_b-3 Y_b^2 X_t-36 \tilde{a}_b^2-12 \tilde{a}_\tau ^2-6 Y_\tau ^2 X_{\tau}
\end{split}\\
\begin{split}
\beta^{(2)}\left[(m_{Q}^2){}_{ii}\right] &=
\frac{199}{75} g_1^4 M_1^2+33 g_2^4M_2^2-\frac{128}{3} g_3^4 M_3^2
+\frac25g_1^2g_2^2 \left(M_1^2+M_1 M_2+M_2^2\right)
 \\&\hspace{50pt}
+\frac{32}{45}g_1^2g_3^2 \left(M_1^2+M_3^2+M_1M_3\right)
+32g_2^2g_3^2 \left(M_2^2+M_3^2+M_2M_3\right)
 \\&\hspace{50pt}
+\frac{2}{5}g_1^2 {S^{(2)}}+\frac{1}{15}g_1^2\sigma_1+3 g_2^2\sigma_2+\frac{16}{3} g_3^2 \sigma _3
\\&\hspace{20pt}+\Bigg\langle\!\Bigg\langle
\frac15g_1^2 \left(4X_t-16 M_1\tilde{a}_t+16 M_1^2 Y_t^2+2X_b-8 M_1\tilde{a}_b+8 M_1^2 Y_b^2\right)
\\&\hspace{50pt}
-10 Y_t^2 X_t-20 \tilde{a}_t^2 -10 Y_b^2 X_b-20 \tilde{a}_b^2-Y_b^2 X_{\tau}-Y_\tau ^2X_b-4 \tilde{a}_b \tilde{a}_\tau\Bigg\rangle\!\Bigg\rangle\stx{for $i=3$}
\end{split}\\
\begin{split}
\beta^{(2)}\left[(m_{\bU}^2){}_{ii}\right] &=
\frac{3424}{75} g_1^4M_1^2-\frac{128}{3} g_3^4 M_3^2
+\frac{512}{45}g_1^2g_3^2 \left(M_1^2+ M_3^2+ M_1 M_3\right)
-\frac{8}{5}g_1^2{{S^{(2)}}}+\frac{16}{15}g_1^2\sigma _1+\frac{16}{3} g_3^2 \sigma _3
\\&\hspace{20pt}+\Bigg\langle\!\Bigg\langle
g_1^2 \left(-\frac{2}{5} X_t+\frac{8}{5} M_1 \tilde{a}_t-\frac{8}{5} M_1^2Y_t^2
\right)+g_2^2 \left(6 X_t-24 M_2 \tilde{a}_t+24 M_2^2Y_t^2\right)
\\&\hspace{120pt}
-16 Y_t^2 X_t-32 \tilde{a}_t^2-2 Y_b^2X_t
-8 \tilde{a}_t \tilde{a}_b-2Y_t^2 X_b
\Bigg\rangle\!\Bigg\rangle\stx{for $i=3$}
\end{split}\\
\begin{split}
\beta^{(2)}\left[(m_{\bD}^2){}_{ii}\right] &=
\frac{808}{75} g_1^4 M_1^2-\frac{128}{3} g_3^4 M_3^2
+\frac{128}{45}g_1^2g_3^2 \left(M_1^2+ M_3^2+ M_1 M_3\right)
+\frac{4}{5} g_1^2 {S^{(2)}}+\frac{4}{15} g_1^2 \sigma_1+\frac{16}{3} g_3^2 \sigma _3
\\&\hspace{20pt}+\Bigg\langle\!\Bigg\langle
g_1^2 \left(\frac{2}{5} X_b-\frac{8}{5} M_1\tilde{a}_b+\frac{8}{5} M_1^2 Y_b^2\right)
+g_2^2\left(6 X_b-24 M_2 \tilde{a}_b+24 M_2^2 Y_b^2\right)
\\&\hspace{50pt}
-8 \tilde{a}_b \tilde{a}_\tau-16 Y_b^2 X_b-32 \tilde{a}_b^2
 -8 \tilde{a}_t \tilde{a}_b-2Y_t^2 X_b-2Y_b^2 X_t
-2 Y_b^2 X_{\tau}-2 Y_\tau ^2X_b\Bigg\rangle\!\Bigg\rangle\stx{for $i=3$}
\end{split}\\
\begin{split}
\beta^{(2)}\left[(m_{L}^2){}_{ii}\right] &=
\frac{621}{25} g_1^4 M_1^2+33 g_2^4 M_2^2
+\frac{18}{5}g_1^2 g_2^2\left(M_1^2+M_2^2+M_1M_2\right)
-\frac{6}{5}g_1^2{S^{(2)}}+\frac{3}{5} g_1^2 \sigma _1+3 g_2^2 \sigma _2
\\&\hspace{20pt}+\Bigg\langle\!\Bigg\langle
g_1^2 \left(\frac{6}{5}X_{\tau}-\frac{24}{5} M_1\tilde{a}_\tau+\frac{24}{5} M_1^2 Y_\tau ^2\right)
-12 \tilde{a}_b \tilde{a}_\tau -3 Y_b^2 X_{\tau}-3 Y_\tau ^2 X_b
-12 \tilde{a}_\tau ^2-6 Y_\tau ^2X_{\tau}\Bigg\rangle\!\Bigg\rangle\stx{for $i=3$}
\end{split}\\
\begin{split}
\beta^{(2)}\left[(m_{\bE}^2){}_{ii}\right] &= \frac{2808}{25} g_1^4 M_1^2+\frac{12}{5} g_1^2 {S^{(2)}}+\frac{12}{5}g_1^2 \sigma _1
\\&\hspace{20pt}+\Bigg\langle\!\Bigg\langle
g_1^2 \left(-\frac{6}{5}X_{\tau}+\frac{24}{5} M_1\tilde{a}_\tau-\frac{24}{5} M_1^2 Y_\tau ^2\right)
+g_2^2 \left(6X_\tau-24 M_2 \tilde{a}_\tau +24 M_2^2 Y_\tau ^2\right)
\\&\hspace{100pt}
 -24 \tilde{a}_b \tilde{a}_\tau -6 Y_b^2 X_{\tau}-6 Y_\tau ^2 X_b+
-16 \tilde{a}_\tau ^2-8 Y_\tau ^2X_{\tau}\Bigg\rangle\!\Bigg\rangle\stx{for $i=3$}
\end{split}\label{eq:MSSM_RGE_2_end}
\end{align}


\allowdisplaybreaks[0]

\newpage
\section{Note on Figures Showing Gauge Coupling Evolution}
\label{app:rge_figures}
Here are documented several remarks on the figures displaying the renormalization group running of coupling constants, i.e., Fig.~\ref{fig:SM_gcu} and Fig~\ref{fig:MSSM_gcu}.

\paragraph{Evaluation for the Standard Model (Fig.~\ref{fig:SM_gcu})}

The gauge coupling running under the Standard Model is employed in the $\MSbar$ scheme~\cite{Bardeen:1978yd}.
The RGEs for the gauge couplings $g_a$ are evaluated at the two-loop level.
The \YUKAWA\ couplings appearing in the RGEs are run under the one-loop level RGEs, which is sufficient to obtain the gauge couplings at the two-loop level accuracy.

Input for the coupling constants are evaluated at the electroweak scale, i.e. at the $Z$-boson mass $m_Z$. The values are~\cite{PDG2012}
\begin{align}
 \alpha_s(m_Z)^{\MSbar}&=0.1184,&
 \sin^2{\theta\s w}(m_Z)^{\MSbar}&=0.2312,&
 \alpha\s{EW}(m_Z)^{\MSbar}&=127.9,
\end{align}
which results in $g_1 = 0.462$, $g_2=0.652$ and  $g_3 = 1.22$. Note that $g_1=\sqrt{5/3}g_Y$.

For precise calculation the \YUKAWA\ couplings should be also evaluated at $m_Z$ with the $\MSbar$ scheme.
However, in drawing Fig.~\ref{fig:SM_gcu}, the constants are evaluated as
\begin{align}
  Y_t    &\approx \frac{m_t(m_t)^{\MSbar}}{v}  \approx\frac{160}{174},&
  Y_b    &\approx \frac{m_b(m_b)^{\MSbar}}{v}  \approx\frac{4.18}{174},&
  Y_\tau &\approx \frac{m_\tau\suprm{pole}}{v} \approx\frac{1.77682}{174}.
\end{align}

Therefore, precisely speaking, the running of Fig.~\ref{fig:SM_gcu} should be understood as a rough estimate; but this simplification would be compensated with the large uncertainty of the $\MSbar$ top mass and the ineffectiveness of $Y_b$ and $Y_\tau$ in the evaluation.

\paragraph{Evaluation for the MSSM (Fig.~\ref{fig:MSSM_gcu})}
As we are not interested in the precise values, several approximations for simplicity are employed in numerical evaluation for Fig.~\ref{fig:MSSM_gcu}.
First the procedure is summarized, followed by the discussion on the approximations.

The ratio $\tan\beta:=\vu/\vd$ and a SUSY scale $M\s{SUSY}$ are set as input values.
The running below $M\s{SUSY}$ is employed with the Standard Model $\MSbar$ RGEs, and that above $M\s{SUSY}$ is with the MSSM $\DRbar$ RGEs.
The exchange at the SUSY scale is employed as
\begin{align}
  g_a\suprm{MSSM}    &= g_a\suprm{SM},&
  Y_t\suprm{MSSM}    &= y_t\suprm{SM}/\sin\beta,&
  Y_b\suprm{MSSM}    &= y_b\suprm{SM}/\cos\beta,&
  Y_\tau\suprm{MSSM} &= y_\tau\suprm{SM}/\cos\beta.
\label{eq:RGEconversion}
\end{align}
Input values at the electroweak scale are taken as the same as the Standard Model case discussed above.

One should first note that the renormalization schemes are different above and below the SUSY scale.
Nevertheless, the RGEs in the two schemes are the same at the one-loop level expression, and thus it is expected that this approximation does not cause serious problem.

More important simplification is that no threshold corrections at $M\s{SUSY}$ are introduced.
Precisely speaking, the matters decouple gradually during the running down of the mass scale $Q$, and thus the RGEs, or the coefficients in the RGEs, depend on the scale $Q$.
However, in usual calculations the RGEs are treated as if they are independent of $Q$, and instead so-called ``threshold corrections'' are introduced at the SUSY scale to compensate the effect from the matter decoupling.
Therefore, ideally, we have to introduce the threshold corrections into the conversion of Eq.~\eqref{eq:RGEconversion}; but as they depend on the SUSY mass spectrum, they are simply not introduced in the numerical evaluation for Fig.~\ref{fig:MSSM_gcu}.

Consequently, the running of Fig.~\ref{fig:MSSM_gcu} should be understood as a rough illustration of the gauge coupling unification.

\starline

Nevertheless, the numerical evaluations in Chapter~\ref{cha:vectorlike}, where we utilize~\withpackage[3.3]{SOFTSUSY}~\cite{SOFTSUSY} to calculate the mass spectrum, the above subtle effects are introduced; i.e. the approximations discussed here are not performed.

\subappendixend


\chapter{The MSSM with Vector-like Matters}
\label{cha:vectorlike}

Now we are ready to discuss the ``V-MSSM,'' an extension of the MSSM with a vector-like pair of supermultiplets.
This model~\cite{Moroi:1991mg} is characterized as\vspace{-5pt}
\begin{equation}
 \text{V-MSSM} = \text{MSSM} + \left(\TEN+\TENbar\right),\vspace{-5pt}
\end{equation}
where $\TEN$ and $\TENbar$ are the decuplet and the anti-decuplet of the $\gSU(5)$ gauge group.

In this model the Higgs mass can be raised by the extra vector-like quarks~\cite{Moroi:1991mg}.
There very large squark masses of $\sim10\TeV$ are not required to realize the $126\GeV$ Higgs boson even with a small mixing parameter $X_t$ as we will see in Sec.~\ref{sec:mssm-h-discussion}.
This feature further allows us to utilize the SUSY as a solution to the muon $g-2$ anomaly even in the simplest SUSY-breaking frameworks.
One should also note that this model respects the underlying $\gSU(5)$ symmetry.

As the gauge-mediated SUSY-breaking (GMSB) scenario is very promising for its freedom from the SUSY $CP$- and flavor problems, the V-MSSM with the GMSB is particularly interesting.
In the GMSB scenario, the slepton soft masses have the same origin as those of the squarks, and the $A$-terms are much smaller than the soft mass terms.
Therefore, under the MSSM framework, the GMSB cannot realize the $126\GeV$ Higgs mass with holding the SUSY explanation of the muon $g-2$ anomaly.
Here, the ``V'' resurrects the GMSB scenario.
As the extra matters raise the Higgs mass, the mass of $126\GeV$ is realized without exploiting heavy squark masses of order 1--$10\TeV$.
It allows the lighter masses of the sleptons and gauginos even under the GMSB framework, and with those masses the muon $g-2$ discrepancy can be explained.

In chis chapter we will examine the V-MSSM, especially focusing on the combination with the simple GMSB scenario, which we call ``V-GMSB model.''

Historically, the V-MSSM was introduced in Ref.~\cite{Moroi:1991mg}, where the increase of the Higgs boson mass was unveiled.
In Ref.~\cite{Martin:2009bg} the model was examined; especially discussed were the infrared fixed point behaviour of the couplings and decays of the extra fermions, which will in this dissertation be explained in Sec.~\ref{sec:vmssm-model} and Sec.~\ref{sec:vector} respectively.

Then author's works follow.\footnote{
This chapter is based on the works by Author, completed in collaboration with Dr.~M.~Endo, Prof.~K.~Hamaguchi, Mr.~K.~Ishikawa and Dr.~N.~Yokozaki~\cite{Endo:2011mc,Endo:2011xq,Endo:2012rd,Endo:2012cc}; a conference article by Author was published in proceedings~\cite{Iwamoto:2012hh}.
}
In Ref.~\cite{Endo:2011mc} the V-MSSM was firstly combined with the muon $g-2$ problem. In Ref.~\cite{Endo:2011xq} the LHC phenomenology is discussed, and Ref.~\cite{Endo:2012rd} was devoted to the vacuum stability condition, which we will see in Sec.~\ref{sec:vac}.
Ref.~\cite{Endo:2012cc} was on constraints from SUSY searches at the LHC, which corresponds to Sec.~\ref{sec:vmssm-lhc} of this dissertation.

\starline

In summary, notable features of this model are:
\begin{itemize}
 \item It respects the underlying $\gSU(5)$.
 \item It goes perturbatively up to the GUT scale.
 \item It allows us to solve the $(g-2)_\mu$ anomaly even with realizing $m_h\simeq 126\GeV$.
\end{itemize}
After defining the model, we will see these advantages, and then discuss current constraints on this model.

\begin{table}[p!]
\catcode`?=\active \def?{\phantom{-}}
\begin{center}
\caption[The field content of the V-MSSM.]{The field content of the V-MSSM. The gauge indices are omitted, and subscripts $i$ denote indices for the three generations running from one to three. The gauge group is the same as the MSSM, and thus the gauge fields are the same.}

\renewcommand{\arraystretch}{1.2}
\vspace{10pt}

 \begin{tabular}[t]{@{\Vrule\ }c|ccc|c@{\ \Vrule}}\Hrule
\multicolumn{5}{@{\Vrule\ }c@{\ \Vrule}}{{\bf Matter fields} (chiral multiplets)}\\\Hrule
 field   & $\gSU(3)\s{color}$& $\gSU(2)\s{weak}$& $\gU(1)_Y$& $R$-parity\\\hline
 $Q_i$               & $\THREE$    & $\TWO$ & $?1/6$ & $-$ \\\hline
 $\bU_i$             & $\THREEbar$ & $\ONE$ & $-2/3$ & $-$ \\\hline
 $\bE_i$             & $\ONE$      & $\ONE$ & $?1$   & $-$ \\\hline
 $\bD_i$             & $\THREEbar$ & $\ONE$ & $?1/3$ & $-$\\\hline
 $L_i$               & $\ONE$      & $\TWO$ & $-1/2$ & $-$\\\hline
 $\Hu$               & $\ONE$      & $\TWO$ & $?1/2$ & $+$\\\hline
 $\Hd$               & $\ONE$      & $\TWO$ & $-1/2$ & $+$\\\Hrule
 $Q'$                & $\THREE$    & $\TWO$ & $?1/6$ & $-$\\\hline
 $\bU'$              & $\THREEbar$ & $\ONE$ & $-2/3$ & $-$\\\hline
 $\bE'$              & $\ONE$      & $\ONE$ & $?1$   & $-$\\\hline
 $\bQ'$              & $\THREEbar$ & $\TWO$ & $-1/6$ & $-$\\\hline
 $U'$                & $\THREE$    & $\ONE$ & $?2/3$ & $-$\\\hline
 $E'$                & $\ONE$      & $\ONE$ & $-1$   & $-$\\\Hrule
\end{tabular}

\vspace{10pt}

\begin{tabular}[b]{@{\Vrule\ }c|c|ccc@{\ \Vrule}}\Hrule
\multicolumn{5}{@{\Vrule\ }c@{\ \Vrule}}{{\bf Gauge fields} (vector multiplets)}\\\Hrule
 gauge group        & field & $\gSU(3)\s{color}$ & $\gSU(2)\s{weak}$ & $\gU(1)_Y$ \\\hline
 $\gSU(3)\s{color}$ & $G$   & $\EIGHT$  & $\ONE$    & $0$      \\\hline
 $\gSU(2)\s{weak}$  & $W$   & $\ONE$    & $\THREE$  & $0$      \\\hline
 $\gU(1)_Y$         & $B$   & $\ONE$    & $\ONE$    & $0$      \\\Hrule
\end{tabular}
\renewcommand{\arraystretch}{1}
\label{tab:FieldContentOfVMSSM}
\end{center}
\end{table}

\clearpage

\section{The V-MSSM Model}
\label{sec:vmssm-model}

\subsection{Definition}
\label{sec:vmssm-definition}
The V-MSSM is an extension of the MSSM, where a vector-like pair of the $\gSU(5)$ decuplets is introduced.
That is, it is defined with the same symmetry as the MSSM,
\begin{equation}
\text{symmetry:}\quad
\Bigl[\gSU(3)\s{strong} \times \gSU(2)\s{weak} \times \gU(1)_Y\Bigr]\times Z_2^{R},
\end{equation}
and the extended field content,
\begin{equation}
\text{fields:}\quad
\text{MSSM} + \TEN + \TENbar \quad =\quad \text{MSSM} + \left(Q, \bU', \bE'\right) 
   +\left(\bQ', U', E'\right).
\end{equation}
The field content is summarized in Table \ref{tab:FieldContentOfVMSSM}.

These ingredients yield the following generic superpotential,
\begin{align}
\label{eq:VMSSMsuperpotential}
 W\s{MSSM}
&= \mu\Hu\Hd
  - \left(Y_u\right)_{ij} \Hu Q_i \bU_j
  + \left(Y_d\right)_{ij} \Hd Q_i \bD_j
  + \left(Y_e\right)_{ij} \Hd L_i \bE_j,
\\
  W\s{extra}
&= - Y'\Hu Q'\bU' + Y'' \Hd\bQ' U' + M_{Q'}Q'\bQ'+M_{U'}\bU'U'+M_{E'}\bE'E',
\\
 W\s{mixing}
&= -\left(\epsilon_u \right)_i \Hu Q_i\bU'
   -\left(\epsilon_u'\right)_i \Hu Q' \bU_i
   +\left(\epsilon_d \right)_i \Hd Q' \bD_i
   +\left(\epsilon_e \right)_i \Hd L_i \bE',
\end{align}
and SUSY-breaking terms
\begin{align}\label{eq:VMSSMsusybreaking}\begin{split}
   -\mathcal{L}_{\rm soft}
 &=
 -\mathcal{L}_{\rm soft}^{\rm MSSM}
 + m_{Q'  }^2 \abssq{Q'}
 + m_{\bQ'}^2 \abssq{\bQ'}
 + m_{\bU'}^2 \abssq{\bU'}
 + m_{U'  }^2 \abssq{U'}
 + m_{\bE'}^2 \abssq{\bE'}
 + m_{E'  }^2 \abssq{E'}
 \\&\qquad\qquad
 + \left(-a' \Hu Q' \bU' + a'' \Hd \bQ' U' + \Hc\right)
 + \left(b_{Q'} Q'\bQ' + b_{U'} \bU'U'+b_{E'} \bE'E'+\Hc\right)
 \\&\qquad\qquad
 +\left[
   -\left(\lambda_u \right)_i \Hu Q_i\bU'
   -\left(\lambda_u'\right)_i \Hu Q' \bU_i
   +\left(\lambda_d \right)_i \Hd Q' \bD_i
   +\left(\lambda_e \right)_i \Hd L_i \bE'
   +\Hc\right].
\end{split}
\end{align}
Note that extra $\mu$-terms, such as $\mu_iQ_i\bQ'$, can be absorbed into the $\epsilon$-terms, and similar absorption can be operated for the SUSY-breaking terms.

However, this general Lagrangian has several problems.
The principal one is on the mixing terms $\epsilon$'s between the Standard Model fermions and the extra fermions.
These mixings must be small enough, especially for $i=1,2$, not to evade current experimental bounds on flavor violating processes.
However, on the other hand, absence of these mixings causes a problem that some of the extra fermions, especially the lightest vector-like quark ($t'_1$, which will be introduced soon), become stable.
This is obviously disfavored because we know our Universe has no heavy stable colored particles.
Quantitative evaluation of these mixing parameters is interesting, and has great importance in search for the extra fermions at the LHC, which is discussed in Sec.~\ref{sec:vector}, because the decay branch of the fermions is determined by these mixings.
Nevertheless, the quantitative discussions are left as future works.
In this dissertation we simply restrict the model to have no mixings between the vector-like matters and the Standard Model fermions in the first and second generations, and to have small mixings between the vector-like matters and the third generation.\footnote{%
Discussions on the mixings with the first and second generation Standard Model quarks are found in Ref.~\cite{Arnold:2010vs} etc.}
Furthermore the SUSY-breaking mixing parameters $\lambda$'s are all ignored.
These terms just govern the mixing between MSSM sfermions and the extra sfermions, and are thus not important in phenomenological discussions.

Later we will impose one more condition that $Y''=0$; this is because the term $Y'' \Hd \bQ' U'$ tends to reduce the Higgs boson mass if $\tan\beta$ is as large as $\sim\Order(10)$. This mechanism is discussed in Sec.~\ref{sec:vmssm-higgs}.
For the SUSY-breaking sector, similarly, $a''=0$ is imposed.

Therefore, in the following discussions, we use the superpotential and the SUSY-breaking terms of
\begin{align}
\begin{split}\label{eq:our-vmssm-spot}
  W\s{VMSSM} &= W\s{MSSM} - Y'\Hu Q'\bU' + M_{Q'}Q'\bQ'+M_{U'}\bU'U'+M_{E'}\bE'E'
 \\&\qquad
 -\epsilon_u  \Hu Q_3\bU'
 -\epsilon_u' \Hu Q' \bU_3
 +\epsilon_d  \Hd Q' \bD_3
 +\epsilon_e  \Hd L_3 \bE',
\end{split}\\
 \begin{split}\label{eq:our-vmssm-sbreaking}
   -\mathcal{L}_{\rm soft}\suprm{VMSSM}
 &=
 -\mathcal{L}_{\rm soft}^{\rm MSSM}
 + m_{Q'  }^2 \abssq{Q'}
 + m_{\bQ'}^2 \abssq{\bQ'}
 + m_{\bU'}^2 \abssq{\bU'}
 + m_{U'  }^2 \abssq{U'}
 + m_{\bE'}^2 \abssq{\bE'}
 + m_{E'  }^2 \abssq{E'}
 \\&\qquad\qquad
 + \left(-a' \Hu Q' \bU' + b_{Q'} Q'\bQ' + b_{U'} \bU'U'+b_{E'} \bE'E'+\Hc\right),
 \end{split}
\end{align}
where $\epsilon$'s are assumed to be tiny; more precisely, they are small enough to avoid disfavored flavor mixings, but not so feeble as the longevity of the extra matters does not cause phenomenological, especially cosmological, problems.

We also define the following variables for later convenience:
\begin{align}
 A' &:=a'/Y',&
 A''&:=a''/Y'',&
 B_{Q'}&:=b_{Q'}/M_{Q'},&
 B_{U'}&:=b_{U'}/M_{U'},&
 B_{E'}&:=b_{E'}/M_{E'}.
\end{align}

\starline

Actually there remains one subtlety: $CP$-violating phases in the parameters.
First, in the supersymmetric sector, one phase cannot be removed from $(Y', Y'', M_{Q'}, M_{U'})$ because we already fixed the phases of $\Hu$ and $\Hd$ to set $\vu$ and $\vd$ positive.
This does not matter under the simplification $Y''=0$, but anyway we just assume all of $(Y', Y'', M_{Q'}, M_{U'})$ are positive for simplicity.
In addition, the SUSY-breaking terms yield many phases as is the case of the MSSM, but we simply neglect all the complex phases in these terms.

\subsection{Extra particles and their masses}
\label{sec:vmssm-mass}
In this model we have eight extra scalar particles and four extra fermions.
Let us discuss the masses of the particles at first.

The starting point of our discussion is Eqs.~\eqref{eq:our-vmssm-spot}--\eqref{eq:our-vmssm-sbreaking}, but we restore $Y''$ for completeness.
After the electroweak symmetry breaking, where $\HuZ$ and $\HdZ$ acquire vacuum expectation values of $\vu=v\sin\beta$ and $\vd=v\cos\beta$, the fermionic mass terms in the Lagrangian are expressed as, with mixing terms,
\begin{equation}
 \begin{split}
-{\mathcal L} &\supset
\pmat{\bar Q'_u&\bU'&\bar t\s R}
\pmat{%
 M_{Q'}         & Y''\vd & 0\\
 Y'\vu  & M_{U'}         & \epsilon_u \vu\\
 \epsilon'_u \vu    & 0              & m_t
}\pmat{Q'_u\\U'\\t\s L}
\\&\qquad
+\pmat{\bar Q'_d&\bar b\s R}
\pmat{
 -M_{Q'} & 0\\
 \epsilon_d \vd & m_b}
\pmat{Q'_d\\b\s L}
+\pmat{\bar E'&\bar \tau\s R}
\pmat{
 M_{E'} & \epsilon_e \vd\\
 0& m_\tau}
\pmat{E'\\\tau\s L}
+\Hc,\end{split}
\end{equation}
where $t_{L,R}$, $b_{L,R}$ and $\tau_{L,R}$ are the Standard Model top, bottom, and tau, and their masses are denoted by $m_t$, $m_b$ and $m_\tau$, respectively.
Similarly, the scalar mass terms are obtained as
\begin{equation}
-{\mathcal L} \supset
\pmat{Q_u'^* \  U'^* \ \bar Q'_u  \ \bar U'}\mathcal{M}_u\pmat{Q'_u \\ U' \\ \bar Q'^*_u \\ \bar U'^*}
+\pmat{Q_d'^* \ \bar Q'_d} \mathcal M_d \pmat{Q'_d \\ \bar Q'^*_d}
+\pmat{E'^* \ \bE'} \mathcal M_e \pmat{E' \\ \bE'^*}
\end{equation}
where the mass matrices are
\begin{align*}
\mathcal M_u &=
\pmat{
 \mathcal M_u^{(1,1)}            & Y'^*\vu M_{U'}+ Y''\vd M_{Q'}^*     &  b_{Q'}^*   & a'^*\vu-\mu Y'^*\vd \\
 Y'\vu M_{Q'}^* + Y''^*\vd M_{Q'}  & \mathcal M_u^{(2,2)}              &  a''^*\vd-\mu Y''^*\vu   & b_{U'}^* \\
 b_{Q'}                          & a''\vd-\mu^*Y''\vu                  & \mathcal M_u^{(3,3)} & Y'^*\vu M_{Q'}+Y''\vd M_{U'}^* \\
 a'\vu-\mu^*Y'\vd                & b_{U'}     &  Y'\vu M_{Q'}^*+Y''^*\vd M_{U'}   & \mathcal M_u^{(4,4)}
},\\
\mathcal M_d&=
\pmat{ \mathcal M_d^{(1,1)}& -b_{Q'}^*\\
-b_{Q'}                    &\mathcal M_d^{(2,2)}
},
\qquad\qquad
\mathcal M_e=
\pmat{ \mathcal M_e^{(1,1)} & b_{E'}^*\\
b_{E'}                       &\mathcal M_e^{(2,2)}
},
\end{align*}
and the diagonal components are
\begin{align*}
 \mathcal M_u^{(1,1)}&=|M_{Q'}|^2+|Y'\vu| ^2+m^2_{Q'}+\left(\frac12-\frac23\sin^2\theta\s w\right)m_Z^2\cos2\beta,\\
 \mathcal M_u^{(2,2)}&=|M_{U'}|^2+|Y''\vd|^2+m^2_{U'}+\left(-\frac23\sin^2\theta\s w\right)m_Z^2\cos2\beta,\\
 \mathcal M_u^{(3,3)}&=|M_{Q'}|^2+|Y''\vd|^2+m^2_{\bQ'}+\left(-\frac12+\frac23\sin^2\theta\s w\right)m_Z^2\cos2\beta,\\
 \mathcal M_u^{(4,4)}&=|M_{U'}|^2+|Y'\vu| ^2+m^2_{\bU'}+\left(\frac23\sin^2\theta\s w\right)m_Z^2\cos2\beta,\\
 \mathcal M_d^{(1,1)}&=|M_{Q'}|^2 + m^2_{Q'} +\left(-\frac12+\frac13\sin^2\theta\s w\right)m_Z^2\cos 2\beta,\\
 \mathcal M_d^{(2,2)}&=|M_{Q'}|^2 + m^2_{\bQ'} +\left(\frac12-\frac13\sin^2\theta\s w\right)m_Z^2\cos 2\beta,\\
 \mathcal M_e^{(1,1)}&=|M_{E'}|^2 + m^2_{E'}   + \left(\sin^2\theta\s w\right)m_Z^2\cos 2\beta,\\
 \mathcal M_e^{(2,2)}&=|M_{E'}|^2 + m^2_{\bE'} - \left(\sin^2\theta\s w\right)m_Z^2\cos 2\beta.
\end{align*}
Note that we use the same symbols for both the scalars and the fermions, but it should not be confusing.

Finally, it is obvious from the above expressions that we have following particles as the extra matters:
\begin{align}
 \text{scalars:} &\quad \Bigl(\tilde t'_1,\tilde t'_2.\tilde t'_3,\tilde t'_4, \tilde b'_1, \tilde b'_2, \tilde \tau'_1, \tilde \tau'_2\Bigr),&
 \text{fermions:}&\quad \Bigl(t'_1, t'_2, b', \tau'\Bigr),
\end{align}
with the definition that $\tilde t'_1 < \tilde t'_2 < \tilde t'_3 < \tilde t'_4$ etc.\ in terms of the masses.

\subsection{Higgs mass increase}
\label{sec:vmssm-higgs}
Let us move on to the most important feature, the Higgs mass in the V-MSSM.
We have discussed much about this topic for the MSSM in Sec.~\ref{sec:MSSM-higgs-1loopmass}, and the discussion given here is completely parallel to that.

As the effective potential has the same form as Eq.~\eqref{eq:MSSMEffPot}, the increase of the Higgs mass specific to the V-MSSM at the one-loop level is given as, in the decoupling limit,
\begin{align}
\Delta m^2_h\Big|\s{VMSSM}\suprm{1-loop}&=
\left[
 \frac{\sin^2\beta}{2}\left(\pdiffn2{}{\vu}-\frac{1}{\vu}\pdiff{}{\vu}\right)
 +\frac{\cos^2\beta}{2}\left(\pdiffn2{}{\vd}-\frac{1}{\vd}\pdiff{}{\vd}\right)
 +\sin\beta\cos\beta\pdiff{^2}{\vu\partial\vd}
\right]\vev{\Delta V^{(1)}\s{VMSSM}},
\end{align}
where
\begin{equation}
\Delta V^{(1)}\s{VMSSM} = \frac{1}{16\pi^2}
\left[
 \sum_{\tilde t_1', \tilde t_2', \tilde t_3', \tilde t_4'}
F\left(m_X^2\right)
-2
 \sum_{t_1', t_2'}
F\left(m_X^2\right)
\right], \qquad  F(x)=\frac{x^2}4\left(\log\frac{x}{Q^2}-\frac32\right).
\end{equation}
Here one should note that the other extra particles, $\tilde b'_i$, $\tilde \tau'_i$, $b'$, $\tau'$, have no contribution to the Higgs mass in the absence of $Y''$.
Therefore, with the masses we calculate in the last section, the increase of the Higgs boson mass can be easily calculated numerically.

\starline

Let us assume $M_{Q'}=M_{U'}(=: M_F)$, $m^2_{Q'}=m^2_{\bar Q'}=m^2_{U'}=m^2_{\bar U'}(=: M_S^2- M_F^2)$, and $b_{Q'}=b_{U'}=0$, in order to obtain an analytic expression of the increase.
With a straightforward computation, one can obtain the following result:
\begin{align}
 \def\fs#1{\frac{M_F^#1}{M_S^#1}}
 \begin{split}
\Delta m^2_h\Big|\s{VMSSM}\suprm{1-loop}
  &= \frac{3 v^2}{4 \pi ^2}\Bigg\{
\left(Y'\sin\beta\right)^4\left[\log\frac{M_S^2}{M_F^2}-\frac16\left(5-\fs2\right)\left(1-\fs2\right)
  +\frac{\Xi'^2}{M_S^2}\left(1-\frac{M_F^2}{3M_S^2}-\frac{\Xi'^2}{12M_S^2}\right)\right]
  \\&\quad\qquad
  -\frac23\left(Y'\sin\beta\right)^3(Y''\cos\beta)
\left[\left(2-\fs2\right)\left(1-\fs2\right)+\fs2\left(\frac{\Xi'^2}{M_S^2}+\frac{\Xi'\Xi''}{2M_S^2}\right)\right]
\\&\quad\qquad
-\left(Y'\sin\beta\right)^2(Y''\cos\beta)^2
\left[\left(1-\fs2\right)^2+\frac{M_F^2\left(\Xi'+\Xi''\right)^2}{3M_S^4}\
\right]
\\&\quad\qquad
  -\frac23\left(Y'\sin\beta\right)(Y''\cos\beta)^3
\left[\left(2-\fs2\right)\left(1-\fs2\right)+\fs2\left(\frac{\Xi''^2}{M_S^2}+\frac{\Xi'\Xi''}{2M_S^2}\right)\right]
\\&\quad\qquad+
\left(Y''\cos\beta\right)^4
\left[\log\frac{M_S^2}{M_F^2}-\frac16\left(5-\fs2\right)\left(1-\fs2\right)
  +\frac{\Xi''^2}{M_S^2}\left(1-\frac{M_F^2}{3M_S^2}-\frac{\Xi''^2}{12M_S^2}\right)\right]
\Bigg\}\label{eq:VMSSM-higgs-full}
 \end{split}
\end{align}
with
\begin{align}
 \Xi' &:= A'-\mu\cot\beta,&
 \Xi''&:= A''-\mu\tan\beta.
\end{align}
Especially, if we set $Y''=0$ as is mentioned above and employed in the following phenomenological discussions, the increase becomes
\begin{equation}
 \def\fs#1{\frac{M_F^#1}{M_S^#1}}
\Delta m^2_h\Big|\s{VMSSM}\suprm{1-loop}\approx
   \frac{3 v^2 Y'^4\sin\beta^4}{4 \pi ^2}
\left[\log\frac{M_S^2}{M_F^2}-\frac16\left(5-\fs2\right)\left(1-\fs2\right)
  +\frac{\Xi'^2}{M_S^2}\left(1-\frac{M_F^2}{3M_S^2}-\frac{\Xi'^2}{12M_S^2}\right)\right].\label{eq:VMSSM-higgs-simp}
\end{equation}
This is similar to the MSSM top--stop contribution~\eqref{eq:MSSM1loopHiggsBound};
the difference simply comes from the fact that the extra top-like quarks are vector-like.
Therefore, the extra contribution is considered to be comparable to that from the MSSM, and thus the Higgs mass is to be increased considerably.

It is interesting that, similarly to the MSSM case, a large ratio of $M_S/M_F$ yields a larger increase in the Higgs mass; the Higgs mass does not raised under $M_S=M_F$.
Therefore, since the parameter $M_S$ is characterized by $M\s{SUSY}$, a lighter $M_F$ gives a large contribution to the Higgs mass for a fixed $M\s{SUSY}$.
We will see this feature in Sec.~\ref{sec:vgmsb-numerical-result}.

 \begin{figure}[t]
 \begin{center}\vspace{\baselineskip}
   \includegraphics[width=0.6\textwidth]{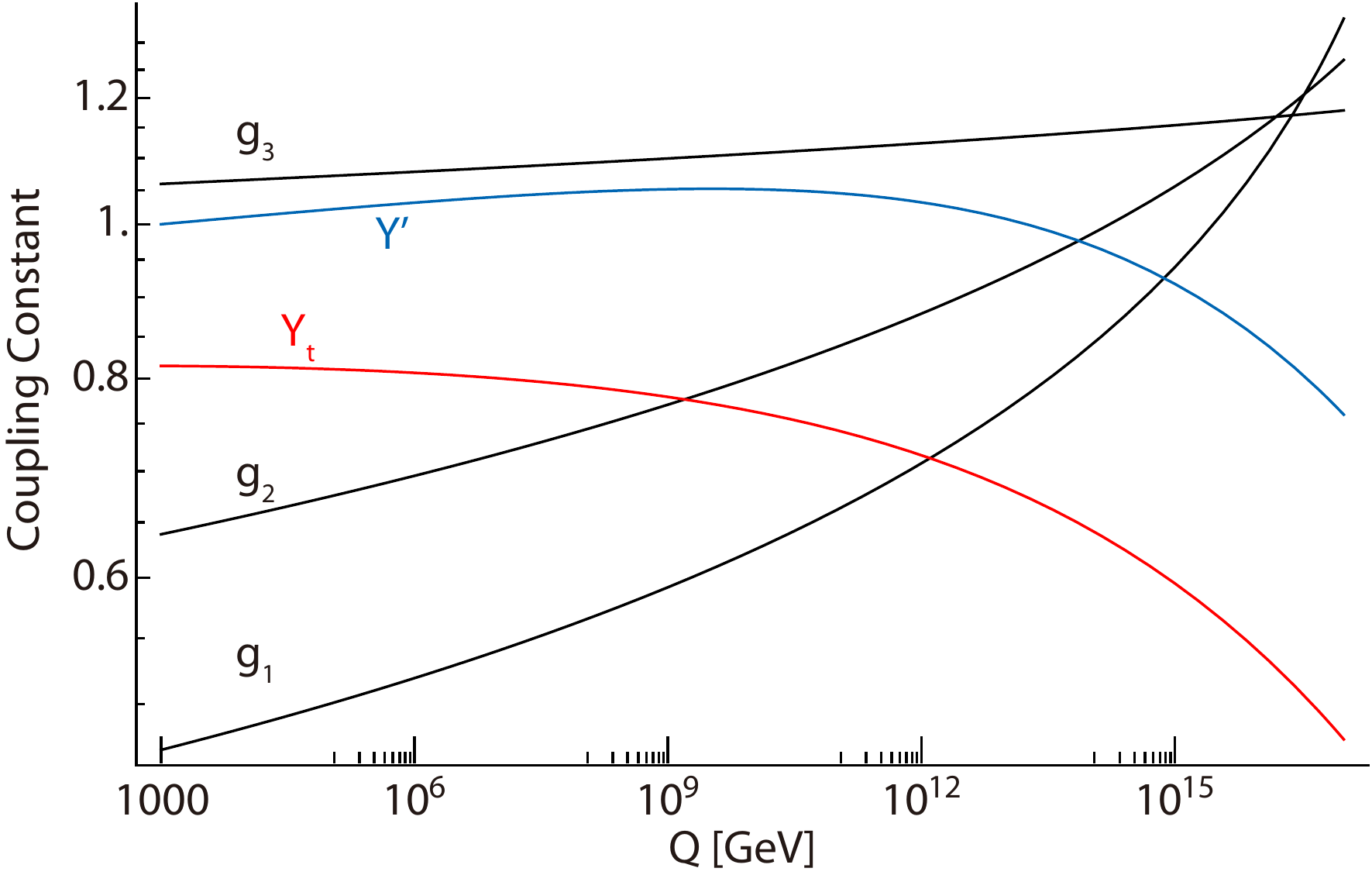}
  \caption[The gauge coupling running under the V-MSSM.]{The renormalization group evolutions of the coupling constants in the V-MSSM.
  Here $Y'$ is set to be $Y'=1$ at the SUSY scale $M\s{SUSY}=1\,{\rm TeV}$. Other parameters are: $Y''=0$, $\tan\beta=20$.
  The drawing procedure is similar to that summarized in the latter half of Appendix~\ref{app:rge_figures}.
  The RGEs are summarized in Appendix~\ref{app:vmssm-rge}.
  The evolutions of $g_3$, $g_2$, and $g_1$ are drawn with black lines (from top to bottom); that for $Y_t$ is with a red line, and for $Y'$ is with a turquoise blue line.
  One should be careful that, as we will observe in Fig.~\ref{fig:vmssm-RGE-flow}, the running of $Y'$ is sensitive to its value at the SUSY scale, $Y'\s{SUSY}$, for its infrared fixed point behaviour.
  Moreover, the value of $Y'\s{SUSY}$ affects the evolutions of the other \YUKAWA\ coupling constants; their flows also significantly depend on $Y'\s{SUSY}$.
  However, as the \YUKAWA\ couplings appear in the RGEs of the gauge coupling constants only at the two-loop level, the renormalization flows of the  gauge couplings are less affected by $Y'\s{SUSY}$.
}
  \label{fig:vmssm-gcu}
 \end{center}
 \end{figure}

\subsection{Renormalization group flow}
\label{sec:vmssm-rgflow}

 \begin{figure}[p]
 \begin{center}
  \includegraphics[width=200pt]{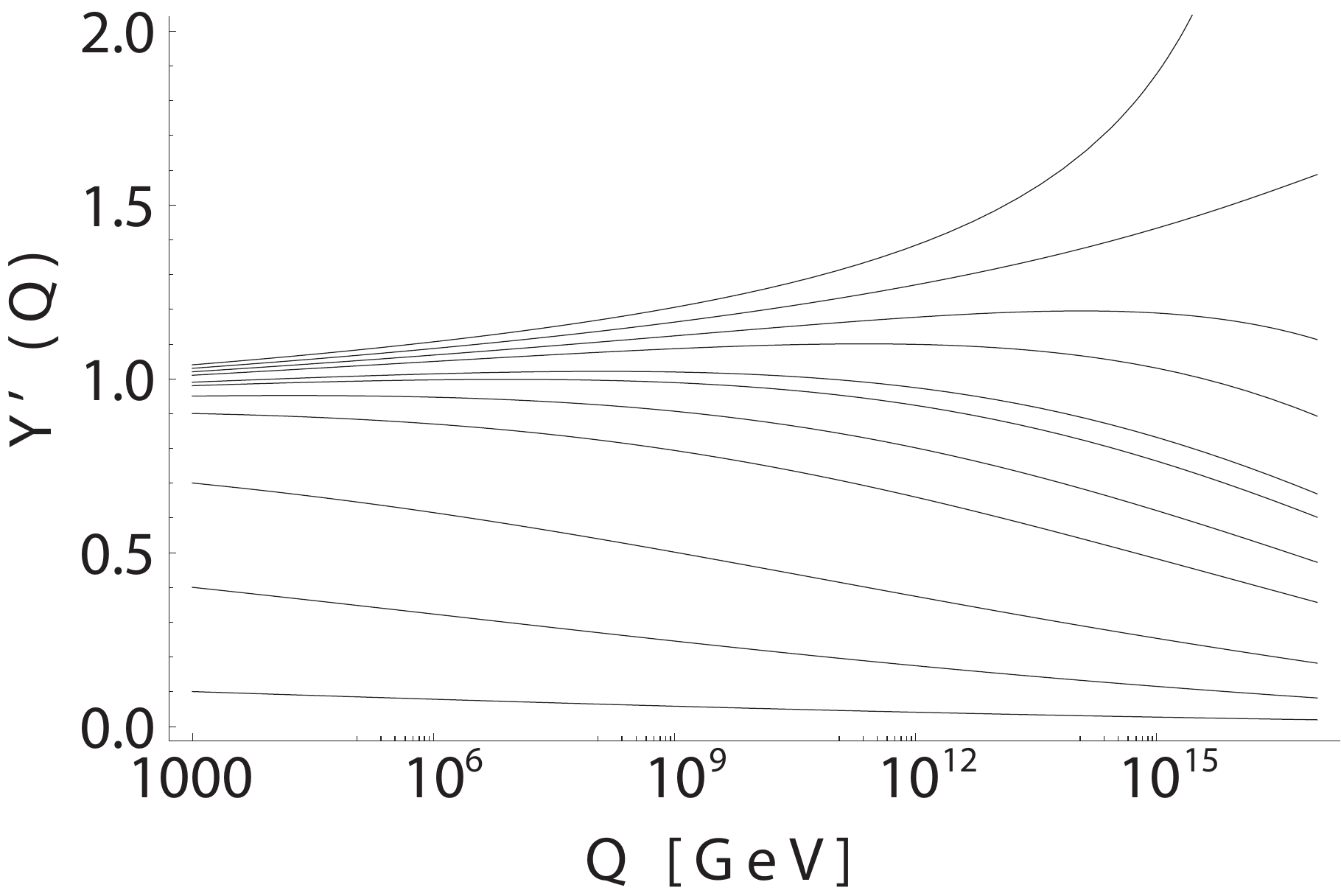}\hspace{10pt}
  \includegraphics[width=200pt]{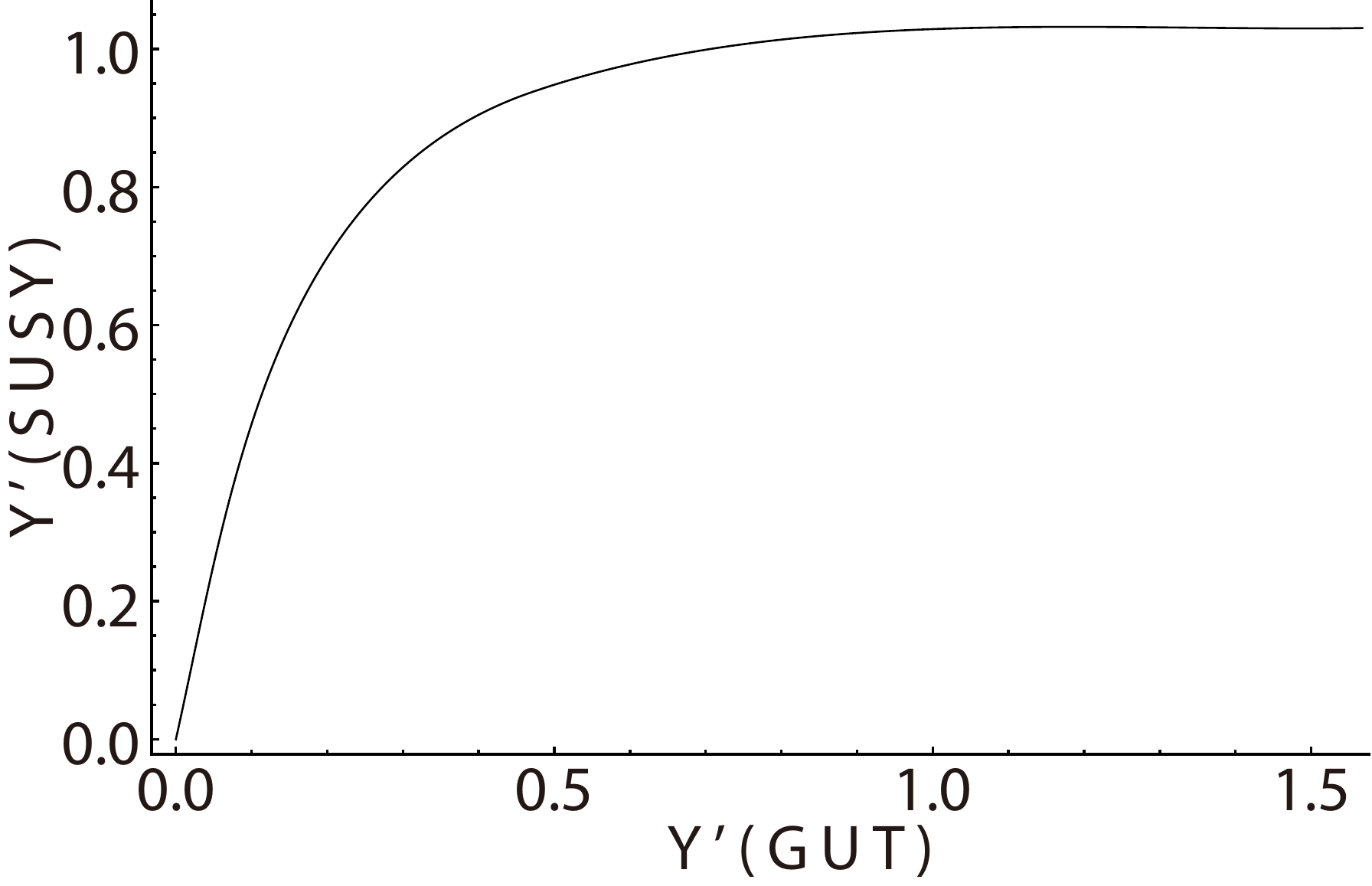}
  \caption[Renormalization group flow of $Y'$, which indicates the quasi infrared fixed point behaviour.]{
    (left) Renormalization group flow of $Y'$ in the V-MSSM.
    We can see that, for a wide range of $Y'$ at the GUT scale (which let us say $Y'\s{GUT}$),  $Y'$ falls into $\sim 1$ at the low-energy scale (which let us say $Y'\s{SUSY}$).
    In other words, the renormalization group flow of $Y'$ severely depends on $Y'\s{SUSY}$.
    This feature, or the quasi infrared fixed point behaviour, can be obvious in the right figure. (right) The value of $Y'\s{SUSY}$ as a function of $Y'\s{GUT}$.
    Here, the GUT scale is defined with the condition $g_1(M\s{GUT})=g_2(M\s{GUT})$.
    In both figures, the SUSY scale is set to be $M\s{SUSY}=1{\,\rm TeV}$, and $\tan\beta = 20$ and $Y''=0$ are used.
    Details of the evaluation procedure are summarized in Appendix~\ref{app:rge_figures}.
 }
  \label{fig:vmssm-RGE-flow}
 \end{center}
 \end{figure}

 \begin{figure}[p]
 \begin{center}
  \includegraphics[width=200pt]{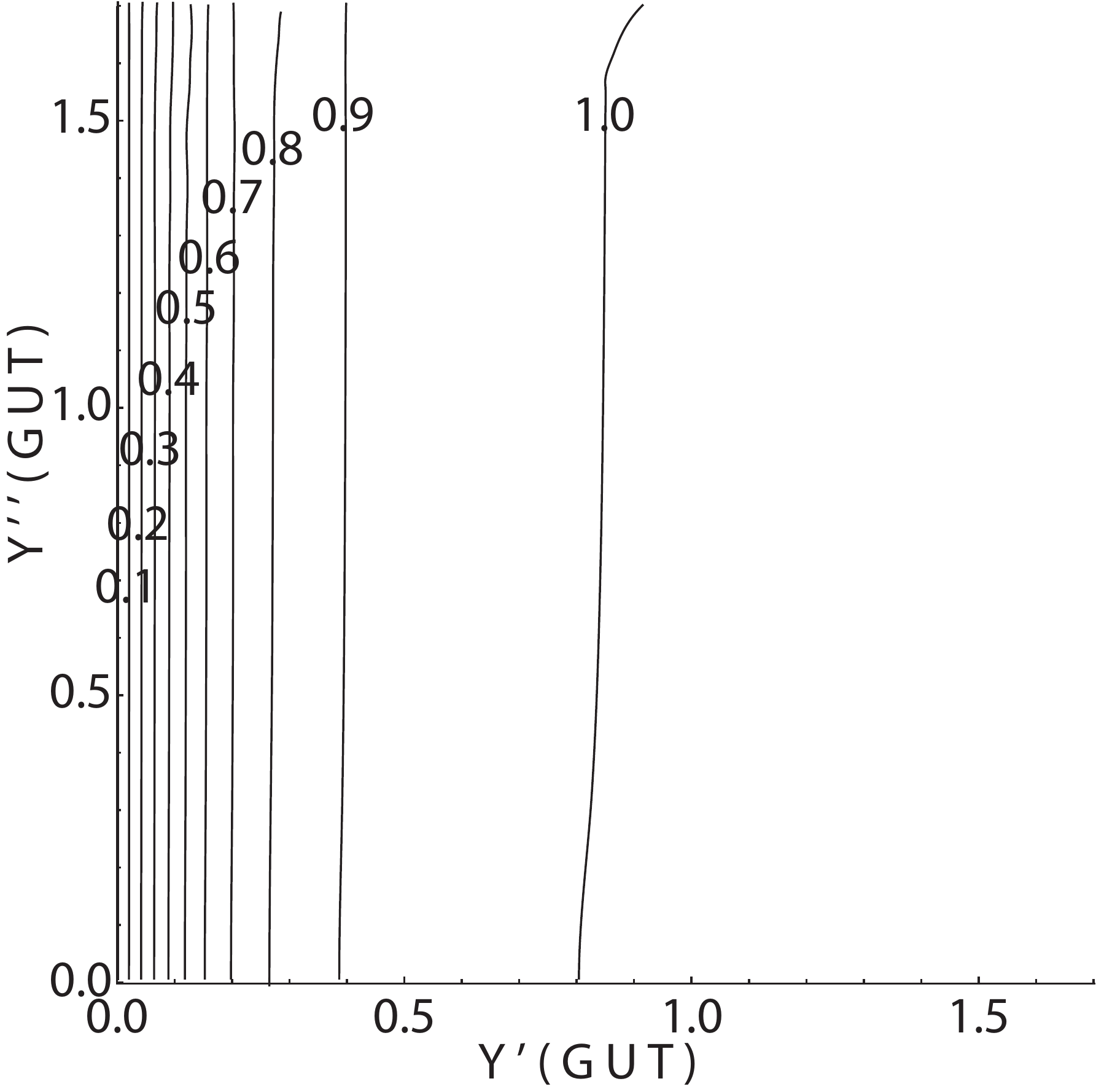}\hspace{10pt}
  \includegraphics[width=200pt]{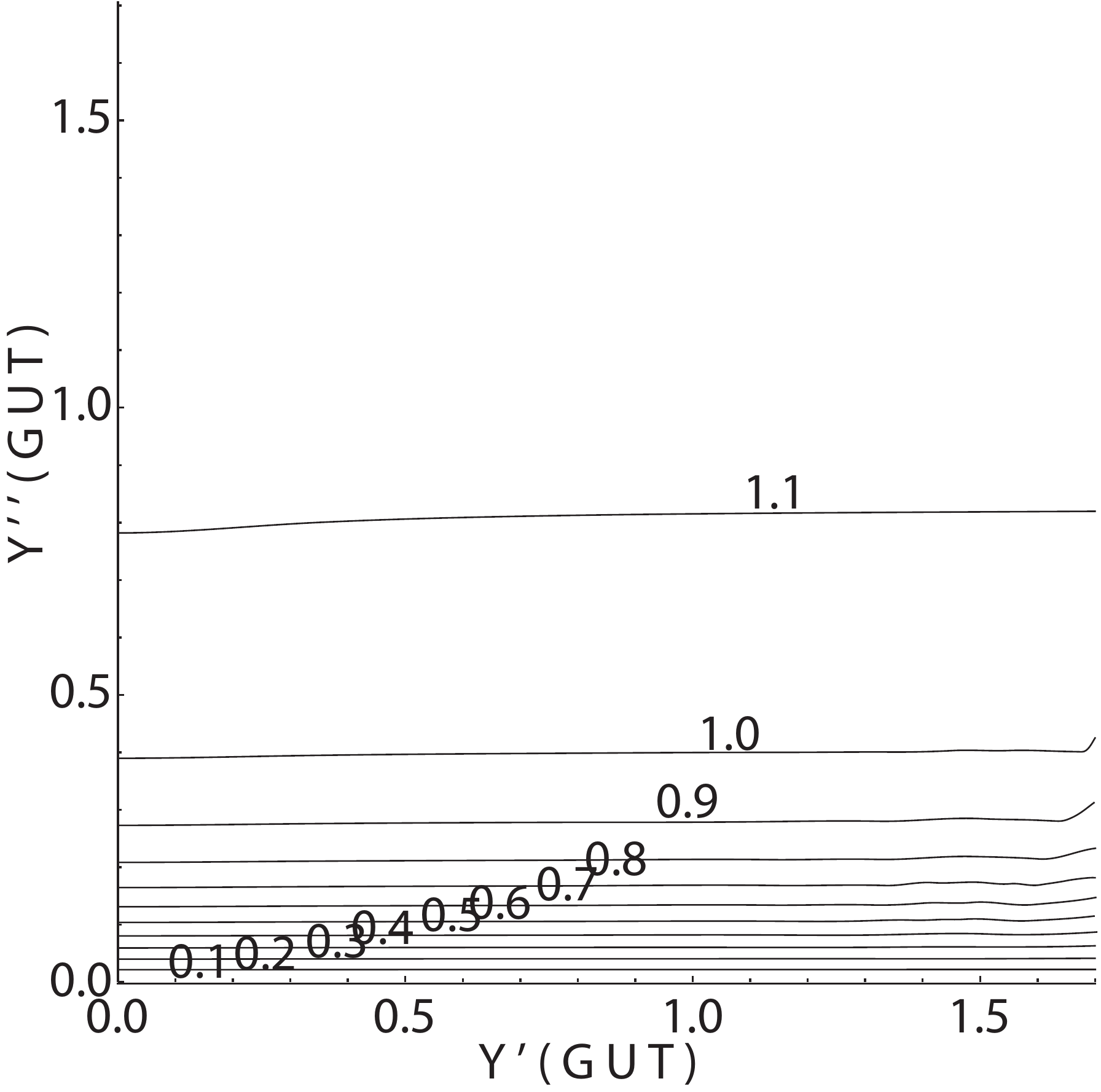}
  \caption[Dependence of $Y'$(SUSY) and $Y''$(SUSY) on $Y'$(GUT) and $Y''$(GUT).]{%
    Contour plots of (left) $Y'\s{SUSY}$ and (right) $Y''\s{SUSY}$ as functions of $(Y'\s{GUT}, Y''\s{GUT})$.
    The parameters are the same as Fig.~\ref{fig:vmssm-RGE-flow}.
    Both $Y'$ and $Y''$ have quasi infrared fixed points $\sim 1$, and they are nearly independent in the plotted region.
}
  \label{fig:vmssm-2dcontour}
 \end{center}
 \end{figure}

\begin{figure}[p]
\begin{center}
\includegraphics[width=200pt]{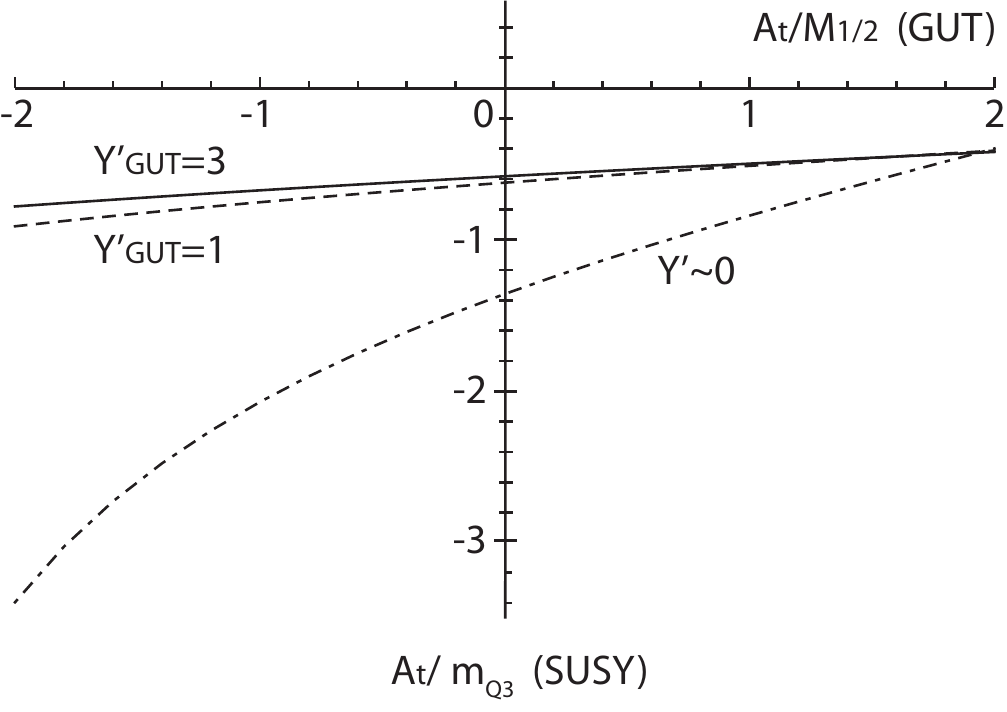}\hspace{10pt}
\includegraphics[width=200pt]{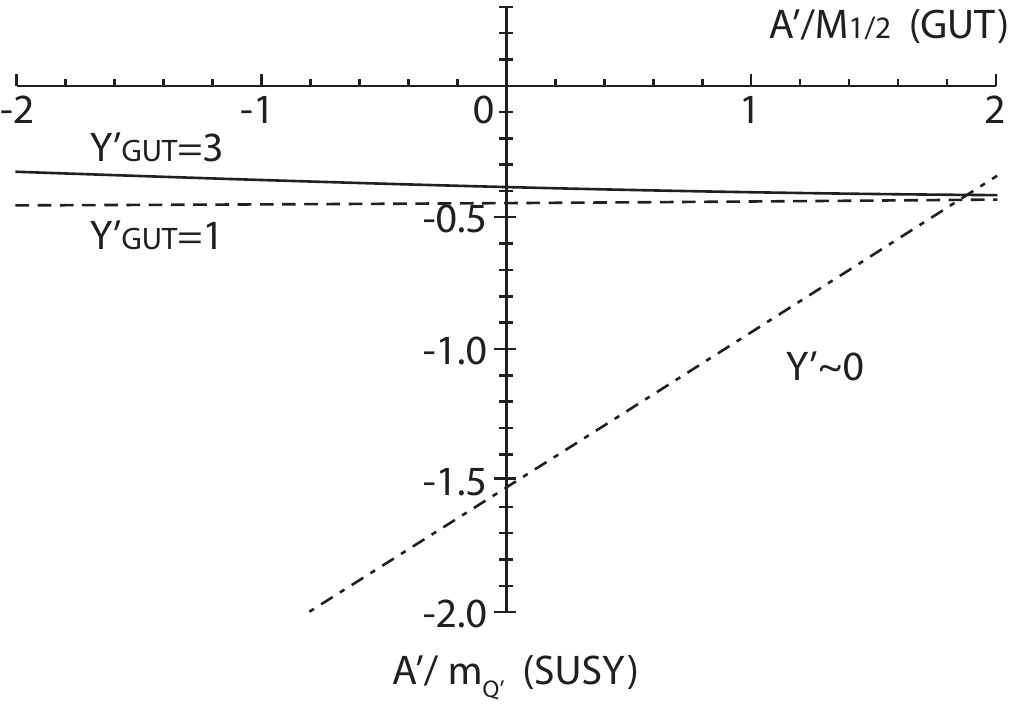}
\caption[Infrared fixed point behaviour of $A_t$ and $A'$.]{
(left) $A_t/m_{Q_3}$ evaluated at the SUSY scale as a function of $A_t/M_{1/2}$, which is an input at the GUT scale.
(right) $A'/m_{Q'}$ evaluated at the SUSY scale as a function of $A'/M_{1/2}$, which is an input at the GUT scale.
 For both figures, the CMSSM scenario is utilized.
 $(m_0,M_{1/2})=(0,1.5\,{\rm TeV})$ at the GUT scale, and $\tan\beta=20$, $\sgn\mu=+1$; $A'$ is set as the same as $A_0$, as well as $A_t$, at the GUT scale.
 $Y'$ is set at the GUT scale as $Y'=3$ (solid), $Y'=1$ (dashed), and $Y'=0$ (dot-dashed). For simplicity, $Y''=A''=0$ is used.
 Note that $m_{Q_3}$ and $m_{Q'}$ is the square-root of the SUSY-breaking soft masses squared.
 From these figure, it is obvious that both $A'$ and $A_t$ perform the quasi infrared fixed point behaviour if $Y'$ works effectively. The infrared fixed points sit at $A_t\sim-0.5m\s{Q_3}$ and $A'\sim-0.5m\s{Q'}$, which means the stop mixing is rather small in the V-MSSM framework (cf.~Sec.~\ref{sec:mssm-h-discussion}).
}
\label{fig:A10AT}
\end{center}
\end{figure}

\begin{figure}[p]
\begin{center}
\includegraphics[width=200pt]{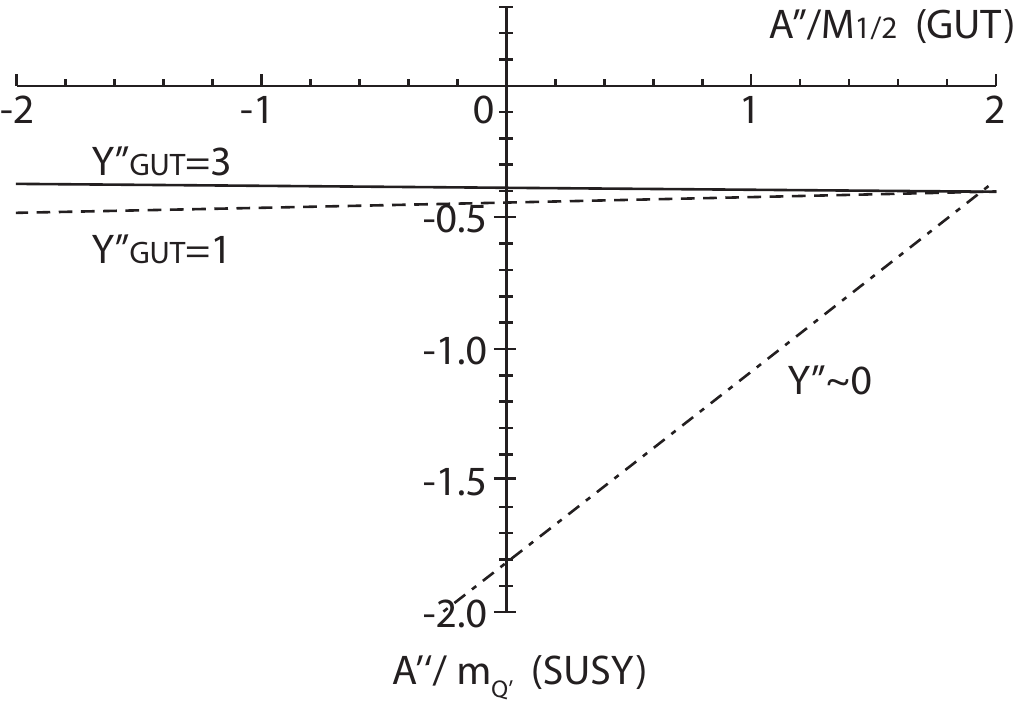}
\caption[Infrared fixed point behaviour of $A''$.]{
 $A''/m_{Q'}$ evaluated at the SUSY scale as a function of $A''/M_{1/2}$, which is an input at the GUT scale.
 Similarly to Fig.~\ref{fig:A10AT}, the CMSSM scenario is utilized with $(m_0,M_{1/2})=(0,1.5\,{\rm TeV})$ at the GUT scale, and $\tan\beta=20$, $\sgn\mu=+1$.
 Here we set $Y'=A'=0$; $A''$ is set as the same as $A_0$ at the GUT scale.
 $Y''$ is set at the GUT scale as $Y''=3$ (solid), $Y''=1$ (dashed), and $Y''=0$ (dot-dashed).
 Here the infrared fixed point behaviour can be seen as Fig.~\ref{fig:A10AT}.}
\label{fig:A10P}
\end{center}
\end{figure}

The unification of the gauge coupling constants is a virtue of the SUSY theories, which are $g_Y=\sqrt{3/5}g_1$ for $\gU(1)_Y$, $g_2$ for $\gSU(2)\s{weak}$ and $g\s s=g_3$ for $\gSU(3)\s{color}$, as is discussed in Sec.~\ref{sec:mssm-gcu}.
Here the vector-like matters modify the renormalization group equations (RGEs).
When we extend the MSSM with $n_5$ copies of $(\FIVE+\FIVEbar)$ and $n_{10}$ copies of $(\TEN+\TENbar)$ supermultiplets, the one-loop level RGEs are modified to be
\begin{align}
 \diff{g_3}{\ln Q} &= \frac{g_3^3}{16\pi^2}\Bigl[-3+\left(3n_{10}+n_5\right)\Bigr],\label{eq:VMSSMg3}\\
 \diff{g_2}{\ln Q} &= \frac{g_2^3}{16\pi^2}\Bigl[1+\left(3n_{10}+n_5\right)\Bigr],\label{eq:VMSSMg2}\\
 \diff{g_1}{\ln Q} &= \frac{g_1^3}{16\pi^2}\Bigl[\frac{33}{5}+\left(3n_{10}+n_5\right)\Bigr].\label{eq:VMSSMg1}
\end{align}

In the V-MSSM, where $(n_{10},n_5)=(1,0)$, it is interesting that $g_3$ does not run at the one-loop level.
Including the two-loop level RGEs, we obtain the renormalization group evolution as Fig.~\ref{fig:vmssm-gcu}, where the V-MSSM parameters are set as: $\left(Y',Y'',\tan\beta\right)=\left(1.0, 0, 20\right)$ at $Q=M\s{SUSY}=1\TeV$.
In the figure the evolutions of $Y_t$ and $Y'$ are drawn, but here one should be careful: the evolution of $Y'$ significantly depends on its value at the SUSY scale $M\s{SUSY}$, which we call $Y'\s{SUSY}$.
It is illustrated in Fig.~\ref{fig:vmssm-RGE-flow}.
In the left figure the renormalization group flow for $Y'$ is drawn; we can see that a wide range of $Y'\s{GUT}$, the value of $Y'$ at the GUT scale $M\s{GUT}$, converges at the low-energy scale to $Y'\s{SUSY}\sim 1$. This is called ``quasi infrared fixed point behaviour,'' and the fixed point is $Y'\sim1$~\cite{Martin:2009bg}.
In the right figure $Y'\s{SUSY}$ is drawn as a function of $Y'\s{GUT}$.

The above discussion assumes $Y''=0$, but actually, in the presence of $Y''$, it behaves in similar manner; moreover, $Y''$ also has a quasi infrared fixed point.
This is shown in Fig.~\ref{fig:vmssm-2dcontour}; $Y'\s{SUSY}$ in the left figure and $Y''\s{SUSY}$ in the right figure are plotted as a function of $(Y'\s{GUT}, Y''\s{GUT})$.
We can see that the both have quasi infrared fixed points, and respective behaviour is almost independent of the value of the other.

It is also known~\cite{Martin:2009bg,Endo:2011mc} that $A_t$, $A'$ and $A''$ among the scalar trilinear couplings perform the infrared fixed point behaviour.
To illustrate this feature, the CMSSM scenario is utilized, which has the following five parameters: $(m_0, M_{1/2}, \tan\beta, A_0, \sgn\mu)$.
Fig.~\ref{fig:A10AT} shows the infrared fixed point behaviour of $A_t$ and $A'$ in the V-MSSM with $Y''=A''=0$.
Here $(m_0,M_{1/2})=(0,1.5\TeV)$ is set at the GUT scale, and $A'(=A_0)$ at the GUT scale is used as an input (horizontal axis). The dependence is drawn for three values of $Y'$.
For any input value of $A_t$ and $A'$ at the GUT scale, $A_t/M\s{Q3}$ and $A'/M\s{Q'}$ converge to $\sim-0.5$ at the SUSY scale $M\s{SUSY}:=1\TeV$ as long as $Y'\gtrsim1$, which is nothing but the infrared fixed point behaviour.
Fig.~\ref{fig:A10P} is that for $A''$ in the V-MSSM; here  $Y'=A'=0$, and $A''=A_0$ are the input values.
We can easily see infrared fixed point behaviour, and it is similar to that of $A'$.

\subsection[Contribution to the muon \texorpdfstring{$g-2$}{g-2}]{Contribution to the muon $\boldsymbol{g-2}$}
\label{sec:vmssm-gm2}
The MSSM parameters relevant to the muon $g-2$ are affected by the extra matters; e.g. from the characteristic behaviour of $g_3$ during the renormalization group running.
Thus the SUSY contribution to the muon $g-2$, which is reviewed in Sec.~\ref{sec:muon-g-2-mssm}, is {\em quantitatively}  modified.

However, it does not change {\em qualitatively}.
Because the extra particles are assumed to have no direct couplings with muons, their contribution appears only as higher-order loop level corrections with the assumed-to-be-small mixing parameters.
Moreover, the particles are expected to be heavy for the SUSY-invariant mass $M_{Q',U',E'}$ and the soft masses.
Therefore, we can use the same formulae as in Sec.~\ref{sec:muon-g-2-mssm} for the V-MSSM.

\section{On the Choices We Have Done}
\label{sec:choices-we-have}
Now we have introduced the V-MSSM.
In the above introduction one might wonder why we have chosen $(\TEN+\TENbar)$ pair, and why we will set $Y''=0$.
For such curious readers, before going into the V-GMSB model, the answers for these questions are provided here.

\subsection[Why did we choose \texorpdfstring{$\TEN+\TENbar$?}{(10) + (10-bar) ?}]{Why did we choose $\TEN\boldsymbol{+}\TENbar$?}
\label{sec:why-ten-tenbar}
Extensions of the MSSM with complete $\gSU(5)$ multiplets are preferred, because there the $\gSU(5)$-GUTs are respected and the gauge coupling unification is realized.
The vector-like insertions, such as $\FIVE+\FIVEbar$ and $\TEN+\TENbar$ have a virtue for vanishment of gauge anomalies.

Then one may consider that we can introduce a vector-like pair $\FIVE+\FIVEbar$ instead as extra matters.
This does not work, because no \YUKAWA\ interactions can be additionally introduced.

Actually an alternate is $\left(\FIVE+\FIVEbar\right)+\left(\ONE+\ONEbar\right)$ model, or $LND$ model~\cite{Martin:2009bg}:
\begin{align}
 \FIVE   &=\left(L'+\bD'\right),&
 \FIVEbar&=\left(\bar L'+D'\right),&
 \ONE    &=\bN',&
 \ONEbar &=N';
\end{align}
\begin{equation}
 W_{LND} = \mathscr M_{L'} L' \bar L' + \mathscr M_{D'} \bD' D' + \mathscr M_{N'} \bN' N' + \mathscr Y' \Hu L' \bN'
             \quad \Bigl(+\mathscr Y'' \Hd \bar L' N'\Bigr).
\end{equation}
As is obvious, the structure of the Lagrangian is almost the same as the V-MSSM.
This model corresponds to $(n_{10},n_5)=(0,1)$ in Eqs.~\eqref{eq:VMSSMg3}--\eqref{eq:VMSSMg1}, and the gauge coupling unification does realize.
One weakness of this model is that the Higgs mass increase is smaller than the V-MSSM.
The effective Higgs potential is almost the same as that in the V-MSSM, but as the leptons, not quarks, form the \YUKAWA\ coupling, the color factor $3$ in the superpotential lacks in this case, i.e.,
\begin{equation}
 \def\fs#1{\frac{M_F^#1}{M_S^#1}}
\Delta m^2_h\Big|_{LND}\suprm{1-loop}=
\frac13\Delta m^2_h\Big|\s{VMSSM}\suprm{1-loop}\approx
   \frac{v^2 \mathscr Y'^4\sin\beta^4}{4 \pi ^2}
\left[\log\frac{M_S^2}{M_F^2}-\frac16\left(5-\fs2\right)\left(1-\fs2\right)
  +\cdots\right].
\end{equation}
Furthermore, it is known that the coupling $\mathscr Y'$ has also a quasi infrared fixed point as well as $Y'$, which however is slightly smaller: $\mathscr Y'\leadsto 0.765$~\cite{Martin:2009bg}. As the mass increase is proportional to $\mathscr Y'^4$, this is crucial.
Therefore, although it is possible to realize $m_h=126\GeV$ with a large $M_S/M_F$, we do not consider this model in this dissertation.

On the other hand, we can consider adding more vector-like pairs: $3n_{10}+n_5 > 3$.
In this case, however, the gauge coupling $g_3$ might blow up before the grand unification;
as is shown in Fig.~\ref{fig:vmssm-gcu}, $g_3$ increases during the running up to the GUT scale in the V-MSSM.
If we add more matters, the RGE for $g_3$ becomes positive even at the one-loop level (See: Eq.~\eqref{eq:VMSSMg3}).
Therefore, this direction is not so promising as is naively expected.
Note that the upper bound on $(3n_{10}+n_5)$ is dependent on other MSSM parameters, in particular the SUSY scale $M\s{SUSY}$ (cf.~Fig.~\ref{fig:MSSM_gcu}).

It should be also mentioned here that the MSSM with four fermion generations, which can be written as ``MSSM$+\vc{10}+\overline{\vc 5}$'', is excluded by the Higgs boson searches~\cite{CMSPASHIG12008} (cf. Ref.~\cite{Ishiwata:2011hr}).

\subsection[Why is \texorpdfstring{$Y''=0$}{Y''=0} required?]{Why is $\boldsymbol{Y''=0}$ required?}
\label{sec:why-ypp-zero}
From the expression \eqref{eq:VMSSM-higgs-full}, one may notice that the contribution from $Y''$ to the Higgs boson mass is similar to that from $Y'$, and thus may wonder why we set $Y''=0$.
This is due to a large $\tan\beta$, which is preferred to keep the SUSY explanation of the muon $g-2$ problem, and a large $\mu$-parameter; as we will see in Sec.~\ref{sec:vgmsb-mass}, the $\mu$-parameter is raised by the vector-like quarks during the renormalization group running and thus tends to be large in the V-MSSM.

Under a large $\tan\beta$, the mass increase is expanded as
\begin{align}
\Delta m^2_h\Big|\s{VMSSM}\suprm{1-loop}
  &= \frac{3 v^2}{4 \pi ^2}\left\{
Y'^4 \log \frac{1}{r^2} - \frac{1}{6} \left[\left(5-r^2\right)\left(1-r^2\right)Y'^4 + 2 (Y'r)^2(Y''\xi)^2 +\frac12(Y''\xi)^4\right] + \Order\left(\frac{1}{\tan\beta}\right)\right\},
\end{align}
where we defined $r:=M_F/M_S$ (which is usually smaller than one) and $\xi:=\mu/M_S$; we set $A'=A''=0$ here for simplicity.
From this expression it is understood that, especially with a large $\mu$-parameter, the interaction $Y''\Hd\bQ'U'$ just decreases the Higgs mass to disgrace the model.

This is the reason we simply employ an unnatural simplification of $Y''=0$;
studies with $Y''\neq 0$ are left as future works.

\section{V-GMSB Model}
\label{sec:vgmsb-model}

Among various SUSY-breaking scenarios, the gauge-mediated SUSY-breaking (GMSB) scenario~\cite{Giudice:1998bp} is notably attractive because of natural suppressions of dangerous flavor-changing processes and $CP$-violations which can appear in soft SUSY-breaking terms.
However, the GMSB under the MSSM faces the difficulty of explaining the mass of the Higgs boson in the range of $125\TO126\GeV$.
This is because the scalar trilinear couplings, $A_{u,d,e}$, arise in two-loop level quantum correction and thus are generally small in the GMSB scenario; consequently the mixing in the stop sector, characterized by $X_t$ in Eq.~\eqref{eq:MSSM1loopHiggsBound}, is suppressed, and we are forced to set the SUSY scale $M\s{SUSY}$ to be considerably large.
Then the SUSY explanation for the muon $g-2$ problem is spoiled, and the SUSY loses one sales point.

In the V-MSSM, the extra vector-like matters efficiently increase the Higgs boson mass to $126\GeV$ without exploiting large $M\s{SUSY}$.
This was qualitatively discussed in Sec.~\ref{sec:vmssm-higgs}, and will be confirmed quantitatively soon.
Thus the GMSB scenario with the V-MSSM is very attractive and phenomenologically viable; the V-MSSM resurrects the GMSB scenario.
We will see that the parameter space has a region where the Higgs boson mass $125\TO126\GeV$ is achieved and the muon $g-2$ is consistent with the experimental value at the 1--2\,$\sigma$ level.

\starline

We adopt the simplest GMSB scenario as the SUSY-breaking framework of the V-GMSB model.
This framework is parameterized by the messenger scale $M_{\rm mess}$, the soft mass scale $\Lambda$, the messenger number $N_5$, the ratio of the Higgs vacuum expectation values $\tan\beta = \vu/\vd$, and the sign of the $\mu$-parameter $\sgn\mu$.

As we saw in Sec.~\ref{sec:vmssm-rgflow}, the gauge coupling constant $g_3$ is marginal to blowing up; as the messengers also worsen the situation, we set $N_5 = 1$ in order to preserve the perturbativity of the gauge coupling constants up to the GUT scale.
We also set $\mu>0$; note that we have taken the convention under which the gaugino masses are positive: $M_a>0$; thus this choice results in positive contribution to the muon $g-2$ (cf.~Sec.~\ref{sec:muon-g-2-mssm}).

Here are summarized the parameters of the V-GMSB model and our setting:
\begin{align}
 \Bigl(\Lambda, \quad M\s{mess}, \quad \tan\beta, \quad \sgn\mu=+1; \qquad M_{Q'} = M_{U'} = M_{E'}(=:M_V), \quad Y'=1.0,\quad Y''=0\Bigr),
\end{align}
where $Y'$ is set at the messenger scale $M\s{mess}$, while $M_V$ is an input at the low-energy scale $M\s{SUSY}$, which is defined as $M\s{SUSY}:=\sqrt{m_{\tilde t_1}m_{\tilde t_2}}$.
Note that the SUSY-breaking parameters specific to the V-GMSB models, $A'$ and $B_{Q',U',E',}$, are not input parameters but yielded by the messengers.

\subsection{SUSY-breaking}
\label{sec:vgmsb-susybreaking}
In the V-GMSB model, the soft SUSY-breaking terms are induced radiatively via the messenger fields as in the ordinary GMSB scenario.
The superpotential of the messenger sector is
\begin{eqnarray}
W_{\rm mess} = ({\mathcal M}_D + {\mathcal F}_D \theta^2) \Psi_D \bar{\Psi}_D + ({\mathcal M}_L + {\mathcal F}_L \theta^2) \Psi_L \bar{\Psi}_L,
\end{eqnarray}
where ${\mathcal M}_D \, ({\mathcal M}_L)$ is the supersymmetric mass, and ${\mathcal F}_D\, ({\mathcal F}_L)$ is the SUSY-breaking $F$-term of the colored (non-colored) messengers.
The messenger scale is taken as $M_{\rm mess} := {\mathcal M}_D = {\mathcal M}_L$, and the SUSY-breaking $F$-terms are as ${\mathcal F}_D={\mathcal F}_L$.
Thus, the soft SUSY-breaking parameters are characterized by the two parameters, $M\s{mess}$ and $\Lambda:={\mathcal F}_D/{\mathcal M}_D={\mathcal F}_L/{\mathcal M}_L$, at the messenger scale.
Precisely speaking, the relations ${\mathcal M}_D={\mathcal M}_L$ and ${\mathcal F}_D={\mathcal F}_L$ do not hold generally at the messenger scale due to the renormalization group evolution, even if we set ${\mathcal M}_D={\mathcal M}_L$ and ${\mathcal F}_D={\mathcal F}_L$ at the GUT scale.
Even though, $\Lambda$ is almost unchanged during the renormalization group running, and hence, the soft mass parameters are not significantly changed under this effect.
Finally, these messenger fields set the soft SUSY-breaking parameters at the messenger scale $M\s{mess}$, and the parameters are evolved down to the weak scale with the RGEs summarized in Appendix~\ref{app:vmssm-rge}.

\subsection{Characteristics of the V-GMSB model}
\label{sec:vgmsb-characteristics}
As is discussed in Sec.~\ref{sec:vmssm-rgflow}, the couplings $Y'$, $A_t$ and $A'$ show quasi infrared fixed point behaviour in the V-MSSM scenario.
However, in the V-GMSB model, the situation is slightly different; since the parameter is input at the messenger $M\s{mess}$, the renormalization group evolution does not work effectively enough to drive them to the fixed points.
This is important for the Higgs boson mass; the Higgs boson mass decreases as $Y'$ is reduced away from $Y' = 1$ at the messenger scale.
Similarly, $A_t$ and $A'$ are much smaller than the fixed point value especially for smaller $M\s{mess}$ as the usual GMSB models.
This disables the Higgs boson mass to be enhanced by the trilinear couplings, but the constraint from the branching ratio of the $b \to s\gamma$ decay is safely satisfied, which is serious in so-called the $m_h$-max scenario of the CMSSM scenarios~\cite{Endo:2011gy}.

Another characteristic feature of the V-GMSB model is the size of the Higgsino mass parameter $\mu$ at the SUSY scale.
It becomes very large compared to the usual GMSB models.
This is because the vector-like matters contribute to the RGE of the up-type Higgs mass squared, which is approximately given as
\begin{eqnarray}
\frac{d m_{\Hu}^2}{d \ln Q}
\simeq
\left.\frac{d m_{H_u}^2}{d \ln Q}\right|_{\rm MSSM}
+
\frac{3}{8 \pi^2} Y'^2 \left(m_{Q'}^2 + m_{\bar{U}'}^2 + m_{H_u}^2 + |A'|^2\right),
\label{eq:RG_mhu}
\end{eqnarray}
where $Q$ is the renormalization scale (cf.~Appendix~\ref{app:vmssm-rge}).
This contribution, which originates in the vector-like matters, gives a large negative contribution to the up-type Higgs mass squared $m_{H_u}^2$.
It is as large as that from the stop, since $Y' \simeq 1$.
Hence, the electroweak symmetry breaking conditions require a large value of $\mu$.
Consequently, the next-to-lightest neutralino $\tilde\chi_2^0$ and the lightest chargino $\tilde\chi_1^\pm$ tend to consist of the wino, and the mass of the lighter stau is likely to be small, because the left-right mixing of the stau mass matrix, which is proportional to $\mu \tan\beta$, becomes large.
Especially when $\tan\beta$ is large,  the lighter stau becomes the NLSP. 
These features can be found in the mass spectrum discussed below.

\subsection{Mass spectrum}
\label{sec:vgmsb-mass}
 \begin{figure}[t]
 \begin{center}
  \includegraphics[height=220pt]{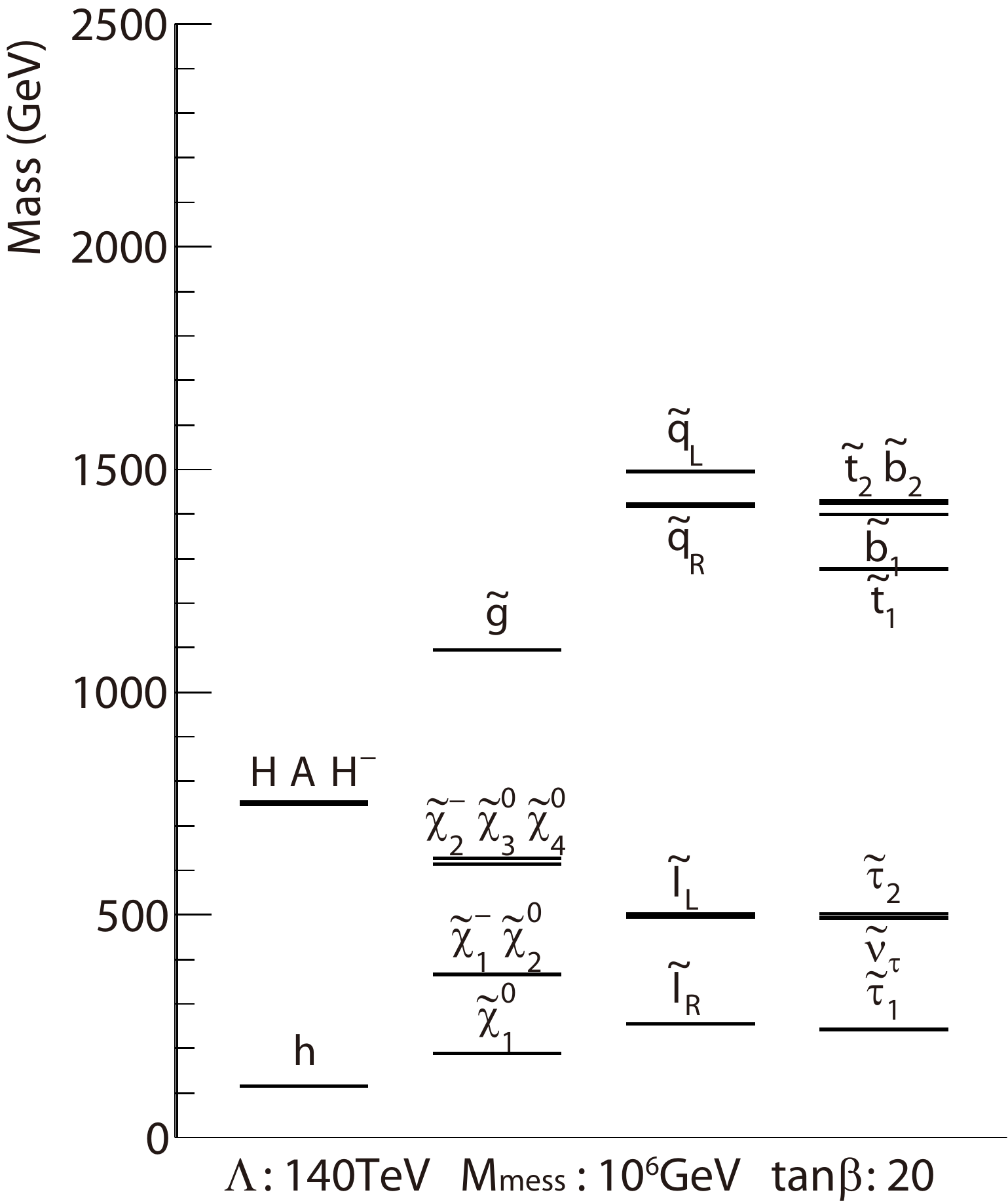}
  \hspace{10pt}
  \includegraphics[height=220pt]{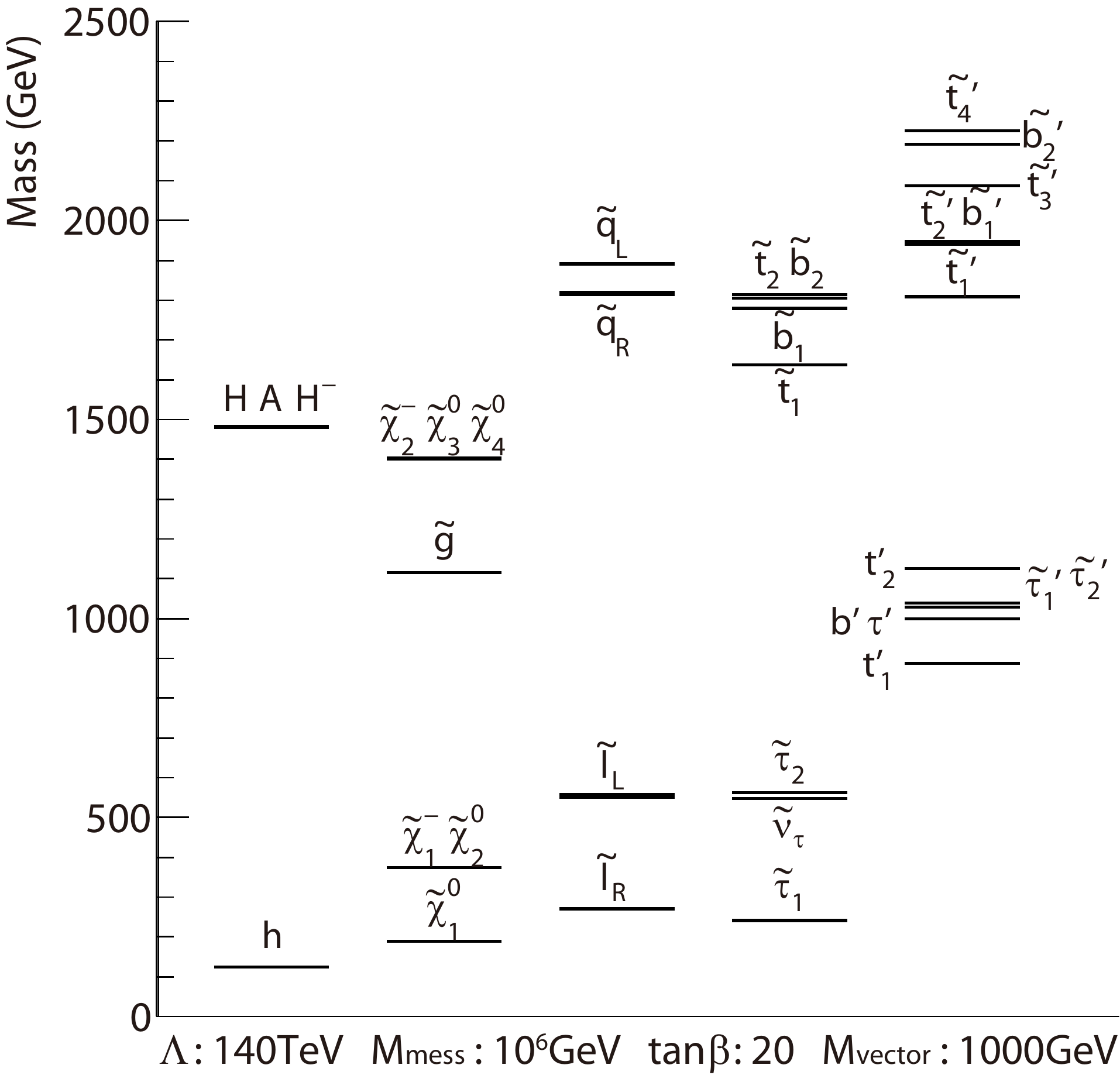}
  \caption[The mass spectra of the GMSB and the V-GMSB.]{%
The mass spectra of the GMSB (left) and the V-GMSB (right). The GMSB parameters are $(\Lambda, M\s{mess},\tan\beta,N_{\rm mess})=(140\,{\rm TeV},10^6\,{\rm GeV},20,1)$ in both cases.
The SUSY-invariant masses of vector-like fields are set as $M_{Q'}=M_{U'}=M_{E'}=1\,\rm TeV$ for the V-GMSB model.
The sfermions of the first and the second generations are labeled with $\tilde q\s L, \tilde q\s R, \tilde l\s L$, and $\tilde l\s R$, where the left--right mixings are ignored.
}
  \label{fig:spectrum}
 \end{center}
 \end{figure}

In this subsection, we discuss the mass spectrum of the V-GMSB.
In the numerical analysis we utilized {\tt SOFTSUSY 3.3}~\cite{SOFTSUSY} and {\tt FeynHiggs 2.9}~\cite{FeynHiggs} to calculate the mass spectrum of the SUSY particles and the Higgs sector.
They were modified to take the effects from the extra vector-like matters into account.

A typical mass spectrum is displayed in Fig.~\ref{fig:spectrum}, where the V-GMSB result is compared to that of the ordinary GMSB.
Here and hereafter, we assume a common SUSY-invariant mass for the vector-like fields, $M_V=M_{Q'}=M_{U'}$ ($=M_{E'}$).
As is explained just above, the heavier chargino $\tilde{\chi}^\pm_2$ and the heavier two neutralinos $\tilde{\chi}^0_{3,4}$ mainly consist of the Higgsino, and are much heavier than the ordinary case.
In addition, it is found that the squarks become heavier; this is because the gauge coupling $g_3$ stays large during the renormalization group evolution.
On the contrary, a ratio of the gaugino masses is less affected by the extra matters.
This is because the ratios of $M_a/\alpha_a$, where $M_a$ are the gaugino masses, are fixed during the renormalization group evolution at the one-loop level, and the gauge coupling constants $g_a$ are (should be) set at the low-energy scale.
The masses of the vector-like squarks are close to those of stops and sbottoms up to the SUSY-invariant mass, because they have the same quantum numbers as the corresponding squarks.

\starline

Let us explain how the mass spectrum is evaluated in {\tt SOFTSUSY} and what kinds of modifications are applied.
First of all, the program estimates the Standard Model gauge couplings $\alpha_a(m_Z)$ and Yukawa couplings $Y_i(m_Z)$ at the scale of the $Z$-boson mass $m_Z$~\cite{Pierce:1996zz}.
They are evolved upwards from $m_Z$ to the messenger scale $M_{\rm mess}$ with solving the V-GMSB RGEs.
In the numerical calculations, the {\tt SOFTSUSY} package was modified to include the two-loop level RGEs of the V-GMSB model.
It is also important to include threshold corrections and self-energy corrections of vector-like matters to the gauge coupling constants at the $m_Z$ scale, since they can give $\sim 10$\% contributions especially to the coupling constant $g_3$.
Those corrections also affect the gaugino masses and the scalar masses squared generated at $M_{\rm mess}$ through the gauge coupling constants.

Next, the soft SUSY-breaking parameters are provided by the messenger loops at $M_{\rm mess}$.
The extra Yukawa couplings are also set at the scale, which are $Y'(M_{\rm mess}) = 1$ and $Y''(M_{\rm mess}) = 0$.
The gauge, Yukawa and soft parameters are evolved with the V-GMSB RGE down to the SUSY scale, $M_{\rm SUSY}$, which is determined by the stop masses as $M_{\rm SUSY}=\sqrt{m_{\tilde{t}_1}m_{\tilde{t}_2}}$.

The masses of the superparticles are evaluated including the whole one-loop level corrections to the self-energies within the MSSM.
The pole mass of the gluino also receives a correction from vector-like matter loops.
However, this turns out to be around a few GeV, since the masses of the vector-like matters are close to $M_{\rm SUSY}$, and their contributions to the self-energy are relatively small.
Similarly, contributions of the vector-like fields to the electroweak gaugino masses are safely neglected.

Then, the Higgs potential is investigated, which determines the Higgs boson mass and the $\mu$-parameter. 
The MSSM part of the Higgs boson mass is evaluated at the two-loop level by the {\tt FeynHiggs} package~\cite{FeynHiggs}.
The contribution from the vector-like matters is estimated with the one-loop level effective potential discussed in Sec.~\ref{sec:vmssm-higgs}, where one-loop level contributions from the vector-like matters are taken into account, and is added to the MSSM Higgs boson mass in quadrature to obtain the V-MSSM Higgs boson mass. 
The two-loop level contribution of the vector-like matters can shift the Higgs boson mass by $\sim 1\text{--}10\GeV$, study for which is reserved for future works.
On the other hand, the $\mu$-parameter is less affected by the vector-like matters except for the renormalization group evolution~\eqref{eq:RG_mhu}, because tree-level contribution to $\mu$ is fairly large in the V-MSSM.

\section{Vacuum Stability Bound}
\label{sec:vac}
One of the most severe constraints on the V-GMSB model is the vacuum stability bound~\cite{Endo:2012rd}.
When $\tan\beta$ is large, the large trilinear coupling of the staus and the Higgs boson can generate charge breaking global minima and destabilize the electroweak symmetry breaking vacuum~\cite{Rattazzi:1996fb}.
Therefore, $\tan\beta$ has a tight upper bound for given $M\s{mess}$  and $\Lambda$ in the GMSB framework from the condition that lifetime of the proper vacuum must be longer than the age of our Universe.

Let us see this condition in detail~\cite{Hisano:2010re}.
The MSSM (and the V-MSSM) Higgs potential together with the stau sector is expressed as (cf.~Eq.~\eqref{eq:MSSMhiggspotTree})
\begin{align}
\begin{split}
 V_{\text{Higgs--stau}}^{\text{tree}+\text{1-loop}}
 &=
   \left(\abssq\mu+m_{\Hu}^2\right) \abssq\HuZ
 + \mstauL^2\abssq{\stauL}
 + \mstauR^2\abssq{\stauR}
 - \left( Y_\tau\mu^*\HuZ^*\stauL\stauR^*+\Hc\right)
 + Y_\tau^2\abssq{\stauL\stauR^*}\\
 &\quad+ \frac{g_2^2}{8}\left(\abssq{\stauL}+\abssq{\HuZ}\right)^2
 + \frac{g_Y^2}{8}\left(\abssq{\stauL}-2\abssq{\stauR}-\abssq{\HuZ}\right)^2
 + \frac{g_2^2 + g_Y^2}{8} \delta_H \left|\HuZ\right|^4.\label{eq:stauhiggspotential}
\end{split}
\end{align}
Note that here the one-loop level correction to the quartic coupling, which actually can be found in Eq.~\eqref{eq:MSSMEffPot}:
\begin{equation}
 \expval{\Delta V^{(1)}} \leadsto\frac{3Y_t^4}{8\pi^2}\log\frac{m_{\tilde t}}{m_t},
\end{equation}
is included as $\delta_H$, whose definition is
\begin{equation}
\delta_H:=\frac{3}{\pi^2}\frac{Y_t^4}{g_Y^2+g_2^2}\log\frac{m_{\tilde{t}}}{m_t}.
\end{equation}
The terms with $\HdZ$ are also ignored because $\expval{\HdZ}$ is tiny in our concerning large $\tan\beta$ region, although they give considerable corrections~\cite{Carena:2012mw}.

Hereafter we simply assume $\mu\in\mathbb R$ as well as $Y_\tau>0$.
Expanding the expression with $\HuZ=v+h$, we obtain the following mass terms for the stau:
\begin{equation}
 V_{\text{Higgs--stau}}^{\text{tree}+\text{1-loop}}
 \supset
 \left(\mstauL^2+\frac{g_2^2-g_Y^2}{4}v^2\right)\abssq{\stauL}-Y_\tau v \mu\left(\stauL\stauR^*+\Hc\right)
+\left(\mstauR^2+\frac{g_Y^2}{2}v^2\right)\abssq{\stauR}.
\end{equation}
The vacuum stability is at classical level understood that the eigenvalues of the stau mass matrix must be positive~\cite{Hisano:2010re}; that is,
\begin{equation}
 \left(m_\tau\cdot\mu\tan\beta\right)^2 < \frac12
\left(2\mstauL^2+2m_W^2-m_Z^2\right)\left(\mstauR^2+m_Z^2-m_W^2\right).
\end{equation}
This condition is certainly not sufficient because it just asserts that ``our'' vacuum, the EWSB vacuum, is a local minimum.
The condition which must hold is that the EWSB vacuum is the global minimum, {\em or}, as a more weaker condition, it is just a local minimum but the transition rate to other minima, especially the global minimum, is longer than the age of our Universe.
For the latter case, the transition rate is estimated by a semi-classical method, searching for so-called bounce solutions~\cite{Coleman:1977py}.
In Ref.~\cite{Hisano:2010re}, an approximate formula for the bound on $\mu\tan\beta$ is  obtained by using multi-dimensional bounce configurations, including top--stop radiative corrections to the Higgs potential.

\starline

In a work of Author\footnote{Done in collaboration with Dr.~M.~Endo, Prof.~K.~Hamaguchi, and Dr.~N.~Yokozaki.}, Ref.~\cite{Endo:2012rd}, this vacuum stability condition on the V-GMSB model is discussed, where a fitting formula given in Ref.~\cite[\href{http://arxiv.org/abs/1011.0260v1}{v1 on arXiv}]{Hisano:2010re} was utilized:
\begin{equation}
  \mu\tan\beta \lesssim 76.9 \sqrt{m_{L_3}m_{\bE_3}} + 38.7 \left(m_{L_3} + m_{\bE_3}\right) - 1.04\EE4\GeV.
  \label{eq:mubound_old}
\end{equation}
Very recently, however, the fitting formula was revised as follows~\cite[\href{http://arxiv.org/abs/1011.0260v1}{v2 on arXiv}]{Hisano:2010re}:
\begin{equation}
 \mu\tan\beta <
 213.5\sqrt{m_{L_3} m_{\bE_3}}
 -17.0 \left(m_{L_3} + m_{\bE_3}\right)
+ 4.52\EE-2 \GeV^{-1} \left(m_{L_3} - m_{\bE_3}\right)^2
 -1.30 \times 10^4\GeV. \label{eq:mubound_new}
\end{equation}
In these expressions, $m_{L_3}$ and $m_{\bE_3}$ are the square-root of the SUSY-breaking soft mass squared for, respectively, the left- and the right-handed stau.
Moreover, also very recently, the vacuum stability bound was reanalyzed with including effects of a radiatively-corrected tau \YUKAWA\ coupling~\cite{Carena:2012mw}.
The results in Ref.~\cite{Carena:2012mw}, however, cannot be directly applied to the V-GMSB model, because the masses and the mixing angle of the staus are different from those in Ref.~\cite{Carena:2012mw}.

The above result \eqref{eq:mubound_new} is obtained in the limit of zero temperature, but thermal effects may tighten the bound~\cite{Endo:2010ya,Endo:2011uw}.
The thermal decay rate of a false vacuum is usually estimated with following the method in Ref.~\cite{Linde:1981zj}.
Evaluating the Higgs potential at the one-loop level with including the thermal effect coming from the top quark and the electroweak gauge bosons, one can find that the stability bound can become more severe by $\sim 10\%$ than the zero temperature result for a small stau mass region~\cite{Endo:2010ya}.
In addition to the thermal corrections, one should note that the proper vacuum may be required to be the global minimum if the vacuum expectation values of the scalar fields stayed away from the ordinary ones in the early universe.

Considering all these various factors together, this dissertation uses the bound of Eq.~\eqref{eq:mubound_new} to draw the upper-bound, and also the bound but weakened by 10\% is drawn for reference use.

\starline

Finally, let us touch on another possibility of the vacuum instability.
In a class of SUSY models, the trilinear coupling of the top squark is predicted to be large in order to raise the Higgs mass.
Such a large trilinear coupling may spoil the vacuum stability in the stop--Higgs plane similarly to the stau--Higgs plane discussed above.
The meta-stability bound was studied in Ref.~\cite{Kusenko:1996jn} and obtained as
\begin{equation}
 A_t^2 + 3\mu^2 < 7.5 \left(m_{\tilde t_L}^2 + m_{\tilde t_R}^2\right).
\end{equation}
However, as the $A$-terms are not so large in the V-GMSB model as ordinary GMSB models, this condition does not appear in our discussion.

\section[Higgs mass, muon \texorpdfstring{$g-2$}{g-2} and vacuum stability in V-GMSB]{Higgs mass, muon $\boldsymbol{g-2}$ and vacuum stability in V-GMSB}
\label{sec:vgmsb-numerical-result}

\begin{figure}[p]
\begin{center}
 \begin{minipage}[t]{0.49\textwidth}\begin{center}
                                     \includegraphics[width=0.99\linewidth]{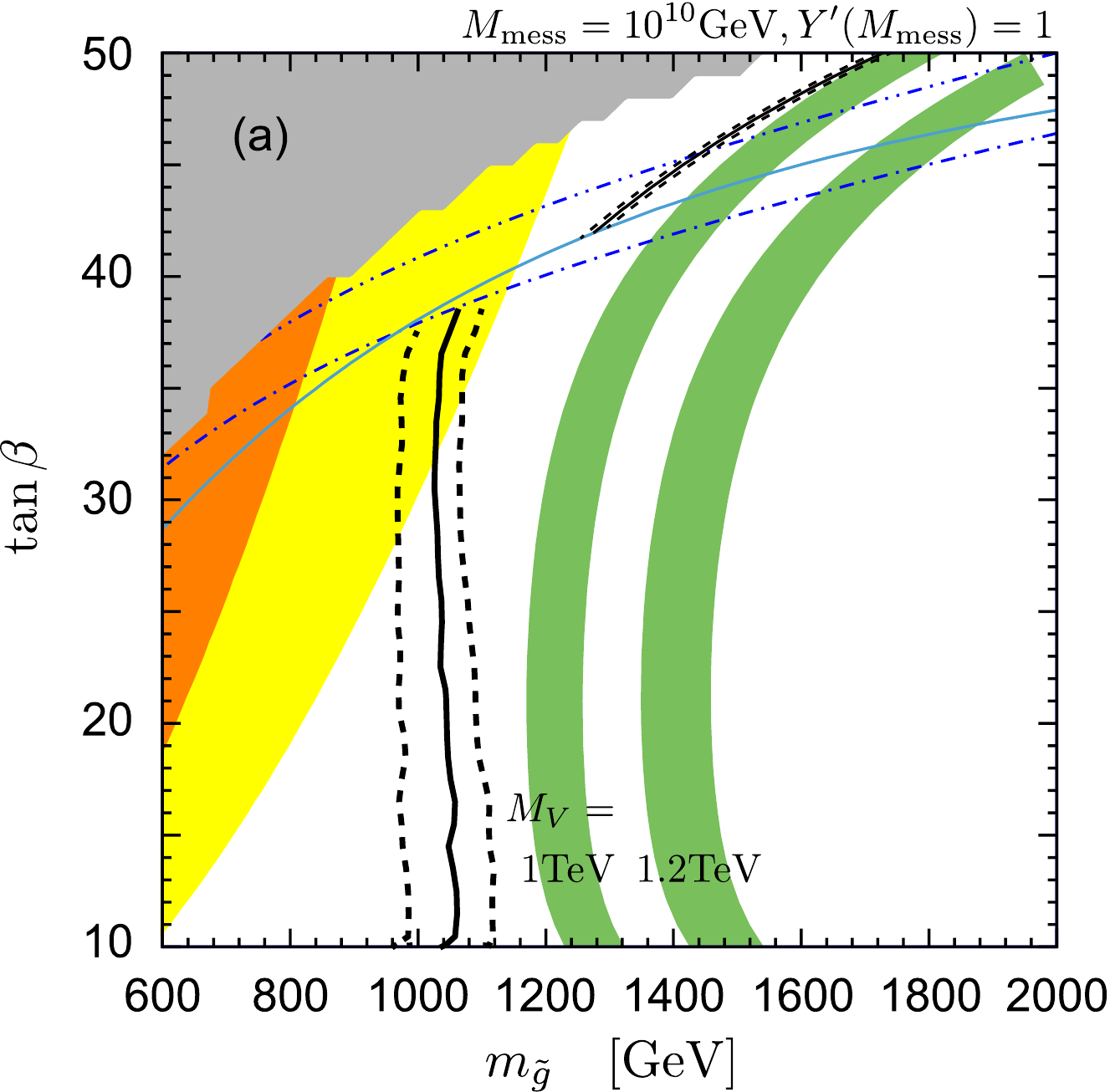}
 \end{center}\end{minipage}
 \begin{minipage}[t]{0.49\textwidth}\begin{center}
  \includegraphics[width=0.99\linewidth]{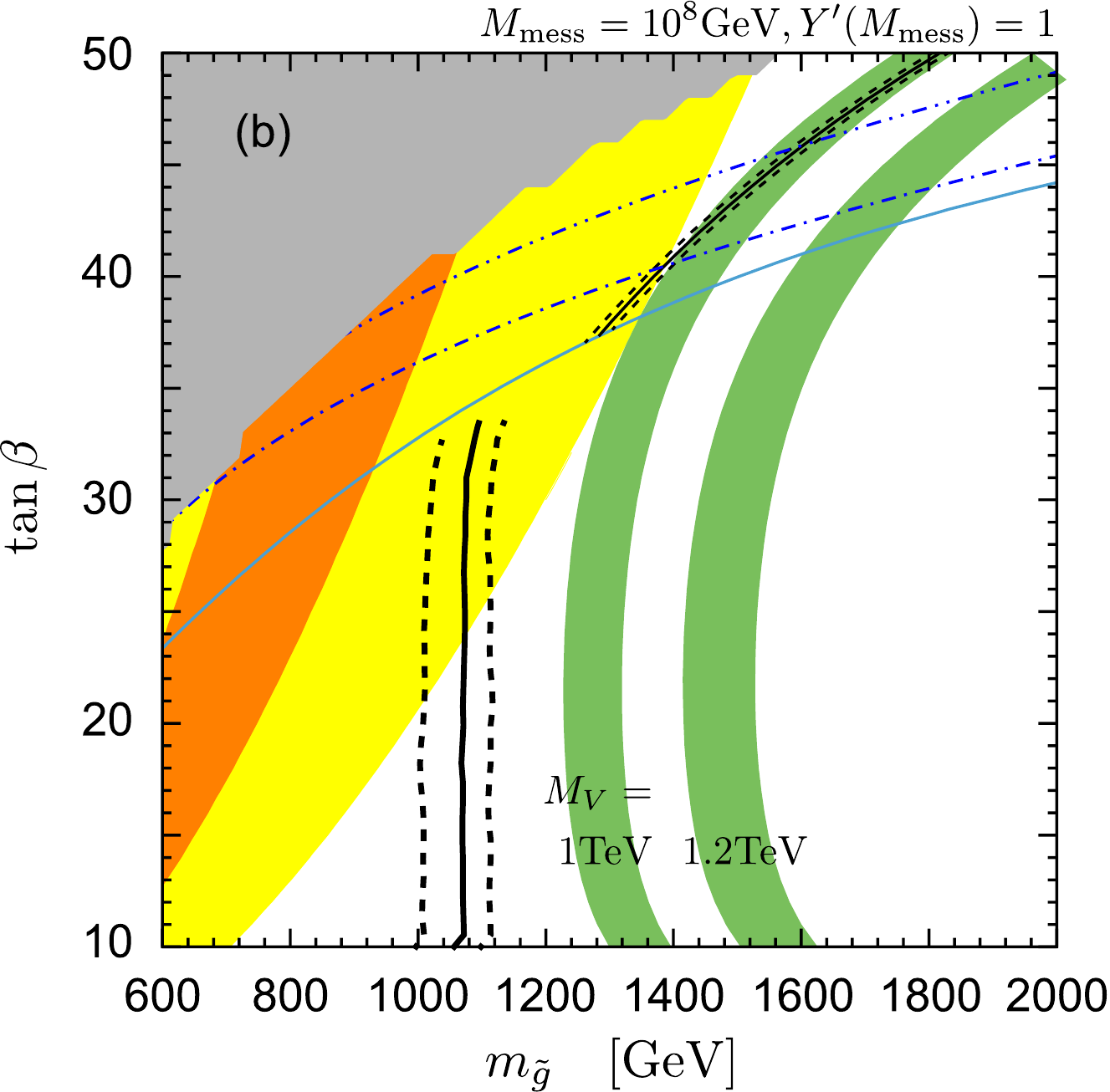}
 \end{center}\end{minipage}

 \begin{minipage}[t]{0.49\textwidth}\begin{center}
  \includegraphics[width=0.99\linewidth]{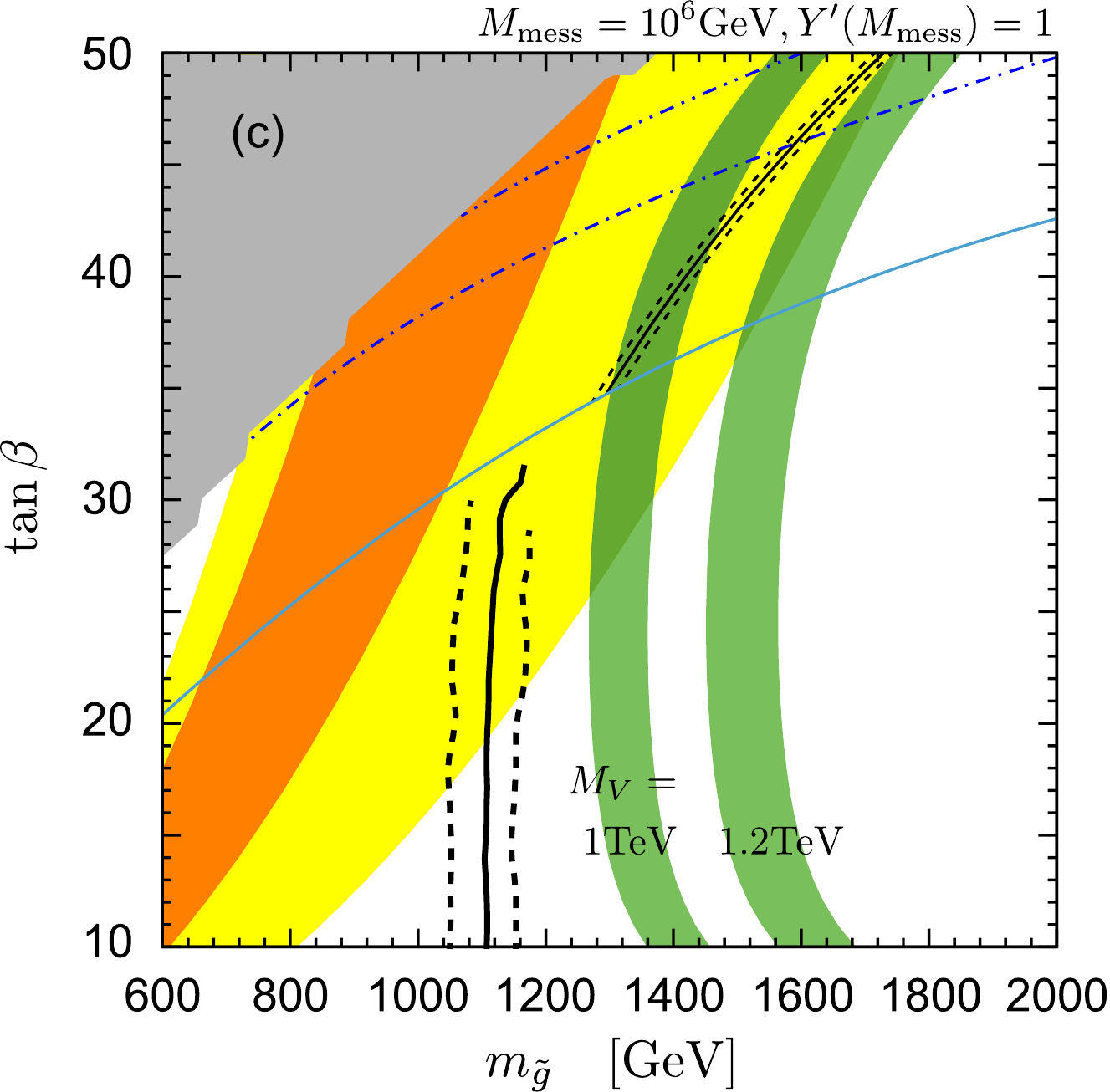}
 \end{center}\end{minipage}
 \begin{minipage}[t]{0.49\textwidth}\begin{center}
  \includegraphics[width=0.99\linewidth]{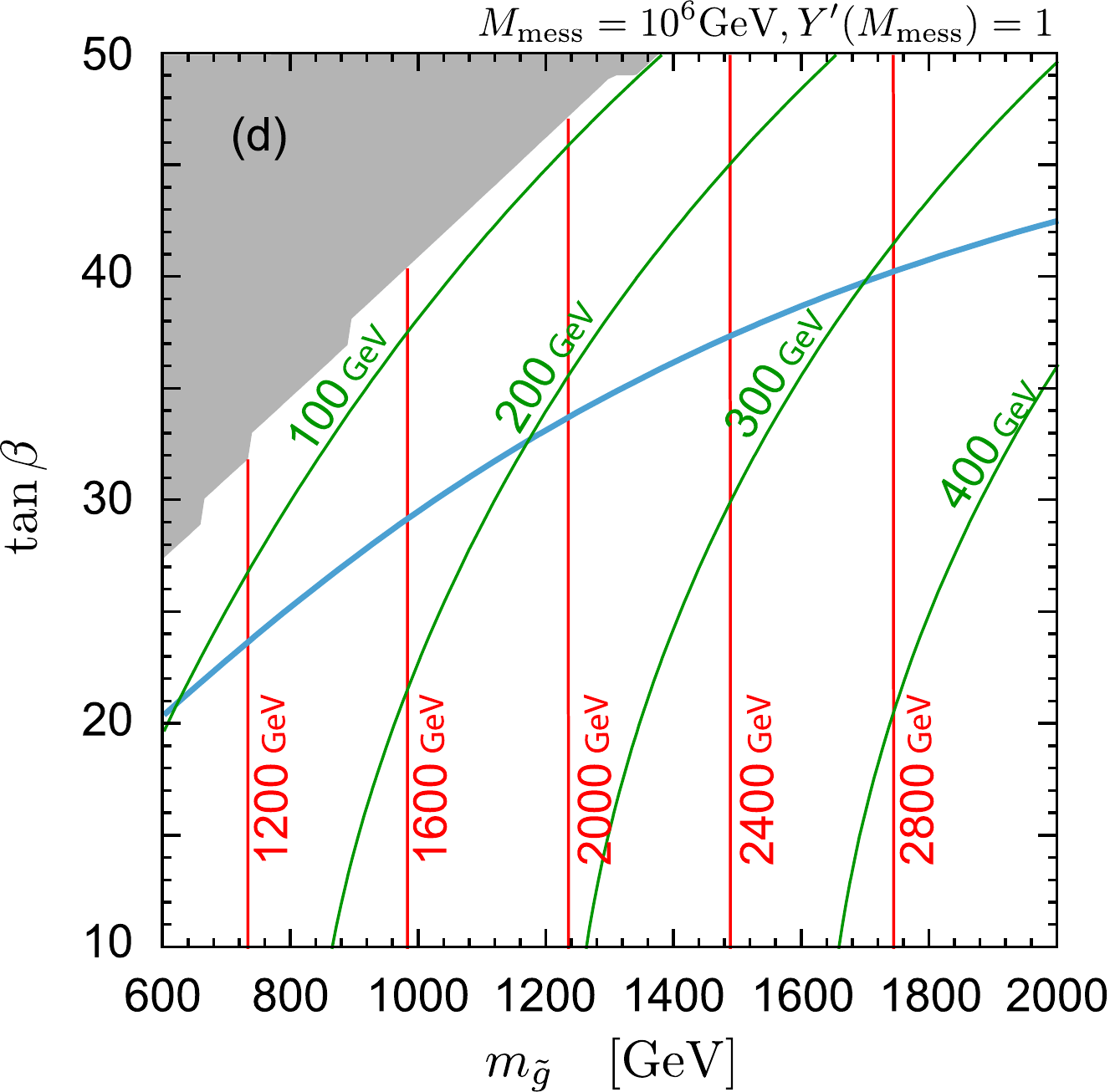}
 \end{center}\end{minipage}
\caption[The conclusive plot of our analysis.]{(a)--(c)
Contours of the Higgs boson mass, the muon $g-2$ and the LHC constraints in the V-GMSB model
are shown for 
(a) $M_{\rm mess}=10^{10}\xGeV$,
(b) $10^{8}\xGeV$, and 
(c) $10^{6}\xGeV$.
In the green bands, the Higgs boson mass is $125\text{--}126\xGeV$ for $M_V=1\xTeV$ and $1.2\xTeV$.
The yellow (orange) regions show the parameter spaces where the muon $g-2$ discrepancy is explained within $2\sigma$ ($1\sigma$)-level by the SUSY contribution.
The light blue lines indicate $m_{\neut}=m_{\tilde\tau_1}$; the NLSP is the lightest neutralino below the lines, and the lighter stau above them.
The regions above the blue dot-dashed lines are constrained by the vacuum stability condition~\eqref{eq:mubound_new}; as a reference the bound which is weakened by 10\% is drawn with the blue double-dotted long-dashed lines.
The LHC constraints, discussed in Sec.~\ref{sec:vmssm-lhc}, are drawn with the black solid lines without theoretical uncertainties, where the regions left to the lines are excluded.
The black dashed lines indicate the LHC bounds with theoretical uncertainties: 35\% error of the production cross section is adopted for the neutralino NLSP, and 2\% error in terms of the stau mass is for the stau NLSP.
The gray shaded regions are excluded by the failure of the {\tt SOFTSUSY} calculation.\quad
(d) Masses of the SUSY particles under $M\s{mess}=10^6\xGeV$.
The green contours are for the lighter stau, and the red lines for the lightest squark among the first two generations. The light blue line indicates $m_{\neut}=m_{\tilde\tau_1}$.
}
\label{fig:results1}
\end{center}
\end{figure}

Now we are ready to discuss numerical results: we are about to see that the V-GMSB model explains the $126\GeV$ Higgs boson mass together with explaining the muon $g-2$ anomaly.

The current status of the V-GMSB model is summarized in Fig.~\ref{fig:results1}, together with the current LHC bounds discussed in Sec.~\ref{sec:vmssm-lhc}. (See also Fig.~\ref{fig:results2}.)
The mass spectrum including the Higgs mass is calculated with the procedure summarized in Sec.~\ref{sec:vmssm-mass}, where \withpackage[3.3]{SOFTSUSY}~\cite{SOFTSUSY} and \withpackage[2.9]{FeynHiggs}~\cite{FeynHiggs} are utilized.
The muon $g-2$ is calculated with \withpackage{FeynHiggs}.
$Y'=1.0$ is set at the messenger scales, $M_{Q'}=M_{U'}=M_{E'}=:M_V$ is set at the low-energy scale, and $Y''=A''=0$ is assumed.
The values of important Standard Model parameters are set as $\alpha_s(m_Z)=0.1184$ and $m_t=173.5\GeV$.

In Fig.~\ref{fig:results1}, contours of the Higgs boson mass and the muon $g-2$ are drawn in the $(m_{\tilde{g}}, \tan\beta)$-plane for the messenger scales of  (a) $M_{\rm mess}=10^{10}\GeV$, (b) $10^{8}\GeV$, and (c) $10^{6}\GeV$.
Here, $m_{\tilde{g}}$ is the gluino pole mass, which is mainly determined by $\Lambda$, and $\tan\beta$ is an input evaluated at the electroweak scale as usual.
As the manner discussed in Sec.~\ref{sec:vac}, the vacuum stability bound is imposed with Eq.~\eqref{eq:mubound_new}.
In each of the figures, it is drawn with the blue dot-dashed line; the region above the line is disfavored.
Above the line, the blue double-dotted long-dashed line is drawn as the vacuum stability condition which is weakened by 10\%.
The black solid and dashed lines show the LHC constraints, which we will discuss in detail in Sec.~\ref{sec:vmssm-lhc}.
The NLSP is the lightest neutralino below the light blue line, and the lighter stau above the line; the LSP is the gravitino in this model.
In the gray shaded region, {\tt SOFTSUSY} fails to calculate the stau mass. Note that such a parameter region is experimentally excluded as long as the NLSP (the lighter stau) is long-lived.

In the green bands in Figs.~\ref{fig:results1}(a)--(c), the Higgs boson mass takes a value of $125\text{--}126\GeV$ for reference values of SUSY-invariant mass of the vector-like matter of  $M_V=1\TeV$ and $1.2\TeV$.
The Higgs boson mass increases as the vector-like matter becomes lighter  (cf.~Sec.~\ref{sec:vmssm-higgs}).
Actually, the Higgs boson mass of $125\text{--}126\GeV$ can be realized in the whole parameter region of Fig.~\ref{fig:results1} by changing the vector-like matter mass.
For fixed $M_V$, the Higgs boson becomes lighter when the gluino mass is smaller, because the radiative corrections to the Higgs potential from stops and vector-like stops decrease.
On the other hand, for higher messenger scale, the stop mass becomes larger during the renormalization group evolution, and thus the Higgs boson mass of $125\text{--}126\GeV$ is realized with a smaller gluino mass.

On the contrary to the mass of the Higgs boson, masses of the MSSM superparticles and the SUSY contributions to the muon $g-2$ are not sensitive to $M_V$.
We have checked that the MSSM superparticle masses change only by $\lesssim 2$\% when $M_V$ is varied from $500\GeV$ to $1\TeV$, and the muon $g-2$ changes by less than 1\%, correspondingly. In the figures, we fix $M_V=1\TeV$ in the calculations of all the quantities except for the Higgs boson mass.

In the yellow (orange) regions in Figs.~\ref{fig:results1}(a)--(c), the muon $g-2$ discrepancy, Eq.~\eqref{eq:muon_g-2}, is explained at the $2\sigma$ ($1\sigma$) level, i.e.,
\begin{align*}
 10.1\EE{-10} &< [\Delta a_\mu]\s{SUSY} < 42.1\EE{-10}\quad \text{in the yellow regions},\\
 18.1\EE{-10} &< [\Delta a_\mu]\s{SUSY} < 34.1\EE{-10}\quad \text{in the orange regions}.
\end{align*}
These regions are sensitive to both the gluino mass, i.e., the soft mass scale $M\s{SUSY}$, and $\tan\beta$~(cf.~Sec.~\ref{sec:muon-g-2-mssm}).
It should be stressed that the gluino mass, and thus $M\s{SUSY}$, has an upper bound for successful explanation of the muon $g-2$ under the vacuum stability condition, which we will discuss again in Sec.~\ref{sec:vmssm-disc} (see Fig.~\ref{fig:results2}).
This bound becomes tighter for a larger $M_{\rm mess}$.
The principal reason is that the renormalization group evolution, which raises the soft scalar masses during the running down from the messenger scale to the low-energy scale, works more significantly under a larger $M\s{mess}$.
In addition, as the gauge couplings are larger at higher energy scales in this model (cf.~Fig.~\ref{fig:vmssm-gcu}), a larger $M\s{mess}$ results in slightly heavier soft scalar masses at $M\s{mess}$.
Thus heavier squarks and sleptons are provided for a larger $M\s{mess}$, and the SUSY contributions to the muon $g-2$ are suppressed, in a fixed value of the gluino mass.

As we shall see in the next section,  LHC searches for the SUSY particles depend on the species of the NLSP.
In Figs.~\ref{fig:results1}(a)--(c), the NLSP is the lightest neutralino (the lighter stau) below (above) the light blue lines.
The stau NLSP region becomes smaller as $M_{\rm mess}$ increases, for the renormalization group running, which increases the sfermion masses relative to those of the gauginos, works more significantly.
For SUSY searches at the LHC, the masses of the gluino, squarks, and the stau are particularly important.
For illustration, we show in Fig.~\ref{fig:results1}(d) the masses of the lighter stau $\tilde\tau_1$ and the lightest squark among the first two generations with green and red lines, respectively, for $M_{\rm mess}=10^6\GeV$.
The squark mass is almost independent of $\tan\beta$, since it is governed by the strong interaction.
On the other hand, the stau mass, which determines the LHC constraint in the parameter region with the stau NLSP, depends on the gluino mass (the soft mass scale), and on $\tan\beta$ through the tau \YUKAWA\ coupling $Y_\tau\approx (m_\tau/v)\tan\beta$.

An interesting feature of the V-GMSB model is that the mass parameter $M_V$ should be $\lesssim1.2\TeV$ in order to explain the muon $g-2$ discrepancy at the $2\sigma$-level together with the $126\GeV$ Higgs boson.
This behaviour can be understood from Eq.~\eqref{eq:VMSSM-higgs-simp}; for larger $M_F(=M_V)$ in the expression the SUSY-breaking mass terms, $M_S$ must be larger to realize enough increase, which means the masses of the sleptons and the gauginos get larger, and the shift of the muon $g-2$ decreases.
Therefore, in this context the vector-like {\em quark} mass should be $\lesssim1.2\TeV$, and this fact makes searches for vector-like quarks very interesting.
We will discuss this topic in Sec.~\ref{sec:vector}, after the discussion on the LHC bounds drawn in the figures.

\section{LHC Bounds on the V-GMSB Model}
\label{sec:vmssm-lhc}
In this section the constraints on the V-GMSB model from LHC SUSY searches are discussed.
Actually the bounds are already shown in Fig.~\ref{fig:results1}; here the procedure to obtain those bounds and interpretations of them are provided.
The V-GMSB model is also constrained by searches for the vector-like quarks, but the constraints are discussed in Sec.~\ref{sec:vector}.

\subsection{Overview}
In the V-GMSB model, the gravitino is the LSP, and collider signatures are determined by species and the lifetime of the NLSP.
As discussed in the previous section, the NLSP is either the lightest neutralino or the lighter stau in the V-GMSB model, which are respectively realized below and above the light blue lines in Fig.~\ref{fig:results1}.
Let us summarize the cases.

\begin{description}
 \item[Long-lived Neutralino NLSP:]
  The promising signature is multi-jets plus large missing energy ($\MET$) yielded from pair-production of colored particles, where the multi-jets are provided by cascade decays of the colored particles and the missing energy is from the NLSPs.

  The ATLAS collaboration reported a search for this channel using the data corresponding to an integrated luminosity of $5.8\invfb$  obtained at $E\s{CM}=8\TeV$~in Ref.~\cite{ATLAS2012109}, and obtained a lower bound on the gluino mass for the case where the squarks are heavy, which is expected in the V-GMSB model, as $m_{\tilde g}\gtrsim 900\GeV$.
  The CMS also analyzed their $(7\TeV, \sim5\invfb)$ data to obtain similar results~\cite{chatrchyan:2012jx,chatrchyan:2012mfa}.

 \item[Neutralino NLSP with prompt decay:]
  In this case the above signature, multi-jets plus large missing energy, is still expected.
  Nevertheless, as the NLSP can decay as $\neut\to\gamma+\tilde G$, di-photon signature with large missing energy is promising.

  For the decoupled squark scenario, the CMS collaboration excluded $m_{\glu}\lesssim1.1\TeV$ for the bino-like NLSP case, and $m_{\glu}\lesssim750\GeV$ for the wino-like NLSP case, with $(8\TeV,4\invfb)$ data~\cite{CMSPASSUS12018}, and also obtained similar results from $(7\TeV,4.3\invfb)$ data~\cite{Chatrchyan:2012bba}; both analyses are based on jet(s) + photon(s) + $\MET$.
  Note that the NLSP is bino-like in the V-GMSB model for the large $\mu$-term.

  The ATLAS collaboration excluded gluinos with $m_{\glu}\lesssim1.1\TeV$ $(\lesssim 950\GeV$) for the bino- (wino-) like NLSP case in the limit where the squarks are decoupled; these bounds are based on jet(s) + di-photon + $\MET$ signature with $(7\TeV,4.8\invfb)$ data~\cite{aad:2012afb}.
  They also employed the di-photon search without jet requirement~\cite{aad:2012afb}; this search focuses the electroweak production of the SUSY particles: $pp\to\tilde\chi\tilde\chi$ etc., and is interesting because it covers the case where the colored particles are extremely heavy and not accessible with the $14\TeV$.

  If the neutralino is quasi-long-lived and decays during the flight in the detector, typical signature at the LHC would be a non-pointing photon~\cite{Kawagoe:2003jv} and a neutralino in-flight decay into a $Z$-boson~\cite{Meade:2010ji}. Since the V-GMSB has an upper bound on the gluino mass $m_{\tilde{g}}\lesssim 1.2\TeV$ (1.8\,TeV) in the neutralino NLSP case once a $125\TO126\GeV$ Higgs and the muon $g-2$ constraints at $1\sigma$ ($2\sigma$) are imposed, it is expected that a large part of the parameter space can be judged also in this case.

 \item[Long-lived stau NLSP:]
  In this case the NLSP does not yield large $\MET$, but is observed as a heavy stable charged particle (HSCP), where the term ``stable'' is used from the experimental viewpoint, i.e.\ it means long-lived enough to escape from the detectors.

  The CMS collaboration reported a lower bound on the mass of the lighter stau as
  \begin{equation}
   m_{\tilde \tau_1}>223\GeV\label{eq:stau223}
  \end{equation}
  with analyzing their data of $(7\TeV,5.0\invfb)$~\cite{Chatrchyan:2012sp}, where the (lighter) stau is assumed to be produced only with the direct production, $pp\to\tilde \tau_1\tilde \tau_1^*$.
  It should be emphasized that this bound is {\em generic}, i.e., is most conservative and can be adopted to any models, because it targets the lighter stau pair-production; one should however note that the production rate depends on the stau mixing.

  The CMS collaboration also published constraints on the stau mass for typical GMSB models, where charginos and neutralinos involve the main production channels~\cite{Chatrchyan:2012sp}.
  However, it cannot be directly applied to the V-GMSB case, because the mass spectra are different  (cf. Fig.~\ref{fig:spectrum}).
  The difference results in a deviation of the velocity distribution of the NLSP staus.
  Once the reconstruction efficiency of the HSCPs is published, the production channels of the charginos and neutralinos can be considered, which would provide much more strict constraints.

  The ATLAS collaboration also published a result on searches for the HSCPs~\cite{aad:2012pra}, but their search {\em seems to} assume direct productions of the whole sleptons, i.e., the production is not limited to the lighter stau, and the reconstruction efficiency of the particle is not provided either.
  Therefore, their results cannot be applied to the V-GMSB.

 \item[Stau NLSP with prompt decay:]
  In this case the promising channel is multi-taus + $\MET$.
  The ATLAS collaboration, utilizing their $(7\TeV,4.7\invfb)$ data, searched for the events with $\ge0$ jet + $\ge1$ tau + 0--1 lepton + $\MET$, and constrained the parameter space of GMSB scenario with $(M\s{mess},N\s{mess},\sgn\mu)=(250\TeV,3,+)$ to obtain the bound of $m_{\glu}\gtrsim1.0\TeV$, which corresponds to $m_{\tilde \tau}>160\GeV$--$70\GeV$ depending on $\tan\beta$ (of 2--60).
  Nevertheless, this result cannot be applied to the V-GMSB model because, especially, the stau is relatively heavier in the V-GMSB model as is already discussed.

  An in-flight-decay of a stau leaves kink signature in the TRT detector, and is expected to be observed if the decay length is $\sim 1\un{m}$ ~\cite{Asai:2011wy}; this feature is the same as the ordinary GMSB scenario.
\end{description}

\starline

In this dissertation we concentrate on the long-lived NLSP scenario.
The LHC exclusion limits in Fig.~\ref{fig:results1} are drawn under this assumption, i.e., they are based on the analysis for the first and the third case in the above listing.
In the following of this section the detailed procedure in the evaluation of the bounds is explained.

\subsection{LHC constraint for the V-GMSB with long-lived neutralino NLSP}
\label{sec:neutralinoNLSP}
For the region with a neutralino NLSP, we interpret the ATLAS result of the search for the superparticles in events with no lepton, 2--6 jets and missing energy, with data obtained at the LHC with $E\s{CM}=8\TeV$ corresponding to an integrated luminosity of $5.8\invfb$~\cite{ATLAS2012109}, as a constraint for the V-GMSB model.
The detailed explanation of the Monte Carlo analysis is given here.

\subsubsection{Data sample and triggering}
The ATLAS collaboration utilized their data taken in 2012 corresponding to a total integrated luminosity of $5.8\invfb$ obtained at the $8\TeV$ LHC.
Their trigger requirement is a jet with $\PT>80\GeV$ and the missing energy $\MET>100\GeV$.\footnote{%
The transverse momentum $\vPT$ and the transverse missing energy are defined as
\begin{equation}
 \vPT:=\left(p_x,p_y\right),\qquad \PT:=\|\vPT\|;\qquad\qquad
 \vMET:=\left(\cancel{E}_x, \cancel{E}_y\right), \qquad \MET=\left\|\vMET\right\|.
\end{equation}
}
Since the event selection, discussed later, includes requirements of a jet with $\PT>130\GeV$ and missing energy of $\MET>160\GeV$, the trigger efficiency is at the efficiency plateau as the ATLAS collaboration wrote ``full efficiency'' in the report.
Therefore, in our analysis the trigger efficiency is not considered.

Our data are obtained with \withpackage[5]{MadGraph}~\cite{MadGraph5} package.
It includes \withpackage[6.4]{Pythia}~\cite{Pythia6.4} to simulate parton shower and initial- and final state radiation (ISR and FSR), and \withpackage[2.0]{Delphes}~\cite{Delphes} for detector simulation; both of them are utilized in our analysis.
The parton distribution functions (PDFs) are obtained from the {\tt CTEQ6L1} set~\cite{PDFCTEQ6}.
{\tt Pythia} setting is based on the {\tt MadGraph} default.

\subsubsection{Simulated events}
The events in $pp\to \tilde g\tilde g, \tilde g\tilde q^{(*)},\tilde q^{(*)}\tilde q^{(*)}$ channels, which are relevant for the SUSY signals, are generated, but among them the production channels with $\tilde t$ and $\tilde b$ are neglected, whose production cross sections are less than a few \% of the total cross section.

The V-GMSB mass spectrum is generated with the customized {\tt SOFTSUSY 3.3}~\cite{SOFTSUSY} as is explained in Sec.~\ref{sec:vmssm-mass}, and passed to {\tt SUSY-HIT 1.3}~\cite{SUSYHIT} to calculate the decay pattern of the SUSY particles and the Higgs bosons.
The generated events are normalized to the NLO cross section obtained with {\tt Prospino 2.1}~\cite{Prospinoweb,ProspinoSG}, where the {\tt CTEQ6L1} and the {\tt CTEQ6.6M} PDFs~\cite{PDFCTEQ6} are used.

\subsubsection{Object reconstruction}
Our detector simulation is based on the \withpackage[2.0]{Delphes} package.
The default parameter set for the ATLAS experiment, which can be found in {\tt DetectorCard\_ATLAS.dat} in {\tt Delphes} package, is used for energy resolutions and calorimeter tile configurations.

Jets are reconstructed using the anti-$k_t$ jet clustering algorithm~\cite{anti-kt} with the distance parameter of $0.4$, where {\tt Delphes} utilizes \withpackage[2.3]{FASTJET}~\cite{FASTJET}.
For the missing energy, the {\tt Delphes} vanilla outputs are used.

In the electron reconstruction, the ATLAS collaboration uses the ``medium'' criterion.
The detection efficiency is reported in Ref.~\cite{Aad:2011mk}, which are already introduced in Sec.~\ref{sec:atlas-object}.
However, the report is based on the data taken at the $7\TeV$ LHC, and no efficiencies at the $8\TeV$ LHC are public.
Therefore, assuming that the efficiencies at the $8\TeV$ LHC are the same as that at the $7\TeV$, the efficiency used in the analysis is taken from Ref.~\cite{Aad:2011mk}.
In principle, the efficiency is, and should be expressed as, a function of $(\PT,\eta)$.
However, such functions are not reported yet due to limitation of the data; they instead reported the efficiency as two separated functions of $\PT$ and of $\eta$ with unignorable uncertainties.
Thus, for simplicity, following efficiency is adopted for the electron reconstruction in our analysis:
\begin{itemize}
 \item The reconstruction efficiency as a constant: $\alpha\s{reco}=0.943$~\cite[Fig.~22(B)]{Aad:2011mk}.
 \item The identification efficiency for ``medium'' as a constant for $\PT>20\GeV$: $\epsilon\s{ID}\suprm{med}=0.942$~\cite[Table~7]{Aad:2011mk}.
\end{itemize}
 \begin{rightnote}
  The identification efficiency significantly drops below $\PT=20\GeV$; for such electrons the efficiency is set as $(\alpha\s{reco},\epsilon\s{ID}\suprm{med})=(0.943,0.8)$ in our simulation set, although $\alpha\s{reco}$ is not reported for this $\PT$ region.
  Note that, however, this efficiency is not used in this analysis, because electrons below $\PT=20\GeV$ are discarded.
 \end{rightnote}

The muon in this analysis seems to be reconstructed with the ``combined'' method, which is explained in Sec.~\ref{sec:atlas-object}.
The corresponding efficiency is found in Ref.~\cite{ATLAS2011063} for muons with $\PT>20\GeV$, which is also based on the $7\TeV$ data.
Our analysis adopts the following simplified efficiency for the muon detection:
\begin{itemize}
 \item The efficiency for inner detectors as a constant: $\epsilon\s{ID}=0.88$~\cite[Fig.~4]{ATLAS2011063}.
 \item The efficiency for muon spectrometer in combined muon method as: $\epsilon\s{MS}\suprm{CB}=0.973$ for $|\eta|>0.25$ and $0.828$ for $|\eta|<0.25$~\cite[Fig.~6]{ATLAS2011063}. This difference comes from the fact that the muon spectrometer is only partially equipped in this region.
\end{itemize}

Consequently, the lepton reconstruction efficiencies are defined as
\begin{equation}
 \epsilon(e)=\alpha\s{reco}\cdot\epsilon\s{ID}\suprm{med}\approx 89\%,\qquad\epsilon(\mu)=\epsilon\s{ID}\cdot\epsilon\s{MS}\suprm{CB}\approx96\%~(82\%) \text{ for $|\eta|>0.25$ ($<0.25$).}
\end{equation}
in our analysis.

 \begin{rightnote}
Actually these efficiencies are prepared for another analysis; in the analysis the events with three leptons are simulated and analyzed, and this set of efficiencies reproduces the ATLAS analysis in Ref.~\cite{ATLAS2012154} very well.
\end{rightnote}

\starline

Only the jets with $\PT>20\GeV$ and $|\eta|<2.8$, the electrons with $\PT>20\GeV$ and $|\eta|<2.47$, and the muons with $\PT>10\GeV$ and $|\eta|<2.4$ are considered.
For overlap removal, jets are rejected if electrons are found within a distance of $\Delta R = 0.2$ from the jet, and then leptons within $\Delta R = 0.4$ of any jets are discarded.
Here, $\PT$ is a missing transverse momentum, $\Delta R := \sqrt{(\Delta \eta)^2 + (\Delta \phi)^2}$ is a distance parameter between two objects, and $\eta$ and $\phi$ are pseudo-rapidity and azimuthal angle around the beam direction, respectively.

 \begin{table}[p]\def\arraystretch{1.4}
\begin{center}
   \caption[Definition of the 12 signal regions for the jets+$\MET$ search.]{The definition of the 12 {\em inclusive} signal regions (SRs), which is the same as the ATLAS original analysis~\cite{ATLAS2012109}.
The effective mass $m\s{eff}^{(n)}$ ($m\s{eff}\suprm{inc}$) is defined as the scalar sum of $\MET$ and $\PT$'s of the leading $n$-jets (all jets with $\PT>40\xGeV$).
Some categories have two or three SRs, which are shown in the final two rows; they are called `tight', `medium', and `loose', respectively.
}
  \label{tab:2012-109}
 \begin{tabular}[t]{|c|c|c|c|c|c|}\hline
       & \multicolumn{5}{|c|}{Signal Regions} \\\cline{2-6}
   & A & B & C & D & E \\
       & ($\ge2$-jets) & ($\ge3$-jets) & ($\ge4$-jets) & ($\ge5$-jets) & ($\ge6$-jets)\\\hline
 $\#$ leptons & \multicolumn{2}{|c}{}&\multicolumn{1}{c}{$=0$}&\multicolumn{2}{c|}{} \\\hline
 $\MET$ [GeV] $>$ & \multicolumn{2}{|c}{}&\multicolumn{1}{c}{160}&\multicolumn{2}{c|}{} \\\hline
 $\PT(j_1)$ [GeV]~~$>$ & \multicolumn{2}{|c}{}&\multicolumn{1}{c}{130}&\multicolumn{2}{c|}{} \\\hline
 $\PT(j_2)$ [GeV]~~$>$ & \multicolumn{2}{|c}{}&\multicolumn{1}{c}{60}&\multicolumn{2}{c|}{} \\\hline
 $\PT(j_3)$ [GeV]~~$>$ & --- & 60 & 60 & 60 & 60 \\\hline
 $\PT(j_4)$ [GeV]~~$>$ & --- &--- & --- & 60 & 60 \\\hline
 $\PT(j_5)$ [GeV]~~$>$ & --- & --- & --- & 60 & 60 \\\hline
 $\PT(j_6)$ [GeV]~~$>$ & --- & --- & --- & --- & 60 \\\hline
 $\Delta\phi(j_i, \vMET)\s{min}$~~$>$ &
    $0.4(i=1,2)$ &     $0.4(i=1,2,3)$ & \multicolumn{3}{|c|}{$0.6~(i\le 3)$;\quad $0.4$ ($\PT>40\GeV$ jets)}\\\hline
 \multirow{2}{*}{$\MET/m\s{eff}^{(n)}$~~$>$}&
   0.3~/~0.4~/~0.4 & 0.25~/~0.3~/~--- & 0.25~/~0.3~/~0.3 & 0.15~/~---~/~--- & 0.15~/~0.25~/~0.3 \\
  & $(n=2)$  & $(n=3)$  & $(n=4)$  & $(n=5)$  & $(n=6)$\\\hline
 $m\s{eff}\suprm{inc}$ [TeV] ~~ $>$&
   1.9~/~1.3~/~1.0 & 1.9~/~1.3~/~--- & 1.9~/~1.3~/~1.0 & 1.7~/~---~/~--- & 1.4~/~1.3~/~1.0 \\\hline
  \end{tabular}
\end{center}
\end{table}

\subsubsection{Signal regions}
Our analysis is employed with the same definition of the 12 inclusive signal regions (SRs) as the original analysis~\cite{ATLAS2012109}, which is summarized in Table~\ref{tab:2012-109}.
However, the selections based on jet quality selection criteria and primary vertex reconstruction are not employed due to the simplified detector simulation.

The events with any reconstructed electrons or muons are vetoed.
At least two jets with $\PT>60\GeV$ are required, and the leading jet must be with $\PT>130\GeV$.
The SRs are classified with the number of jets having $\PT>60\GeV$.
The definition is inclusive; for example, if an event has four jets with $\PT>60\GeV$, it can be a candidate for the events in the SRs A, B, and C.

Two requirements designed to reduce the background events from multi-jet processes are imposed.
One is on the azimuthal separations between $\vMET$ and jets, which is expressed as $\Delta\phi(j_i,\vMET)\s{min}$ in Table~\ref{tab:2012-109}.
For the SR A (B), leading two (three) jets are considered for this selection.
For the others, leading three jets must have a separation of at least 0.4, and all the other jets with $\PT>40\GeV$ must at least $0.2$.
The other requirement is on the ratio between $\MET$ and $m\s{eff}^{(n)}$, which is defined as
\begin{equation*}
 m\s{eff}^{(n)} := \MET + \sum_{i=1}^{n}\left\|\vPT{}\text{\footnotesize~of $i$-th leading jet}\right\|.
\end{equation*}

Finally, the inclusive effective mass is utilized, which is defined as
$ m\s{eff}\suprm{inc} :=m\s{eff}^{(N)}$, where $N$ is the number of jets with $\PT>40\GeV$.

 \begin{table}[t]\def\arraystretch{1.3}\catcode`?=\active \def?{\phantom{.}}\catcode`@=\active \def@{\phantom{0}}
\begin{center}
 \caption[The result of the ATLAS jet+$\MET$ search.]{The result of the ATLAS analysis in Ref.~\cite{ATLAS2012109}. The last two columns show the 95\% upper limits (UL) on the excess number of events, $N\s{BSM}$, and that on the cross section of the new physics, $\sigma\s{BSM}$.
 One should note that the experimental uncertainties are already considered in the calculation of the upper limits.}
 \label{tab:atlasresult}
   \begin{tabular}[t]{|c|c|c|c|c|}\hline
   SR    & Background & Observed & UL on $N\s{BSM}$ & UL on $\sigma\s{BSM}$ [fb] \\\hline
 A-loose & $ 650\pm 130$ & 643 & 224.8 & 38.8@ \\
 A-medium& $ 140\pm @33$ & 111 & @33.9 & @5.84 \\
 B-medium& $ 115\pm @30$ & 106 & @43.8 & @7.55 \\
 C-loose & $ 155\pm @31$ & 156 & @65.7 & 11.3@ \\
 C-medium& $ @33\pm @@8$ & @31 & @17.9 & @3.09 \\
 E-loose & $@ 5.7\pm @1.7$ & @@9 & @10.4 & @1.79 \\
 E-medium& $@ 3.5\pm @1.7$ & @@7 & @@9.9 & @1.71 \\\hline
 A-tight & $ @14\pm @@5$ & @10 & @@8.9 & @1.53 \\
 B-tight & $@ 8.7\pm @3.4$ & @@7 & @@7.3 & @1.26 \\
 C-tight & $@ 2.8\pm @1.2$ & @@1 & @@3.3 & @0.57 \\
 D-tight & $@ 6.3\pm @2.1$ & @@5 & @@6.0 & @1.03 \\
 E-tight & $ @10\pm @@4$ & @@9 & @@9.3 & @1.60 \\\hline
 \end{tabular}
\end{center}
\end{table}

\subsubsection{Analysis procedure}
The ATLAS collaboration found no significant excesses; therefore the cross section of models beyond the Standard Model, $\sigma\s{BSM}$, receives an upper bound.
The expected number of the background events, the observed number, and the upper limits are summarized in Table~\ref{tab:atlasresult}.
This result is interpreted to obtain the constraints (shown in Fig.~\ref{fig:results1}) with the following procedure, so-called $\CL{s}$-method.

First, the {\em expected} sensitivity $\CLexp{s}$ is defined for each of the 12 SRs.
It is defined as
\begin{align}
  \CLexp{s}:= &~\frac{\CLexp{s+b}}{\CLexp{b}};\\
  \CLexp{b}&:=1-\left(\text{the probability that a random variable which obeys $f(N\s b,\sigma\s{b})$ exceeds $N\s{b}$}\right),\notag\\
  \CLexp{s+b}&:=1-\left(\text{the probability that  a random variable which obeys $f(N\s{s+b},\sigma\s{s+b})$ exceeds $N\s{b}$}\right),\notag\\
  N\s{b}&  := \left(\text{expected number of background events}\right),\notag\\
  N\s{s+b}&:= N_b + \left(\text{expected number of SUSY events}\right),\notag\\
  \sigma\s{b}&  := \left(\text{uncertainty of $N\s{b}$}\right),\notag\\
  \sigma\s{s+b}&  := \sqrt{\sigma\s{b}^2+\sigma\s{s}^2}, \text{where $\sigma\s{s}$ is the uncertainty of the number of SUSY events},\notag\\
  f(\mu,\sigma)&:=\mathop{\mathrm{Poisson}}\left(\mathop{\mathrm{Normal}}(\mu,\sigma)\right),\notag
\end{align}
where $\mathop{\mathrm{Poisson}}(x)$ is the Poisson distribution with the mean $x$, and $\mathop{\mathrm{Normal}}(\mu,\sigma)$ is the normal distribution with the mean $\mu$ and the variance $\sigma^2$.
As the uncertainties in the cross section of the SUSY events are not included in the procedure to obtain the upper limits (shown in Table~\ref{tab:atlasresult}) but are considered afterwards (cf.~Fig.~\ref{fig:msugracompare} etc.), our procedure uses $\sigma\s{s}=0$.
Here one should note that $\CLexp{s}$ is determined for each model point, and that the number of background events is {\em assumed for simplicity} to distribute with the normal distribution.
 \begin{rightnote}
$\CLexp{b}$ and $\CLexp{s+b}$ can be interpreted as the {\em expected} confidence level of the background-only and the signal-plus-background hypothesis, respectively, or the {\em expected} probability that the respective hypothesis describes the data.
Here, since the ``expected data'' are nothing but the expected number of background events, $\CLexp{b}$ and $\CLexp{s+b}$ are defined as the probability that the respective hypothesis describes the number of background.
 \end{rightnote}

The SR which gives the smallest $\CLexp{s}$ is expected to give the most stringent constraints on the model point, and therefore adopted as ``the SR {\em for the point}.''
In Figs.~\ref{fig:msugracompare} and \ref{fig:results1-reg}, which will be introduced later, which SR is selected for each model point is displayed.
Then, the {\rm observed} confidence level $\CL{s}$ is calculated for the selected SR, which is defined as
\begin{align}
  \CL{s}:= &~\frac{\CL{s+b}}{\CL{b}};\\
  \CL{b}&:=1-\left(\text{the probability that a random variable which obeys $f(N\s b,\sigma\s{b})$ exceeds $N\s{data}$}\right),\notag\\
  \CL{s+b}&:=1-\left(\text{the probability that  a random variable which obeys $f(N\s{s+b},\sigma\s{s+b})$ exceeds $N\s{data}$}\right),\notag\\
  N\s{data}&  := \left(\text{observed number of events}\right).\notag
\end{align}
If the $\CL{s}$, calculated for the selected SR, satisfies $\CL{s}<0.05$, the point is excluded at 95\% confidence level (CL). Otherwise, the point is not excluded.
It should be important that, although we have 12 SRs, the analysis based on the data is employed for a sole SR, and thus the condition $\CL{s}<0.05$ results in 95\% CL limit.
 \begin{rightnote}
 It is known that the $\CL{s}$-method generally gives a conservative limit, i.e., the obtained constraint is looser than the ``true'' one.
 \end{rightnote}

\subsubsection{Verification of our analysis}
\begin{figure}[p!]
\begin{center}
   \includegraphics[width=0.9\textwidth]{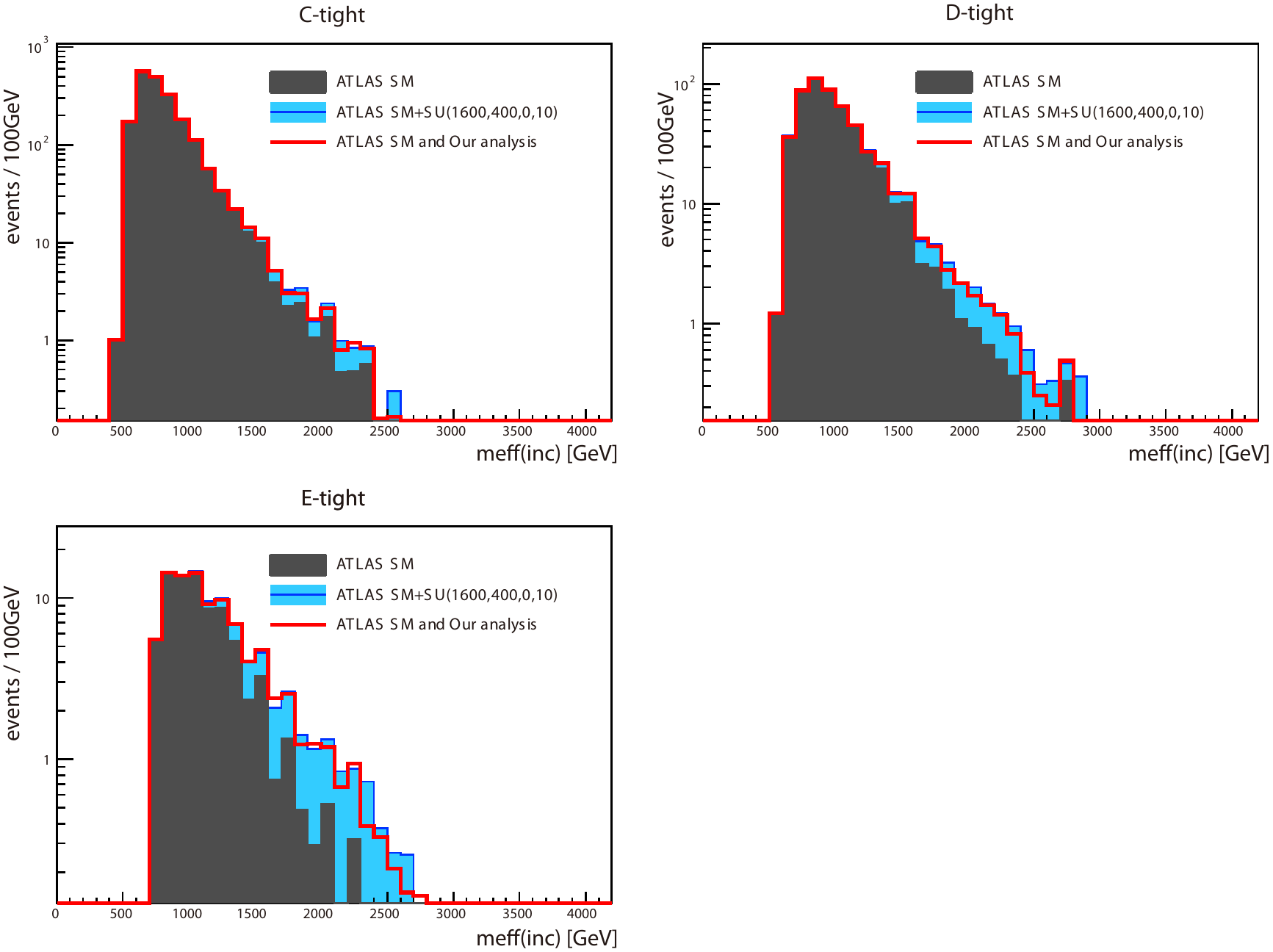}
  \caption[The distributions of the effective mass, used to certify our analysis.]{The $m\s{eff}\suprm{inc}$ distributions before the selection on itself.
 Here the lepton efficiencies are taken into account (cf.\ Fig.~\ref{fig:histocompare-noleff}).
As a benchmark model, the CMSSM with $(m_0, M_{1/2},A_0,\tan\beta,\sgn\mu)=(1600\xGeV, 400\xGeV, 0, 10, +)$ is chosen.
 Only the distributions for the SRs C-tight, D-tight, and E-tight are shown here.
 The results from the ATLAS Monte Carlo analysis~\cite{ATLAS2012109} are shown as the histograms; the gray are the events from Standard Model background, and the blue, stacked on the gray, are from the SUSY signal.
 The results based on our analysis are drawn with red lines, but since our analysis does not simulate background processes, the background results reported by the ATLAS collaboration are used as the background histogram for our analysis.
 These two results are compared to certificate our analysis.
  Note that the observed data are not shown here.
 }\label{fig:histocompare-leff}
\end{center} 
\end{figure}

To verify our analysis, the distributions of the inclusive effective mass, $m\s{eff}\suprm{inc}$, are compared.
The distributions for the SRs C-tight, D-tight and E-tight are displayed in Fig.~\ref{fig:histocompare-leff}, where the results from ATLAS Monte Carlo simulations and our Monte Carlo simulation are shown, in which all the selections but on $m\s{eff}\suprm{inc}$ itself are employed.
As the ATLAS collaboration provides only the histograms which include both the Standard Model and the SUSY contributions, the comparison was done between the ATLAS Standard Model simulation plus the ATLAS SUSY simulation (displayed with gray plus blue histograms), and, the ATLAS Standard Model simulation plus our SUSY simulation (displayed with red lines).
The results do agree well with slight deficit in our analysis.

\begin{figure}[p!]
\begin{center}
   \includegraphics[width=0.9\textwidth]{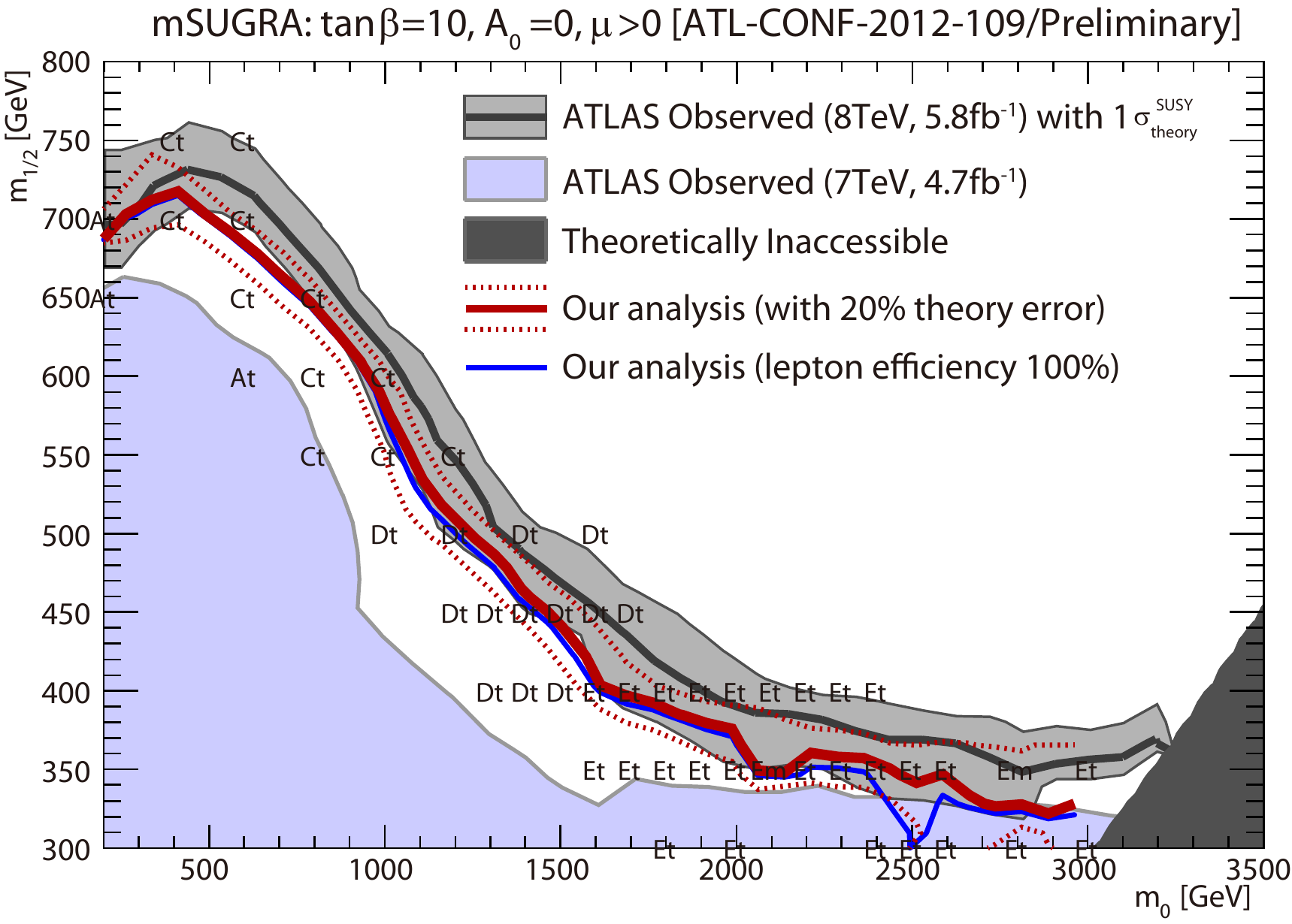}
  \caption[The CMSSM exclusion limit from ATLAS jets+$\MET$ search, used to certify our analysis.]{95\% CL exclusion limit for the CMSSM scenario with $(A_0,\tan\beta,\sgn\mu)=(0, 10, +)$ presented in the $(m_0,M_{1/2})$-plane.
 The black line with a gray band is the result reported by the ATLAS collaboration in Ref.~\cite{ATLAS2012109} with $\pm1\sigma$ uncertainty on the calculation of SUSY cross section.
 The red line is the result from our analysis, and the letters denote the SR which is adopted for the point. (See text for detail.)
 As a reference, the limits obtained in our analysis with $\pm20\%$ theoretical uncertainty are shown as the red dotted lines.
 Interpolation to obtain the curve is based on $\log\CL{s}$.
 The matching is not complete, but a decision was made to accept this result since it is conservative.
 The blue line describes the result from our analysis but with setting the lepton detection efficiencies as 100\%. It tightens the lepton veto, and slightly loosens the bound.
 {\bf This figure is based on a figure produced by the ATLAS collaboration and used in Ref.~\cite{ATLAS2012109}. (ATLAS Experiment \copyright 2012 CERN)}
 }
 \label{fig:msugracompare}
\end{center} 
\end{figure}

The exclusion limit on the CMSSM $(m_0,M_{1/2})$ plane from our analysis is also shown in Fig.~\ref{fig:msugracompare}.
This is based on the procedure, the $\CL{s}$-method, described just above; the letters in the parameter spaces denote the expected-to-be-most-sensitive SR, that is, the SR which is selected for the point.
Interpolation to obtain the curve is based on $\log\CL{s}$.
The constraint is slightly looser than that reported by the ATLAS collaboration, but as ours is still conservative, we decided to accept this analysis.

\subsubsection{Result}
The results are already shown in Fig.~\ref{fig:results1}: the black solid lines in the neutralino NLSP region, i.e., below the light blue lines.
The exclusion limits with considering theoretical uncertainties are also drawn as the black dashed lines.
The theoretical uncertainties mainly originate in the evaluation of renormalization and factorization scales, and choice of the PDFs~\cite{Aad:2011ib}.
For the uncertainties we adopt $\pm 35\%$ in the production cross section.

The left sides of the black lines are excluded.
Consequently, the gluino mass is required to be larger than approximately $1100\GeV$ for $M_{\rm mess}=10^6\GeV$, if the theoretical uncertainties are not included.
The theoretical uncertainties can shift the mass bound by $\lesssim 100\GeV$.
When $M_{\rm mess}$ is larger, the exclusion becomes weaker for the fixed gluino mass.
This is because the squarks become heavier, and the production cross section of the squark, especially that of $pp\to \tilde g\tilde q$, becomes smaller. When the messenger scale is as large as $10^{10}\GeV$, the bound becomes $m_{\tilde g} \gtrsim 1030\GeV$.

\begin{figure}[p]
\begin{center}
 \begin{minipage}[t]{0.49\textwidth}\begin{center}
                                     \includegraphics[width=0.99\linewidth]{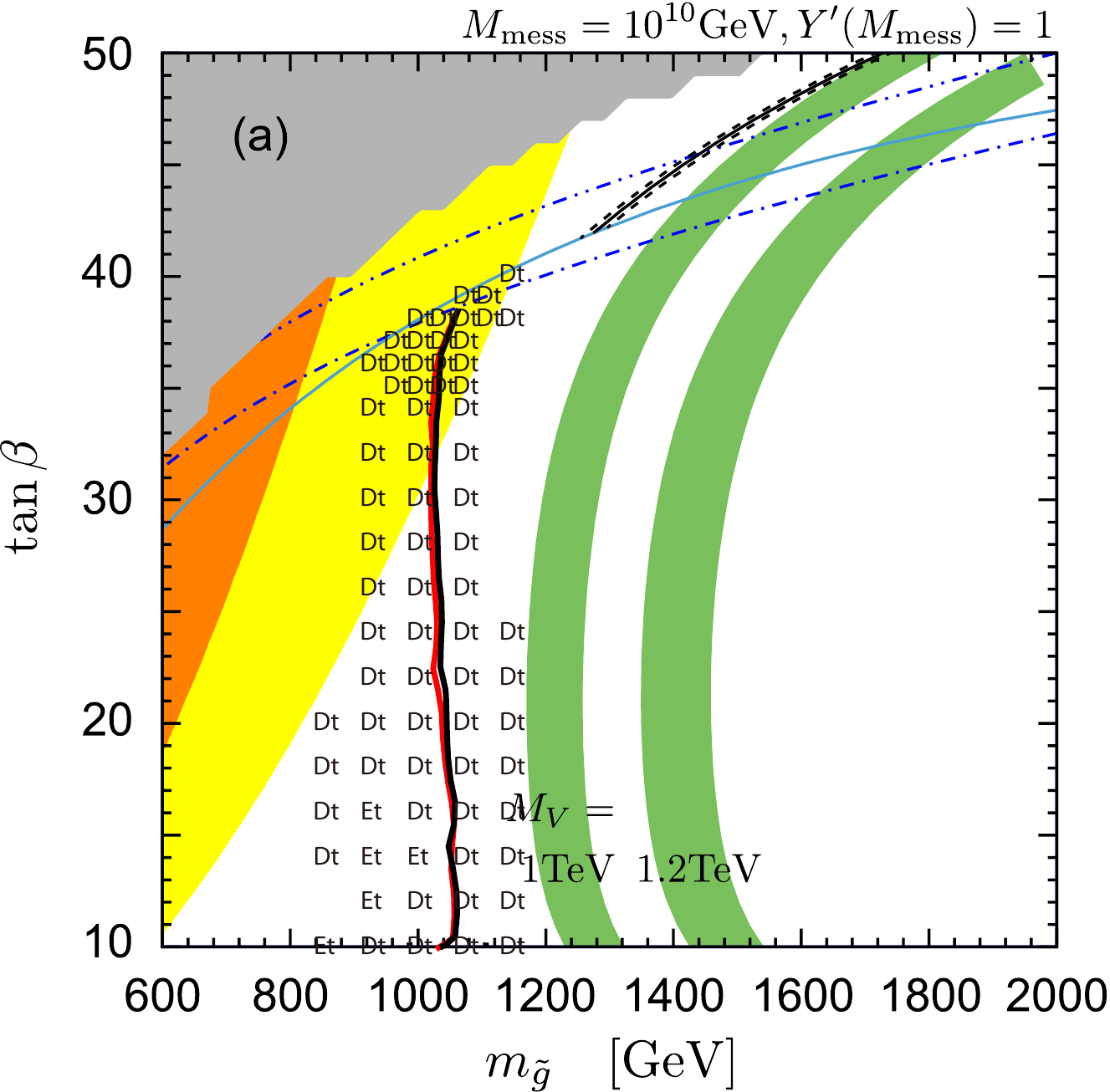}
 \end{center}\end{minipage}
 \begin{minipage}[t]{0.49\textwidth}\begin{center}
  \includegraphics[width=0.99\linewidth]{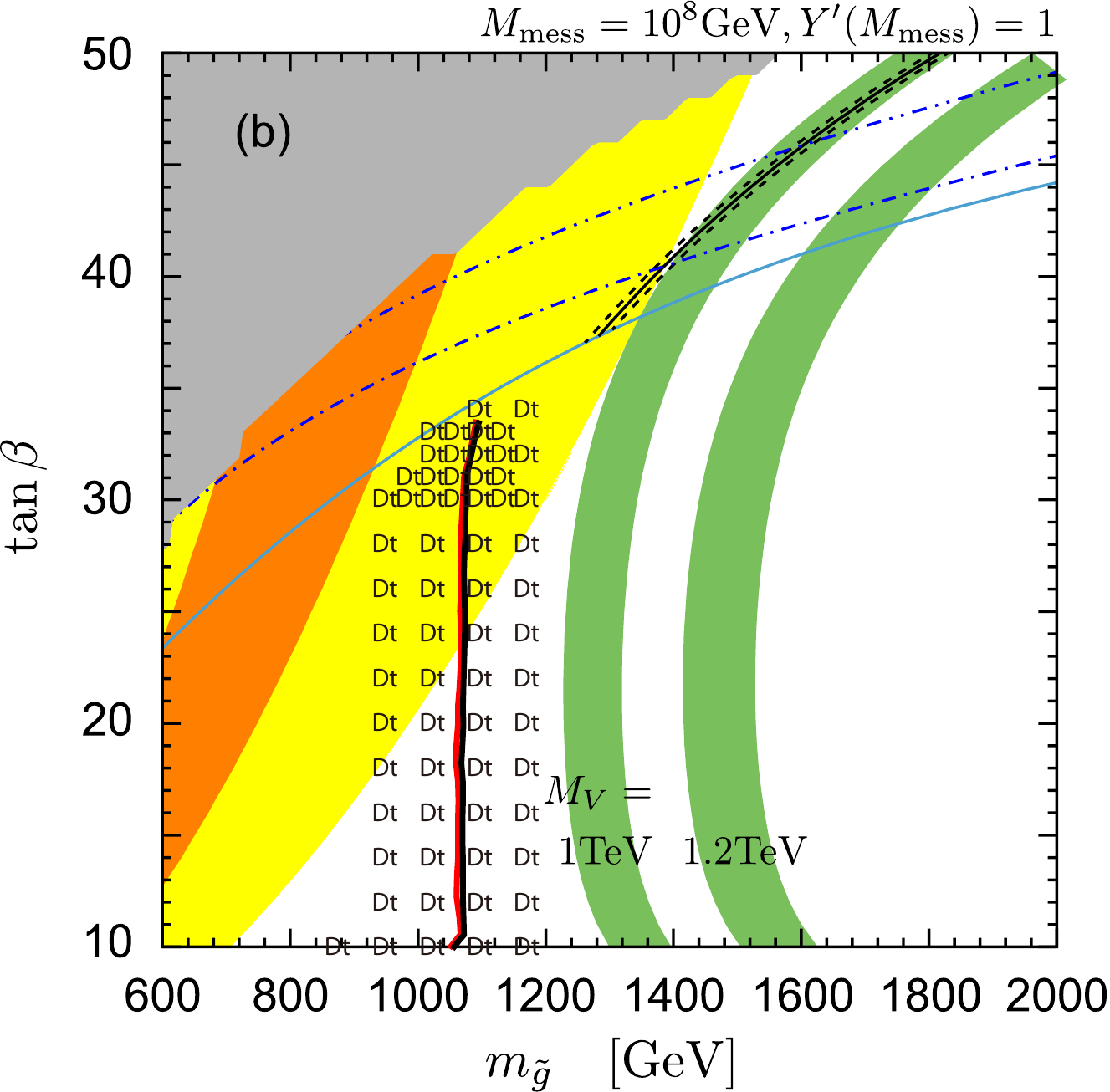}
 \end{center}\end{minipage}

 \begin{minipage}[t]{0.49\textwidth}\begin{center}
  \includegraphics[width=0.99\linewidth]{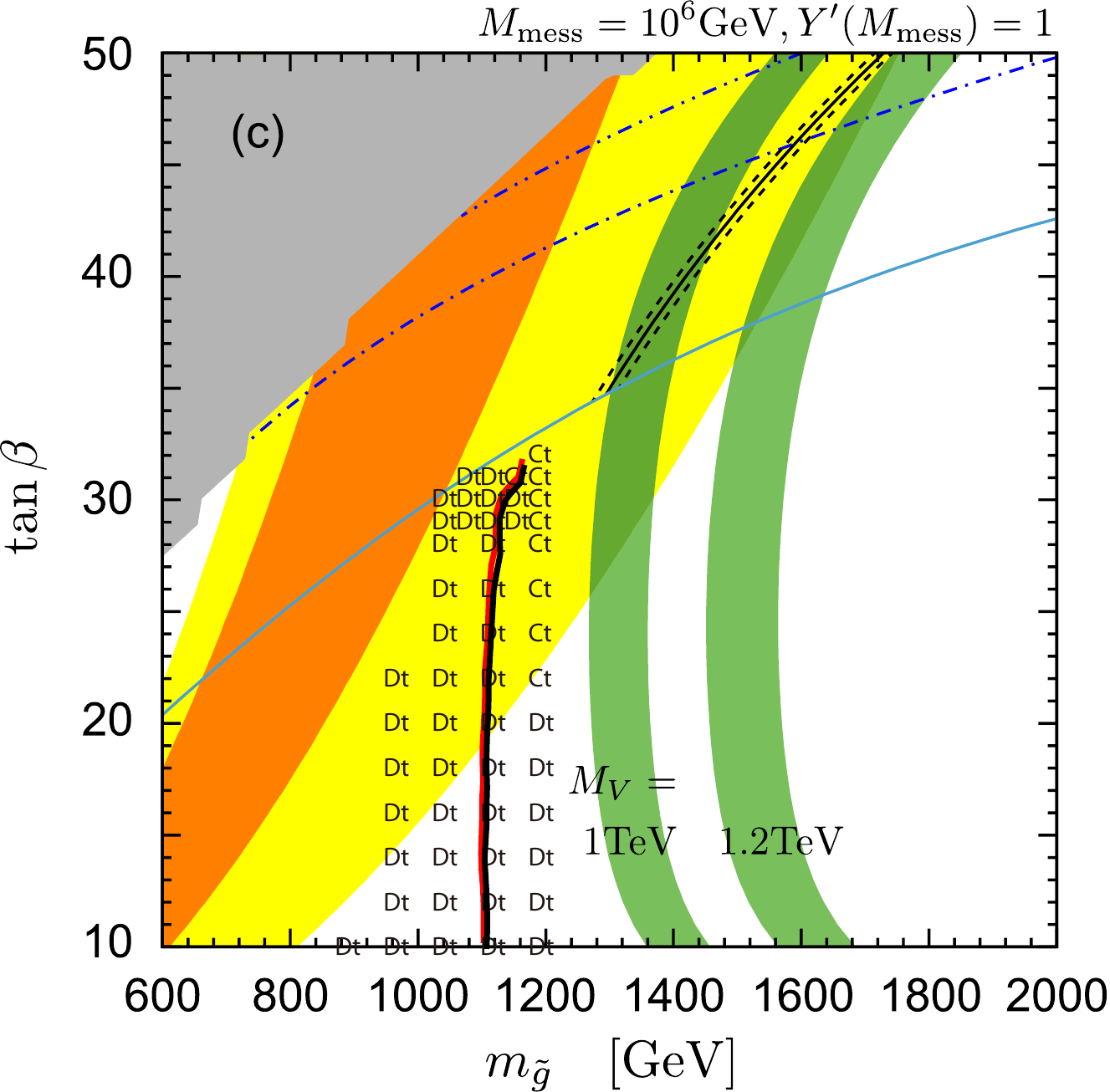}
 \end{center}\end{minipage}
 \begin{minipage}[t]{0.49\textwidth}\hfill\end{minipage}
\caption[Figure for discussion on the lepton efficiencies. Also the adopted SRs are shown.]{
The same as Fig.~\ref{fig:results1}, but here, for the cases where the neutralino is the NLSP, the SRs used for the limit calculation are shown instead of the upper limits with uncertainties.
In addition, the limits obtained with assuming the lepton efficiencies as 100\% are shown as red lines to see the effect of the lepton efficiencies.
}
\label{fig:results1-reg}
\end{center}
\end{figure}

The SRs used in the limit calculation, i.e., the expected-to-be-most-stringent SRs, are shown in Fig.~\ref{fig:results1-reg}.
In the figure the red lines show the exclusion limits under assuming full lepton reconstruction, i.e., setting the lepton efficiencies as $100\%$. Discussion on this topic is performed later.
What is important is that for almost all points the ``D-tight'' SR is selected.
The ``D-tight'' SR requires five hard jets and large missing energy.
It is easy to understand the origin of the four hard jets: the gluino is lightest among the colored superparticles, and it generates at least two hard jets in its cascade decay.
The origin of the fifth jet can be understood as follows.
In the channel of $pp\to \tilde g\tilde q$, the hard fifth jet is produced in decays of the squarks into the gluino. In the $pp\to \tilde g\tilde g$ channel, the largest fraction of events which pass the ``D-tight'' cut have additional hard jet(s) from ISR and FSR (cf.~Ref.~\cite{Alwall:2009zu}).
The ISR can yield additional hard jet(s) in the $pp\to \tilde g\tilde q$ channel as well.
In the rest of the events the fifth jet comes from decays of $W^\pm$ and $\tau$.
Here $W^\pm$ is generated in the decay of the top quark, and $\tau$ comes from the cascade decay of $\chgPM(\neut[2])\to \tau \tilde \tau \to \tau \tau \neut$, where $\chgPM(\neut[2])$ is provided from decays of colored superparticles.

One might worry that the hardness of the ISR and FSR jets might not be simulated correctly in {\tt Pythia}.
In order to resolve this uncertainty, the analysis for the ``D-tight'' SR with the MLM-matching scheme~\cite{Mangano:2006rw}, which is implemented in {\tt MadGraph}, is performed at several model points.
The analysis employs a shower $k_t$ clustering with avoiding double counting between the gluino and squarks~\cite{Alwall:2008qv}.
With this analysis it is checked that the results from {\tt Pythia} agree with those with the MLM-matching.

The main production channels of SUSY events are $pp\to \tilde g\tilde g, \tilde g\tilde q,\tilde q\tilde q$.
Although the squarks are relatively heavy compared to the gluino as shown in Fig.~\ref{fig:spectrum} and Fig.~\ref{fig:results1}(d), they are not decoupled from the productions.
Since the squark masses are almost independent of $\tan\beta$, so are the LHC exclusion lines in Figs.~\ref{fig:results1}(a)-(c).
For illustration, in Table~\ref{tab:gg_gq_qq} we show the cross section of each channel at two model points,
$(\Lambda, M_{\rm mess}, \tan\beta,M_V)=(140\TeV, 10^6\GeV, 20,1\TeV)$
and 
$(130\TeV, 10^{10}\GeV, 20,1\TeV)$.
These two points are, as shown in Fig.~\ref{fig:results1-reg}, analyzed with the ``D-tight'' SR, and close to the LHC exclusion limit.
It is found that the channels of $pp\to \tilde g\tilde g$ and $\tilde g\tilde q$ comparably contribute to the SUSY searches for a low messenger scale (i.e., for a relatively light squark), while $pp\to \tilde g\tilde g$ dominates for a high messenger scale (a heavy squark).

\begin{table}[t]
\begin{center}
 \caption[The cross section, the acceptance, and the expected number of events of our analysis.]{The cross section, the acceptance, and the expected number of events, on the V-GMSB model points
$(\Lambda, M_{\rm mess}, \tan\beta,M_V)=(140\xTeV, 10^6\xGeV, 20,1\xTeV)$
and 
$(130\xTeV, 10^{10}\xGeV, 20,1\xTeV)$.
The gluino mass, $m_{\tilde{g}}$, and the mass of the lightest squark among the first and the second generations, $m_{\tilde{q}}$,
are also displayed. See Fig.~\ref{fig:spectrum} for the full mass spectrum of the former point. 
For both points, the ``D-tight'' SR is selected as the expected-to-be-most-promising SR, and thus adopted for the evaluation.
For each process 5000 events are generated.
As can be seen in Fig.~\ref{fig:results1}, the former model point is not excluded since $N<6.0$ (cf.~Table~\ref{tab:atlasresult}), while the latter is excluded.
}
\label{tab:gg_gq_qq}

\vspace{\baselineskip}\def\arraystretch{1.2}
\begin{tabular}{|c|c|c|c|c|c|c|}
\hline
$(\Lambda, M_{\rm mess}, \tan\beta)$
& \multicolumn{3}{|c|}{$(140\TeV, 10^6\GeV, 20)$}
& \multicolumn{3}{|c|}{$(130\TeV, 10^{10}\GeV, 20)$}
\\\hline
$m_{\tilde{g}}$
& 
\multicolumn{3}{|c|}{1116 GeV} &\multicolumn{3}{|c|}{1002 GeV}
\\\hline
$m_{\tilde{q}_{\rm R}}$
& 
\multicolumn{3}{|c|}{1813 GeV} &\multicolumn{3}{|c|}{1887 GeV}
\\\hline
production channel 
& $\tilde{g}\tilde{g}$& $\tilde{g}\tilde{q}$ & $\tilde{q}\tilde{q}$ 
& $\tilde{g}\tilde{g}$& $\tilde{g}\tilde{q}$ & $\tilde{q}\tilde{q}$ 
\\ \hline
cross section $\sigma$ (fb) 
& 5.54  & 2.75  & 0.238 & 16.2  & 3.31  & 0.119
\\ \hline
acceptance $A$ 
& 0.0758 & 0.184 & 0.201& 0.0488 & 0.155 & 0.164
\\ \hline
$N=\sigma\times A\times 5.8{\rm fb}^{-1}$
& 2.44  & 2.93  & 0.28  & 4.59  & 2.98  & 0.11
\\\hline
$N$ (total)
& 
\multicolumn{3}{|c|}{5.65}
& 
\multicolumn{3}{|c|}{7.68}
\\ \hline
\end{tabular}
\end{center}
\end{table}

 \begin{figure}[t]
\begin{center}
   \includegraphics[width=0.9\textwidth]{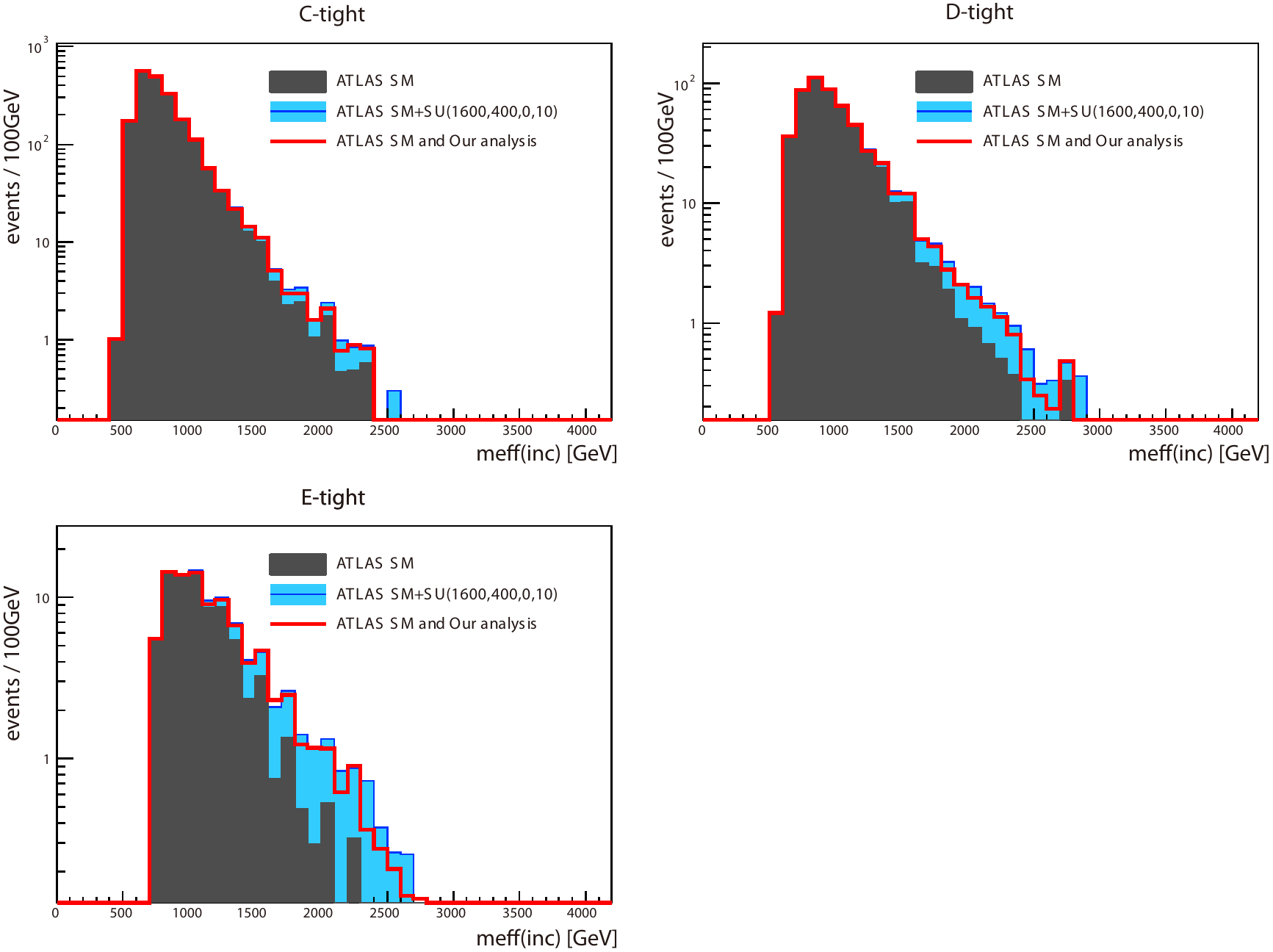}
  \caption[The distributions of the effective mass, with assuming $100\%$ lepton efficiencies.]{The same as Fig.~\ref{fig:histocompare-leff}, but here the lepton efficiencies are set to be $100\%$, which tightens the lepton veto and makes event yields decrease by $\sim5\%$.
}\label{fig:histocompare-noleff}
\end{center} 
\end{figure}

\subsubsection{Appendix --- On lepton efficiencies}
To check the effect of the lepton efficiencies, the above analysis is also performed with setting all the lepton efficiencies as $100\%$.
Here the leptons are perfectly detected, and it is expected to tighten the lepton veto.

As is expected, the $100\%$ lepton efficiencies result in slightly looser limits, which are illustrated in Figs.~\ref{fig:histocompare-noleff} and \ref{fig:results1-reg}, and also in Fig.~\ref{fig:msugracompare}.
In Fig.~\ref{fig:histocompare-noleff}, the histogram of Fig.~\ref{fig:histocompare-leff} is reproduced with assuming $100\%$ lepton detection.
The event yields decreases by $\sim5\%$.
In Fig.~\ref{fig:results1-reg}, the exclusion limits with nominal cross section is drawn with red lines.
It is observed that the bound is imperceptibly looser than the original one (the black solid lines).
This comparison is also performed in Fig.~\ref{fig:msugracompare}; there the bound with $100\%$ efficiencies are drawn with a blue line.

\starline

Now everything what Author wanted to explain is fully shown.
Let us move on to the stau NLSP case.

\subsection{LHC constraint for the V-GMSB with long-lived stau NLSP}
\label{sec:stauNLSP}
The LHC constraints for this case are also already drawn in Figs.~\ref{fig:results1}(a)--(c); the black solid lines above the light blue lines. The left-side regions of the lines are excluded.

To obtain this limit we just utilize the CMS bound of $223\GeV$, Eq.~\eqref{eq:stau223}, corresponding to the 95\% CL exclusion, obtained from searches for the heavy long-lived staus via the direct production~\cite{Chatrchyan:2012sp}.
The CMS collaboration evaluated the theoretical uncertainties as 3--7\% in the cross section calculations due to renormalization and factorization scales, $\alpha_s$, and choice of PDFs.
In the figure, we assigned theoretical uncertainty of 8\% for the cross section, which corresponds to 2\% uncertainty for the stau mass bound, shown by black dashed lines. 
Consequently, the regions with the gluino mass lighter than $1200\GeV$ are excluded; the bound becomes more severe for a larger $\tan\beta$.

\begin{figure}[t]
\begin{center}
\begin{tabular}{cc}
\includegraphics[width=0.46\linewidth]{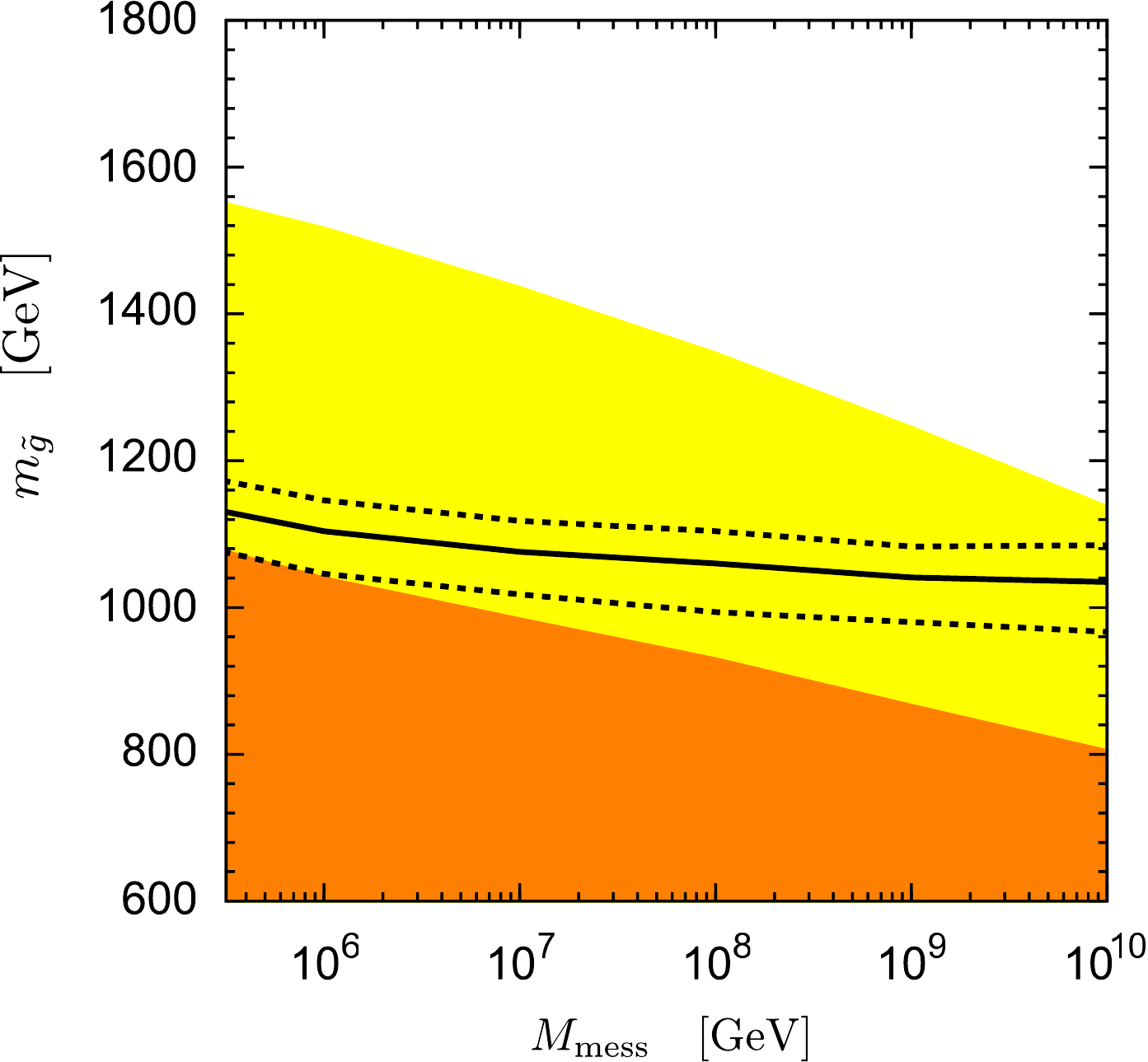}  &
\includegraphics[width=0.46\textwidth]{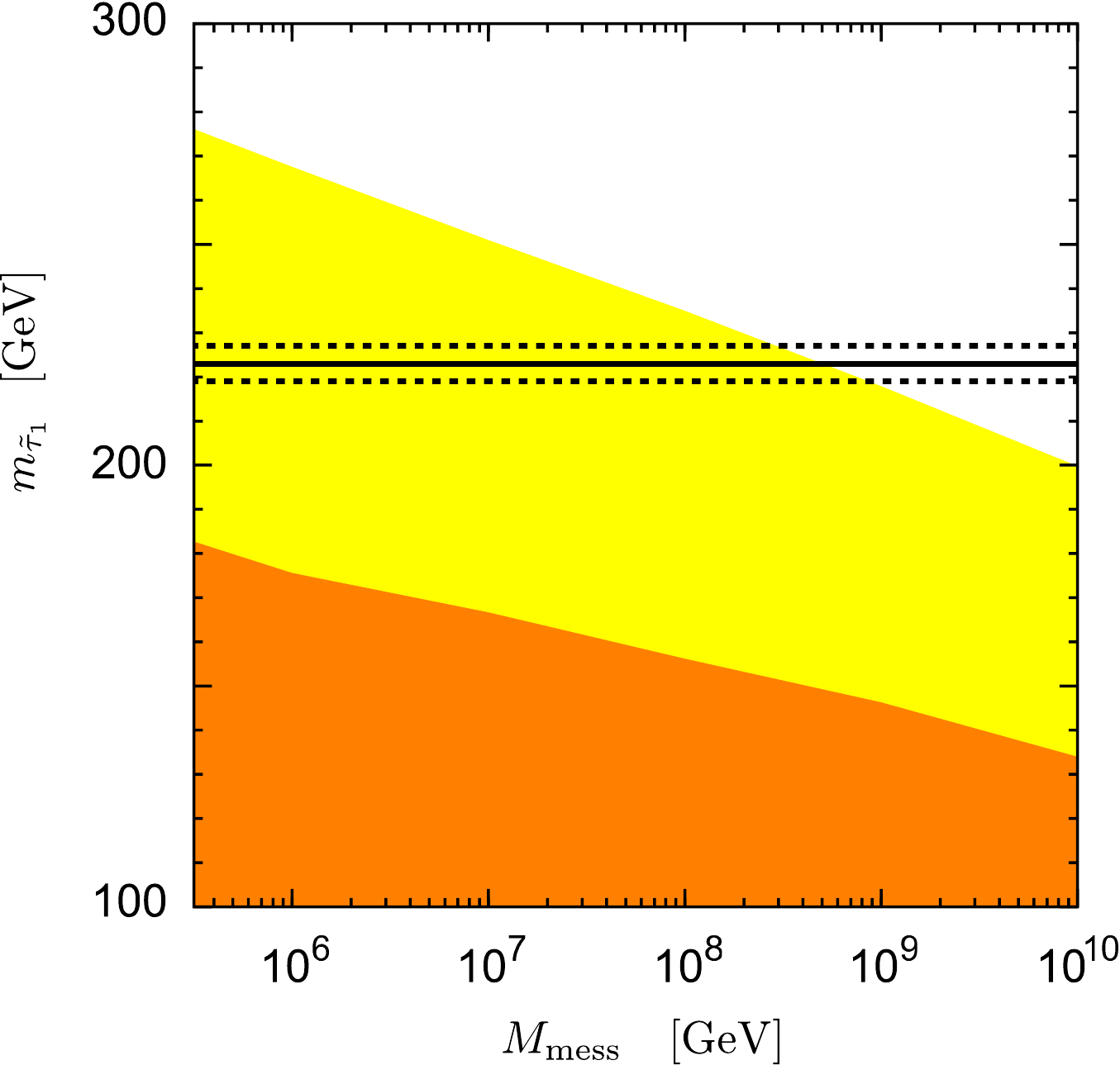}  
\end{tabular}
 \caption[The LHC bound as a function of $M\s{mess}$.]{(left) The lower bound, obtained from the LHC SUSY searches, on the gluino mass when the NLSP is the neutralino. (right) That on the lighter stau mass for the cases where the NLSP is the stau.
The yellow (orange) regions show the parameter spaces where the muon $g-2$ discrepancy can be explained within $2\sigma$- ($1\sigma$-) level.}
\label{fig:results2}
\end{center}
\end{figure}

\subsection{Discussion}
\label{sec:vmssm-disc}
The LHC constraint can be interpreted as a function of the messenger scale in each category of the NLSP.
The results are shown in Fig.~\ref{fig:results2}.
The left and right panels correspond to the cases with the neutralino NLSP and the stau NLSP, respectively.
Note that, throughout this section, the NLSP is assumed to be long-lived.
For a given $M_{\rm mess}$, the muon $g-2$ discrepancy can be explained at the $1\sigma$ ($2\sigma$) level when the gluino/stau mass is within the orange (yellow) region.
The upper ends of these regions represent the upper bounds on the gluino/stau mass in order to explain the muon $g-2$ at the $1\sigma$ ($2\sigma$) level.

In the case with the neutralino NLSP (left panel), the upper bound is determined by the requirement that the neutralino is lighter than the stau.
In fact, for a fixed value of the muon $g-2$, the gluino mass is maximized when $\tan\beta$ comes on the light blue lines in Fig.~\ref{fig:results1}.
When $M_{\rm mess}$ is as large as $10^{10}\GeV$, the vacuum stability bound can give a more severe bound, but the result is almost unchanged as can be noticed from Fig.~\ref{fig:results1}(a). 
This upper bound should be compared with the lower bound on the gluino mass from the LHC SUSY search, drawn with the black solid line.
Here, the exclusion limit at $\tan\beta=20$ is used for illustration, since it does not depend much on $\tan\beta$.
The black dashed lines show the theoretical uncertainties. (See discussion above.)
It is found that the whole region where the muon $g-2$ discrepancy is explained at the 1$\sigma$ level is already excluded by the direct searches for the superparticles at LHC.
The region with the $2\sigma$ explanation is still viable, and expected to be covered with the $14\TeV$ LHC.

For the stau NLSP case, the upper bound on the stau mass is shown in the right panel of Fig.~\ref{fig:results2}.
The upper ends of the yellow and orange regions are obtained from the requirement that the stau is the NLSP.
According to Fig.~\ref{fig:results1}(c) and  (d), the stau mass is maximized for a fixed muon $g-2$ when it is close to the neutralino mass.
The black solid line is the lower bound on the stau mass from the LHC. 
It does not include the theoretical uncertainties; they are taken into account in the black dashed lines.
Although the vacuum stability condition can give a tighter bound for $M_{\rm mess} \sim 10^{9}\text{--}10^{10}\GeV$, such parameter regions are already constrained by the LHC (see Fig.~\ref{fig:results1}(a)). 
Consequently, it is found that the region where the muon $g-2$ is explained at $1\sigma$-level is fully excluded by the searches for the heavy long-lived charged particles at the LHC.
It is even expected to become more severe, for instance by analyzing the data obtained at the $8\TeV$ LHC.

\clearpage

\section{Searches for the Vector-like Quark}
\label{sec:vector}
Finally, let us discuss searches for the extra vector-like quarks.
The LHC experiments have not discovered any extra quarks, and thus the V-MSSM is constrained.
What is important is that what is constrained is the SUSY-invariant mass terms $M_{Q'}$ and $M_{U'}$ together with $Y'$, and therefore the constraint is independent on the SUSY-breaking scenario.
Therefore, the discussion performed in this section is not specific to the V-GMSB model, but generic for the V-MSSM scenario.

The extra vector-like quarks are produced with considerable cross sections at the LHC, a hadron collider.
Searches for these particles are of great interest and importance because the existence of these particles would be a direct evidence of the V-MSSM, and besides, because these particles should be near the TeV-scale as we have seen in Sec.~\ref{sec:vgmsb-numerical-result},  they are expected to be within the reach of the LHC.

The vector-like quarks have characteristic decay channels, which are different from the ordinal fourth generation quarks. We first check their masses and decay modes, and then current experimental bounds on the masses of those particles are discussed. After that we will mention prospects of further searches.

\starline

In the following discussions we use $Y'=1.05$, the infrared fixed point value~\cite{Martin:2009bg}, as a reference value.
The approximation $M_{Q'}=M_{U'}(=:M_V)$ is also exploited; one should understand $M_{Q'}=M_{U'}$ when we refer to $M_V$.

As the Standard Model parameters relevant to the decay branch, the following are used:
\begin{align*}
 m_t&=173.5\GeV,  & m_Z&=91.2\GeV, & v&=174\GeV,  & g_2&=0.652, \\
 m_b&=4.78\GeV,   & m_W&=80.4\GeV, & m_h&=126\GeV,& g_Z&=0.743.
\end{align*}

 \begin{figure}[p!]
\begin{center}
   \includegraphics[width=0.6\textwidth]{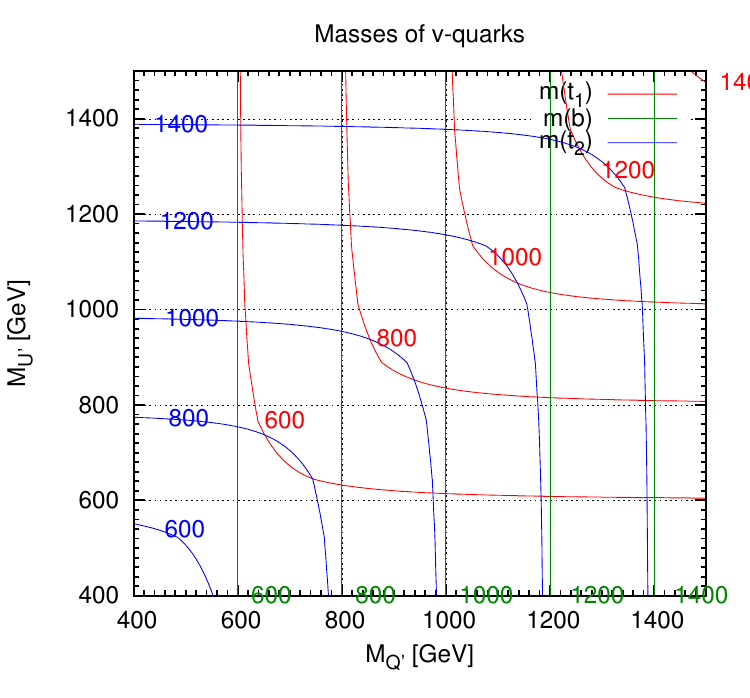}
  \caption[Masses of the vector-like quarks.]{Masses of the vector-like quarks as functions of $(M_{Q'}, M_{U'})$. The mixing term is taken as $Y'\vu=1.05\times174\xGeV$.}
 \label{fig:vq_mass_2d}
\end{center} 
\end{figure}

\begin{table}[p!]
\begin{center}\def\arraystretch{1.2}
\caption[Benchmark points for the mixings between Standard Model and extra quarks.]{Benchmark points for the mixing parameters.
The shown values as the branching ratios are calculated at $m_{t'_1}=400\,\rm GeV$ ($M_{Q'}=M_{U'}\simeq 483\,\rm GeV$), $Y'=1.05$, $\tan\beta=30$ and $m_h=125\,\rm GeV$;
they have nontrivial dependence on $m_{t'_1}$, but are almost stable under changes of the absolute values of the mixing parameters ($\epsilon$'s) as long as the mixing parameters are much smaller than $\Order(1)$.
}
\label{tab:vmssm-mixingsample}
 \begin{tabular}[t]{|c|c|c|c|c|}\hline
       & \multirow{2}{*}{$\epsilon_U : \epsilon_U' : \epsilon_D$}  & \multicolumn{3}{|c|}{Branching Ratios for $m_{t'_1}=400\GeV$}\\\cline{3-5} &
 & $\Br\bigl(t'_1\to bW\bigr)$ & $\Br\bigl(t'_1\to tZ\bigr)$ & $\Br\bigl(t'_1\to th\bigr)$\\\hline
(A)    & $0:0:1$       & 1 & 0 & 0\\
(B)    & $1:1:1$       & 0.51 & 0.44 & 0.05\\
(C)    & $1:0:0$       & 0.48 & 0.13 & 0.39\\
(D)    & $0:1:0$       & 0.15 & 0.19 & 0.65\\
(E)    & $1:2:0$     & 0.01 & 0.48 & 0.51\\\hline
 \end{tabular}
\end{center}
\end{table}

\subsection{Masses and decay modes}
The masses of the vector-like quarks are already reviewed in Sec.~\ref{sec:vmssm-mass}.
When we set $M_{Q'}=M_{U'}=:M_V$, the masses are approximately given as
\begin{align}
 m_{t'_1,t'_2}&=
\sqrt{\frac{M_{Q'}^2+M_{U'}^2+Y'^2\vu^2}{2}
 \pm\sqrt{\left(\frac{M_{Q'}^2+M_{U'}^2+Y'^2\vu^2}{2}\right)^2- M_{Q'}^2M_{U'}^2}}
\approx
M_{V}\left(1\pm\alpha+\frac{\alpha^2}{2}\right),
\\
m_{b'} &=M_{Q'}=M_V,
\end{align}
where one can see the mass hierarchy
\begin{equation}
 m_{t_1'}<m_{b'}<m_{t'},
\end{equation}
and the mass splittings among them are characterized by
\begin{equation}
 \alpha:=\frac{Y' v\sin\beta}{2M_{Q'}} \approx \frac{91\GeV}{M_V}\times \left(\frac{Y'}{1.05}\right)
\end{equation}
with a large $\tan\beta$.
As a reference, Fig.~\ref{fig:vq_mass_2d} is provided, which shows the masses of the vector-like quarks as functions of $(M_{Q'}, M_{U'})$.

As we assume that the vector-like quarks are only mixed with the third generation Standard Model quarks ($Q_3$, $\bU_3$, and $\bD_3$), the possible decay channels are summarized as
\begin{align}
 t'_2&\to b'W,\ t'_1 h,\ t'_1Z,
&
 b'&\to t'_1W,
&
 t'_1&\to bW,\ th,\ tZ,
\end{align}
where some of them may be kinematically forbidden if the mass separation is smaller.

The lightest one, $t'_1$, is the most important for the vector-like quark search because of its large production cross section.
As we assume very tiny mixing between the Standard Model quarks and the vector-like quarks, the pair-production $pp\to t_1' \bar t_1'$ is the most promising channel, and it leaves characteristic final state particles with multi-$b$-jets plus leptons depending on the decay pattern.
Thus the decay branching ratio of $t'_1$ is crucial for the vector-like quark search.

The decay widths~\cite{Martin:2009bg} are summarized in Appendix~\ref{sec:decay-rates-t_1}, and the branching ratio can be calculated straightforwardly.
Then one finds that the ratio has nontrivial dependence on the mass $m_{t'_1}$, and {\em the ratio among} the mixing parameters $\epsilon_U$, $\epsilon_U'$, and $\epsilon_D$~\cite{Martin:2009bg}.
Here note that the branching ratio is insensitive to the absolute value of the mixings as long as they are tiny, and also that the dependence on $\tan\beta$ appears only with a form $\epsilon_D/\tan\beta$ as we focus on the cases with a large $\tan\beta$.

In order to proceed the discussion, we pick up several mixing patterns shown in Table ~\ref{tab:vmssm-mixingsample} as benchmark points.
At the benchmark point (A), $t'_1$ exclusively decays into $bW$, or $\Br\bigl(t'_1\to bW\bigr)=1$, for any $m_{t'_1}$.
However, except for that point, the decay branching ratio of $t'_1$ has nontrivial dependence on the mass of $t'_1$. (cf. Figs.~\ref{fig:CMSboundsTZ}--\ref{fig:TEVATRONboundsBW}.)
This is mainly because the $t'_1\to tZ$ and $t'_1\to th$ channels are closed if the mass of $t'_1$ is below ``thresholds.''
Especially, if $m_{t'_1} < m_t + m_Z \simeq 264\GeV$, only the $t'_1\to bW$ channel is open regardless of the mixing parameters; this case corresponds to $M_V\lesssim 316\GeV$ under the $M_{Q'}=M_{U'}$ approximation.

\subsection{Current experimental bounds}

\begin{table}[t]
\begin{center}\def\arraystretch{1.3}
  \caption[Summary of current results from vector-like quark searches.]{Summary of current vector-like quark searches.
The displayed limits are at 95\% CL. Several obsoleted analyses are also shown with parentheses as a reference.
   For $t'\to bW$ search, ``$2l$'' denotes the di-lepton channel $bbWW\to bbll\nu\nu$, while ``$l+j$'' does the lepton plus jets channel $bbWW\to bbl\nu jj$.}
  \label{tab:vecsearchlist}
 \begin{tabular}[t]{|c|c|c|c|c|}\hline
 Decay channel & EXP. & Analyzed data & Obtained limit & References\\\hline
 \multirow{4}{*}{$\Br\bigl(t'_1\to bW\bigr)=1$} & ATLAS & $7\TeV, 4.7\invfb$ ($l+j$)& $m_{t'_1}>656\GeV$ & \cite{ATLAS:2012qe} (\cite{Aad:2012xc}) \\\cline{2-5}
        & \multirow{2}{*}{CMS} & $7\TeV, \sim5\invfb$ ($l+j$)& $m_{t'_1}>570\GeV$ & \cite{Chatrchyan:2012vu} (\cite{CMSPASEXO11051}) \\\cline{3-5}
        &        & $7\TeV, 5.0\invfb$ ($2l$)& $m_{t'_1}>557\GeV$ & \cite{CMS:2012ab}\\\cline{2-5}
        &  CDF   & $1.96\TeV, 5.6\invfb$ ($2l$)& $m_{t'_1}>358\GeV$ & \cite{Aaltonen:2011tq}\\\hline
 \multirow{3}{*}{$\Br\bigl(t'_1\to qW\bigr)=1$} &  ATLAS  & $7\TeV, 1.04\invfb$ ($2l$)& $m_{t'_1}>350\GeV$ & \cite{Aad:2012bt}\\\cline{2-5}
        &  CDF   & $1.96\TeV, 5.6\invfb$ ($l+j$)& $m_{t'_1}>340\GeV$ & \cite{Aaltonen:2011tq}\\\cline{2-5}
        &  D0    & $1.96\TeV, 5.3\invfb$ ($l+j$)& $m_{t'_1}>285\GeV$ & \cite{Abazov:2011vy}   \\\hline
 $\Br\bigl(t'_1\to tZ\bigr)=1$ &  CMS  & $7\TeV, 5.0\invfb$& $m_{t'_1}>625\GeV$ & \cite{Chatrchyan:2012af} (\cite{Chatrchyan:2011ay})\\\hline
 $t'_1\to bW,tZ,th$ &  ATLAS & $7\TeV, 4.7\invfb$  & See Fig.~\ref{fig:vqbound_2d}. &\cite{ATLAS:2012qe}\\\hline
 \end{tabular}
\end{center} 
\end{table}

\begin{figure}[p]
\begin{center}
 \includegraphics[width=0.65\textwidth]{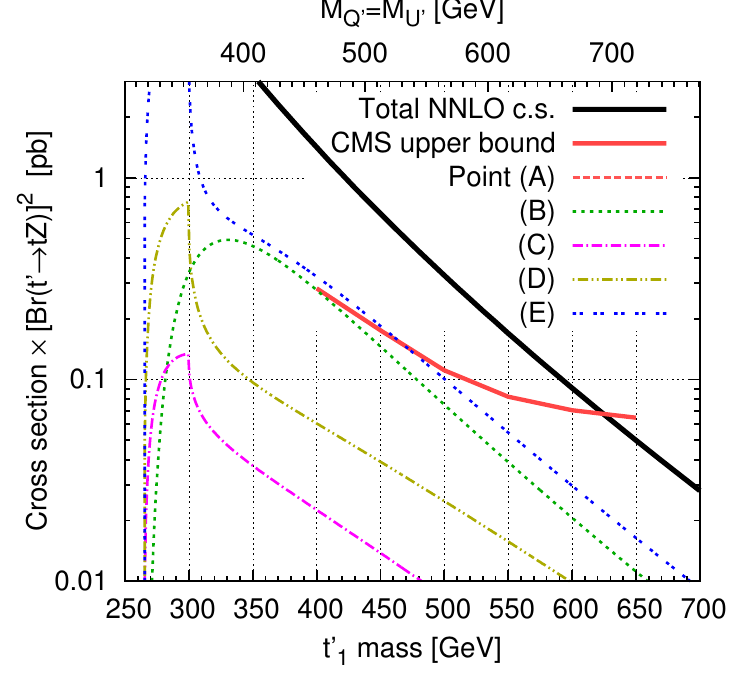}\vspace{-2em}
 \caption[The CMS bound on the vector-like quark with assuming $\Br\bigl(t'_1\to tZ\bigr)=1$.]{The CMS experimental 95\% CL upper limit (red line) on the $t'_1 \bar t'_1$ pair-production cross section with $5.0\rm{\,fb^{-1}}$ data, assuming $t'_1$ exclusively decays via $t'_1\to tZ$~\cite{Chatrchyan:2012af}. Their limit is obtained for $400\xGeV<m_{t'_1}<650\xGeV$.
The black solid line is the NNLO {\em total cross section} of $t' \bar t'$ production, calculated with {\tt HATHOR}~\cite{HATHORNNLO}.
Considering the branching ratio $\Br\bigl(t'_1\to tZ\bigr)$, this limit can be applied to the vector-like quark with generic decay branch. We show the corresponding $t'_1\bar t'_1$ cross section with the branching effect at the benchmark points (A)--(E) as dashed and dotted lines. Note that the line corresponding to the point (A) is not shown since $\Br\bigl(t'_1\to tZ\bigr)=0$ at the point.
}
 \label{fig:CMSboundsTZ}
\end{center}\vspace{-20pt}
\end{figure}

\begin{figure}[p]
\begin{center}
 \includegraphics[width=0.65\textwidth]{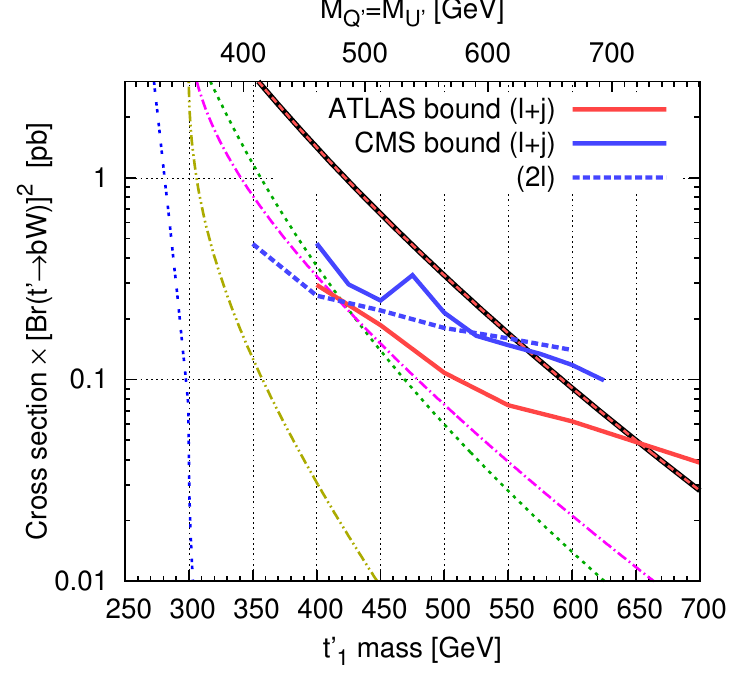}\vspace{-2em}
 \caption[The LHC bounds on the vector-like quark with assuming $\Br\bigl(t'_1\to bW\bigr)=1$.]{%
The same as Fig.~\ref{fig:CMSboundsTZ}, but here $t'_1$ is assumed to decay exclusively via $t'_1\to bW$ channel.
The red line denotes the upper limit from the ATLAS~\cite{ATLAS:2012qe}, while the blue solid and the blue dashed lines are from the CMS~\cite{CMS:2012ab,Chatrchyan:2012vu}.
The line for (A) overlaps the total cross section line since $\Br\bigl(t'_1\to bW\bigr)=1$.}
 \label{fig:CMSboundsBW}
\end{center}
\end{figure}

\begin{figure}[t]
\begin{center}
 \includegraphics[width=0.65\textwidth]{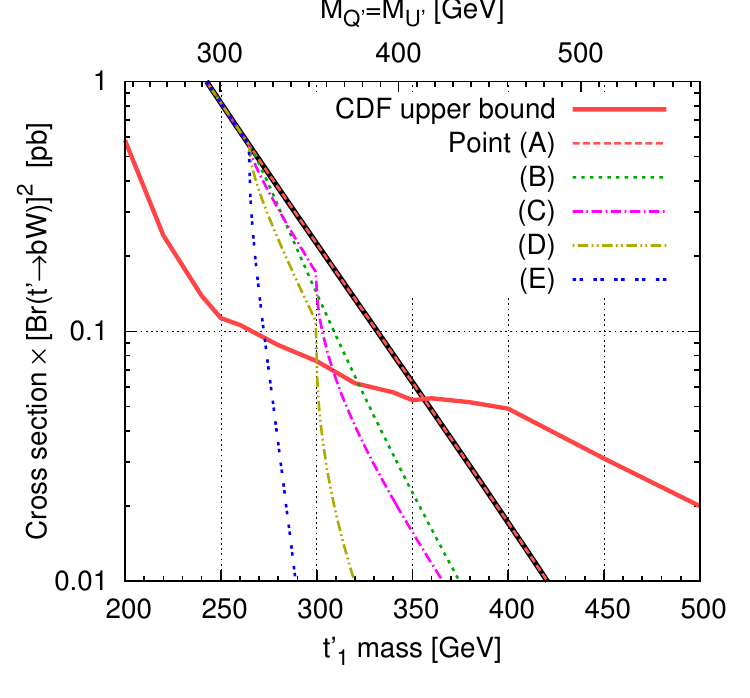}
 \caption[The CDF bounds on the vector-like quark with assuming $\Br\bigl(t'_1\to bW\bigr)=1$.]{%
The same as Fig.~\ref{fig:CMSboundsBW}, but from the CDF experiment at the Tevatron~\cite{Aaltonen:2011tq}.
$t'_1$ is assumed to decay via $t'_1\to bW$ channel, and the cross section corresponds to $1.96\xTeV$ $p\bar p$ collision.
The red solid line denotes the upper bounds on the cross section.
The other lines are the same as Figs.~\ref{fig:CMSboundsTZ}--\ref{fig:CMSboundsBW}, but the cross sections are for $1.96\xTeV$ $p\bar p$ collision.
Note again that the line for (A) overlaps the total cross section line.}
 \label{fig:TEVATRONboundsBW}
\end{center}
\end{figure}

Searches for the vector-like quarks can be employed in a similar way to those for the fourth generation quarks.
The current reports from the LHC experiments are summarized in Table~\ref{tab:vecsearchlist}, together with results from the Tevatron collider.
If $t'_1$ decays exclusively into $bW$, i.e., $\Br\bigl(t'_1\to bW\bigr)=1$, results from searches for the fourth generation up-type quark can be applied, and the tightest bound is $m_{t'_1}>656\GeV$~\cite{ATLAS:2012qe}.
The ATLAS collaboration also obtained a bound for the case in which $\Br\bigl(t'_1\to qW\bigr)=1$, where $q$ is a generic down-type quark: $q=(d,s,b)$, of $m_{t'_1}>350\GeV$~\cite{Aad:2012bt}.
For the case where $\Br\bigl(t'_1\to tZ\bigr)=1$, the CMS obtained a limit of $m_{t'_1}>475\GeV$~\cite{Chatrchyan:2011ay}.

Those bounds can be applied to the above model points (A)--(E).
The result is shown in Fig.~\ref{fig:CMSboundsTZ}--\ref{fig:TEVATRONboundsBW}.
In Fig.~\ref{fig:CMSboundsTZ} the 7\,TeV LHC bound for the $t'\to tZ$ channel reported by the CMS collaboration~\cite{Chatrchyan:2012af} is displayed as a red solid line.
They reported upper bounds on the production cross section in a mass range of $400\GeV\le m_{t'_1}\le 650\GeV$.
The black solid line is the total cross section of $pp\to t'_1\bar t'_1$, and the other (dotted and dashed) lines are the ``effective'' cross section of $pp\to t'_1\bar t'_1 \to (bW)(\bar bW)$.
It is observed that the model points (B) and (E) have lower bounds of $m_{t'_1}\gtrsim400\GeV$ and $460\GeV$, respectively.
Note that the line for (A) does not appear in this figure because $\Br\bigl(t'_1\to tZ\bigr)=0$.

The results for the $t'_1\to bW$ channel from the LHC experiments are shown in Figs.~\ref{fig:CMSboundsBW}; here one result from the ATLAS collaboration~\cite{ATLAS:2012qe} and two from the CMS~\cite{CMS:2012ab,Chatrchyan:2012vu} are displayed (cf.~Table~\ref{tab:vecsearchlist}).
Benchmark points (A), (B), and (C) are constrained as, respectively, $m_{t'_1}\gtrsim650\GeV,420\GeV$, and $420\GeV$.

The point (D) is not constrained by the above results; one reason is that the LHC results are reported only for $m_{t'_1}>350\GeV$.
To give constraint for this benchmark point the analysis by the CDF collaboration~\cite{Aaltonen:2011tq} can be utilized.
Their result obtained at the Tevatron collider with $E\s{CM}=1.96\TeV$ $p\bar p$ collision is displayed in Fig.~\ref{fig:TEVATRONboundsBW}, and yields a limit of $m_{t'_1}\gtrsim 300\GeV$ for the benchmark point (D).

In this analysis the production cross sections are calculated with \withpackage[1.3]{HATHOR} at the NNLO level. The \withpackage{CT10nnlo} PDFs~\cite{PDFCT10} are used.

\starline

\begin{figure}[p]
\begin{center}
 \includegraphics[width=0.96\textwidth]{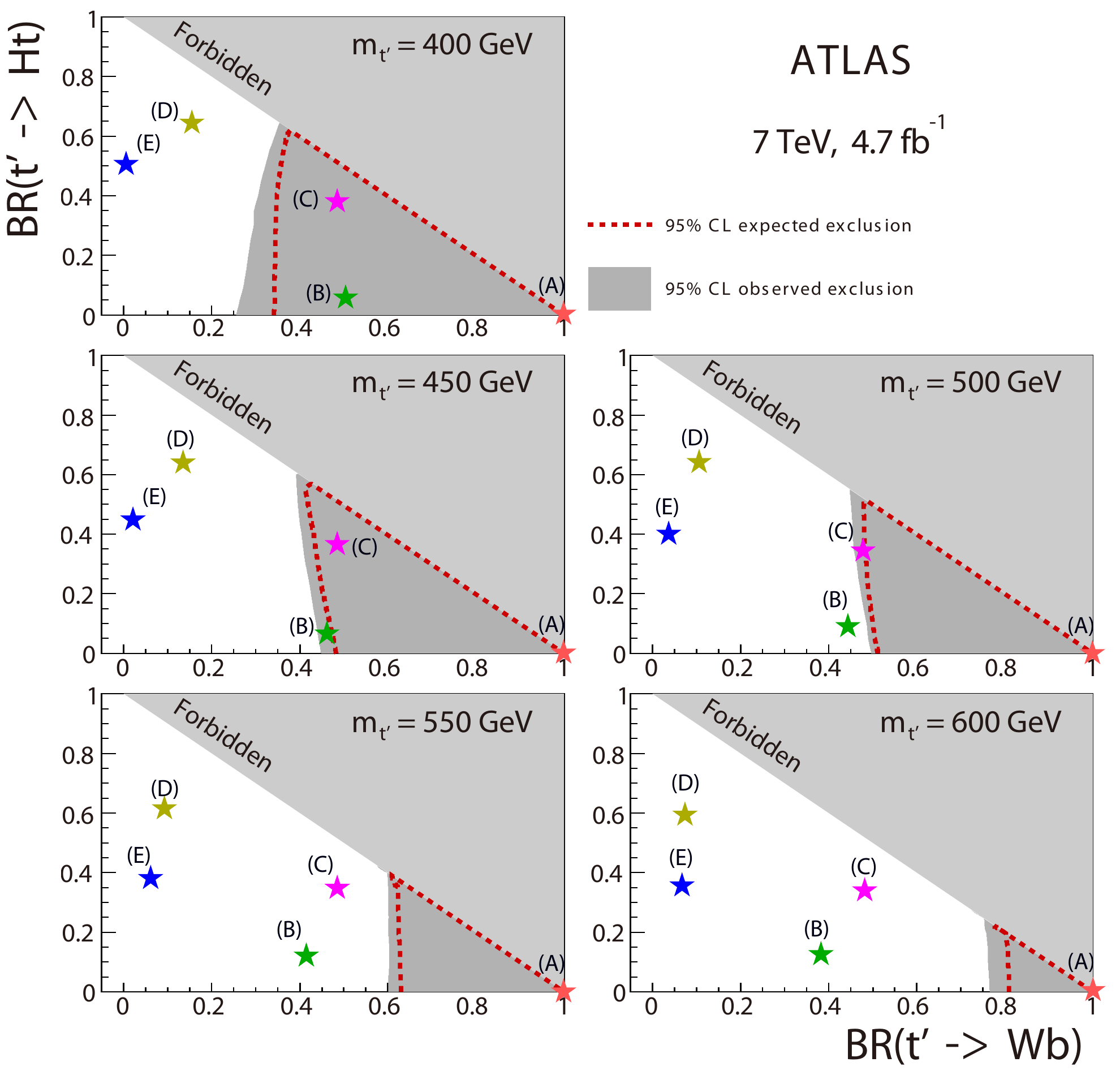}
 \caption[The ATLAS general bound on the vector-like quark.]{%
The result of the {\em inclusive} search for the vector-like quark $t'_1$ reported by the ATLAS collaboration~\cite{ATLAS:2012qe}.
In this analysis the decay branch of $t'_1$ is not specified, but the assumption that the vector-like quarks only mix with the third generation Standard Model quarks is employed.
Therefore, this search gives an inclusive study for the three decay patterns: $t'_1\to bW, tZ, th$.
The result is shown in the plane of the decay branch; here one should note that $\Br\bigl(t'_1\to tZ\bigr)$, which is not displayed explicitly in this figure, is equal to $1-\Br\bigl(t'_1\to bW\bigr)-\Br\bigl(t'_1\to th\bigr)$, and thus implicitly shown.
This report gives tighter bounds for our benchmark points (B) and (C) than the analyses focusing on a specific decay branch, i.e., of Figs.~\ref{fig:CMSboundsTZ} and \ref{fig:CMSboundsBW}.
\quad
{\bf Original version of this figure is produced by the ATLAS collaboration and used in Ref.~\cite{ATLAS:2012qe}. (ATLAS Experiment \copyright 2012 CERN)}}
 \label{fig:vqbound_2d}
\end{center}
\end{figure}

An interesting analysis was reported by the ATLAS collaboration~\cite{ATLAS:2012qe}.
They employed an inclusive search for the vector-like quark $t'_1$ in the events with at least three jets, at least one of which should be tagged as a $b$-jet, and at least one lepton ($e$ or $\mu$), and reported the excluded regions on the parameter space of $\bigl[\Br\bigl(t'\to bW\bigr),\Br\bigl(t'\to th\bigr)\bigr]$-plane with the mass $m_{t'_1}$ fixed.
The result is shown in Fig.~\ref{fig:vqbound_2d} with a modification that our benchmark points (A)--(E) are plotted instead of theirs.
Interpreting the figure, we obtain tighter bounds for the benchmark point (B) as $m_{t'_1}\gtrsim 450\GeV$ and (C) as $m_{t'_1}\gtrsim 500\GeV$.

\subsection{Prospects of further searches}

\begin{figure}[t]
\begin{center}-
 \includegraphics[width=0.6\textwidth]{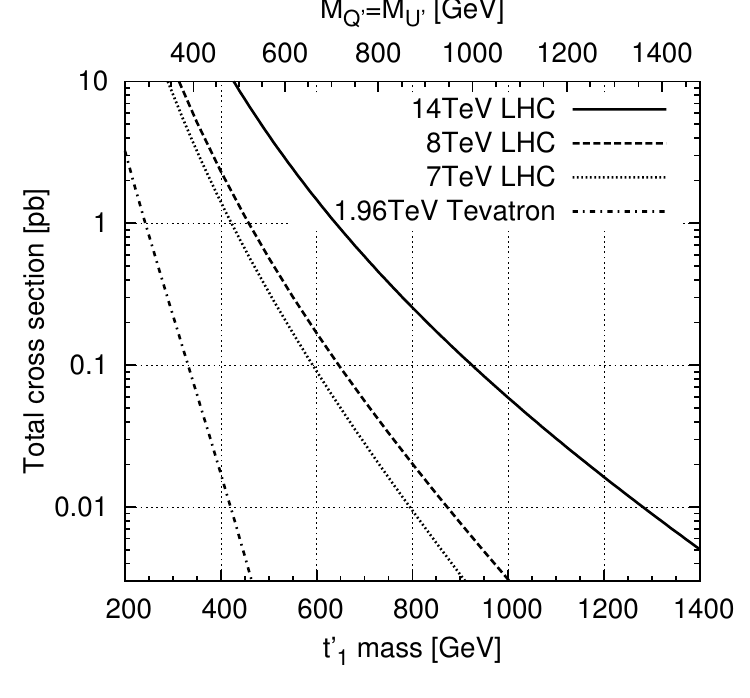}
 \caption[The production cross sections of $pp (p\bar p)\to t'_1 \bar t'_1$ as functions of $m\s{t_1'}$.]{%
The production cross sections of $pp (p\bar p)\to t'_1 \bar t'_1$ as functions of the $t'_1$ quark mass, calculated with {\tt HATHOR\,1.3}~\cite{HATHORNNLO} at the NNLO level.}
 \label{fig:ttbarCS}
\end{center}
\end{figure}

In order is discussion on future prospects to conclude this section.
Fig.~\ref{fig:ttbarCS} shows the production cross sections of $pp (p\bar p)\to t'_1 \bar t'_1$ as functions of the $t'_1$ quark mass, together with an extra axis of $M_V(=M_{Q'}=M_{U'})$.
The cross sections are calculated with \withpackage[1.3]{HATHOR}~\cite{HATHORNNLO} and the \withpackage{CT10nnlo} PDFs~\cite{PDFCT10} at the NNLO level.
The cross section doubles at the $8\TeV$ LHC, and is ten times larger at the $14\TeV$, than that at $E\s{CM}=7\TeV$.

As we have just seen, the inclusive search for $pp\to t't'\to(bW,th,tZ)+(bW,th,tZ)$ by the ATLAS collaboration~\cite{ATLAS:2012qe} is of great importance and interest.
Especially, because the pair-production cross section drastically improves at the $14\TeV$ LHC, we can expect that the $t'_1$ quark with $\sim1\TO1.2\TeV$ will be in our experimental reach.
As the V-GMSB model requires $M_V\lesssim 1.2\TeV$ to explain the muon $g-2$ anomaly, the $14\TeV$ LHC would be the court for the V-GMSB model; that is, the fate of the V-GMSB scenario, and thus the GMSB framework, with the explanation of the muon $g-2$ problem will be determined there.

Here it should be emphasized that the ATLAS analysis utilizes the event with at least {\em one} $b$-jet.
As the $t'_1$ decay generally yields 1--3 $b$-quarks, its pair-production would be searched more efficiently with multiple $b$-jets requirement~\cite{Endo:2011xq,Harigaya:2012ir}.
In Ref.~\cite{Harigaya:2012ir} the prospects of this strategy are discussed with Monte Carlo fast simulations, and it is found that the requirement of $1\,\text{lepton}+\ge3\,b\text{-jets}$ would yield complementary exclusion regions on the parameter spaces of Fig.~\ref{fig:vqbound_2d}.
This analysis would also be an important key for searches for the heavier vector-like quarks, $b'$ and $t'_2$.

On the experimental side, we now realize that $b$-tagging, discussed briefly in Chapter \ref{cha:atlas}, is acquiring importance from the fact that the Higgs boson is expected to decay mainly into $b$-quarks.
Improvements on the efficiency and the mis-tag rate, and deep understanding of the algorithms, are now much awaited for searches of new physics beyond the Standard Model.

\section{Summary}
\label{sec:vgmsb-summary}

Now the main chapter of this dissertation is finishing.
Let us summarize what we learned.

We in this section considered the V-MSSM, an extension of the MSSM with vector-like matters.
It has three virtues.
First, it respects the daydream of the $\gSU(5)$-GUTs.
Also, the Higgs boson mass of $126\GeV$ is realized with help of the vector-like quarks without exploiting the heavy stops of $\sim1$--$10\TeV$.
Then squarks need not to be so heavy, and thus sleptons either.
It allows us to explain the muon $g-2$ discrepancy with the SUSY contributions, even under the GMSB scenario.

These features are summarized in Fig.~\ref{fig:results1} for the V-GMSB model.
Even after imposing the LHC limits, the yellow regions, where the muon $g-2$ anomaly is explained within $2\sigma$-level, survive in all the three plots.
With the LHC limits, the gluino mass is constrained as $m_{\tilde g}\gtrsim 1.0\TeV$ under the V-GMSB model.

Another important point in this figure is that the mass parameter $M_V$ is also constrained for successful explanation of the muon $g-2$ anomaly.
For (c)$M\s{mess}=10^6\GeV$ case, the mass parameter $M_V$ should be $\sim800\TO1200\GeV$; for (a) and (b), it should be lighter as $\sim700\TO1000\GeV$.
Therefore, the vector-like quark should be lighter than $\lesssim1.1\TeV$, and this is obviously within the reach of the $14\TeV$ LHC.

As for the vector-like quark searches, the one by the ATLAS collaboration~\cite{ATLAS:2012qe} is very interesting and promising (cf.~Fig.~\ref{fig:vqbound_2d}).
This analysis is based on the requirement of at least one $b$-jet, but since more $b$-quarks are expected from the decays of $t'_1$, much more parameter regions is expected to be excluded with multi-$b$-jets analysis~\cite{Harigaya:2012ir}.
The importance of $b$-tagging should be, therefore, emphasized now.

Several future works are remained just for the vector-like quark searches.
First, in this dissertation we simply restrict the model to have no mixings between the vector-like matters and the Standard Model fermions in the first and second generations, and to have small mixings between the vector-like matters and the third generation.
Actually the $b$-jet signature is provided by this assumption; if the vector-like quarks mix with the lighter quarks, other strategies are mandatory for vector-like quark searches.
In this context, quantitative evaluation of current experimental bounds on the mixing parameters is awaited.

Analyses for the cases with a short-lived NLSP are, although we briefly discussed, left as future works.

\starline

From a theoretical viewpoint, the V-MSSM scenario has several interesting features.
One of them that the large $\mu$-term.
This feature actually worsens the argument of the little hierarchy problem.
Therefore, it is still difficult to settle the collision between the little hierarchy and the $126\GeV$ Higgs mass, and hence, with examining this argument, another window towards fundamental theories might arise, as we have seen that the muon $g-2$ anomaly bore the V-GMSB scenario.

The mechanism to suppress the coupling $Y''$ is also worth investigated, for which a model with Peccei--Quinn symmetry was proposed~\cite{Nakayama:2012zc}.

Cosmological discussion, which is not performed in this dissertation at all, is also left as future works, such as the effect of the vector-like matters to the history of our Universe, and consequences of the large strong coupling $g_3$ in the early universe (cf.~Fig.~\ref{fig:vmssm-gcu}).

\subappendix
\section{Renormalization Group Equations for the V-MSSM}
\label{app:vmssm-rge}
The two-loop level renormalization group equations of the V-MSSM are shown here, which are used in our numerical evaluation in this chapter.
As emphasized in Ref.~\cite{Martin:2009bg}, the two-loop level effect is significant especially for the running of the gaugino masses.

For completeness all the $\beta$-functions up to two-loop level of the renormalization group equations are included. The $\beta$-functions of the equations are defined as
\begin{equation}
 \diff{X(Q)}{\log Q} = \frac{1}{16\pi^2} \beta^{(1)}\left[X\right] + \frac{1}{\left(16\pi^2\right)^2} \beta^{(2)}\left[X\right],
\end{equation}
where $X$ are one of the parameters and $Q$ is the renormalization scale.
Shown are not only the $\beta$-functions for the extra parameters but also corrections to the MSSM $\beta$-functions $\beta\stx{MSSM}$,  i.e.
\begin{equation}
\beta^{(i)}\stx{V-MSSM} = \beta^{(i)}\stx{MSSM} + \Delta\beta^{(i)}.
\end{equation}
Note that the definition of the parameters is shown in Eqs.~\eqref{eq:VMSSMsuperpotential} and \eqref{eq:VMSSMsusybreaking}, and the MSSM $\beta$-functions are listed in Appendix~\ref{sec:MSSM_RGE}.

The $\beta$-functions are calculated with \withpackage[1.1]{Susyno}~\cite{Susyno}.

\subsection{Restriction and notation}
The following assumptions are employed, as is done for the MSSM $\beta$-functions (in Appendix \ref{sec:MSSM_RGE}).
\begin{itemize}
 \item The $R$-parity is conserved.
 \item The scalar soft mass terms $m^2_X$ are diagonal,
 \item For the $A$-terms $a_X$ and the \YUKAWA~coupling $Y_X$, all the components but the $(3,3)$ are neglected.
 \item The gaugino masses $M_a$ are real.
\end{itemize}
The $\DRbarPrime$ scheme~\cite{Jack:1994rk} is chosen as the renormalization scheme.

\starline

\allowdisplaybreaks[1]
The following variables are used in the expressions of the $\beta$-functions.
\begin{align*}
 X_t    &:= 2a_t^2    + 2Y_t^2    \left(m^2_{\Hu}+(m^2_{Q})_{33}+(m^2_{\bU})_{33}\right)\\
 X_b    &:= 2a_b^2    + 2Y_b^2    \left(m^2_{\Hd}+(m^2_{Q})_{33}+(m^2_{\bD})_{33}\right)\\
 X_\tau &:= 2a_\tau^2 + 2Y_\tau^2 \left(m^2_{\Hd}+(m^2_{L})_{33}+(m^2_{\bE})_{33}\right)\\
 X'     &:= 2a'^2     + 2Y'^2     \left(m^2_{\Hu}+m^2_{Q'}+m^2_{\bU'}\right)\\
 X''    &:= 2a''^2    + 2Y''^2    \left(m^2_{\Hd}+m^2_{\bQ'}+m^2_{U'}\right)\\
 \tilde a_{(t,b,\tau)} &:= Y_{(t,b,\tau)}a_{(t,b,\tau)}\\
 \tilde a^{\prime(\prime)} &:= Y^{\prime(\prime)}a^{\prime(\prime)}\\
 S      &:= m^2_{\Hu}-m^2_{\Hd} +\sum_{i=1}^3\left[ (m^2_{Q})_{ii} -2 (m^2_{\bU})_{ii}+ (m^2_{\bD})_{ii}- (m^2_{L})_{ii}+ (m^2_{\bE})_{ii}\right]\\
 S'     &:= \left(m^2_{Q'} - 2m^2_{\bU'} + m^2_{\bE'}\right)-\left(m^2_{\bQ'} - 2m^2_{U'} + m^2_{E'}\right)\\
 \begin{split}
 S^{(2)}&:=
-Y_t^2\left(3m^2_{\Hu}+(m^2_Q)_{33}-4(m^2_{\bU})_{33}\right)
+Y_b^2\left(3m^2_{\Hd}-(m^2_Q)_{33}-2(m^2_{\bD})_{33}\right)
+Y_\tau^2\left(m^2_{\Hd}+(m^2_L)_{33}-2(m^2_{\bE})_{33}\right)
 \\&\qquad
+\left(\frac3{10}{g_1^2}+\frac32g_2^2\right)\left(m^2_{\Hu}-m^2_{\Hd}-\sum_{i=1}^3 (m^2_{L})_{ii}\right)
+\left(\frac1{30}g_1^2+\frac32g_2^2+\frac83g_3^2\right)
\sum_{i=1}^3 (m^2_{Q})_{ii}
 \\&\qquad
-\left(\frac{16}{15}g_1^2+\frac{16}3g_3^2\right)\sum_{i=1}^3 (m^2_{\bU})_{ii}
+\left(\frac{2}{15}g_1^2+\frac83g_3^2\right)\sum_{i=1}^3 (m^2_{\bD})_{ii}
+\frac65g_1^2\sum_{i=1}^3 (m^2_{\bE})_{ii}
 \end{split}\\
\sigma_1&:=\frac15 g_1^2\left[ 3m^2_{\Hu}+3m^2_{\Hd}+\sum_{i=1}^3\left((m^2_Q)_{ii}+8(m^2_{\bU})_{ii}+2(m^2_{\bD})_{ii}+3(m^2_L)_{ii}+6(m^2_{\bE})_{ii}\right)\right]\\
\sigma_2&:=g_2^2\left[m^2_{\Hu}+m^2_{\Hd}+\sum_{i=1}^3\left((3m^2_Q)_{ii}+(m^2_L)_{ii}\right)\right]\\
\sigma_3&:=g_3^2\sum_{i=1}^3\left(2(m^2_Q)_{ii}+(m^2_{\bU})_{ii}+(m^2_{\bD})_{ii}\right)\\
 S^{(2)}\s{tot}&:=S^{(2)}
-Y'^2\left(3m^2_{\Hu}+m^2_{Q'}-4m^2_{\bU'}\right)
+Y''^2\left(3m^2_{\Hd}+m^2_{\bQ'}-4m^2_{U'}\right)\\
&\qquad\qquad
+\left(\frac1{30}g_1^2+\frac32g_2^2+\frac83g_3^2\right)\left(m^2_{Q'}-m^2_{\bQ'}\right)
-\left(\frac{16}{15}g_1^2+\frac{16}3g_3^2\right)\left(m^2_{\bU'}-m^2_{U'}\right)
+\frac65g_1^2\left(m^2_{\bE'}-m^2_{E'}\right)\\
\sigma_{{\rm tot};1}&:=\sigma_1+\frac15 g_1^2\left(m^2_{Q'} + m^2_{\bQ'} + 8m^2_{\bU'}+ 8m^2_{U'} + 6m^2_{\bE'}+6m^2_{E'}\right)\\
\sigma_{{\rm tot};2}&:=\sigma_2+g_2^2\left(3m^2_{Q'}+3m^2_{\bQ'}\right)\\
\sigma_{{\rm tot};3}&:=\sigma_3+g_3^2\left(2m^2_{Q'}+2m^2_{\bQ'}+m^2_{\bU'}+m^2_{U'}\right)
\end{align*}
\allowdisplaybreaks[0]

\allowdisplaybreaks[4]
\subsection[One-loop level \texorpdfstring{$\beta$}{beta}-functions]{One-loop level $\boldsymbol \beta$-functions}
\paragraph{Corrections to the evolution of the MSSM parameters}
\begin{align}
\Delta\beta^{(1)}\left[g_i\right] &= 3 g_i^3\\
\Delta\beta^{(1)}\left[M_i\right] &= 6 g_i^2 M_i\\
\Delta\beta^{(1)}\left[Y_{t}\right] &= 3 Y'^2 Y_{t}\\
\Delta\beta^{(1)}\left[Y_{b}\right] &= 3 Y''^2 Y_{b}\\
\Delta\beta^{(1)}\left[Y_{\tau}\right] &= 3 Y''^2 Y_{\tau}\\
\Delta\beta^{(1)}\left[a_{t}\right] &= 3 Y'\left(Y' a_{t}+2 a' Y_{t}\right)\\
\Delta\beta^{(1)}\left[a_{b}\right] &= 3 Y''\left(Y'' a_{b}+2 a'' Y_{b}\right)\\
\Delta\beta^{(1)}\left[a_{\tau}\right] &= 3 Y''\left(Y'' a_{\tau}+2 a'' Y_{\tau}\right)\\
\Delta\beta^{(1)}\left[\mu\right] &= 3 \mu  \left(Y'^2+Y''^2\right)\\
\Delta\beta^{(1)}\left[b\right] &= 6\mu\left(\tilde a' +\tilde a''\right)+3 b \left(Y'^2+Y''^2\right)\\
\Delta\beta^{(1)}\left[m_{\Hu}^2\right] &= \frac{3}{5} g_1^2 S'+3 X'\\
\Delta\beta^{(1)}\left[m_{\Hd}^2\right] &= 3 X'-\frac{3}{5} g_1^2 S'\\
\Delta\beta^{(1)}\left[(m_{Q}^2){}_{ii}\right] &= \frac15 g_1^2 S'\\
\Delta\beta^{(1)}\left[(m_{\bU}^2){}_{ii}\right] &= -\frac{4}{5} g_1^2 S'\\
\Delta\beta^{(1)}\left[(m_{\bD}^2){}_{ii}\right] &= \frac{2}{5} g_1^2 S'\\
\Delta\beta^{(1)}\left[(m_{L}^2){}_{ii}\right] &= -\frac{3}{5} g_1^2 S'\\
\Delta\beta^{(1)}\left[(m_{\bE}^2){}_{ii}\right] &= \frac{6}{5} g_1^2 S'
\end{align}

\paragraph{Evolution of the extra parameters}
\begin{align}
\beta^{(1)}\left[Y'\right] &= \left(3 Y_{t}^2+6Y'^2-\frac{13}{15} g_1^2-3 g_2^2-\frac{16}{3}g_3^2\right)Y'\\
\beta^{(1)}\left[Y''\right] &= \left(3 Y_b^2+Y_{\tau}^2+6Y''^2-\frac{13}{15}g_1^2-3 g_2^2-\frac{16}{3} g_3^2\right)Y''\\
\beta^{(1)}\left[M_{Q'}\right] &= \left(Y'^2+Y''^2-\frac{1}{15}g_1^2-3 g_2^2-\frac{16}{3}g_3^2\right)M_{Q'} \\
\beta^{(1)}\left[M_{U'}\right] &= \left(2Y'^2+2Y''^2-\frac{16}{15} g_1^2-\frac{16}{3}g_3^2\right)M_{U'} \\
\beta^{(1)}\left[M_{E'}\right] &= -\frac{12}{5} g_1^2 M_{E'}\\
\beta^{(1)}\left[a'\right] &=  \left(6 \tilde a_{t}+\frac{26}{15} g_1^2 M_1+6 g_2^2M_2+\frac{32}{3} g_3^2 M_3\right)Y'
+ \left(3 Y_{t}^2+18 Y'^2-\frac{13}{15} g_1^2-3g_2^2-\frac{16}{3} g_3^2\right)a'\\
 \begin{split}
\beta^{(1)}\left[a''\right] &= \left(6 \tilde a_{b} +2 \tilde a_{\tau}+\frac{26}{15} g_1^2 M_1+6 g_2^2 M_2+\frac{32}{3} g_3^2 M_3\right)Y''
\\&\hspace{100pt}
+\left(3 Y_b^2+Y_{\tau }^2+18 Y''^2-\frac{13}{15} g_1^2-3 g_2^2-\frac{16}{3}g_3^2\right)a''
\end{split}\\
\begin{split}
\beta^{(1)}\left[b_{Q'}\right] &=
\left(2 \tilde a' +2 \tilde a''+\frac{2}{15} g_1^2M_1+6 g_2^2 M_2+\frac{32}{3} g_3^2 M_3\right)M_{Q'}
\\&\hspace{100pt}
+\left(-\frac{1}{15}g_1^2-3 g_2^2-\frac{16}{3} g_3^2+Y'^2+Y''^2\right)b_{Q'}
\end{split}\\
 \beta^{(1)}\left[b_{U'}\right] &=  \left(4 \tilde a' +4 \tilde a'' +\frac{32}{15} g_1^2M_1+\frac{32}{3} g_3^2 M_3\right)M_{U'}
 +\left(-\frac{16}{15} g_1^2-\frac{16}{3}g_3^2+2 Y'^2+2 Y''^2\right)b_{U'} \\
\\
\beta^{(1)}\left[b_{E'}\right] &= \frac{24}{5} g_1^2 M_1 M_{E'}-\frac{12}{5} g_1^2b_{E'}\\
\beta^{(1)}\left[m_{Q'}^2\right] &= -\frac{2}{15} g_1^2 M_1^2-6 g_2^2M_2^2-\frac{32}{3} g_3^2 M_3^2+\frac15g_1^2(S+S')+X'\\
\beta^{(1)}\left[m_{\bU'}^2\right] &= -\frac{32}{15} g_1^2 M_1^2-\frac{32}{3} g_3^2M_3^2-\frac{4}{5} g_1^2 (S+S')+2 X'\\
\beta^{(1)}\left[m_{\bE'}^2\right] &= -\frac{24}{5} g_1^2 M_1^2
+\frac{6}{5} g_1^2(S+S')\\
\beta^{(1)}\left[m_{\bQ'}^2\right] &= -\frac{2}{15} g_1^2 M_1^2-6 g_2^2M_2^2-\frac{32}{3} g_3^2 M_3^2-\frac15g_1^2(S+S')+X'\\
\beta^{(1)}\left[m_{U'}^2\right] &= -\frac{32}{15} g_1^2 M_1^2-\frac{32}{3} g_3^2M_3^2+\frac{4}{5} g_1^2 (S+S')+2 X'\\
\beta^{(1)}\left[m_{E'}^2\right] &= -\frac{24}{5} g_1^2 M_1^2-\frac{6}{5} g_1^2(S+S')
\end{align}


\subsection[Two-loop level \texorpdfstring{$\beta$}{beta}-functions]{Two-loop level $\boldsymbol \beta$-functions}
\paragraph{Corrections to the evolution of the MSSM parameters}
\begin{align}
\Delta\beta^{(2)}\left[g_1\right] &= \left(\frac{23}{5} g_1^2+\frac{3}{5}g_2^2+\frac{48}{5} g_3^2-\frac{26}{5} Y'^2-\frac{26}{5} Y''^2\right)g_1^3\\
\Delta\beta^{(2)}\left[g_2\right] &= \left(\frac{1}{5}g_1^2+21 g_2^2+16 g_3^2-6 Y'^2-6Y''^2\right)g_2^3\\
\Delta\beta^{(2)}\left[g_3\right] &=  \left(\frac{6}{5} g_1^2+6 g_2^2+34 g_3^2-4 Y'^2-4Y''^2\right)g_3^3\\
\Delta\beta^{(2)}\left[M_1\right] &= \left[\frac{52}{5} \left(\tilde a'-Y'^2M_1+ \tilde a''-Y''^2M_1\right)+\frac{92}{5} g_1^2 M_1+\frac{6}{5} g_2^2 \left(M_1+M_2\right)+\frac{96}{5} g_3^2\left(M_1+M_3\right)\right]g_1^2\\
\Delta\beta^{(2)}\left[M_2\right] &= \left[12\left(\tilde a'-Y'^2M_2+\tilde a''-Y''^2M_2\right)+\frac{2}{5} g_1^2\left(M_1+M_2\right)+84 g_2^2 M_2+32 g_3^2 \left(M_2+M_3\right)\right]g_2^2\\
\Delta\beta^{(2)}\left[M_3\right] &= \left[8\left(\tilde a'-Y'^2M_3+\tilde a''-Y''^2M_3\right)+\frac{12}{5} g_1^2\left(M_1+M_3\right)+12 g_2^2 \left(M_2+M_3\right)+136 g_3^2 M_3\right]g_3^2\\
\Delta\beta^{(2)}\left[Y_t\right] &=\left[
\frac{13}{5} g_1^4 +9 g_2^4 +16g_3^4
+\left(\frac{4}{5} g_1^2 +16 g_3^2-9Y'^2-9Y_t^2\right)Y'^2 -3 Y_b^2Y''^2
\right]y_t
\\
\Delta\beta^{(2)}\left[Y_b\right] &=\left[
\frac{7}{5} g_1^4 +9 g_2^4 +16g_3^4
 -3 Y_t^2Y'^2 +\left(\frac{4}{5} g_1^2 +16 g_3^2 -9 Y_b^2 -9 Y''^2\right)Y''^2
\right]Y_b
\\
\Delta\beta^{(2)}\left[Y_\tau \right] &=\left[ \frac{27}{5} g_1^4  +9 g_2^4+\left( \frac{4}{5} g_1^2 +16g_3^2   -9 Y_\tau^2-9 Y''^2\right)Y''^2 \right]Y_\tau\\
\begin{split}
 \Delta\beta^{(2)}\left[a_t\right] &=
\frac{13}{5} g_1^4\left( a_t-4Y_tM_1\right)+9g_2^4 \left(a_t-4Y_tM_2\right)+16g_3^4 \left(a_t-4Y_tM_3\right)
 \\&\hspace{20pt}
+\frac45 g_1^2 Y'\left(Y'a_t+2a' Y_t-2M_1 Y' Y_t\right)
+16g_3^2Y'\left(Y'a_t+2 a' Y_t-2 M_3 Y' Y_t\right)
 \\&\hspace{20pt}
 +Y'\left(-27Y'a_t Y_t^2-9 Y'^3 a_t-36 a' Y'^2 Y_t-18 a' Y_t^3\right)
 \\&\hspace{20pt}
 +Y_bY''\left(-3 Y'' a_t Y_b-6 Y'' Y_t a_b-6 a'' Y_t Y_b\right)
\end{split}\\
\begin{split}
\Delta\beta^{(2)}\left[a_b\right] &=
\frac75g_1^4 \left(a_b-4Y_bM_1\right)+9g_2^4 \left( a_b-4Y_bM_2\right)+16g_3^4 \left(a_b-4 Y_bM_3\right)
 \\&\hspace{20pt}
+\frac45g_1^2Y'' \left(Y''a_b+2a''Y_b-2 M_1 Y''Y_b\right)+16g_3^2 Y''\left(Y'' a_b+2 a'' Y_b-2 M_3 Y''Y_b\right)
 \\&\hspace{20pt}
+Y_tY'\left(-3 Y' Y_t a_b-6 Y' a_t Y_b-6 a' Y_t Y_b\right)
 \\&\hspace{20pt}
+Y''\left(-27 Y'' a_b Y_b^2-9 Y''^3 a_b-36 a''Y''^2 Y_b-18 a'' Y_b^3\right)
\end{split}\\
\begin{split}
\Delta\beta^{(2)}\left[a_\tau \right] &=
\frac{27}{5}g_1^4 \left(a_\tau-4Y_\tau M_1\right)+9g_2^4 \left(a_\tau -4 Y_\tau M_2\right)
 \\&\hspace{20pt}
+ \frac45g_1^2Y'' \left(Y'' a_\tau+2 a'' Y_\tau -2 M_1 Y'' Y_\tau \right)
+16g_3^2Y'' \left(Y'' a_\tau +2 a'' Y_\tau -2 M_3Y'' Y_\tau \right)
 \\&\hspace{20pt}
+Y''\left(-27 Y'' a_\tau  Y_\tau ^2-9 Y''^3 a_\tau -36 a''Y''^2 Y_\tau -18 a'' Y_\tau ^3\right)
\end{split}\\
\Delta\beta^{(2)}\left[\mu\right] &= \left[
\frac{9}{5} g_1^4+9 g_2^4+\frac{4}{5} g_1^2 \left(Y'^2+Y''^2\right)+16 g_3^2\left(Y'^2+Y''^2\right)-9\left(Y'^4+Y''^4\right)\right]\mu\\
\begin{split}
 \Delta\beta^{(2)}\left[b\right] &= 
 \Bigg[
-\frac{36}{5} g_1^4 M_1-36 g_2^4 M_2
 -36\tilde a' Y'^2-36 \tilde a'' Y''^2
 \\&\hspace{50pt}
+\frac{8}{5} g_1^2\left(\tilde a'+\tilde a''-\left(Y'^2+Y''^2\right)M_1\right)
+32 g_3^2\left(\tilde a'+\tilde a''-\left(Y'^2+Y''^2\right)M_3\right)
\Bigg]\mu
 \\&\hspace{20pt}
+\left[
\frac{9}{5} g_1^4+9 g_2^4+\frac{4}{5} g_1^2 \left(Y'^2+Y''^2\right)+16 g_3^2\left(Y'^2+Y''^2\right)-9\left(Y'^4+Y''^4\right)\right]b
\end{split}\\
\begin{split}
 \Delta\beta^{(2)}\left[m_{\Hu}^2\right] &= \frac{54}{5} g_1^4 M_1^2+54 g_2^4 M_2^2+\frac{6}{5} g_1^2S'+\frac{3}{5}g_1^2 \sigma'_1+3 g_2^2 \sigma'_2
\\&\hspace{20pt}
+\frac{4}{5}g_1^2 \left(X'-4 M_1\tilde{a}'+4 M_1^2 Y'^2\right)
+16g_3^2 \left(X'-4 M_3 \tilde{a}'+4 M_3^2 Y'^2\right)
-36\tilde{a}'^2-18 Y'^2 X'
\end{split}\\
\begin{split}
 \Delta\beta^{(2)}\left[m_{\Hd}^2\right] &=\frac{54}{5} g_1^4 M_1^2+54 g_2^4 M_2^2-\frac{6}{5}g_1^2 S'+\frac{3}{5}g_1^2 \sigma'_1+3 g_2^2 \sigma'_2
\\&\hspace{20pt}+\frac{4}{5} g_1^2 \left(X''-4 M_1\tilde{a}''+4 M_1^2 Y''^2\right)
+16g_3^2 \left(X''-4 M_3 \tilde{a}''+4 M_3^2Y''^2\right)-36 \tilde{a}''^2-18 Y''^2 X''
\end{split}\\
\begin{split}
\Delta\beta^{(2)}\left[(m_{Q}^2){}_{ii}\right] &= \frac{6}{5} g_1^4 M_1^2+54 g_2^4 M_2^2+96g_3^4 M_3^2+g_1^2 \left(\frac{2}{5} S'+\frac{\sigma'_1}{15}\right)+3 g_2^2 \sigma'_2+\frac{16}{3} g_3^2 \sigma'_3
\\&\hspace{20px}+\Bigg\langle\!\Bigg\langle -12 \tilde{a}' \tilde{a}_t-3Y_t^2 X'-12 \tilde{a}'' \tilde{a}_b-3 Y_b^2 X''-3 Y''^2 X_b-3 Y'^2X_t\Bigg\rangle\!\Bigg\rangle\stx{for $i=3$}
\end{split}\\
\begin{split}
\Delta\beta^{(2)}\left[(m_{\bU}^2){}_{ii}\right] &= \frac{96}{5} g_1^4 M_1^2+96 g_3^4M_3^2-\frac{8}{5}g_1^2S'+\frac{16}{15}g_1^2 \sigma'_1+\frac{16}{3} g_3^2 \sigma'_3
\\&\hspace{100px}+\Bigg\langle\!\Bigg\langle -24 \tilde{a}' \tilde{a}_t-6Y_t^2 X'-6 Y'^2 X_t\Bigg\rangle\!\Bigg\rangle\stx{for $i=3$}
\end{split}\\
\begin{split}
\Delta\beta^{(2)}\left[(m_{\bD}^2){}_{ii}\right] &= \frac{24}{5} g_1^4 M_1^2+96 g_3^4M_3^2+\frac{4}{5} g_1^2 S'+\frac{4}{15} g_1^2 \sigma'_1+\frac{16}{3} g_3^2 \sigma'_3
\\&\hspace{100px}
+\Bigg\langle\!\Bigg\langle -24 \tilde{a}'' \tilde{a}_b-6Y_b^2 X''-6 Y''^2 X_b\Bigg\rangle\!\Bigg\rangle\stx{for $i=3$}
 \end{split}\\
\begin{split}
\Delta\beta^{(2)}\left[(m_{L}^2){}_{ii}\right] &= \frac{54}{5} g_1^4 M_1^2+54 g_2^4M_2^2-\frac{6}{5} g_1^2 S'+\frac{3}{5} g_1^2 \sigma'_1+3 g_2^2\sigma'_2
\\&\hspace{100px}
+\Bigg\langle\!\Bigg\langle -12 \tilde{a}'' \tilde{a}_\tau -3Y_\tau ^2 X''-3 Y''^2 X_{\tau}\Bigg\rangle\!\Bigg\rangle\stx{for $i=3$}
\end{split}\\
\begin{split}
\Delta\beta^{(2)}\left[(m_{\bE}^2){}_{ii}\right] &= \frac{216}{5} g_1^4 M_1^2+\frac{12}{5} g_1^2S'+\frac{12}{5} g_1^2 \sigma'_1
+\Bigg\langle\!\Bigg\langle -24 \tilde{a}'' \tilde{a}_\tau -6Y_\tau ^2 X''-6 Y''^2 X_{\tau}\Bigg\rangle\!\Bigg\rangle\stx{for $i=3$}
\end{split}
\end{align}
\paragraph{Evolution of the extra parameters}
\begin{align}
\begin{split}
 \beta^{(2)}\left[Y'\right] &= \Bigg[
 \frac{3913}{450} g_1^4+\frac{33}{2} g_2^4+\frac{128}{9} g_3^4
 +g_1^2g_2^2+\frac{136}{45} g_1^2g_3^2+8g_2^2 g_3^2
 \\&\hspace{20pt}
+\left(\frac{4}{5}Y_t^2+\frac{6}{5}Y'^2\right)g_1^2
+6 g_2^2Y'^2
+16g_3^2(Y_t^2+Y'^2)
 -3 Y_t^2 Y_b^2
-9Y_t^4
 -9 Y_t^2Y'^2
-22 Y'^4
 \Bigg]Y'
\end{split}
\\
\begin{split}
 \beta^{(2)}\left[Y''\right] &= \Bigg[
\frac{3913}{450} g_1^4+\frac{33}{2}g_2^4+\frac{128}{9} g_3^4
 +g_1^2g_2^2+\frac{136}{45} g_1^2g_3^2+8g_2^2 g_3^2
 \\&\hspace{20pt}
+g_1^2 \left(-\frac{2}{5}Y_b^2+\frac{6}{5} Y_\tau ^2+\frac65Y''^2\right)+6 g_2^2Y''^2+16g_3^2 (Y_b^2+Y''^2)
-9 Y_b^4-3 Y_\tau ^4
 \\&\hspace{20pt}
-3 Y_t^2 Y_b^2
+Y''^2 \left(-9 Y_b^2-3Y_\tau ^2-22Y''^2\right)\Bigg]Y''
\end{split}\\
\begin{split}
 \beta^{(2)}\left[M_{Q'}\right] &= \Bigg[
\frac{289}{450} g_1^4 +\frac{33}{2} g_2^4 +\frac{128}{9}g_3^4 +\frac{1}{5} g_1^2 g_2^2 +\frac{16}{45} g_1^2 g_3^2+16 g_2^2 g_3^2
 \\&\hspace{50pt}
+\frac{4}{5} g_1^2 (Y'^2+Y''^2) -3 Y'^2 Y_t^2 -3 Y''^2 Y_b^2 -Y''^2Y_\tau ^2 
-5 Y'^4 -5 Y''^4
\Bigg]M_{Q'}
\end{split}\\
\begin{split}
 \beta^{(2)}\left[M_{U'}\right] &= \Bigg[\frac{2432}{225}g_1^4 +\frac{128}{9} g_3^4 +\frac{256}{45} g_1^2 g_3^2
-\frac{2}{5} g_1^2 \left(Y'^2+Y''^2\right) +6 g_2^2 \left(Y'^2+Y''^2\right)
 \\&\hspace{50pt}
-6 Y'^2 Y_t^2 -6 Y''^2 Y_b^2 -2Y''^2 Y_\tau ^2 -8Y'^4 -8 Y''^4 \Bigg]M_{U'}
\end{split}\\
\beta^{(2)}\left[M_{E'}\right] &= \frac{648}{25} g_1^4 M_{E'}\\
\begin{split}
\beta^{(2)}\left[a'\right] &=
\frac{3913}{450}g_1^4 \left(a'-4Y'M_1\right)+\frac{33}{2}g_2^4 \left(a'-4Y' M_2\right)+\frac{128}{9}g_3^4 \left(a'-4Y'M_3\right)
 \\&\hspace{20pt}
+8g_2^2g_3^2 \left(a'-2Y'M_2-2Y'M_3\right)
+g_1^2g_2^2 \left(a'-2 Y'M_1-2 Y'M_2\right)
 \\&\hspace{20pt}
+\frac{136}{45}g_1^2g_3^2 \left(a'-2Y' M_1 -2Y'M_3\right)
+g_1^2\left(\frac{8}{5} Y' \tilde a_t+\frac{4}{5} a' Y_t^2-\frac{8}{5} M_1Y' Y_t^2+\frac{18}{5} \tilde a'Y'-\frac{12}{5} M_1 Y'^3\right)
 \\&\hspace{20pt}
+g_2^2 \left(18 \tilde a' Y'-12 M_2 Y'^3\right)
+g_3^2 \left(32 Y' \tilde a_t+16 a'Y_t^2-32 M_3 Y' Y_t^2+48 \tilde a' Y'-32 M_3 Y'^3\right)
 \\&\hspace{20pt}
-6 Y' \tilde a_t Y_b^2-6 Y' Y_t^2 \tilde a_b-3 a' Y_t^2 Y_b^2
-18 Y'^3 \tilde a_t-36 Y' \tilde a_tY_t^2-27 \tilde a' Y' Y_t^2-9 \tilde a' Y_t^3
-110 a'Y'^4
\end{split}
\\
\begin{split}
\beta^{(2)}\left[a''\right] &= 
\frac{3913}{450}g_1^4 \left(a''-4Y''M_1\right)+\frac{33}{2}g_2^4 \left(a''-4Y'' M_2\right)+\frac{128}{9}g_3^4 \left(a''-4Y''M_3\right)
 \\&\hspace{20pt}
+8g_2^2g_3^2 \left(a''-2Y''M_2-2Y''M_3\right)
+g_1^2g_2^2 \left(a''-2 Y''M_1-2 Y''M_2\right)
 \\&\hspace{20pt}
+\frac{136}{45}g_1^2g_3^2 \left(a''-2Y'' M_1 -2Y''M_3\right)
 \\&\hspace{20pt}
+
g_1^2 \left(
\frac{4}{5} \left(Y_b^2-3Y_\tau ^2-3Y''^2\right)Y''M_1
-\frac{4}{5} Y'' a_b Y_b
+\frac{12}{5} Y'' a_\tau Y_\tau
+\left(\frac{18}{5}Y''-\frac25Y_b^2+\frac65Y_\tau^2\right)a''
\right)
 \\&\hspace{20pt}
+g_2^2\left(18 Y'' \tilde a''-12 M_2 Y''^3\right)
+16g_3^2 \left(3 Y'' \tilde a''+2 Y'' a_bY_b+ a'' Y_b^2-2 M_3 Y'' Y_b^2-2 M_3 Y''^3\right)
 \\&\hspace{20pt}
-27 Y''\tilde a'' Y_b^2-9 Y'' \tilde a'' Y_\tau ^2
-110 Y''^3 \tilde a''-6 Y'' a_tY_t Y_b^2-6 Y'' Y_t^2 a_b Y_b-3 a'' Y_t^2 Y_b^2
 \\&\hspace{20pt}
-18Y''^3 a_b Y_b-36 Y'' a_b Y_b^3-9 a'' Y_b^4-6 Y''^3 a_\tau  Y_\tau -12 Y'' a_\tau  Y_\tau ^3-3 a'' Y_\tau ^4
\end{split}\\
\begin{split}
 \beta^{(2)}\left[b_{Q'}\right] &= \Bigg[
-\frac{578}{225}g_1^4 M_1-66 g_2^4 M_2-\frac{512}{9}g_3^4 M_3
-\frac25g_1^2g_2^2 \left(M_1+M_2\right)-\frac{32}{45}g_1^2g_3^2\left(M_1+M_3\right)
\\&\hspace{20pt}
 -32g_2^2 g_3^2 \left( M_2+M_3\right)+\frac85g_1^2 \left(\tilde a'+\tilde a''- M_1Y'^2-M_1 Y''^2\right)
\\&\hspace{20pt}
-6 Y'^2 \tilde a_t-6 \tilde a' Y_t^2-6Y''^2 \tilde a_b -6 \tilde a'' Y_b^2-2 Y''^2 \tilde a_\tau -2\tilde a'' Y_\tau ^2
-20 \tilde a' Y'^2-20 \tilde a'' Y''^2\Bigg]M_{Q'}
\\&\quad
+\Bigg[
\frac{289}{450} g_1^4 +\frac{33}{2} g_2^4 +\frac{128}{9}g_3^4 +\frac{1}{5} g_1^2 g_2^2 +\frac{16}{45} g_1^2 g_3^2+16 g_2^2 g_3^2
 \\&\hspace{20pt}
+\frac{4}{5} g_1^2 (Y'^2+Y''^2) -3 Y'^2 Y_t^2 -3 Y''^2 Y_b^2 -Y''^2Y_\tau ^2 
-5 Y'^4 -5 Y''^4
\Bigg]b_{Q'}
\end{split}\\
\begin{split}
\beta^{(2)}\left[b_{U'}\right] &= \Bigg[
-\frac{9728}{225} g_1^4 M_1-\frac{512}{9} g_3^4M_3-\frac{512}{45}g_1^2g_3^2\left(M_1+M_3\right)
 \\&\hspace{20pt}
-\frac45g_1^2\left(\tilde a'+\tilde a''-M_1Y'^2-M_1 Y''^2\right)+12g_2^2 \left(\tilde a'+\tilde a''- M_2 Y'^2- M_2 Y''^2\right)
 \\&\hspace{20pt}
-32 Y'^2 \tilde a'-32Y''^2 \tilde a''-12 Y'^2\tilde a_t-12 Y''^2\tilde a_b-4 Y''^2\tilde a_\tau
-12 \tilde a' Y_t^2-12\tilde a'' Y_b^2-4 \tilde a'' Y_\tau ^2
\Bigg]M_{U'}
 \\&\quad
+\Bigg[\frac{2432}{225}g_1^4 +\frac{128}{9} g_3^4 +\frac{256}{45} g_1^2 g_3^2
-\frac{2}{5} g_1^2 \left(Y'^2+Y''^2\right) +6 g_2^2 \left(Y'^2+Y''^2\right)
 \\&\hspace{20pt}
-6 Y'^2 Y_t^2 -6 Y''^2 Y_b^2 -2Y''^2 Y_\tau ^2 -8Y'^4 -8 Y''^4 \Bigg]b_{U'}
\end{split}\\
\beta^{(2)}\left[b_{E'}\right] &= \frac{648}{25} g_1^4 b_{E'}-\frac{2592}{25} g_1^4 M_1M_{E'}\\
\begin{split}
\beta^{(2)}\left[m_{Q'}^2\right] &=
\frac{289}{75} g_1^4 M_1^2+87 g_2^4M_2^2+\frac{160}{3} g_3^4 M_3^2
+\frac25g_1^2g_2^2 \left(M_1^2+M_2^2+ M_1M_2\right)
 \\&\hspace{20pt}
+\frac{32}{45}g_1^2g_3^2 \left(M_1^2+M_3^2+M_1M_3\right)
+ 32g_2^2g_3^2 \left(M_2^2+M_3^2+M_2M_3\right)
 \\&\hspace{20pt}
+\frac{2}{5}g_1^2 S^{(2)}\s{tot}+\frac{1}{15}g_1^2\sigma_{{\rm tot};1}
+3g_2^2\sigma_{{\rm tot};2}
+\frac{16}{3} g_3^2 \sigma_{{\rm tot};3}
 \\&\hspace{20pt}
+\frac45g_1^2\left(X'-4M_1\tilde a'+4M_1^2Y'^2\right)
-12 \tilde{a}' \tilde{a}_t-3 Y_t^2 X'-3 Y'^2X_t-20 \tilde{a}'^2-10 Y'^2 X'
\end{split}\\
\begin{split}
\beta^{(2)}\left[m_{\bU'}^2\right] &=
\frac{4864}{75} g_1^4M_1^2+\frac{160}{3} g_3^4 M_3^2
+\frac{512}{45}g_1^2g_3^2 \left(M_1^2+M_3^2+M_1M_3\right)
\\&\hspace{20pt}
-\frac{8}{5}g_1^2S^{(2)}\s{tot}+\frac{16}{15} g_1^2\sigma_{{\rm tot};1}
+\frac{16}{3} g_3^2 \sigma_{{\rm tot};3}
-\frac25g_1^2\left(X'-4M_1 \tilde{a}'+4 M_1^2 Y'^2\right)
\\&\hspace{20pt}
+6g_2^2\left(X'-4M_2 \tilde{a}'+4 M_2^2 Y'^2\right)
 -24 \tilde{a}' \tilde{a}_t-6 Y_t^2 X'-6 Y'^2X_t-32 \tilde{a}'^2-16 Y'^2 X'
\end{split}\\
\begin{split}
\beta^{(2)}\left[m_{\bE'}^2\right] &= \frac{3888}{25} g_1^4 M_1^2+\frac{12}{5}g_1^2 \left(S^{(2)}\s{tot}+\sigma_{{\rm tot};1}\right)
\end{split}\\
\begin{split}
\beta^{(2)}\left[m_{\bQ'}^2\right] &=
\frac{289}{75} g_1^4 M_1^2+87 g_2^4 M_2^2+\frac{160}{3} g_3^4M_3^2
+\frac25g_1^2g_2^2 \left(M_1^2+M_2^2+ M_1M_2\right)
 \\&\hspace{20pt}
+\frac{32}{45}g_1^2g_3^2 \left(M_1^2+M_3^2+M_1M_3\right)
+ 32g_2^2g_3^2 \left(M_2^2+M_3^2+M_2M_3\right)
\\&\hspace{20pt}
-\frac{2}{5}g_1^2S^{(2)}\s{tot}+\frac1{15}g_1^2{\sigma_{{\rm tot}; 1}}
+3g_2^2 \sigma_{{\rm tot}; 2}+\frac{16}{3} g_3^2 \sigma_{{\rm tot}; 3}
+\frac45g_1^2 \left(X''-4M_1 \tilde{a}''+4M_1^2 Y''^2\right)
\\&\hspace{20pt}
-4\tilde{a}'' (3\tilde{a}_b+\tilde{a}_\tau)
-(3 Y_b^2+Y_\tau^2) X''
-3 Y''^2 (X_b+X_\tau)
-20 \tilde{a}''^2
-10 Y''^2 X''
\end{split}\\
\begin{split}
\beta^{(2)}\left[m_{U'}^2\right] &=
\frac{4864}{75} g_1^4M_1^2+\frac{160}{3} g_3^4 M_3^2
+\frac{512}{45}g_1^2g_3^2\left(M_1^2+M_3^2+M_1M_3\right)
\\&\hspace{20pt}
+\frac{8}{5}g_1^2S^{(2)}\s{tot}+\frac{16}{15} g_1^2\sigma_{{\rm tot};1}
+\frac{16}{3} g_3^2 \sigma_{{\rm tot};3}
-\frac25g_1^2\left(X''-4M_1 \tilde{a}''+4 M_1^2 Y''^2\right)
\\&\hspace{20pt}
+6g_2^2 \left(X''-4M_2 \tilde{a}''+4 M_2^2 Y''^2\right)
\\&\hspace{20pt}
-8 \tilde{a}'' \left(3\tilde{a}_b+\tilde a_\tau\right)-2 (3Y_b^2+Y_\tau ^2) X''-2 Y''^2 (3X_b+X_\tau)-32\tilde{a}''^2-16 Y''^2 X''
\\&\hspace{20pt}
\end{split}\\
 \beta^{(2)}\left[m_{E'}^2\right] &= \frac{3888}{25} g_1^4 M_1^2+\frac{12}{5}g_1^2 \left(-S^{(2)}\s{tot}+\sigma_{{\rm tot};1}\right)
\end{align}

\allowdisplaybreaks[0]

\newpage

\section[Decay Rates of \texorpdfstring{$t'_1$}{t'\_1}]{Decay Rates of $\boldsymbol{t'_1}$}
\label{sec:decay-rates-t_1}
Here are the decay rates of the lightest vector-like quark in the V-MSSM, $t_1'$, summarized, which are cited from Ref.~\cite{Martin:2009bg}.
The relevant mixing parameters and the mass matrix are, as already referred,
\begin{align}
  W &= - Y'\Hu Q'\bU' + M_{Q'}Q'\bQ'+M_{U'}\bU'U'
 -\epsilon_u  \Hu Q_3\bU'
 -\epsilon_u' \Hu Q' \bU_3
 +\epsilon_d  \Hd Q' \bD_3;
\\
-{\mathcal L} &\supset
\pmat{Q'_u&U'&t\s L}
\mathcal M_t
\pmat{\bar Q'_u\\\bU'\\\bar t\s R}
+\pmat{Q'_d&b\s L}
\mathcal M_b
\pmat{\bar Q'_d\\\bar b\s R}+\Hc;
\\
&\hspace{0.18\textwidth}
\mathcal M_t :=\pmat{%
 M_{Q'}         & Y'\vu & \epsilon'_u \vu\\
 0  & M_{U'}         & 0\\
 0  & \epsilon_u\vu              & Y_t\vu},
\qquad
\mathcal M_b :=
\pmat{
 -M_{Q'} & \epsilon_d\vd\\
 0 & Y_b\vd}.
\end{align}
Utilizing the singular value decomposition method, we can diagonalize these two matrices with unitary matrices $R, R, L', R'$ as
\begin{align}
  L^* \mathcal M_t R^\dagger &= \pmat{m_t & 0 & 0\\ 0 & m_{t'_1} & 0 \\ 0&0&m_{t'_2}},
&
  L'^*\mathcal M_b R'^\dagger &= \pmat{m_b&0\\ 0&m_{b'}}.
\end{align}
The gauge interactions are extracted to be
\begin{align}
\begin{split}
  \Lag&\supset
 \slashed{W}^-
 \left(
 \frac{g_2}{\sqrt2}Q'^\dagger_d Q'_u
 + \frac{g_2}{\sqrt2}\bQ'^\dagger_d \bQ'_u
 + \frac{g_2}{\sqrt2}\bL^\dagger \tL
 \right)
 +
 \slashed{W}^+
 \left(
  \frac{g_2}{\sqrt2}Q'^\dagger_u Q'_d
 + \frac{g_2}{\sqrt2}\bQ'^\dagger_u \bQ'_d
 + \frac{g_2}{\sqrt2}\tL^\dagger \bL
 \right)\\
 &+ \slashed{A}\left[
  \frac23|e|\left( Q_u'^\dagger Q'_u + U'^\dagger U' +\tL^\dagger\tL
                   - \bQ_u'^\dagger \bQ'_u - \bU'^\dagger \bU' - \bar t\s R^\dagger \bar t\s R\right)
 - \frac13|e|\left( Q_d'^\dagger Q'_d + \bL^\dagger\bL - \bQ_d'^\dagger \bQ'_d -\bar b\s R^\dagger \bar b\s R\right)
 \right]%
 \\&n
 + \slashed{Z}\left[
 \left(\frac{|e|}{2t}-\frac{t|e|}{6}\right)\left(Q_u'^\dagger Q_u' + \tL^\dagger\tL - \bQ_u'^\dagger\bQ_u'\right)
 +
 \left(-\frac{|e|}{2t}-\frac{t|e|}{6}\right)\left(Q_d'^\dagger Q_d' + \bL^\dagger\bL - \bQ_d'^\dagger\bQ_d'\right)
 \right.\\&\qquad\qquad\left.
 -\frac{2t|e|}{3}\left(U'^\dagger U' - \bar t\s R^\dagger\bar t\s R- \bU'^\dagger \bU'\right)
 -\frac{t|e|}{3}\bar b\s R^\dagger\bar b\s R
 \right]
 \\&
 +\left(
-Y_t\HuZ \bar t\s R\tL - Y_b\HdZ \bar b\s R\bL - Y'\HuZ Q'_u \bU' -\epsilon_u\HuZ \tL\bU' -\epsilon_u'\HuZ Q'_u \bar t\s R - \epsilon_d \HdZ Q_u'\bR+\Hc\right)
\end{split}\nonumber
\\
 \begin{split}
&=
 \frac{g_2}{\sqrt2}\left(L'^*_{11} L_{21}+L'^*_{12}L_{23}\right)\slashed{W}^-b^\dagger t'_1
 + \frac{g_2}{\sqrt2}R'^*_{11}R_{21}\cdot\slashed{W}^- \bar b^\dagger\bar t_1'
\\&\quad
 + \frac{g_Z}{2}\left(L_{21}^* L_{11}+L^*_{23}L_{13}\right)\slashed{Z}t'^\dagger_1 t
-\frac{g_Z}{2}R^*_{11}R_{21}\cdot\slashed{Z}\bar t^\dagger\bar t_1'
\\&\quad
- \left(Y_t L_{23}L_{13} + Y' L_{21}R_{12} + \epsilon_u L_{23}R_{12} + \epsilon_u' L_{21}R_{13}\right)\HuZ\bar t t'_1
\\&\quad
 - \left(Y_t L_{13} R_{23} + Y' L_{11}R_{22} + \epsilon_u L_{13}R_{22} + \epsilon_u' L_{11}R_{23}\right)\HuZ\bar t_1't
 +\Hc
 \end{split}
\end{align}
Therefore, we obtain the decay rates as, defining the following coupling:
\begin{align}
g^W_{t_1'b^\dagger}&:= \frac{g_2}{\sqrt2}\left(L'^*_{11} L_{21}+L'^*_{12}L_{23}\right),
\qquad\qquad
g^W_{\bar t_1'\bar b^\dagger}= \frac{g_2}{\sqrt2}R'^*_{11}R_{21},
\\
g^Z_{t'_1 t^\dagger}&:= \frac{g_Z}{2}\left(L_{21}^* L_{11}+L^*_{23}L_{13}\right),
\qquad\qquad\ 
g^Z_{\bar t'_1\bar t^\dagger}:=-\frac{g_Z}{2}R^*_{11}R_{21},
\\
y^{h}_{t_1'\bar t}&:=\frac{\cos\alpha}{\sqrt2} \left(Y_t L_{23}L_{13} + Y' L_{21}R_{12} + \epsilon_u L_{23}R_{12} + \epsilon_u' L_{21}R_{13}\right),
\\
y^h_{\bar t'_1 t}&:=\frac{\cos\alpha}{\sqrt2} \left(Y_t L_{13} R_{23} + Y' L_{11}R_{22} + \epsilon_u L_{13}R_{22} + \epsilon_u' L_{11}R_{23}\right),
\end{align}
\begin{align}
 \Gamma\left(t'_1\to W^+b\right)&=\frac{m_{t_1'}}{32\pi}
\left(1-r_W\right)^2\left(2+r_W^{-1}\right)
\left(
\abssq{g^W_{t_1'b^\dagger}}+\abssq{g^W_{\bar t_1'\bar b^\dagger}}
\right),
\\
 \Gamma\left(t'_1\to Zt\right)&=\frac{m_{t_1'}}{32\pi}\sqrt{\lambda(1,r_Z,r_t)}
\left[
\left(1+r_t-2r_Z+\frac{(1-r_t)^2}{r_Z}\right)\left(\abssq{g^Z_{t'_1 t^\dagger}}+\abssq{g^Z_{\bar t'_1\bar t^\dagger}}\right)
+12\sqrt{r_t}\Re\left(g^Z_{t'_1 t^\dagger}g^Z_{\bar t'_1\bar t^\dagger}\right)
\right],
\\
 \Gamma\left(t'_1\to ht\right)&=\frac{m_{t_1'}}{32\pi}\sqrt{\lambda(1,r_h,r_t)}
\left[
 (1+r_t-r_h)\left(\abssq{y^h_{t_1'\bar t}}+\abssq{y^h_{\bar t'_1 t}}\right)
+4\sqrt{r_t}\Re\left(y^h_{t_1'\bar t}y^h_{\bar t'_1 t}\right)
\right],
\end{align}
where
\begin{equation}
\lambda(x,y,z):=x^2+y^2+z^2-2xy-2yz-2zx, \qquad r_X=\frac{m_X^2}{m^2_{t'_1}}.
\end{equation}

\subappendixend


\chapter{Coda}
\label{cha:coda}

In this coda, the {\em theme} of this dissertation is repeated.

\starline

On 4th July 2012, the ATLAS and the CMS collaborations claimed that they respectively observed a new boson with a mass approximately $126\GeV$ in the search for the Standard Model Higgs boson~\cite{20120704CMS,20120704ATLAS}, which completes the Standard Model~\cite{StandardModel,HiggsBoson}.
This model has the electroweak symmetry breaking as its heart, which is governed by the Higgs mechanism, and explains almost all of Nature.

However, we already know that the Standard Model is not the ultimate theory.
We have the Dark Matter problem, the fine-tuning problem, and the muon $g-2$ discrepancy.
Also the mechanism which generated current baryon asymmetry of our Universe is still unknown.
Moreover, we have four forces, not one.
To achieve the grand unification, the slight mismatch on gauge coupling unification should be resolved.
These topics were discussed in Chapter~\ref{cha:foundation}.

Now we have the LHC experiments as a powerful tool to investigate the next theory beyond the Standard Model.
In this dissertation the ATLAS detector~\cite{aad:2008zzm,Aad:2009wy}, a general-purpose detector for the LHC, is examined in Chapter~\ref{cha:atlas}.
There we saw that the discovery of the Higgs boson was realized with help from the increases of the energy and the collision rate in the 2012 run, and also the importance of $b$-tagging was emphasized.

Then we reviewed the SUSY, a silver bullet for the problems in the Standard Model.
It solves the hierarchy problem in a miraculous manner, and as we saw in Chapter~\ref{cha:mssm}, the muon $g-2$ problem and the slight mismatch on gauge coupling unification can be solved with the SUSY.
Moreover, a promising candidate for the Dark Matter is provided.

However, we have not detected any hints of the SUSY at the LHC.
Now the colored superparticles are highly constrained as $m_{\glu}\gtrsim900\GeV$ and $m_{\squ}\gtrsim1400\GeV$, which indicates that the SUSY is heavier than expected.
We saw that this indication is supported also by the Higgs boson having a mass of $126\GeV$.
Under the no-mixing scenario, the stop mass is required to be $\Order(10)\TeV$.
Even with the maximal-mixing of $X_t\sim\pm\sqrt6m_{\tilde t}$, it should be $\sim1$--$2\TeV$.
There the argument of the little hierarchy is not fulfilled.

Two possibilities were introduced there.
One was the heavy-colored light-non-colored scenario, which can be investigated with searches focusing on pair-production of charginos, neutralinos, and sleptons via electroweak interactions.
Such searches do not need higher energy, but require a higher luminosity; the data expected in the $13\text{--}14\TeV$ run, corresponding to $\Order(100)\invfb$, would help this direction, and also the HL-LHC is of great importance (cf.\ Sec.~\ref{sec:atlas-concluding-remark}).

However, this scenario is not preferred from a theoretical viewpoint because it does not fully respect the $\gSU(5)$-GUTs.
As a possibility which respects the $\gSU(5)$-GUTs daydream, there the V-MSSM scenario, an extension of the MSSM with vector-like matters, was introduced, and was investigated in Chapter~\ref{cha:vectorlike} as the main dish of this dissertation.

We saw that in this model the Higgs mass can be raised by the extra vector-like quarks.
It allows us to explain the muon $g-2$ anomaly even under the GMSB framework, which is a very promising framework for its freedom from the SUSY $CP$- and flavor problems, but disfavored because it cannot simultaneously realize the $126\GeV$ Higgs and the muon $g-2$ explanation.
There, the ``V'' resurrects the GMSB.

Fig.~\ref{fig:results1} was the conclusive figure of this dissertation.
There it was obviously shown that the $126\GeV$ Higgs boson can be realized with keeping the SUSY explanation of the muon $g-2$ at the $2\sigma$-level, while the $1\sigma$-level explanation was excluded by the LHC SUSY searches.
Also it was emphasized that the mass parameter $M_V$, which governs the masses of the vector-like quarks and the vector-like lepton, is constrained as $M_V\lesssim 1.2\TeV$.
This means the lightest vector-like quark $t'_1$ should be lighter than $\lesssim1.1\TeV$, which is within the reach at the $14\TeV$ LHC.

Then collider searches for the vector-like quarks were investigated.
We saw that the search performed by the ATLAS collaboration~\cite{ATLAS:2012qe} was very interesting and promising, and that improvements and better understanding of $b$-tagging algorithms is awaited for further progress.

\starline

On 17 December 2012, the LHC was shut down to prepare for collisions with $E\s{CM}=14\TeV$, leaving the following message on the LHC monitor:
 \begin{quote}
  *** End of operation for 2012! ***\\
  See you again briefly for p-Pb in 2013.\\
  High energy proton proton physics\\
  will be resumed in 2015.\\
  So long and thanks for all the fish.
 \end{quote}
In 2015, we will obtain a much more powerful tool to investigate new physics beyond the Standard Model.
There SUSY searches would strikingly proceed, and we expect we will obtain many clues for the next theories.
Especially, the fate of the V-GMSB scenario is expected to be determined.

Now we have to wait for three years, but now, contrary to that after the accident in 2009, we have data, which are enough to allow us to find the ``tail'' of the physics beyond the Standard Model buried inside them.
We should exhaust the obtained data to be ready for the $14\TeV$ run starting in, hopefully, 2015.

\backmatter
\pdfbookmark[0]{Bibliography}{bookmark:bib}
\addtocontents{toc}{\protect\contentsline{chapter}{Bibliography}{\thepage}{bookmark:bib.0}}

\itemsep=30pt
\bibliography{AstroPhysics,Cosmology,ExpDetector,ExpLHC,ExpNeutrino,ExpResLHC,ExpResTevatron,ExpTevatron,FlavorSym,Gravitino,HEP,HEPComputing,Higgs,LFV,LHCPhenom,LinearCollider,MagneticMoment,Nuclear,PhotonConversion,Positronium,ProtonDecay,QED,QFT,RPV,SUGRA,SUSY,SUSYHiggs,TopAFB,axion,sphaleron,vectorlike,ExpResAstro}
\end{document}